\documentclass{aa}  

\usepackage{graphicx}
\usepackage{txfonts}
\usepackage{color}
\usepackage{hyperref}

\usepackage{ulem}
\usepackage[below]{placeins}

\begin{document}

   \title{Eight new, triply eclipsing triple star systems}

   \author{T. Borkovits\inst{1,2,3}  
	  \and
	  S. A. Rappaport\inst{1,4}
          \and
	  G. Torres\inst{5}
	  \and
	  A. Tokovinin\inst{6}
	  \and
	  T. Mitnyan\inst{1,7}
	  \and
	  R. Gagliano\inst{8}
	  \and
	  T. Jacobs\inst{9}
	  \and
	  B. Powell\inst{10}
	  \and
	  V. Kostov\inst{10,11}
	  \and
	  M. Omohundro\inst{12}
	  \and
	  M. H. Kristiansen\inst{13,14}
	  \and
	  I. Terentev\inst{15}
	  \and
	  H. M. Schwengeler\inst{12}
	  \and
	  D. LaCourse\inst{16}
	  \and
	  Z. Garai\inst{17,18}
	  \and
	  T. Pribulla\inst{17}
	  \and
	  I. B. B\'\i r\'o\inst{1,2}
          \and
          I. Cs\'anyi\inst{2}
	  \and
	  Z. Dencs\inst{3,18}
	  \and
	  D. W. Latham\inst{5}
          \and
          R. Uhla\v{r}\inst{19}
	  \and
	  A. P\'al\inst{3}
          }

   \institute{HUN--REN--SZTE Stellar Astrophysics Research Group, H-6500 Baja, Szegedi \'ut, Kt. 766, Hungary 
         \and
   	   Baja Astronomical Observatory of University of Szeged, H-6500 Baja, Szegedi \'ut, Kt. 766, Hungary.\\
             \email{borko@bajaobs.hu} 
	\and
	   Konkoly Observatory, Research Centre for Astronomy and Earth Sciences,  H-1121 Budapest, Konkoly Thege Mikl\'os \'ut 15-17, Hungary	   
	\and
	  Department of Physics, Kavli Institute for Astrophysics and Space Research, M.I.T., Cambridge, MA 02139, USA \\
	   \email{sar@mit.edu}	
	\and
	  Center for Astrophysics $\vert$ Harvard \& Smithsonian, 60 Garden Street, Cambridge, MA 02138, USA 
	\and
          Cerro Tololo Inter-American Observatory---NFSs NOIRLab Casilla 603, La Serena, Chile
	\and
	   Department of Experimental Physics, University of Szeged, H-6720, Szeged, D\'om t\'er 9, Hungary
	\and
	   Amateur Astronomer, Glendale, AZ 85308 
	\and
	   Amateur Astronomer, Missouri City, Texas 77459 USA
	\and
	   NASA Goddard Space Flight Center, 8800 Greenbelt Road, Greenbelt, MD 20771, USA 
	\and
	   SETI Institute, 189 Bernardo Avenue, Suite 200, Mountain View, CA 94043, USA
	\and
	   Citizen Scientist, c/o Zooniverse, Dept,~of Physics, University of Oxford, Denys Wilkinson Building, Keble Road, Oxford, OX1 3RH, UK 
	\and
	   Brorfelde Observatory, Observator Gyldenkernes Vej 7, DK-4340 T\o ll\o se, Denmark 
	\and 
	   National Space Institute, Technical University of Denmark, Elektrovej 327, DK-2800 Lyngby, Denmark 
        \and
           Citizen Scientist, Planet Hunter, Petrozavodsk, Russia
	\and
	   Amateur Astronomer, 7507 52nd Place NE Marysville, WA 98270, USA 
	\and
	   Astronomical Institute, Slovak Academy of Sciences, 05960 Tatransk\'a Lomnica, Slovakia
        \and
	   ELTE Gothard Astrophysical Observatory, Szent Imre herceg u. 112, H-9700 Szombathely, Hungary
        \and
           Private Observatory, Poho\v{r}\'{\i} 71, CZ-254 01 J\'{\i}lov\'e u Prahy, Czech Republic
       }

   \date{Received  .., 2026; accepted ... , ...}

  \abstract
  {}
{We have identified numerous triply eclipsing hierarchical triple star systems from observations made with the space telescope TESS. For many of these systems we can potentially determine their dynamical and astrophysical parameters with high precision, in some cases even without the use of radial velocity data. In the present paper we report the comprehensive study of eight new, triply eclipsing triple star systems, taken from this larger sample: TICs 77177148, 220456364, 232874003, 306690194, 343763382, 358795247, 379084660, 385933895.}
   {Besides the photometric data from the TESS observations, we obtained supplementary ground-based photometric measurements, as well as radial velocity measurements for some of the systems. The eclipse timing variation curves extracted from the TESS data and from the ground-based follow up data, the photometric light curves, radial velocity data, and the spectral energy distribution are combined in a complex photodynamical analysis to yield the stellar and orbital parameters for all of the investigated systems.}
   {The outer periods are in the range of 63-294 days, including the currently sixth tightest known triple system, TIC\,385933895 with an outer to inner period ratio of 9.82. We find that all of the triples are nearly coplanar, with mutual inclination angles of $\lesssim2.5\degr$. Despite such flatness, we still detect eclipse depth variations in TIC\,220456364 that are induced by orbital plane precession. Moreover, we find that TIC\,358795247 is likely a (2+1)+1-type hierarchical quadruple system. Finally we make some statistical assessments of a number of the orbital elements, stellar parameters, and distances of these systems.}
   {}


   \maketitle
   
\nolinenumbers

\section{Introduction}  
\label{sec:intro}

 It was only 15 years ago, when the discoveries of the first two triply eclipsing triple star systems were reported \citep{carter11,derekasetal11}. These novel findings were based on investigations of the photometric data from the \textit{Kepler}-spacecraft \citep{borucki10}. Later, further photometric observations with the \textit{Kepler} space telescope led to the discovery of more than a dozen additional triply eclipsing systems.  While a few subsequent discoveries of triply eclipsing triples were reported sporadically with the use of other space-borne and ground-based instruments \citep[see the list of][]{borkovits22}, the discovery of triply eclipsing triples became routine only with the advent of the TESS space telescope \citep{ricker15}.  Due to its monitoring of the entire sky over the past eight years\footnote{This monitoring covers some 5\% of the sky for a month before the field of regard is moved. The higher the ecliptic latitude of a source, the longer it continues to be monitored.}, by now TESS has found more than a hundred triply eclipsing triple star candidates.

Triply eclipsing triple star systems, in general, consist of at least one close\footnote{In general, but somewhat arbitrarily, we call a triple system `close', when $P_\mathrm{out}\lesssim1000$\,d.} and tight\footnote{That is, $P_\mathrm{out}/P_\mathrm{in}\lesssim50$, since in this case one can expect readily measurable dynamical perturbations, even on timescales of $P_\mathrm{out}$} triple stellar subsystem, where, moreover, the innermost pair is generally an eclipsing binary (EB). It is not necessary for a triply eclipsing system to be close and/or tight, but the closeness of the third component substantially increases the chance of mutual extra eclipsing events. Similarly it is also not required that the inner pair should be an EB (though, in the absence of regular, inner eclipses, the name `triply eclipsing' would not be exactly correct); however, the discovery of extra eclipsing events, and the investigations of such triples are generally much easier when inner, regular eclipses are also present.

In several former papers, we investigated in a homogeneous manner TESS-discovered triply eclipsing stellar systems using a spectro-photodynamical analysis.  Apart from the very first two TESS-discovered triply eclipsing triples (TIC 209409435 - \citealp{borkovitsetal20b}; TIC 278825952 - \citealp{mitnyanetal20}), we have subsequently tended to group systems into sets of between 3 and 10 triples, and then investigated these subsets in detail \citep{borkovitsetal22a,borkovitsetal25b,rappaportetal22,rappaportetal23,rappaportetal24,czavalingaetal23}. In some cases, however,  when a particular system was found to be exceptionally interesting and/or a record holder in some way, we dedicated individual papers to such objects \citep{powelletal22,kostovetal24,borkovitsetal26a,borkovitsetal26c,rappaportetal26}. Here we also note  the paper of \citet{powelletal26} which was dedicated to only a particular triply eclipsing triple stellar system, but where the method of analysis followed a somewhat different approach.

This current paper also joins the former category of the study of a modest-sized group of eight triply eclipsing triple star systems analysed with our complex photodynamical code.  For these particular systems, such an analysis has become possible only this year (2026), thanks to the latest TESS observations, as well as to some new additional, ground-based photometric and/or spectroscopic follow-up observations.

In Section \ref{sec:8triples} we describe the collection of systems that we have selected for this detailed study. We explain briefly how these sources were discovered.  In Section~\ref{sec:archivaldata} we analyse the archival data for these triples. In Section \ref{sec:lightcurves}, the light curves exhibiting third body eclipses are introduced along with model fits, and briefly discussed.  The eclipse timing variation curves (ETVs) are introduced and discussed in general in Section \ref{sec:etvs}.  The photodynamical code, with which the system parameters are extracted, is briefly reviewed in Section \ref{sec:photodynamics}. In Section \ref{sec:results} we summarize the system parameters in a set of comprehensive tables, and the results for each individual system are presented. Finally, some conclusions and statistical assessments are made in Sect.~\ref{sec:discuss}.
   
\section{The eight triply eclipsing triples}  
\label{sec:8triples}

\subsection{Catalogued Information}

For this collection of eight formerly unanalysed triply eclipsing triple systems, we started by gathering their main catalogued properties (see Table\,\ref{tbl:mags8}).  Five of the eight triples are located in the southern ecliptic hemisphere, including one system (TIC 220456364) which is located in the Southern Continuous Viewing Zone (SCVZ), while two systems are located in the northern hemisphere. The eighth triple, TIC 306690194, is also in the northern hemisphere, and is located near to the ecliptic; hence, it was observed in some of the ecliptic sectors of the TESS satellite (see in Table\,\ref{tbl:sectors}).

\subsection{Discovery}

Before the advent of the eight-year-long monitoring of nearly the entire sky with TESS, the triply eclipsing triple nature of all eight of the currently analysed systems was unknown.  Moreover, in the case of the majority of the current systems, the eclipsing binary (EB) nature of the inner pair was either not known or it was discovered only in the last few years due to such ground-based surveys as, e.g. the All-Sky Automated Survey for Supernovae \citep[ASAS-SN;][]{shappee14}.  Regarding the current eight systems, seven of them were identified by our `Visual Survey Group' \citep[VSG;][]{kristiansen22} in the manner described in detail, e.g., in \citet{rappaportetal24}. Note, however, that one of these seven (TIC\,343763382 = ASASSN-V J181441.19-755530.2) was also independently discovered and reported by \citet{zascheetal22}, who also gave the correct outer period. Moreover, the triply eclipsing nature of TIC\,232874003 was found, and communicated personally by Petr Zasche.

\subsection{Follow-up photometric observations}

Since the majority of the currently investigated eight systems are located far in the southern hemisphere, they are mostly unavailable to European and North-American observers.  Therefore, we organized ground-based photometric follow-up campaigns only for the three targets which are located in the northern hemisphere (TICs\,306690194, 358795247 and 379084660). We obtained usable ground-based measurements for all of them. The ground-based photometric observations of TIC\,3587956247 were carried out with two identical 80-cm telescopes in Gothard Astrophysical Observatory, Szombathely (GAO80) and Baja Astronomical Observatory of Szeged University, Baja (BAO80), both in Hungary. The GAO80 observations were carried out in Sloan $z'$-band, while for the BAO80 ones a Sloan $r'$-band filter was used. One regular eclipse observation of TIC\,379084660 was also carried out with the BAO80 instrument, using a Sloan $r'$-band filter. The details of both these instruments, and the way in which the observations were made, were described briefly in \citet{borkovitsetal22a}.  These observations were carried out over 13+1 nights (for TICs\,358795247 and 379084660, respectively) and resulted in 9+1 further regular eclipse observations as well as, in  the case of TIC\,358795247, some parts of one additional third-body eclipse event. 

Finally, only after completing the spectro-photodynamical analyses, we received 8+16 nights of additional observations of TICs\,306690194 and 358795247, respectively. These latter data sets were obtained from the small private observatory of R.U. in J\'{\i}lov\'e u Prahy (Czech Republic), as well as from some remote telescopes in southern Switzerland, in Carona, operated by the same observer. These observations were carried out with three different telescopes, namely, an unfiltered 34-mm refractor and 150-mm and 200-mm reflectors in Cousins $R_\mathrm{C}$-band. When we plotted these photometric light curves against the formerly accepted models, we found that they agreed with the emulated model light curves both in timings and shapes. Therefore, we decided not to rerun our fitting processes, including these new data, but to use them simply to verify the former results. The eclipse times determined from all the ground-based observations (including those that were derived from these late ground-based observations) are listed together with the eclipse times extracted from the TESS-light curve in Tables of H.1-H.8.)

\subsection{Follow-up spectroscopic observations}
\label{Subsect: spectroscopy}

Three of the eight targets were bright enough for ground-based spectroscopic follow-up observations with moderate-sized telescopes to obtain radial velocities (RVs). In the case of the brighter northern targets, i.e., TIC 306690194 and TIC 358795247, we obtained high-resolution spectra using the Tillinghast Reflector Echelle Spectrograph \citep[TRES;][]{Furesz:2008, Szentgyorgyi:2007} on the 1.5m Tillinghast reflector at the Fred L.\ Whipple Observatory (Mount Hopkins, Arizona, USA), which is operated by the Smithsonian Astrophysical Observatory. This instrument has a resolving power of about 44,000, and observes the wavelength range from 390 to 910 nm. For TIC 306690194 we obtained six spectra, from which we were able to extract RVs for the two members of the inner double-lined spectroscopic EB (stars Aa and Ab). The velocities were measured using {\tt TODCOR}, a two-dimensional cross-correlation technique \citep{Zucker:1994}. In the case of TIC 358795247, which the ETV data indicate is a quadruple system (see Sect.~\ref{sec:etvs} below), we were able to identify the spectral lines of all four components, and measure their velocities in most of our 29 spectra. The measurements were made with {\tt QUADCOR}, which is an extension of {\tt TODCOR} to four dimensions \citep{Torres:2007}. The velocities for these two systems are shown in Appendix~\ref{app:RVplots} and tabulated in Appendix~\ref{app:RVs}.

The third system for which we carried out spectroscopic observations and, hence, obtained RV data, was TIC\,343763382. In the case of this far-southern system we obtained six RV points for all the three stellar components (that is, stars Aa, Ab and B).  The data were obtained with the CHIRON fiber-fee optical spectrometer \citep{tokovininetal13} which is mounted on an 1.5-m telescope at Cerro Tololo Inter-American Observatory (CTIO) in Chile and operated by the SMARTS consortium.\footnote{\url{http://www.astro.gsu.edu/~thenry/SMARTS/}} Spectra with a resolution of 28,000 were processed by the standard pipeline \citep{Paredes2021} and the RVs were measured by cross-correlation with a binary template and fitting three dips. These RV data are also shown and tabulated in Appendices~\ref{app:RVplots} and \ref{app:RVs}, respectively.

\section{Archival photometric data}
\label{sec:archivaldata}

As is the usual case in our analysis of triple systems, just after the discovery of any triply eclipsing triple, we gather all the available archival data for the given target and, carry out an analysis of these data. The two main purposes of such an analysis are (i) to strengthen or refute the triple nature of the object, and (ii) to obtain the outer orbital period, if possible. The latter is necessary for a further, more detailed analysis of the system. In some cases, however, at the time of discovery, the available archival data are insufficient to yield a clear detection of the outer eclipses or for a determination of the outer period. In such cases we iterate our analysis every couple of years as more and more archival data have been acquired. Here, we present the results of our current archival data analysis for these eight systems.

The outer period is a crucial parameter to determine in triple-star systems.  This can be measured via RV measurements, ETVs of the inner binary, and the timing of the outer eclipses. Some difficulties are that we often do not have good (or any) RV coverage, and both the ETVs and outer eclipses observed with TESS (our main source of photometry for triples) are typically sparsely sampled over the eight years of TESS observations.  This can lead to multiple solutions for the outer orbital period. 


We therefore make use of archival data which has thousands of samples spread over years to more than a decade.  If the third-body eclipses last for $\sim$0.3-1 day, and the outer periods are $\sim$ 30-300 days, then the eclipsing intervals occupy of order 1/2\% ($\pm$ a factor of $\sim$5) of the outer period. Therefore, with some 3000 typical photometric samples in an archival data set, we might expect about 15 low points due to outer eclipses. If these can be detected (i.e., separated from low noise points) via a box least squares (BLS; \citealt{kovacs02}), then the outer orbital period can be determined independently from the TESS data.

In this work, we have made use of archival photometry from the ASAS-SN \cite{shappee14,kochanek17}, the Asteroid Terrestrial-impact Last Alert System (ATLAS; \citealt{tonry18}), and the Zwicky Transient Facility (ZTF; \citealt{masci19}, \citealt{bellm19}) to search for the outer periods of the eight triples studied in this work. This procedure was successful for six of the eight sources in our study, and the results are summarized in Fig.~\ref{fig:outer_folds}.

If the area under the EB light curve dips in flux is substantially greater than under the outer eclipse dips (as is typically the case), then the EB eclipses must be removed in order to have a realistic chance of detecting the outer eclipses in a BLS transform. To this end, we first remove any obvious high outliers or those below what could reasonably be expected from inner or outer eclipses.  The photometric measurements from different filter bands within a given archival data set are median normalized to each other.  If we combine more than one archival data set, the magnitudes from each set are, in turn, median normalized to each other.  The final data preparation step is to force-fit a harmonic series, starting at the base frequency of the EB orbit, and subtract it from the data.  We typically use 75 orbital harmonics, and the resultant linear fit then requires inverting a $151 \times 151$ matrix, essentially an instantaneous calculation on a modern laptop.  

The archival data set, prepared in the manner described above, is then subjected to a BLS transform\footnote{\url{https://exoplanetarchive.ipac.caltech.edu/cgi-bin/Pgram/nph-pgram}}. If the highest BLS peak (or a harmonic of comparable amplitude) turns out to represent the ultimately correct outer orbital period, then we report this as an independent and robust detection of the outer period. The pointwise folds of the outer orbital periods for the six sources in this study for which this criterion was met are shown in Fig.~\ref{fig:outer_folds}.  In all cases, phase 0 of the folds is taken to be the approximate mid-time of one of the TESS third-body eclipses.  Collections of low flux points at common phases are highlighted by a red curve that represents an arbitrary eclipse shape, width, and depth.  These are not fits to the data.  In some cases, the secondary outer eclipse is not centered between the two primary outer eclipses, and their locations, if there are no obvious low points to guide us, are informed by the outer orbital parameters ($e_{\rm out}$ and $\omega_{\rm out}$) given in Tables~\ref{tab:syntheticfit_TIC077177148220456364} -- \ref{tab:syntheticfit_TIC379084660385933895}. 

In four of the six panels of Fig.~\ref{fig:outer_folds}, the deepest outer eclipse is in agreement with the phasing given by the TESS outer eclipse in that it appears at phase 0.  For TIC 343763382, both outer eclipses are of about the same average depth, so either one could have been phased to zero.  In the case of TIC 306690194, the deepest outer eclipse actually comes at a phase very near 1.5.  The smaller red dip at phase zero corresponds to the outer eclipse times detected by TESS (and too shallow to be picked up by the archival data).  The times of the deeper outer eclipses were not observed by TESS, and thus not yet detected.  However, the model light curve for this source does indeed show that such outer eclipses will eventually be seen with TESS.  

\section{Light curve and model fits}  
\label{sec:lightcurves}

\begin{figure*}
   \centering
\includegraphics[width=0.32\textwidth]{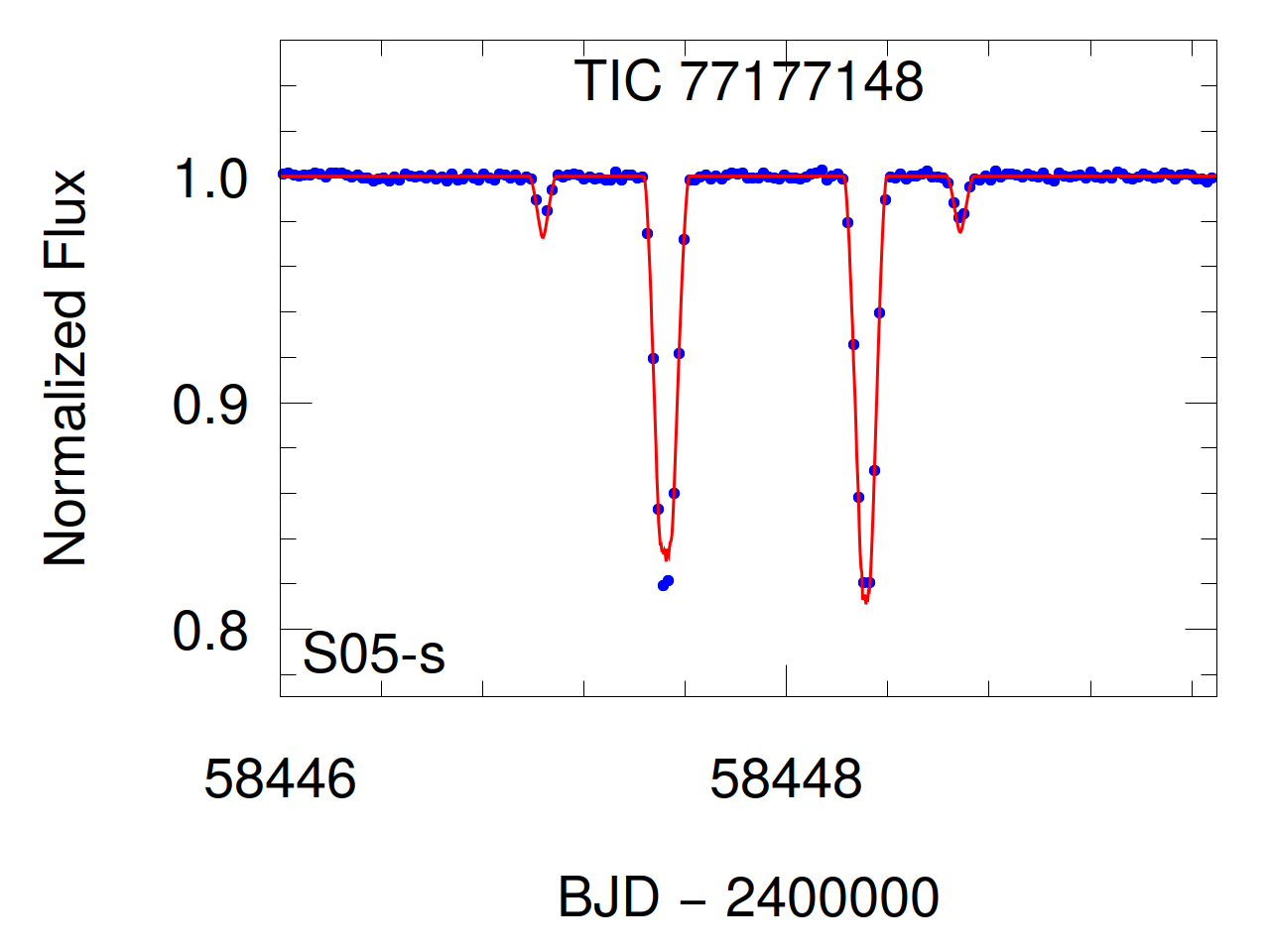}\includegraphics[width=0.32\textwidth]{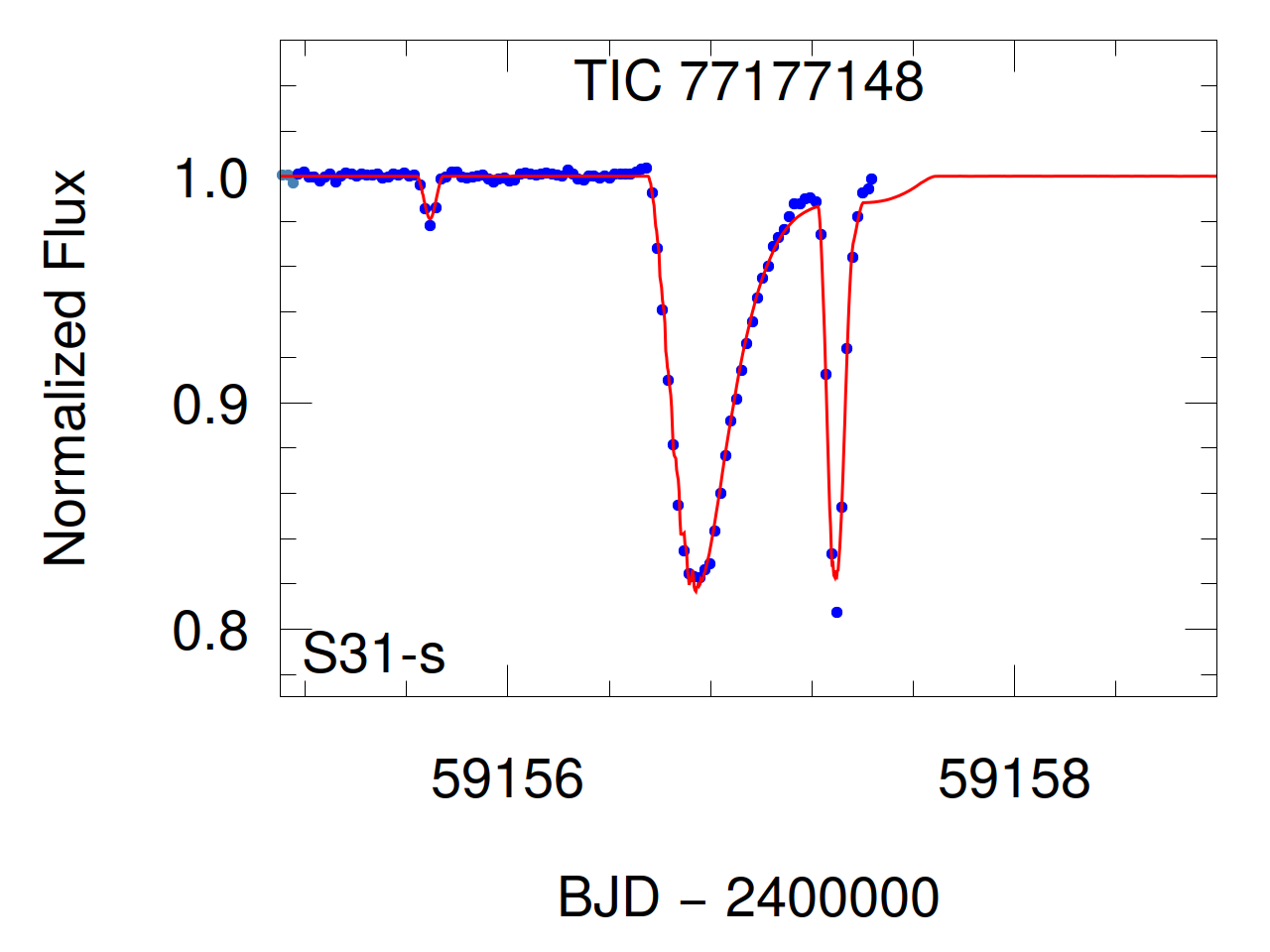}\includegraphics[width=0.32\textwidth]{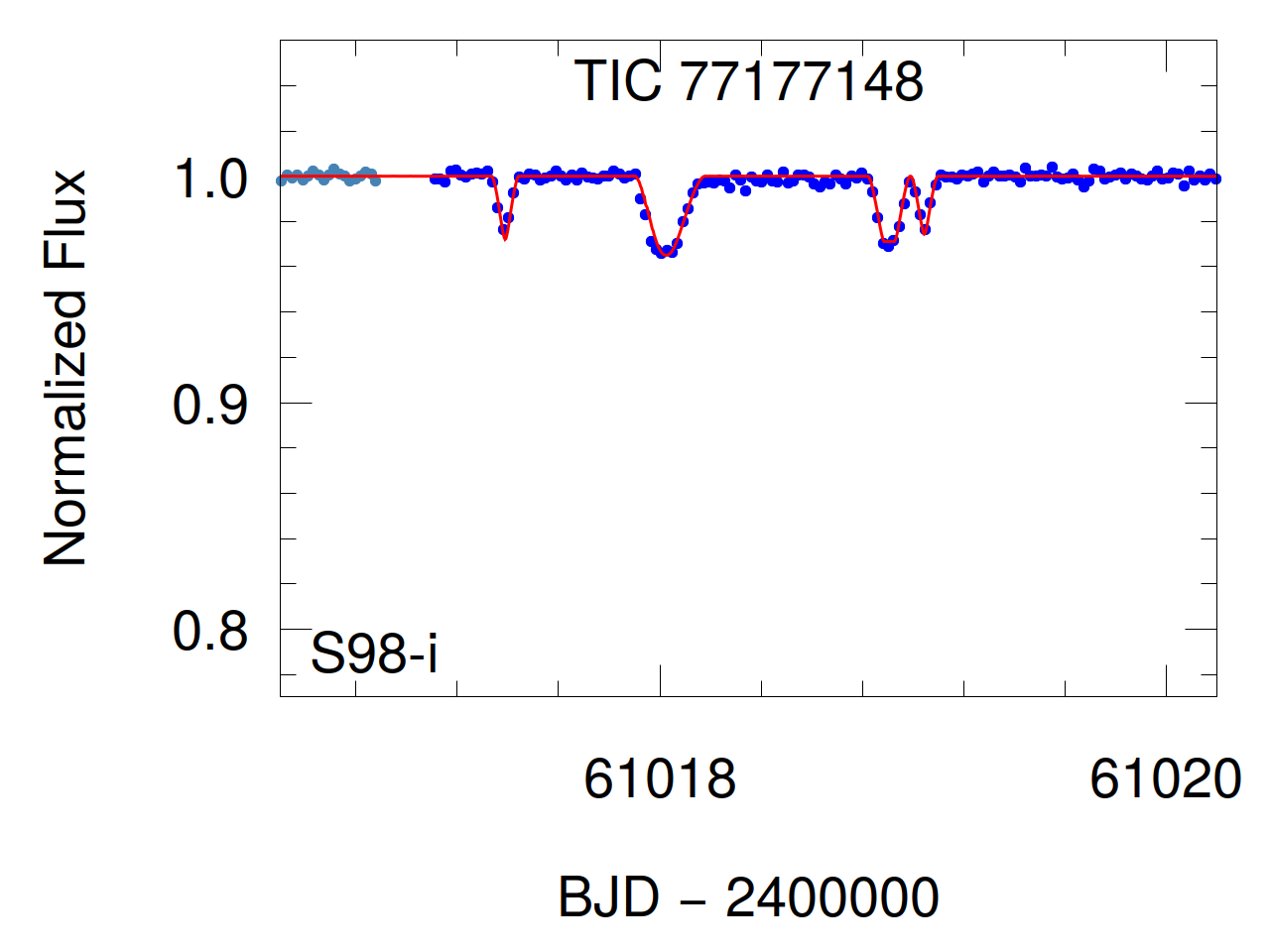}
   \caption{Observed light curves (blue points) and model fits (smooth red curves) near three illustrative third-body eclipses of TIC\,77177148. Dark/pale blue points are for those light curve sections that were used/not-used for the photodynamical solution. The sector numbers are indicated in the lower left corner of each panel. Letters `i' or `s' after the sector numbers refer to an inferior or superior conjunction of the third star, respectively.}
   \label{fig:77177148lc}
\end{figure*}

The TESS space telescope has observed a total of 39 triply eclipsing events (i.e., outer eclipses) in these eight triple star systems. From among these events, for illustrative purposes, we show 23 of them in  Figs.\,\ref{fig:77177148lc}, \,\ref{fig:lcs1} and \ref{fig:lcs2} (three for each system, except for TIC 306690194 which has only two).  We note that each outer eclipsing `event' really should be referred to as `event'. This is so, because, as one can see, there are third-body eclipsing events which are multiply dipped.  In general, all outer eclipses are unique, even in the same triple star system. This results from the fact that the shapes and the depths, as well as the number of the dips in flux, even in the same triple system, depend very strongly on the relative positions of the three stars.  Moreover, we intentionally selected for display third-body eclipsing events from both the superior and inferior conjunctions of the third-body.  Third-body eclipsing events at the superior conjunction of the third star (i.e., when the third star is eclipsed by the members of the inner EB either together or separately) or at the inferior conjunction (when the third star eclipses member[s] of the EB) may differ not only due to the varying relative stellar positions and the different surface brightness ratios, but also due to not atypical significant outer eccentricities. The sole exception is TIC\,306690194 where TESS observed only two third-body eclipsing events, and both of them occurred at the inferior conjunction of the third star. (In this regard see also our corresponding discussion above, in Sect.\,\ref{sec:archivaldata}.) 

We show in Figs.\,\ref{fig:77177148lc}, \ref{fig:lcs1} and \ref{fig:lcs2} the best-fit spectro-photodynamical model light curves (as red curves). superimposed on the TESS flux data points (blue dots) with 1800-sec cadence. The spectro-photodynamical models and results will be discussed below in Sects.\,\ref{sec:photodynamics}, \ref{sec:results}. For now, however, we simply make some general comments about these plots. As one can see, there are some panels where the accepted model eclipse depths depart from the observed ones. Some examples can be seen, e.g. in the first two panels of Fig.\,\ref{fig:77177148lc}, where some of the third-body (i.e., outer) eclipses have lower depths in the model than in the the observed  eclipses. But, further examples can also be seen in some panels of Figs.\,\ref{fig:lcs1} and \ref{fig:lcs2}. For example, in the case of the last panel of the second line of Fig.\,\ref{fig:lcs1} (third-body eclipse event during the Sector 104 superior conjunction of TIC\,232874003) not only the third-body eclipses, but even the regular two-body eclipses, have smaller model depths than the observed ones. But, oppositely, in the other two panels of the same  row (that is, other third-body eclipses in the very same triple, but during other TESS sectors), the regular eclipses have the correct depths, while the modeled third-body eclipses show a bit greater depths than the observed ones. In our opinion, the main reason for these discrepancies comes from the contaminating fluxes and stray-light, which vary slightly sector-by-sector during the TESS observations, that do not allow a more perfect modeling. (In our experience, some tenth of a degree in the mutual inclination of the inner and outer orbits, and/or in the locations of the relative ascending node -- that is, $\Omega_\mathrm{out}-\Omega_\mathrm{in}$ -- may result in such discrepancies. Despite this, we are convinced that these slightly varying contaminating fluxes and stray light affect our solution certainly within its 1-$\sigma$ uncertainty.) Naturally, these statements about the varying contaminating flux do not explain the in-sector discrepancies in the light curves of TIC\,358795247 (upper row in Fig.\,\ref{fig:lcs2}), where the departures between the observed and model light curves may arise from some unmodelled chromospheric activity that causes real, and rapidly varying, irregular light curve distortions.

\section{ETV curves}
  \label{sec:etvs}
  
After the detection of a third-body eclipsing event in any of TESS light curves, in addition to gathering the archival observations, the other immediate task is the determination of the mid-minima times of all the regular eclipses and the production and analysis of the ETV curve. In fortuitous cases, the ETV curve itself can reveal the third-body period. This can happen because a cyclic behaviour with the outer period ($P_\mathrm{out}$) must be present in the ETV of any compact and/or tight triple system. The more continuous the data train, the easier it is to uniquely determine the outer period via the ETV curve itself.

The $P_\mathrm{out}$-period cycles seen in an ETV curve, in general, have two basic origins. One of them is a purely geometrical effect. This is the well-known R{\o}mer-delay, which is also referred to as the light-travel-time effect \citep[LTTE;][]{roemer1677}. The LTTE results from the finite speed of light ($c$), and occurs in every triple system (as long as it is not observed pole-on), where the distance of the inner EB from the observer varies cyclically as the EB revolves (with period $P_\mathrm{out}$) around the  common centre of mass of the triple system. It is easy to show that the amplitude of this clearly sinusoidal effect is 
\begin{equation}
\mathcal{A}_\mathrm{LTTE}=\frac{M_\mathrm{B}}{M_\mathrm{AB}}\frac{a_\mathrm{out}\sin i_\mathrm{out}}{c}\sqrt{1-e_\mathrm{out}^2\cos^2\omega_\mathrm{out}}
\end{equation} 
or, 
\begin{equation}
\mathcal{A}_\mathrm{LTTE}\approx1.1\times10^{-4}\frac{M_\mathrm{B}\sin i_\mathrm{out}}{M_\mathrm{AB}^{2/3}}P_\mathrm{out}^{2/3}\sqrt{1-e_\mathrm{out}^2\cos^2\omega_\mathrm{out}},
\end{equation}
where, in this latter form, the numerical constant is valid only when the masses are expressed in solar masses, the period is in days, and the amplitude itself is also in days. (Here, and hereafter, $M_\mathrm{B}$ indicates the mass of the tertiary component [i.e., star B], while $M_\mathrm{AB}$ stands for the total mass of the system [i.e., $M_\mathrm{AB}=M_\mathrm{A}+M_\mathrm{B}$, where $M_\mathrm{A}=M_\mathrm{Aa}+M_\mathrm{Ab}$], and the orbital elements have their usual meanings. See also these same notations in Table\,\ref{tbl:definitions}.)

The other effect, which has the same period of $P_\mathrm{out}$, arises from perturbations by the third star that result in departures from a pure Keplerian two-body orbital motion of the inner binary.  Hence, we call these `dynamical effects' (DEs). Such perturbations operate on different timescales, but one of them, the so-called `medium-timescale perturbations' has a characteristic timescales of $P_\mathrm{out}$.  Currently, the most-detailed analytic investigations of such perturbations for the ETV curves can be found in \citet{borkovitsetal15}.  When the inner and outer orbits are nearly coplanar, then the characteristic, theoretical amplitude of this approximately periodic effect is
\begin{equation} \mathcal{A}_\mathrm{DE}^\mathrm{coplanar}\approx\frac{3}{2\pi}\frac{M_\mathrm{B}}{M_\mathrm{AB}}\frac{P_\mathrm{in}^2}{P_\mathrm{out}}\frac{e_\mathrm{out}}{\left(1-e_\mathrm{out}^2\right)^{3/2}}
\end{equation}
\citep{rappaportetal13}.

In most cases, however, a clear detection of the outer orbital period from the ETV results requires several observing sessions with TESS of a given target. This is so because, in general, one TESS observing session is shorter (in many cases, much shorter) than the outer period itself. In this circumstance, the only things we can detect from a single or few sectors of TESS-observed ETVs, are that either the curves exhibit some curvature, or ETV sections which were taken a few years apart have different slopes. Both of these phenomena are evidence of rapid period variations. Moreover, this is the chief reason why we need photometric ground-based follow-up monitoring of the regular, i.e., inner EB eclipses.

\begin{figure*}
\centering
  \includegraphics[width=0.39\textwidth]{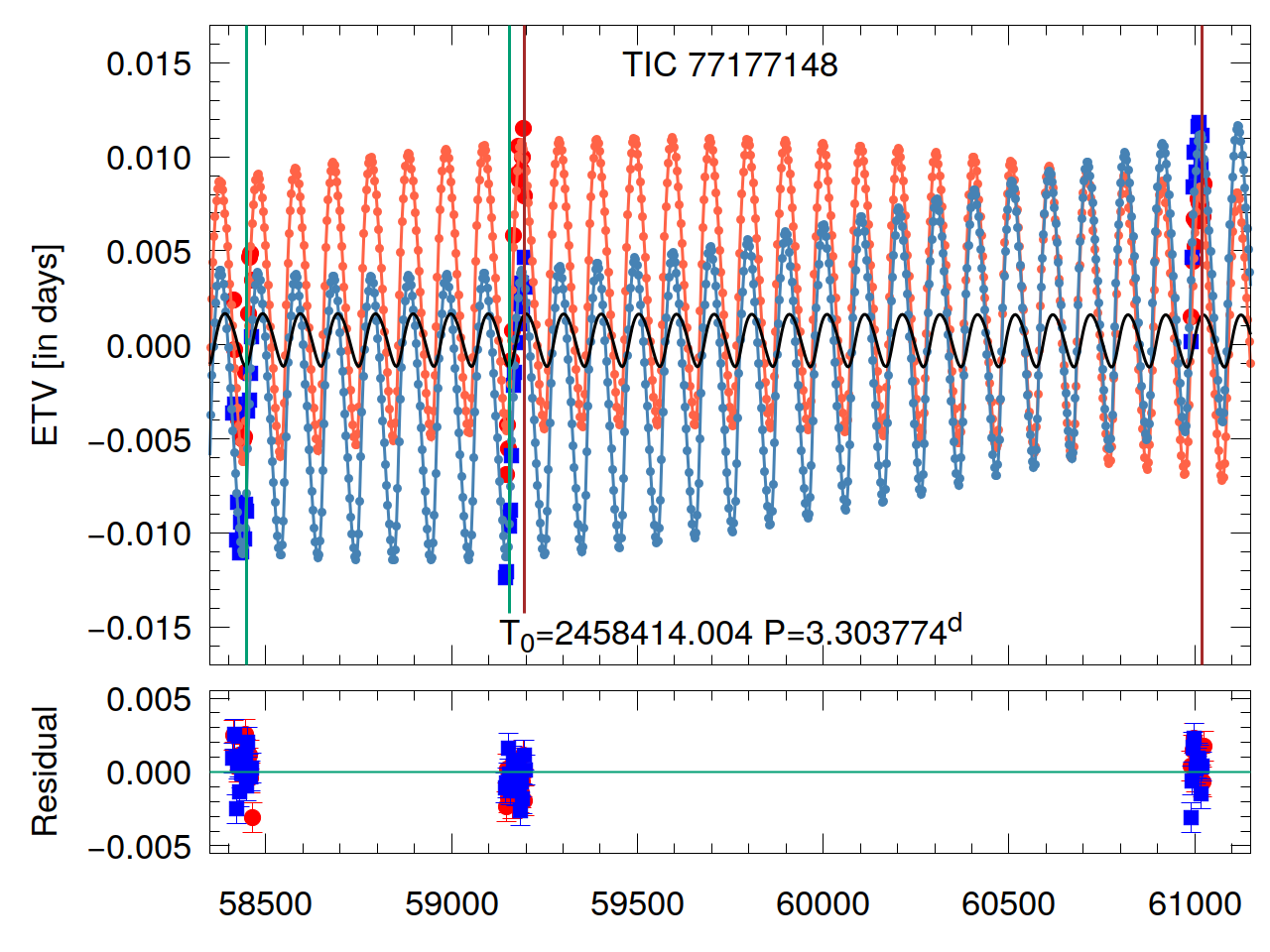}\includegraphics[width=0.39\textwidth]{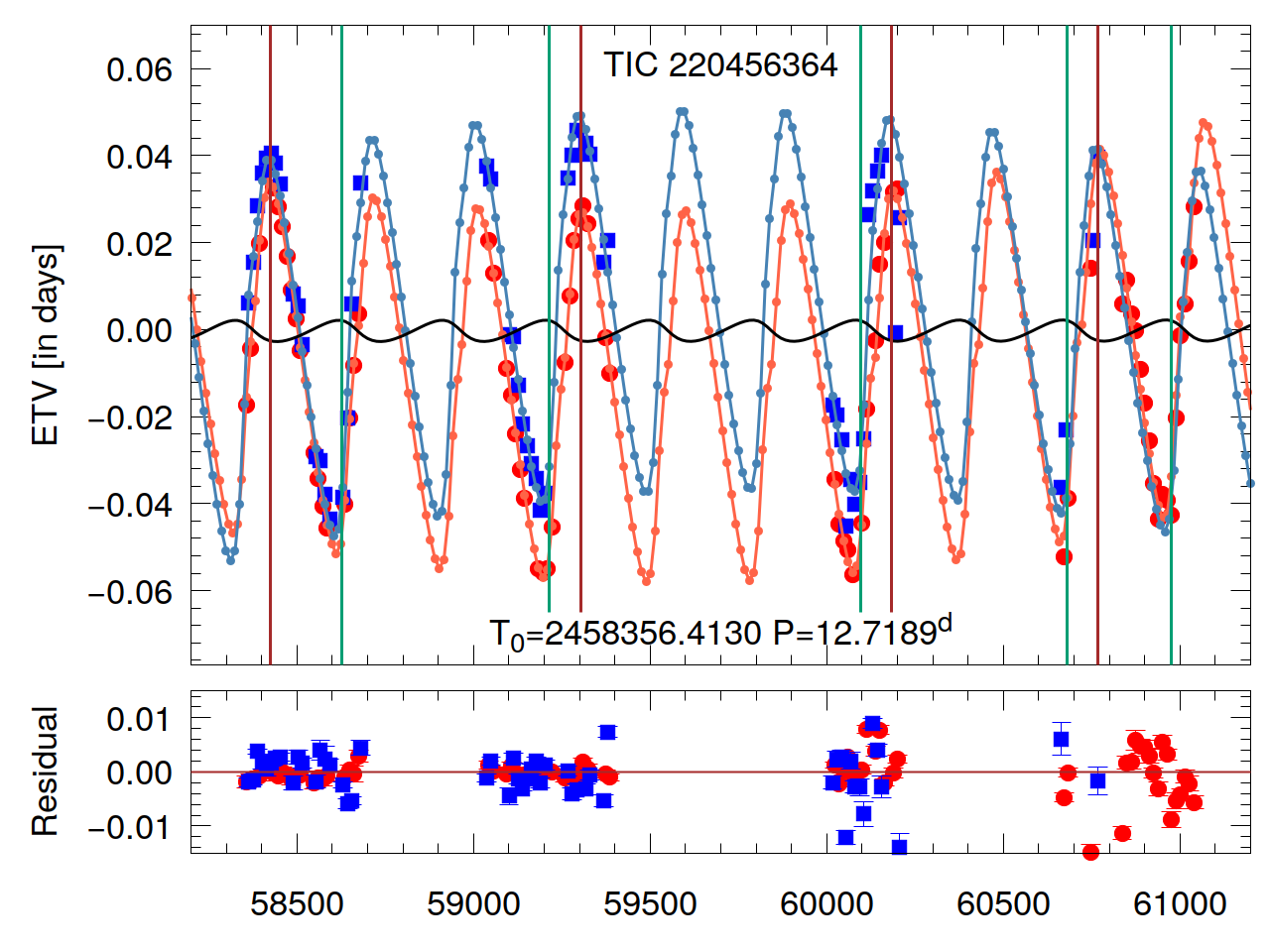}
  \includegraphics[width=0.39\textwidth]{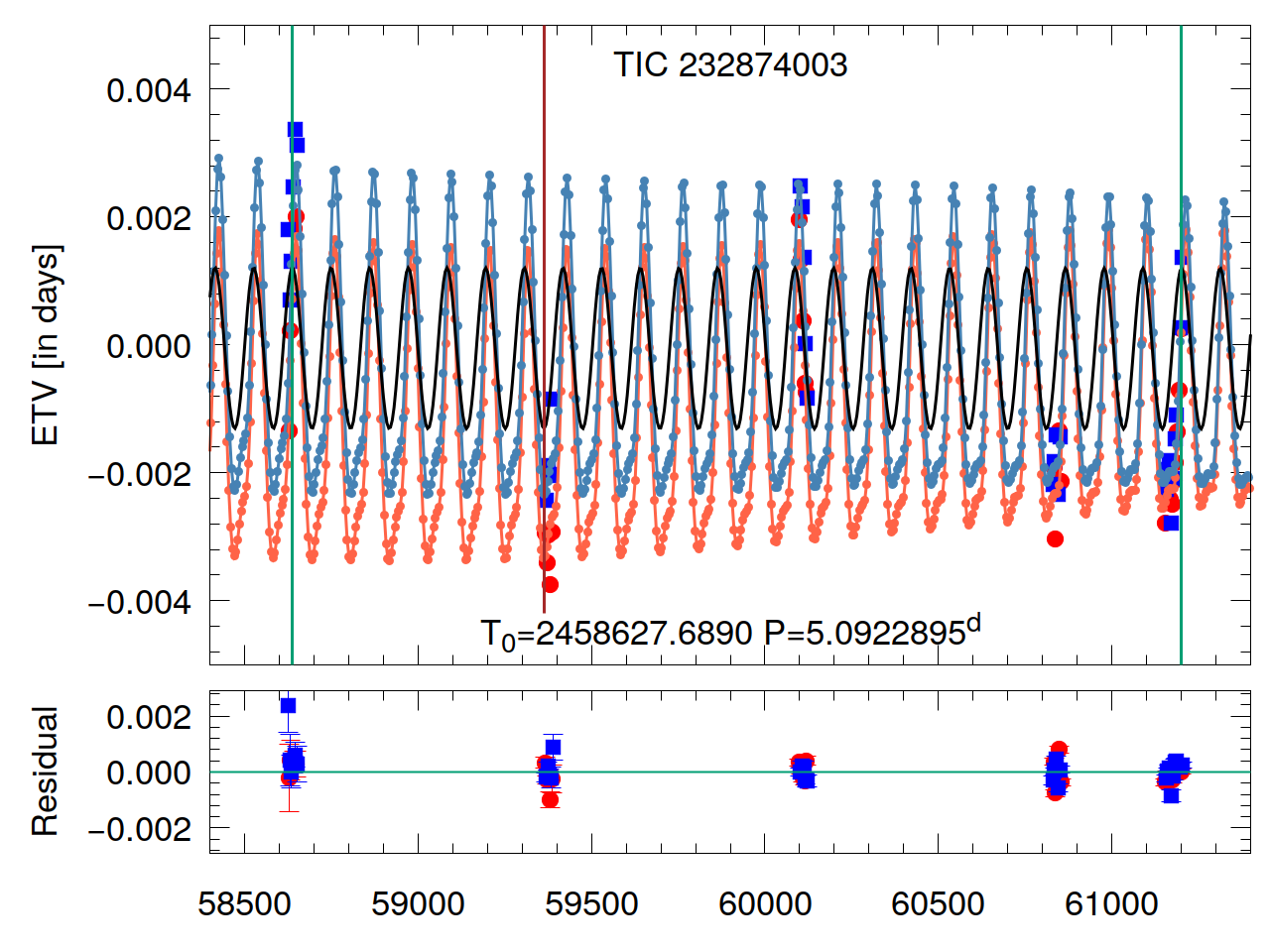}\includegraphics[width=0.39\textwidth]{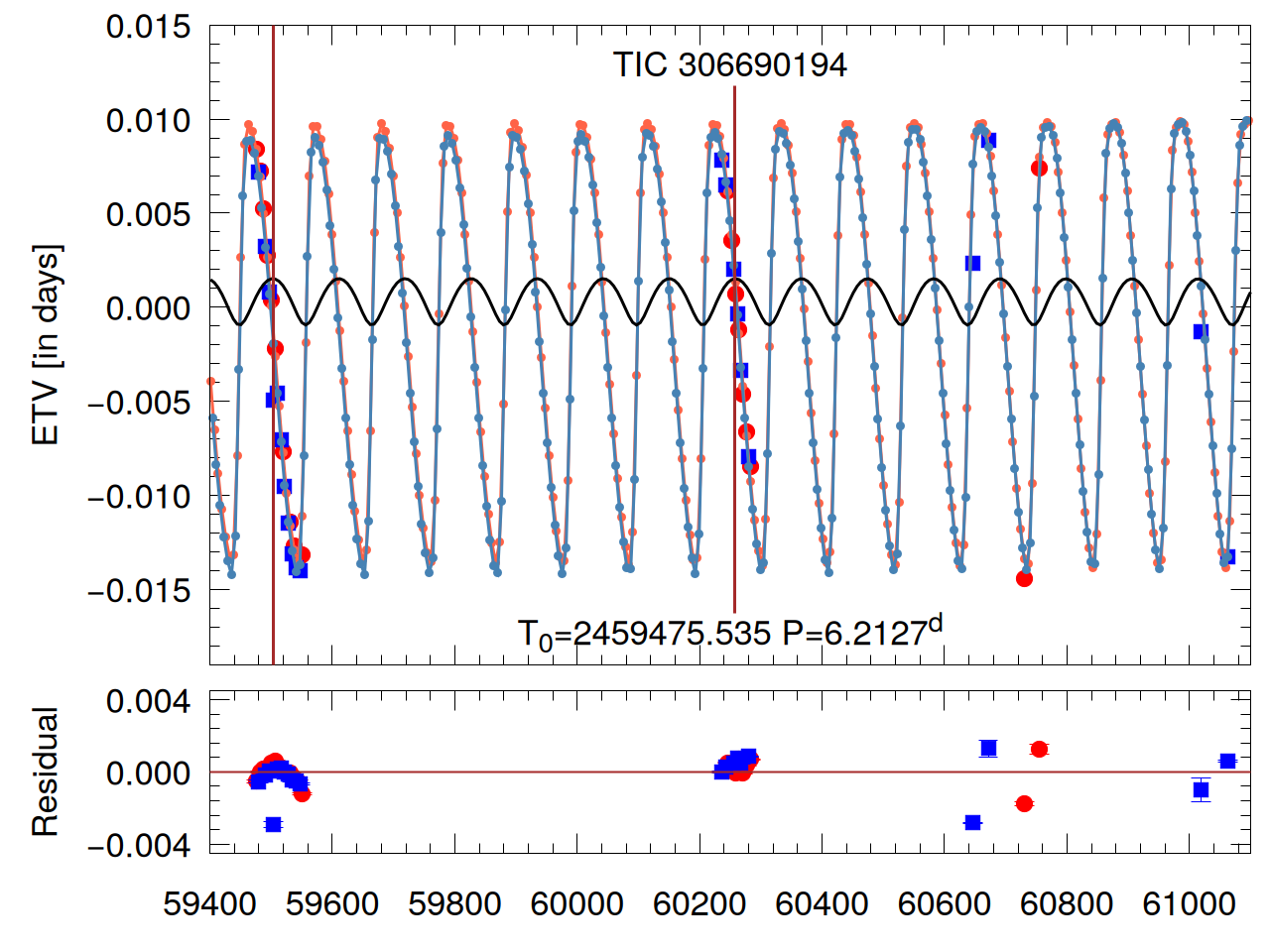}
  \includegraphics[width=0.39\textwidth]{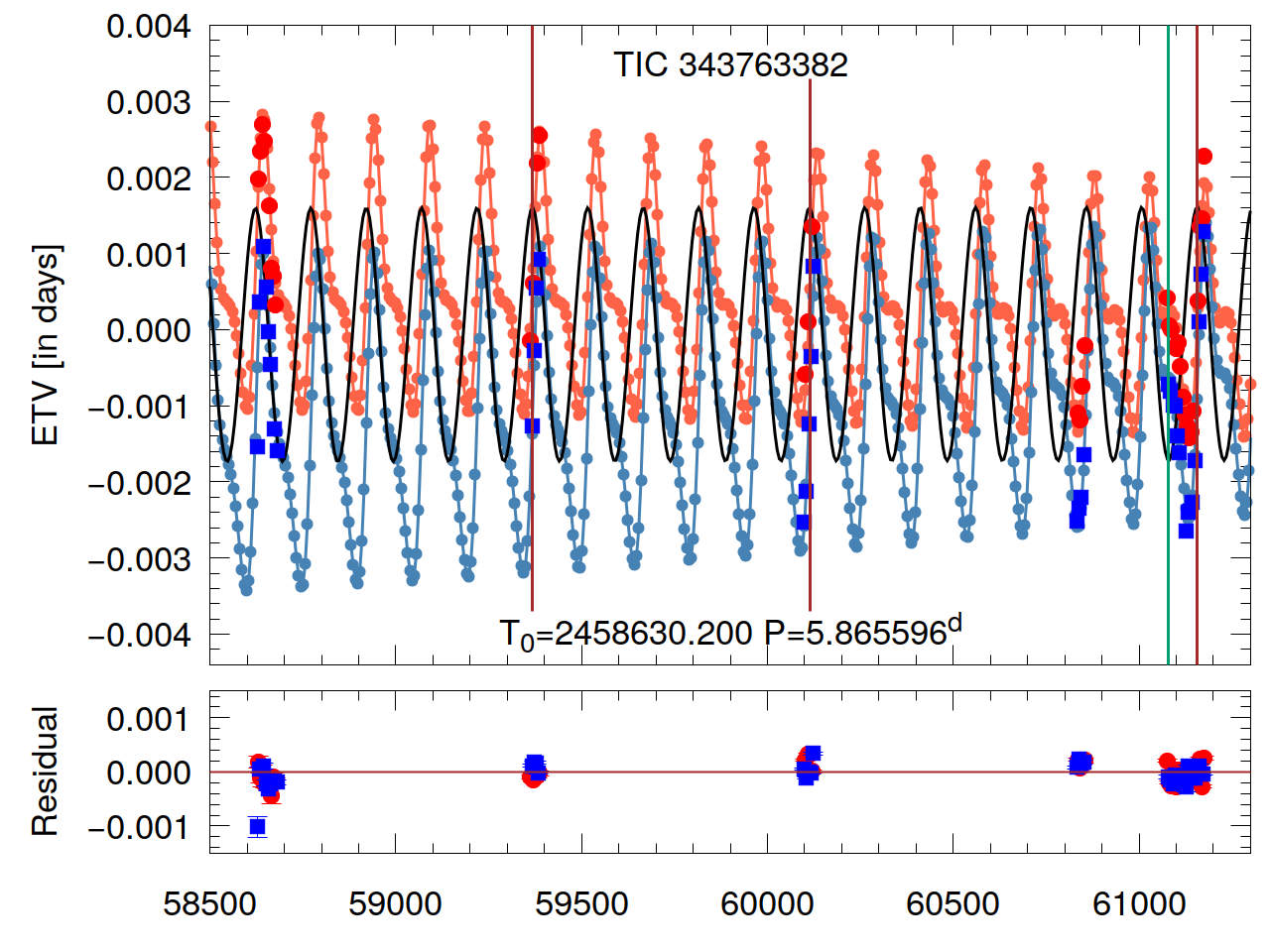} \includegraphics[width=0.39\textwidth]{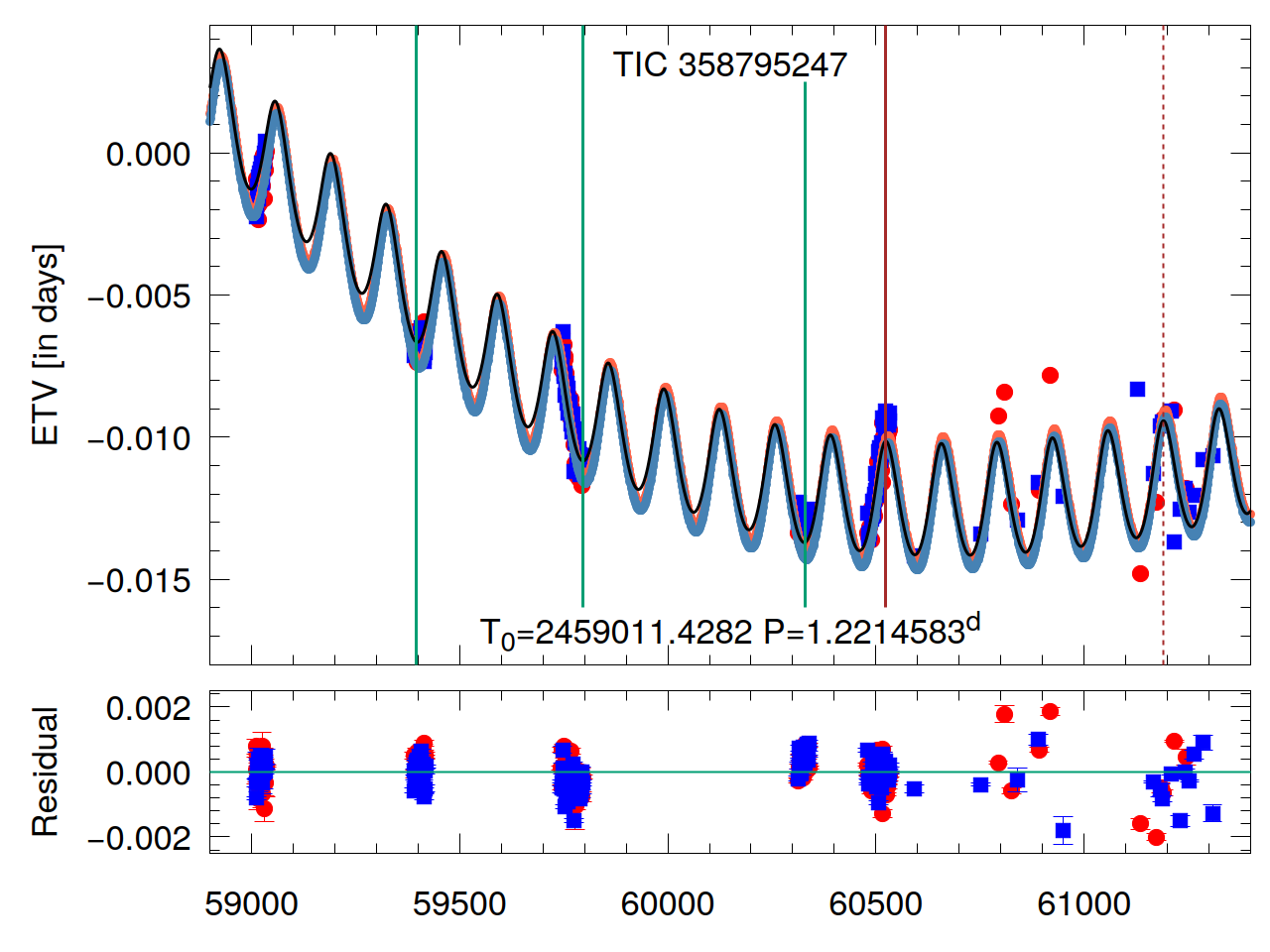}
  \includegraphics[width=0.39\textwidth]{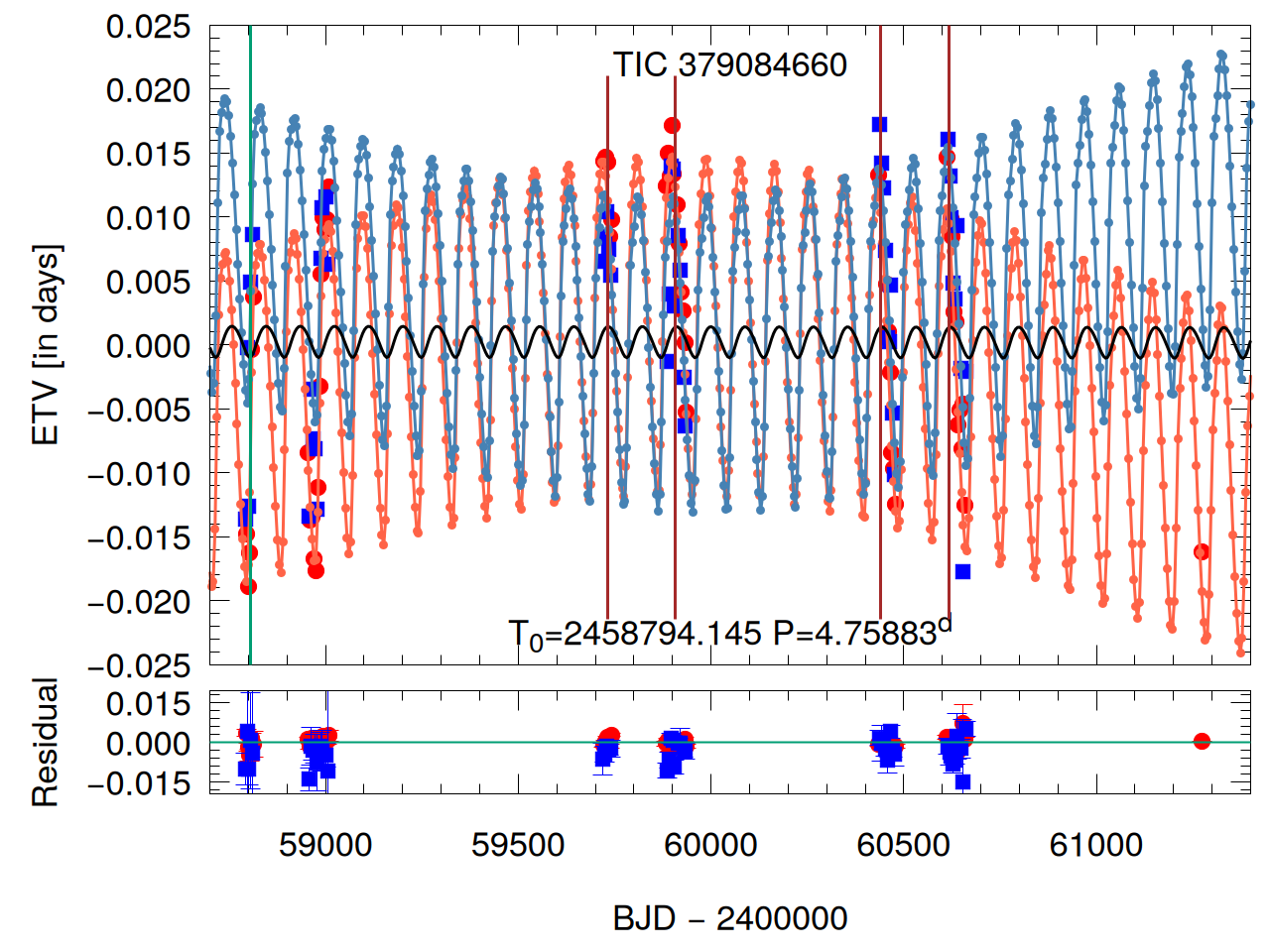}\includegraphics[width=0.39\textwidth]{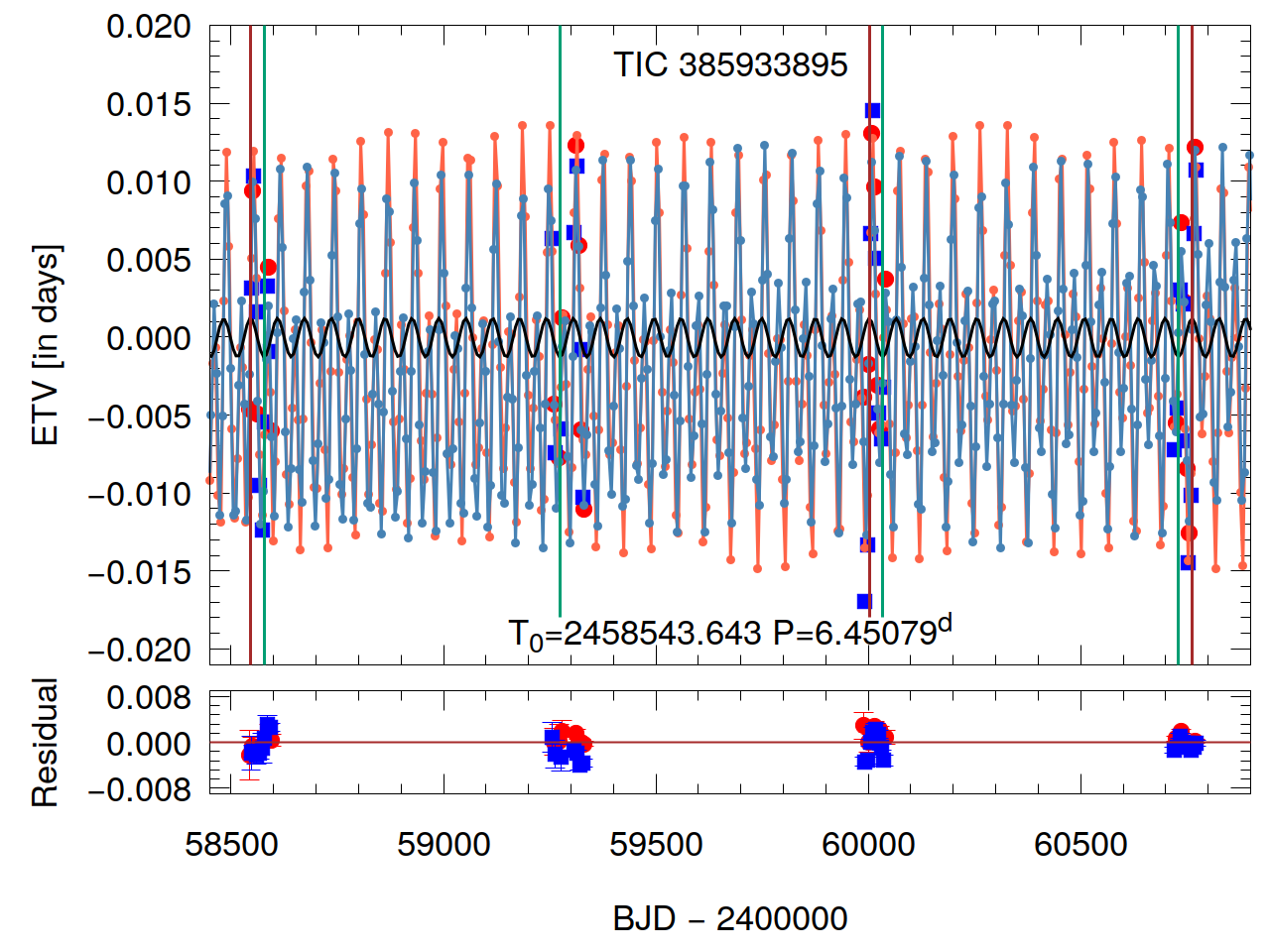}
 \caption{Primary and secondary ETV curves (heavy red circles and blue boxes, respectively) formed from the TESS observations, with the best-fit photodynamical solution superposed, for the eight triples in the work. The black curves represent the pure LTTE contributions. Apart from TIC\,358795247, which was found to be a quadruple system, this black curve is horizontally centred, which means there is no other source of any LTTE. Moreover, this black curve is clearly dominant in the case of the mentioned TIC\,358795247 and is quite large in the triples of TICs 232874003 and 343763382, while being almost negligible relative to the dynamical components in the remaining five very tight systems. Vertical lines mark the times of the observed outer eclipses (green -- the binary occulting the tertiary star and, brown -- vice versa). (In the case of TIC\,358795247 the dashed line refers to the location of the third-body eclipse that was observed during the ground-based, follow-up observations.)}
\label{fig:etvs}
\end{figure*}

In Fig.\,\ref{fig:etvs} we illustrate both LTTE and DE dominated ETVs for the current eight systems. The larger red circles and blue boxes represent those mid-eclipse times that were calculated from TESS as well as ground-based primary and secondary eclipse observations. Moreover, the connected, smaller red and blue dots are numerically generated mid-eclipse times, and represent the best spectro-photodynamical model (see Sect.\,\ref{sec:photodynamics}). While these red and blue curves come mainly from the sum of the LTTE and DE contributions to the ETV, we also show separately the pure LTTE contribution as the continuous black curve. As one can see, in most cases, the total amplitude of the ETV is much higher than the amplitude of the pure LTTE-term (in black). This means that most of the investigated ETV curves are mainly dominated by dynamical terms.  Moreover, the fact that the average of the LTTE contribution, with the sole exception of the case of TIC\,358795247, is horizontal means that the average distance of the considered triple system is constant, or varies only linearly (at least on the timescale of the observations). Opposite to this scenario, in the case of TIC\,358795247 (i.e. right panel, third row) one can see that the pure LTTE contribution (in black) is practically equal to the total ETV. That is, here the DEs are negligible, and what is seen is actually the pure classical LTTE. Moreover, one can see a longer-term  non-linearity in the ETV, as well. Before considering the spectroscopic RV data, all that we could conclude is that some additional, longer term periodicity is present in this system. Later, as was mentioned already in Sect.\,\ref{sec:8triples} and will also be discussed below in Sect.\,\ref{Sect:discussion_TIC358795247}, we spectroscopically detected the lines of a fourth star, whose RVs vary linearly and, in accord with the ETV. Therefore, we strongly suspect that the longer term variation is the consequence of a longer period LTTE. Consequently, we model TIC\,358795247 as a 2+1+1-type quadruple system.

We also show, as green and brown vertical lines, the times of those third-body eclipsing events that were observed by TESS (or, in the case of the last dashed horizontal line in TIC\,358795247, by the BAO80 instrument). One can see that the green lines in every case are located close to the lower extrema of the black LTTE curves, while the brown ones are near to the upper extrema of the LTTE. This is so, because third-body eclipsing events may occur either during the superior or the inferior conjunctions of the third star, that is, when the distance of the EB from the observer is close to its minimum or maximum. (Exact equality would hold only for a circular outer orbit.)

Regarding the ETV curves and the spectro-photodynamical solutions for the other systems (see the different panels of Fig.\,\ref{fig:etvs}), one can readily see that none of the other systems is clearly LTTE dominated. In the cases of TICs\,232874003 and 343763382 the LTTE contributions are as large as the DE terms, but the remaining five systems are clearly dynamically dominated in the sense that the LTTE terms are almost negligible relative to the DE contributions.

Finally, we note that the ETV curves obtained from the primary (red) and secondary (blue) eclipses do not exactly overlap each other, for example, in the case of TIC\,77177148 (first row, left panel of Fig.\,\ref{fig:etvs}), or in some other ETVs shown in the different panels. When the two ETV curves do not overlap, this directly indicates that the difference between the mid-eclipse times of the primary and the consecutive secondary eclipses differ from half of the current eclipsing period. Since the vertical separation between a primary and the consecutive secondary ETV points for small inner eccentricities is primarily related to $\Delta\propto e_\mathrm{in}\cos\omega_\mathrm{in}$, one can readily see that this effect is a direct consequence of the non-zero eccentricities of the inner EB orbits. Moreover, when this vertical separation varies, or even reverses sign (e.g., the blue curve can be either above or below the red curve), this is a direct consequence of apsidal motion (AM; i.e., the variation of $\omega_\mathrm{in}$)\footnote{Naturally, the vertical separation in ETV curves may also vary due to third-body-induced eccentricity ($e_\mathrm{in}$) peturbations, and we found such tight systems where this latter variation is also present \citep[see, e.g,][]{borkovitsmitnyan23,borkovitsetal25b,rappaportetal26}. But, the effect of the perturbation-induced variations in $\omega_\mathrm{in}$ on the AM induced ETV, even in these cases, clearly dominate over that of the eccentricity variations.} which, in the current systems, are mostly also dynamically dominated.

The mid-eclipse times used for the derivation of these ETV curves are tabulated in Appendix~\ref{app:ToMs}. These mid-eclipse times were calculated with the use of template eclipses in the very same manner as was described already in several former papers, for example in \citet{borkovitsetal16}.
  
\section{Photodynamical models}
\label{sec:photodynamics}

We carried out detailed (spectro-)photodynamical analysis of the current eight triply eclipsing triples. This analysis was done with the sofware package {\sc Lightcurvefactory} that we have used previously for numerous other multi-stellar systems, as were listed in the Introduction.  Therefore, our current analyses are practically the same as we have done with more than 40 previous triple-star systems that we analyzed.  This tends to result in a homogeneously analyzed set of parameters for all these systems. The details of the entire analysis process, including the input data sets, the input parameters (which might be either adjusted, fixed or constrained), as well as the different adjusting strategies have all be extensively explained in several former papers \citep[like, e.g.][]{borkovitsetal18,borkovitsetal19a,borkovitsetal19b,borkovitsetal20a,borkovitsetal20b,borkovitsetal21,mitnyanetal20} and, hence, we did not repeat them in detail. 

Here we note only the basics of {\sc Lightcurvefactory}.  In general, the code is able to handle the motions of the stars using a seventh order Runge-Kutte-Nystr\"om integrator\footnote{There is also the option for computing the motions analytically, assuming pure Keplerian two-body motions for each binary subsystem.}, as well as the coupled spin rotations of all the bodies. It also emulates the light curves for rotationally and tidally distorted stars with any kind of eclipses that may occur in the following hierarchical stellar (or, of course, planetary) systems: binaries, triples, both kinds of hierarchical quadruples (i.e., both 2+2 and (2+1)+1 type hierarchies) and, moreover, some kinds of higher-order hierarchies up to six stars.  The code is able to simultaneously analyze photometric light curves (with a maximum of five different passbands), RV data, ETV curves, as well as composite stellar energy distribution (SED) data (i.e., observed passband magnitudes).  Moreover, the code contains different sets of pre-built \texttt{PARSEC} isochrone tables \citep{PARSEC}, from which the physical radii, effective temperatures, as well as different internal passband magnitudes of any given stars can be tri-linearly iterated and, hence, constrained. The latter assumes that a given star has evolved as a single object and that there has been no previous mass transfer.  Finally, to solve the inverse problem, the software package contains a Markov Chain Monte Carlo (MCMC)-based search routine, which is used not only to determine the best-fit system parameters, but also to obtain the statistical uncertainties.  This feature uses our own implementation of the generic Metropolis-Hastings algorithm \citep[see, e.g., ][]{ford05}.

Regarding the eight current triple systems, we numerically integrated the motions of the stars for all of them. In seven cases we used a 2+1-type hierarchical triple star configuration, The exception is TIC\,358795247, where both the ETV curve and the RV data prefer the presence of a fourth, more distant, component, for which we applied a (2+1)+1-type hierarchical quadruple configuration. We used the TESS light curves for all the systems (see below). In the case of TIC\,358795247, we also incorporated two additional light curves from ground-based follow up photometry. The GAO80 observations were carried out in Sloan $z'$-band, while the BAO80 photometry was taken in Sloan $r'$-band. For TIC\,379084660, due to the very different eclipse depths (see below), we also included a second light curve in passband Sloan $r'$, despite the fact that this latter light curve contained only one ground-based regular eclipse observation. For all eight systems we used both the primary and the secondary ETV curves in the simultaneous, complex analysis. 

Regarding fitting the ETVs as an quasi-independent data set, one may object that the timings of the regular eclipse minima are already inherent in the light curves.  However, we have found empirically that a fit to the system parameters is far more sensitive when the ETVs are fit independently from, but simultaneously with, the light curve itself. Each eclipse yields its own precise mid-minimum time by fitting effectively for only a single parameter -- namely, the time. When mixed in with the 30 or so other parameters to be adjusted or constrained in the fit to the light curves, the time of minimum becomes unnecessarily entangled with the other parameters and the results are not as precise or accurate.

We had RV data for only three of the systems and, therefore, the photodynamical analysis for just these three targets included RV information. These are TICs\,306690194, 343763382 and 358795247, where RV curves for two, three and four stars are available, respectively. All these multiple RV curves were included in our analysis. Here we mention that when both stars' RV curves are known in a binary, that determines the two stellar masses (if the inclination angle is known). Similarly, in the case of the latter two systems, the presence of the RVs of all the stars strongly constrains all the stellar masses.  In the case of TIC\,306690194, in theory, only the minimum mass of the tertiary is strictly constrained. For the other five targets, however, where no RV curves are available, we had to use some other methods to estimate the stellar masses and, hence, the physical dimensions of the stars. 

Fortunately, we have found in our previous studies that the composite SED together with the \texttt{PARSEC} isochrones can be used very effectively as proxies for the masses. Our experience and results in this latter regard are described in detail in \citet{borkovitsetal20a} and \citet{borkovitsetal22a}. Note that, for the sake of a fully homogeneous investigation, this latter method, i.e., the use of built-in tabulated \texttt{PARSEC} stellar ischrones, as well as composite system SED data, to constrain the fundamental stellar parameters, was also used in the case of the three systems for which RVs were available. These three targets were thereby used as `control' systems, since the two methods (using RVs vs.~isochrones for constraining masses and other stellar parameters) can result in different stellar masses. The way this works is that if isochrone-deduced masses turn out to be inconsistent with the RV-inferred masses, this will result in unusually high values of $\chi^2$. We do not find this to be the case. Further discussion in this regard can be found in Appendix\,\ref{app:proxy}.

In conclusion, in most cases, where no RV data were available, we adjusted and constrained 16 and 15 variables, as follows.  The adjusted variables include the primary's (i.e., star $Aa$) mass and the other two mass ratios; three and six orbital elements or other parameters of the inner and outer orbits, respectively; extra flux in the TESS-band; and age, metallicity and interstellar extinction of the system. The constrained parameters are the stellar radii and effective temperatures from the built-in \texttt{PARSEC} evolutionary tracks of the three stars, as well as the period and inferior conjunction time of the inner orbit, the (photometric) distance of the triple star, and the passband-dependent limb-darkening coefficients of the stars.  Where RV data were available, the systematic RV of the entire system ($V_\gamma$) is an additional constrained parameter. We suggest that the interested reader consult with the two schematic flow-charts given as Fig.\,5 of \citet{borkovitsetal20a} and Fig.\,3 of \citet{borkovitsetal25b} for further information about this whole process.

Turning now to the observational material that we used our analysis, we summarize these as follows: 
\begin{itemize}
\item[(i)] We binned all the available TESS light curves to 1800-sec cadence times. Similar to our previous triples' studies, we did this mainly because the cadence times of the exclusively available full-frame-image (FFI)-type TESS photometric data were different for the various sectors. In Years 1 and 2 (Sectors 1--26) the cadence time of the FFIs was 1800-sec. Then, in Years 3 and 4 (Sectors 27--55) the FFIs had a cadence of 600-sec, while from Year 5, i.e., after Sector 56, 200-sec cadence FFI data are available. By binning all the data to 1800 s, we effectively obtained a  homogeneous cadence for all the data used in our light curve analysis. (Though, naturally, we always determined the mid-eclipse times from the original TESS dataset and, hence, in most cases, we used finer cadenced data.)  For six of the eight systems we made use of the {\sc FITSH} package of \citet{pal12} for the photometry of the FFI data. This pipeline, however, was only available until the Sector 77 data. Photometry of the newer data was carried out with the freely available {\sc Lightkurve} \citep{2018ascl.soft12013L} software. Moreover, in the cases of TICs\,220456234 and 379084660 all the FFI photometry was carried out with this latter software. 
\item[(ii)] In order to save computational time and also to give higher weight to the most valuable sections of the light curves, we used only data from the vicinity of those light curve sections where eclipses were observed. Therefore, we excluded the out-of-eclipse light curve sections, and retained only those parts where the photometric phase was within the $\pm0\fp10$-phase domains of the inner primary and secondary eclipses.  Moreover, we kept 5-10 day-long light curve sections around each of the third-body eclipsing events (including those light-curve sections near to the (outer) inferior and/or superior conjunctions where no eclipses were observed due to the lower and/or varying outer inclinations)\footnote{Here we note that, due to the precession of the inner orbital plane in TIC\,220456364, even the inner, i.e., regular EB secondary eclipses disappeared during the most recent TESS sectors. Despite this, naturally we still included those light curve segments in our analysis where the secondary EB eclipses would have occurred during higher inner inclinations.}. 
\item[(iii)] In the cases of TIC\,358795247 (TIC\,379084660), as was mentioned above, we also used two (one) additional light curves that were made from the ground-based follow-up observations. Since the original exposure times were different from night-to-night, and even for the same instrument under different weather conditions, we binned these data to 900-sec, and used these as two further datasets. Also, as mentioned earlier, the GAO80 ground-based photometry was carried out in the Sloan $z'$-band, while the BAO80 data were in the Sloan $r'$-band. These ground-based observations were strongly down-weighted, and were mostly used as a check for times in between the sparse TESS-observed data sections. The latter was accomplished by penalizing (in a $\chi^2$ sense) only really bad fits to the ground-based data.
\item[(iv)]  As mentioned earlier, especially in Sect.~\ref{sec:etvs}, we computed mid-eclipse times from the original TESS and ground-based datasets. These times of minima, again, are listed in Appendix~\ref{app:ToMs}. The use and necessity of ETVs in our spectro-photodynamical analysis were explained above.
\item[(v)] For three of the sources, we also included RV data in our complex analysis. In these cases the systemic velocity $V_\gamma$ was also constrained with an a-posteriori minimisation of the $\chi^2_\mathrm{RV}$ value at the end of each trial step, and the result obtained for $V_\gamma$ is also tabulated.
\item[(vi)] Finally, in all cases, a simultaneous SED fit was also included. The observed catalog magnitudes used for this fitting are listed in Table\,\ref{tbl:mags8}.
\end{itemize}

\section{System parameters}
\label{sec:results}

\subsection{Tables of fitted parameters}

We tabulate the results of the above described (spectro-) photodynamical runs in the tables of Appendix~\ref{app:resulttables}. The meanings of most of the parameters are explained in Table~\ref{tbl:definitions}, while the results on the eight investigated triple systems are tabulated in the following Tables~\ref{tab:syntheticfit_TIC077177148220456364}--\ref{tab:syntheticfit_TIC379084660385933895}. We emphasize again that almost all of the orbital parameters, except for the first two lines, give the instantaneous osculating orbital elements (or are computed from them). These are valid strictly only at the epoch $t_0$ (given in the first line of each table). The osculating parameters are computed at any given time from the instantaneous (Jacobian) coordinates and velocities of the given object\footnote{The first Jacobian position and velocity vectors give the position and velocity of the secondary, i.e., star Ab, relative to star Aa, and hence, provide the orbital elements of the inner orbit, while the second Jacobians, which are rendered for star B, give the position and velocity of the third star relative to the centre of mass of the inner pair, i.e., can be used to derive the elements of the outer orbit}, and give the elements of that two-body Keplerian orbit on which the object would move in the future---if all the perturbations were to disappear at that moment. Naturally, the true motion of each object lies on the envelope curve of the instantaneous osculating orbits. Therefore, the instantaneous osculating elements are not identical with the sets of orbital elements that appear, e.g., in the theoretical perturbation equations. Importantly, they cannot be observed directly. For example, the period, $P$, tabulated in the third line of each results table is the osculating anomalous period that belongs to the inner and the outer orbits at instant $t_0$. This period should not be confused with the period that can be inferred, e.g., from an ETV or, RV curve, or simply, by observing the time between the consecutive eclipses.

In order to offer a natural comparison with other observational studies, and also for purpose of predicting future eclipses in these systems, we give the average observing periods, denoted by $P_\mathrm{ETV}^\mathrm{avg}$ (second line of each table). We calculate these quantities in the following manner. First, we emulate a model light curve of the system using the parameters for the best-fit (spectro-)photodynamical model corresponding to the time interval 2010-01-01 to 2030-01-01. Then, we calculate model mid-eclipse times from these emulated light curves for the 20-year-long interval in a manner similar to what we used to compute mid-eclipse times  from the observational data. Naturally, {\sc Lightcurvefactory} directly generates mid-eclipse times by itself; but instead of those, we use the indirect step of determining mid-eclipse times from the emulated light curves for a double-check on the results. Finally, the dense ETV curves, calculated in such a manner for this 20-year-long interval, are analysed analytically with the same code used and described, e.g., in \citet{borkovitsetal15,borkovitsetal16,borkovitsetal25a}. In this way, we obtain the inner and outer observable average periods together with their uncertainties, and we tabulate these values in the line for $P_\mathrm{ETV}^\mathrm{avg}$. We also note that these periods have clear geometrical meanings. The value obtained for the inner orbit gives the period that should be used for an ETV curve with zero average slope (i.e., horizontal). The value valid for the outer orbit, however, is the average period of the cycles caused by the third-body on the ETV curve. We stress that this is not necessarily equal to the time interval between two `primary' (or `secondary') outer eclipses, as the time between consecutive, similar types of outer eclipses also depends on the locations of the three stars in their orbits, and can vary by even more than one inner orbital cycle. 

Turning back the instantaneous osculating orbital elements, and also to the other quantities given in Tables~\ref{tab:syntheticfit_TIC077177148220456364}--\ref{tab:syntheticfit_TIC379084660385933895}, we list there the median of all the posteriors obtained in all the individual runs of the MCMC chains.  Moreover, we also give the $1-\sigma$ upper and lower statistical uncertainties. We do not necessarily give the adjusted (or constrained) quantities directly but, in some cases, we list only parameters that are computed from the adjusted (or constrained) parameters. For example, though in each MCMC run the adjusted parameters are $e\cos\omega$ and $e\sin\omega$ for each orbit, in the Tables we directly list $e$ and $\omega$. We also note that several derived quantities are listed as well. For example, the code directly calculates the RV amplitudes ($K_\mathrm{pri/sec}$) from other model parameters. Therefore, we stress that the given $K$-s are always derived and not fitted, independent of whether RV data are available or not for a given system. 

Besides the tabulated results, in what follows we discuss our results for all eight individual triply eclipsing triple systems. First, however, we make some general remarks. In \citet{borkovitsetal25b} we considered ultra-compact hierarchical triples (UCHTs), that is, such systems where the outer period is less than (or equal to) 100\,days. In the current work, we did not apply such a restriction, and only two of the eight investigated systems can be considered UCHTs. (Or, if we allow borderline cases, we may add a third system, TIC\,77177148 as a marginal UCHT, given that its outer period is $P_\mathrm{out}=101\fd4$\,days.) For the current systems, however, all but one of them have another interesting characterization for triples, i.e., they are `tight', i.e., $P_{\rm out}/P_{\rm in}$ is relatively small (see the last line of Table\,\ref{tbl:mags8} for the tightness parameter). Apart from TIC\,358795247 (which is, however, a quite rare, four-lined spectroscopic quadruple), the tightness parameter of the systems is below 31. Interestingly, in the case of TIC\,385933895 (which is one of the two or three UCHTs in the current study) this ratio is smaller than 10, making this triple one of the tightest systems known. Therefore, it is far from surprising that the currently investigated systems are dynamically quite active, where the third-body perturbations already play important roles on monthly timescales.  These can be seen nicely in the ETV curves, that is, in Fig~\ref{fig:etvs}. (We will return to this question below in connection with TIC\,220456364, in Subsect.\,\ref{Sect:discussion_TIC220456364}.) 

What we can also say globally for the investigated systems is that all eight triples contain only main-sequence stars. Apart from TIC\,358795247, all the inner periods have  $P_\mathrm{in}>3\fd3$ day. Or, in other words, there are only two inner binaries from the current eight where the fractional radii of the stars exceed $r=R/a>0.1$. And, even the maximum of the fractional radii (which is $r=0.141$ in the case of star Aa of the above mentioned TIC\,358795247) is quite small and, therefore, all the stars can be considered as nearly spherical (though we did not apply such an approximation during our analyses). Due to the small $r=R/a$ values, there are no noticeable proximity effects on the light curves---this can nicely be seen in the light curve plots, i.e., in Figs.\,\ref{fig:77177148lc} and \ref{fig:lcs1}--\ref{fig:lcs2}). Therefore, any tidal effects can safely be neglected in all eight systems. This is also the reason why $\Delta\omega_\mathrm{3b}>>\Delta\omega_\mathrm{tide}$ for all but one of the inner binaries (see the appropriate lines of Tables\,\ref{tab:syntheticfit_TIC077177148220456364}--\ref{tab:syntheticfit_TIC379084660385933895}) or, in other words, the AM in all cases where it is present, is clearly dynamically dominated. 

Finally, note that the light curve sections of the (spectro-)photodynamical model fits can be seen in the above mentioned Figs.\,\ref{fig:77177148lc} and \ref{fig:lcs1}--\ref{fig:lcs2}); the model ETVs are plotted against the observed points in Fig.\,\ref{fig:etvs}; while the model RVs are shown against the observed data both in the time and phase domains in Figs.\,\ref{fig:T306690194RV}--\ref{fig:T358795247RV}.

\subsection{Results for the individual systems}

\subsubsection{TIC 77177148}
\label{Sect:discussion_TIC077177148}

TESS observed a two-dipped, $\approx20\%$-deep third-body eclipse event back in November, 2018 (in TESS Sector 5, see the left panel of Fig.\,\ref{fig:77177148lc}). While this system was one of the very first triply eclipsing triples discovered in the TESS data, unfortunately we were unable to determine the outer period or to find any working photodynamical models. This was mainly due to the fact that the regular eclipses are quite shallow, and do not reach even $\approx2\%$ deep. This is a consequence of the fact that the vast majority ($\approx88\%$) of the system's total light comes from the third, most massive stellar component (see in Table\,~\ref{tab:syntheticfit_TIC077177148220456364}), which strongly dilutes the depths of the regular eclipses. (Moreover, note that our solution also resulted in a high-contamination flux ratio of $\ell_\mathrm{x}=0.46\pm0.02$ that is an additional source of the light dilution. The source of this contamination is likely TIC\,77177146 that is only $\approx28''$ away from our target, and which is, in the TESS-band, fainter only by $\Delta T\approx0.9$\, mag. According to Gaia DR3, and the derived distance of \citealt{bailer-jonesetal21}, however, the distance between the two stars is about 28\,pc or even larger. Therefore, this object is likely only a visual companion, and does not belong to the triple system.)  

In the end, we find that this triple is formed by three low-mass, cool stars.  The total mass of the inner binary is less than the mass of our Sun, being $M_\mathrm{Aa}=0.49\pm0.04\,\mathrm{M}_\sun$ and $M_\mathrm{Ab}=0.46\pm0.03\,\mathrm{M}_\sun$ respectively, while the tertiary component is more massive with $M_\mathrm{B}=0.86\pm0.05\,\mathrm{M}_\sun$. Considering the more certain mass ratios, these are $q_\mathrm{in}=0.94\pm0.01$ and $q_\mathrm{out}=0.90\pm0.02$, respectively. We found the inner orbit of the two M dwarfs to be only slightly eccentric with $e_\mathrm{in}=0.0029\pm0.0003$, which is reflected most strongly in the dynamically induced AM. The latter AM effect manifests itself in the low-amplitude, anti-correlated variations of the primary (red) and secondary (blue) ETV curves on the $\simeq 8$-year timescale of the TESS observations, as can be seen in the first row, left panel of Fig.\,\ref{fig:etvs}.  On the other hand, the outer orbit has a moderate eccentricity of $e_\mathrm{out}=0.244\pm0.001$. However, in spite of this substantial outer eccentricity, the mutual inclination angle was found to be quite small at $2\fdg2\pm0\fdg4$, which, in turn, implies a practically flat triple system\footnote{Interestingly, in spite of how flat this system is, most of the other systems in this study are even flatter, i.e., have smaller $i_{\rm mut}$.}. 

\subsubsection{TIC 220456364}
\label{Sect:discussion_TIC220456364}

This system is located in the SCVZ, hence, TESS monitored it nearly continuously during all the southern eclipsing sectors (see in Table\,\ref{tbl:sectors}). Similar to the previous triple, this one is also formed by three low-mass, cool stars, of masses $M_\mathrm{Aa}=0.63\pm0.02\,\mathrm{M}_\sun$, $M_\mathrm{Ab}=0.32\pm0.02\,\mathrm{M}_\sun$ and, $M_\mathrm{B}=0.75\pm0.03\,\mathrm{M}_\sun$, respectively. That is, the total mass ($M_\mathrm{A}$) of the inner pair, again, is lower than the mass of our Sun and, moreover, the third star is the most massive. On the other hand, the inner mass ratio here is the smallest amongst all of our eight systems, while the difference between stars Aa and B is not so significant, as in the case of the previous triple ($q_\mathrm{in}=0.495\pm0.004$, $q_\mathrm{out}=0.79\pm0.01$).  Moreover, while this system was found to be quite tight with $P_\mathrm{out}/P_\mathrm{in}\approx23.07$, both the inner and the outer periods are by far the longest, respectively, among all the eight triples in our sample, being $P_\mathrm{in}=12\fd72$ and $P_\mathrm{out}=293\fd46\pm0\fd06$. Primarily, this is the reason why we see here definitive eclipse depth variations (EDVs) in this system in spite of the fact that the mutual inclination is only $i_\mathrm{mut}=1\fdg7\pm0\fdg1$, that is, somewhat, but not significantly, smaller than $i_{\rm mut}$ for TIC\,77177148, where no EDVs were observed. Strictly speaking, this is a consequence of the fact that, due to the larger separations and, hence, to the much smaller sum of the fractional radii (where, in the current system this is $r_\mathrm{Aa}+r_\mathrm{Ab}=0.0416\pm0.0007$, in contrast to that of TIC\,77177148, where $r_\mathrm{Aa}+r_\mathrm{Ab}=0.101\pm0.004$), the eclipse depths are much more sensitive to any small variations in the inner orbital inclination.

\begin{figure*}
\centering
     \includegraphics[width=0.5\hsize]{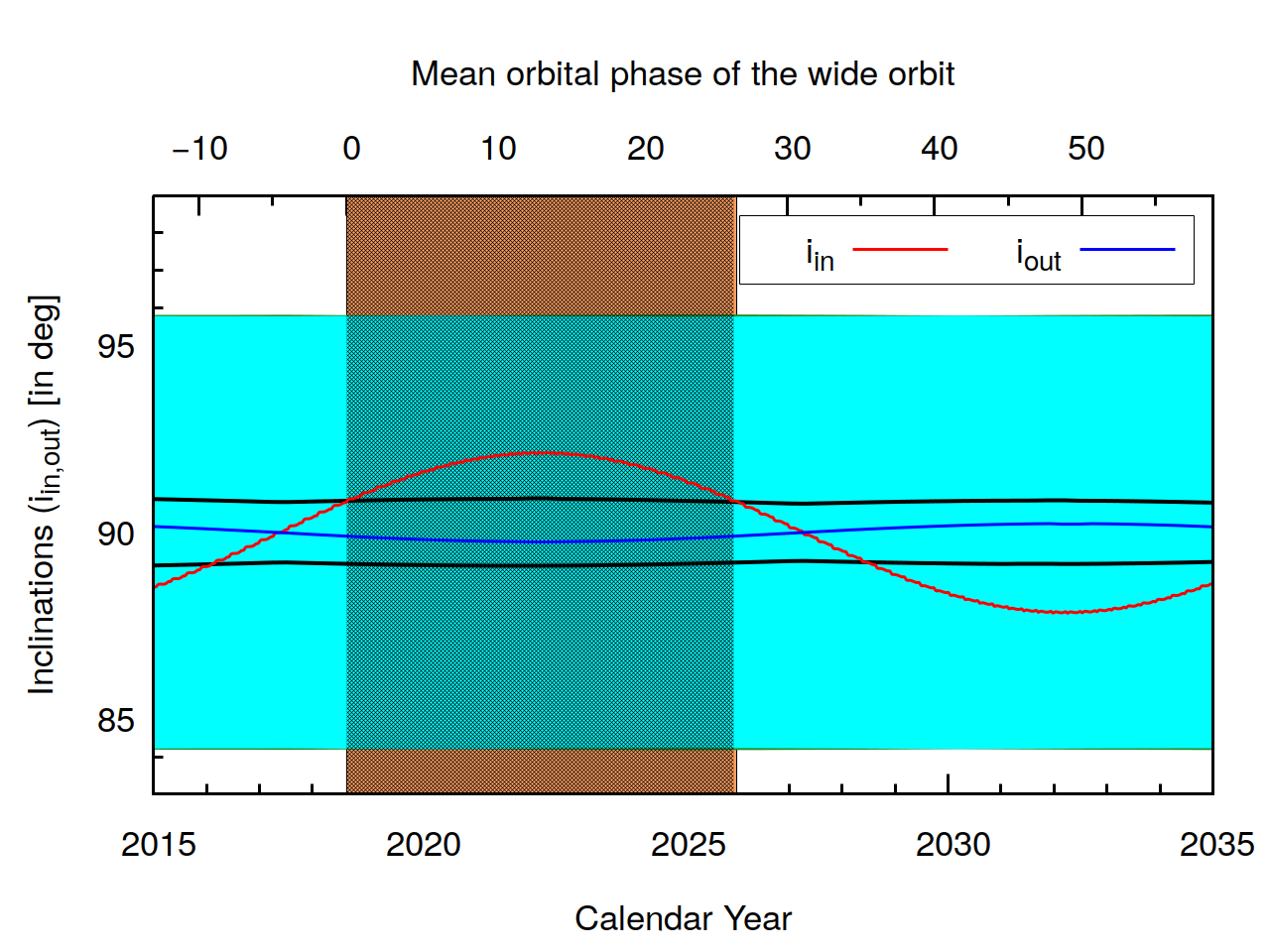}\includegraphics[width=0.5\hsize]{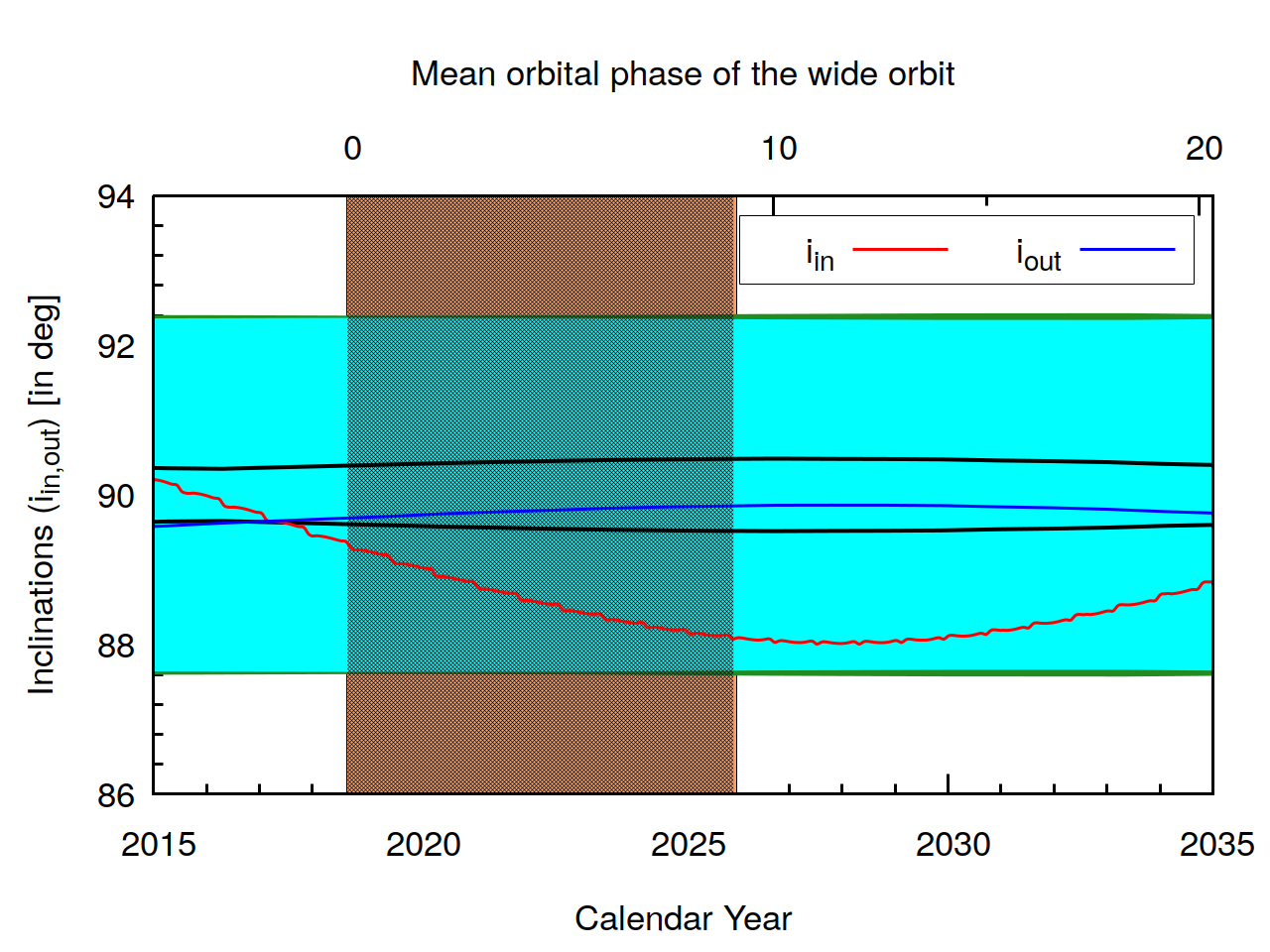}
     \caption{Variations of the inner and outer inclinations (red and blue curves, respectively) for the systems TICs\,77177148 (left) and 220456364 (right). The horizontal cyan regions denote the inclination regime where the inner pair should show two eclipses. The black, nearly horizontal curves denote the area within which third-body eclipsing events should be (and are observed to be) present. The vertical, shaded regions denote the interval between TESS Sectors 1 and 98, that is the the total time span which contains the various TESS observations of the SCVZ. (Note the different vertical scales on the two panels.) See text for further details.}
\label{fig:inclvar}
\end{figure*}

In the two panels of Figure \ref{fig:inclvar} we plot the inclination angle variations and also the domains of the regular and third body eclipses for these two systems, namely, TICs\,077177148 (left) and 220456364 (right). As can be seen readily, the inclinations of both triples are located within areas where both regular and third-body eclipses must and may occur\footnote{Within these areas, regular eclipses should certainly occur, but in the case of third-body events, realizations of these eclipses also depends on the configuration of the three bodies; therefore, in this latter case the inclination-criterion is only a necessary, but not sufficient, condition.}, respectively. On the other hand, however, in the case of TIC\,220456364, the domain of regular eclipses is fairly narrow and, moreover, for the last TESS observing session, $i_\mathrm{in}$ is very close to the lower border of the domain of the regular eclipses. This is the reason why we observe significant EDVs in this system and also why there are very shallow secondary eclipses. While the latter should be present, at least according to this figure, they get lost in the photometric noise.

\begin{figure}
\centering
     \includegraphics[width=0.99\hsize]{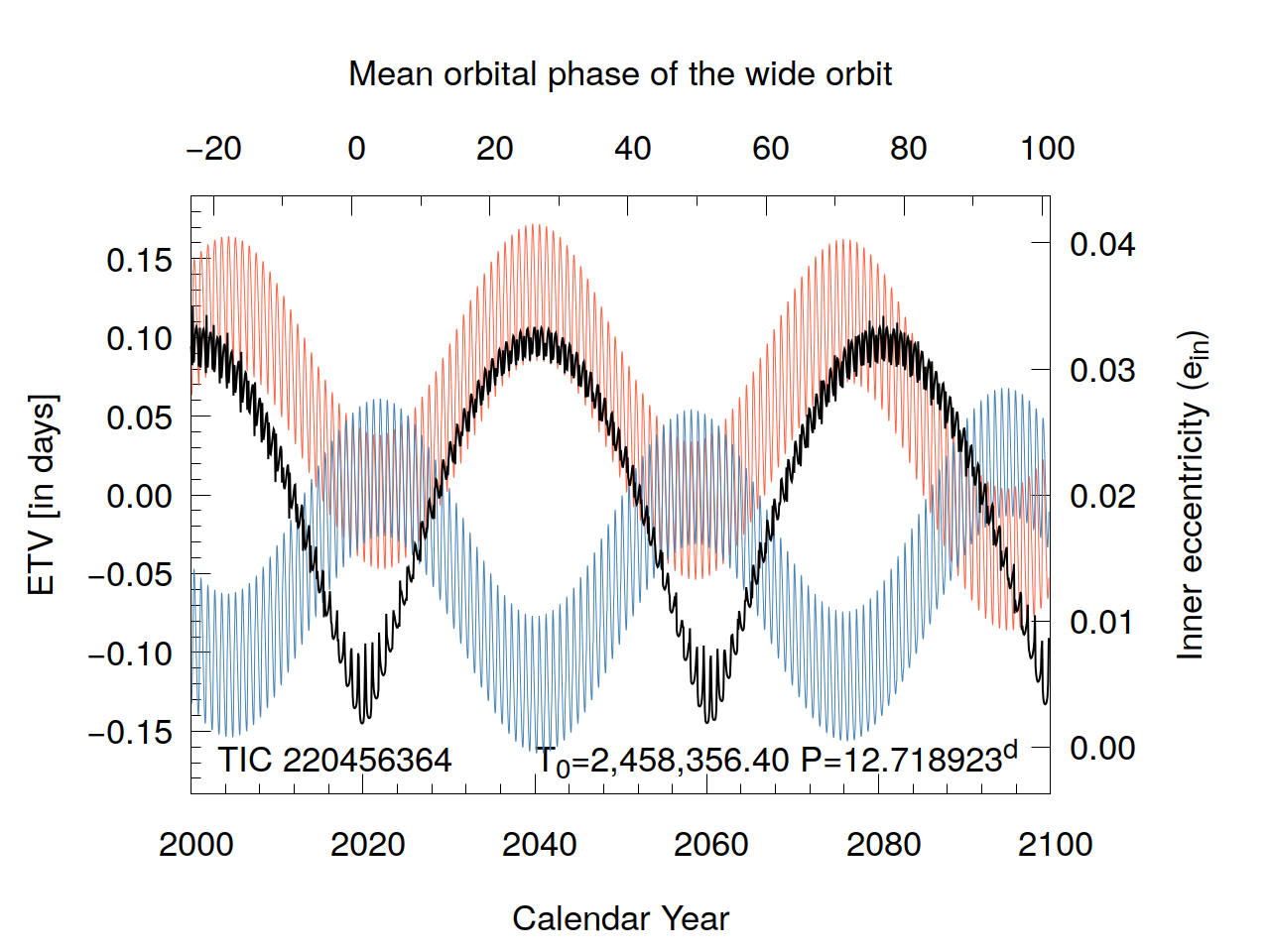}
     \caption{The primary (red) and secondary (blue) ETV curve for TIC\,2204563643 during the current century, according to the best-fit photodynamical model. The black curve represents the variations of the inner eccentricity ($e_\mathrm{in}$) during the same 100 years. As one can see, around the minima of the eccentricity curve, the primary and secondary ETV curve just about overlap each other. Oppositely, when the eccentricity is close to its maxima, the difference between the primary and secondary ETV curves and, hence, the AM is quite manifest. See further explanations in the text.}
\label{fig:ETVecc100yr}
\end{figure}

The inner orbit of TIC\,220456364 is currently nearly circular $e_\mathrm{in}=0.0099\pm0.0002$. Due to the third-body perturbations, however, according to our numerical integrations, the inner eccentricity varies cyclically between 0.002 and 0.033 on a timescale of $\approx14,800$~days. Such a huge eccentricity variation in the forthcoming years will be easily detectable even without spectroscopy, using only simple, photometric methods. In order to illustrate this, we plot in Fig.\,\ref{fig:ETVecc100yr} both modelled ETV curves, as well as the $e_\mathrm{in}$ curve of the system until the end of the current century. As one can see, currently, due to the almost negligible inner eccentricity, the primary and secondary ETV curves nearly overlap (see also the first row, right panel of Fig.\,\ref{fig:etvs}). In other words, this means that the secondary eclipses occur nearly at the midpoint between two consecutive primary eclipses. In the next decade, however, the eccentricity is expected to grow significantly and, therefore, the secondary eclipses will arrive earlier and earlier with respect to the primary eclipses. Therefore, the situation that we will observe in the next years or decades, will be some mixture of the usual AM, where only $\omega_\mathrm{in}$ varies but $e_\mathrm{in}$ remains constant,  and such an anomalous AM, where not only $\omega_\mathrm{in}$ varies, but also $e_\mathrm{in}$ changes substantially. Thus, the vertical reversals of the red and blue curves (dictated by whether the secondary eclipses occur before or after the midtimes of the consecutive primary eclipses), are determined by the classic apsidal motion, that is, by the variation of the line of the apsides (i.e. $\omega_\mathrm{in}$). By contrast, the maximum vertical displacements between the red and blue curves (which are $\Delta\propto\pm e_\mathrm{in}$, assuming that $\cos\omega_\mathrm{in}=\pm1$), are the result of the variations in $e_\mathrm{in}$.

Finally, we note that this is the only triple in our sample, for which a Gaia astrometric solution is available for the outer orbit (see near the end of Sect.\,\ref{sec:discuss}). The angular separation of the outer orbit (which is the highest in our sample), however, even in this case is not larger than 2.3\,mas that somewhat exceeds the resolution limit of the current largest optical interferometers.

\subsubsection{TIC 232874003}
\label{Sect:discussion_TIC232874003}

In contrast to the previous two cool and low-mass systems, TIC\,232874003 consists of more massive and hot stars. Their masses, radii and temperatures are $M_\mathrm{Aa}=2.4\pm0.2\,\mathrm{M}_\sun$, $M_\mathrm{Ab}=2.2\pm0.2\,\mathrm{M}_\sun$, $M_\mathrm{B}=1.7\pm0.1\,\mathrm{M}_\sun$; $R_\mathrm{Aa}=2.33\pm0.05\,\mathrm{R}_\sun$, $R_\mathrm{Ab}=2.09\pm0.09\,\mathrm{R}_\sun$, $R_\mathrm{B}=1.57\pm0.08\,\mathrm{R}_\sun$; $T_\mathrm{Aa}=9600\pm500\,\mathrm{K}$, $T_\mathrm{Ab}=9300\pm500\,\mathrm{K}$, $T_\mathrm{B}=7800\pm500\,\mathrm{K}$.  Note that, in the absence of RV measurements for this system, we could determine the masses only to an accuracy of $\sim10\%$, which is the weakest among our sample. Nonetheless,  in terms of the relative, or dimensionless, quantities that are constrained more strongly by the light curve and the ETV curve, we were able to obtain much better accuracies.  In particular, for the inner mass ratio we find $q_\mathrm{in}=0.934\pm0.007$ (that is, better than 1\% accuracy).  On the other hand, for the outer mass ratio, our accuracy is not better, being $q_\mathrm{out}=0.36\pm0.03$. Moreover, considering the fractional radii, our photodynamical solutions resulted in $r_\mathrm{Aa}=0.113\pm0.001$, $r_\mathrm{Ab}=0.101\pm0.001$ and, $r_\mathrm{B}=0.0088\pm0.0003$, respectively.  At least for the inner EB stars, these fractional radii are again accurate to around 1\%.

This triple was found to be one of the youngest among our sample, with an age of $400\pm170$\,Myr. The metallicity is slightly (but not significantly) enhanced compared to our Sun, with [M/H]$=0.06\pm0.07$ dex. Regarding the somewhat larger inaccuracies in the parameters for this system, we note that the Gaia DR3 gives a (geometric) distance of $d=2221\pm2175$\,pc \citet{bailer-jonesetal21}, which, naturally, can be interpreted only as a large upper limit. Instead, our photometric distance of $d=1184\pm66$\,pc appears to be more secure.

A fact that makes this nearly flat ($i_\mathrm{mut}=2\fdg3\pm0\fdg2$) triple rather uncommon from a dynamical point of view, is that not only is the system flat, but both the inner and outer orbits have low eccentricities ($e_\mathrm{in}=0.00126\pm0.00008$, and $e_\mathrm{out}=0.049\pm0.003$). Such a low outer eccentricity is likely primordial in this triple, as we know of no mechanism that would circularise the outer orbit in such a relatively wide, but tight system.

\subsubsection{TIC 306690194}
\label{Sect:discussion_TIC306690194}

This triple system is located close to the ecliptic and, hence, it was observed only during some of the ecliptic sectors of the TESS spacecraft  (see in Table\,\ref{tbl:sectors}). Unfortunately, during the infrequent TESS visits to this source, the same sections of the outer orbit were observed.  One can see this nicely in the second row, right panel of Fig.\,\ref{fig:etvs}, and notice the complete lack of data points in the ascending branches of the ETV curves. Therefore, TESS observed only two third-body eclipsing events, and both events belong to the same superior conjunction of the third star. Thus, only one kind of outer eclipse was observed.

Fortunately, this system is quite bright for spectroscopic follow-up observations (see in Table\,\ref{tbl:306690194_RVs}) and, therefore, despite the above mentioned obstacles, we were able to determine the astrophysical and dynamical parameters of this triple with good accuracy. Thus, while the stellar masses of the current system are very similar to those of the previous system, we obtained the system parameters with factors of 3-10 better accuracy, leading to $M_\mathrm{Aa}=2.25\pm0.06\,\mathrm{M}_\sun$, $M_\mathrm{Ab}=2.19\pm0.05\,\mathrm{M}_\sun$, $M_\mathrm{B}=1.61\pm0.03\,\mathrm{M}_\sun$; $R_\mathrm{Aa}=1.81\pm0.01\,\mathrm{R}_\sun$, $R_\mathrm{Ab}=1.78\pm0.01\,\mathrm{R}_\sun$, $R_\mathrm{B}=1.49\pm0.01\,\mathrm{R}_\sun$; $T_\mathrm{Aa}=9900\pm300\,\mathrm{K}$, $T_\mathrm{Ab}=9750\pm250\,\mathrm{K}$, $T_\mathrm{B}=7650\pm250\,\mathrm{K}$.

Regarding the relative dimensionless quantities, these are $q_\mathrm{in}=0.975\pm0.004$ and $q_\mathrm{out}=0.362\pm0.003$ (in particular, even the outer mass ratio was found with an accuracy better than 1\%), while for the fractional radii we obtained $r_\mathrm{Aa}=0.0776\pm0.0003$, $r_\mathrm{Ab}=0.0760\pm0.0003$ and, $r_\mathrm{B}=0.00850\pm0.00008$, respectively. Despite the very similar masses of this and the previous system, as well as the similar separations in the inner EBs (being $a_\mathrm{in}=23.4\pm0.2\,\mathrm{R}_\sun$ instead of $a_\mathrm{in}=20.6\pm0.6\,\mathrm{R}_\sun$ in the former EB), the fractional radii differ substantially from the values in the previous system.  This results from the fact that TIC\,306690194 is found to be significantly younger than the previously investigated TIC\,232874003 with an age of $168\pm17$\,Myr.  Its age makes this triple likely to be the youngest among the current sample. Therefore, the physical stellar radii were found to be substantially smaller than in the previous system. This latter conclusion naturally holds only because we found a similar, slightly super-solar metallicity ([M/H]$=0.06\pm0.04$) for the current triple as was obtained for the previous system.

What is also remarkable about this system is that it is found to be almost totally flat, with $i_\mathrm{mut}=0\fdg10\pm0\fdg06$. However, we should note that this triple, being located near to the ecliptic, was observed only very sparsely with TESS. Therefore, some of the true physical uncertainties might be much larger than the formal statistical ones. Connected to this fact, we call the attention of the reader to the ETV residuals in the lower part of the second row, right panel of Fig.\,\ref{fig:etvs}. In the case of the first TESS observations (in Sectors 43--45), the ETV residuals (apart from one clear outlier point) show a systematic trend that may suggest our fit and solution should be slightly refined. This cannot be done in this moment, as no further, high accuracy TESS measurements are available.
  
For TIC\,306690194 we obtained a photometric distance of $d=990_{-15}^{+11}$\,pc which is in excellent agreement with the geometric distance of $d_\mathrm{DR3}=984\pm15$\,pc \citep{bailer-jonesetal21}.

\subsubsection{TIC 343763382}
\label{Sect:discussion_TIC343763382}

This is a triple lined spectroscopic triple star. We obtained five RVs for star Aa, and six data points for stars Ab and B with the  CHIRON spectrometer, the data for which are plotted in Fig.\,\ref{fig:T343763382RV}, and listed in Table\,\ref{tbl:343763382_RVs}. According to our results, this triple is formed by three quite similar stars, that are slightly more massive than our Sun. The derived masses, radii and temperatures are as follows: $M_\mathrm{Aa}=1.27\pm0.02\,\mathrm{M}_\sun$, $M_\mathrm{Ab}=1.13\pm0.02\,\mathrm{M}_\sun$, $M_\mathrm{B}=1.26\pm0.02\,\mathrm{M}_\sun$; $R_\mathrm{Aa}=1.31\pm0.01\,\mathrm{R}_\sun$, $R_\mathrm{Ab}=1.08\pm0.01\,\mathrm{R}_\sun$, $R_\mathrm{B}=1.29\pm0.02\,\mathrm{R}_\sun$; $T_\mathrm{Aa}=6290\pm40\,\mathrm{K}$, $T_\mathrm{Ab}=5950\pm40\,\mathrm{K}$, $T_\mathrm{B}=6270\pm50\,\mathrm{K}$.

As usual, the dimensionless quantities are more accurate, being $q_\mathrm{in}=0.890\pm0.003$, $q_\mathrm{out}=0.525\pm0.006$, $r_\mathrm{Aa}=0.0715\pm0.0005$, $r_\mathrm{Ab}=0.0591\pm0.0004$ and, $r_\mathrm{B}=0.0071\pm0.0001$, respectively.

From dynamical point of view, this is the second such triple in our sample where we have found a quite low outer eccentricity: $e_\mathrm{out}=0.0575\pm0.0003$. The inner orbit also has a small, but significantly non-zero, eccentricity with $e_\mathrm{in}=0.001640\pm0.000009$.  This small, but non-zero inner eccentricity manifests itself in that fact that the primary and secondary ETV curves are slightly displaced (see the third row, left panel in Fig.\,\ref{fig:etvs}). Note also, that the mutual inclination was found to be $i_\mathrm{mut}=0\fdg6\pm0\fdg2$, that is, the flatness of this triple remains under $1\degr$.

Finally note that similar to the previous triple, our photometrically deduced distance ($d=492\pm6$\,pc) is in good accord with the geometrical distance calculated by \citet{bailer-jonesetal21}. Their value, which is based on the Gaia DR3 data gives $d_\mathrm{DR3}=482\pm4$\,pc.

\subsubsection{TIC 358795247}
\label{Sect:discussion_TIC358795247}

This relatively bright northern system is likely a (2+1)+1-type hierarchical quadruple stellar system. We discovered this as a usual triply eclipsing triple star, after the detection of the first third-body eclipsing event in Sector 40 (see the upper left panel in Fig.\,\ref{fig:lcs2}). The most up to date ETV curves, augmented by the latest TESS observations, however, suggested that, besides some month-long cycles caused by the third-body, a longer-term non-linear trend is also present (see third row, right panel of Fig.\,\ref{fig:etvs}). Such a long-term non-linear trend may signify a number of different kinds of period variations. The system, however, is fortunately bright enough for spectroscopic follow-up observations. Therefore, we carried out extensive spectroscopic follow-up observations with the TRES spectrograph (see in Sect.\,\ref{Subsect: spectroscopy}), and detected the lines of four stars (see in Table\,\ref{tbl:358795247_RVs}). Three of the stars were moving in accord with the photometric observations of the inner triple subsystem. Regarding the lines of the fourth star, it showed linearly varying RV values that are in perfect agreement with the longer-term non-linear variations in the ETV curve.  Hence, we successfully modelled the system in our spectro-photodynamical runs as a (2+1)+1-type hierarchical quadruple system. We made additional extensive photometric follow-up observations to obtain the orbital period of the most distant, fourth body. But, to date, we are unable to obtain a certain period for the fourth body. Hence, similar to our previous works on the likely quadruple systems TICs\,229785001 \citep{rappaportetal23} and 321978218 \citep{rappaportetal24} and, in contrast to TIC\,120362137 \citep{borkovitsetal26a}, the parameters of the outermost orbit should not to be taken too seriously.

TIC\,358795247 is the only system in our sample that cannot be considered to be a tight triple with $P_\mathrm{mid}/P_\mathrm{in}=109.07$. Therefore, in this triple subsystem, we do not expect strong gravitational perturbations. This can be seen nicely directly in the (above mentioned) third row, right panel of Fig.\,\ref{fig:etvs}, where one can see that the observed ETV is nearly identical with the (black) LTTE contribution, therefore, there are no visible dynamical contributions to the ETV curves.

This quadruple system consists of four Sun-like, i.e., FGK-type stars ($M_\mathrm{Aa}=0.91\pm0.02\,\mathrm{M}_\sun$, $M_\mathrm{Ab}=0.87\pm0.02\,\mathrm{M}_\sun$, $M_\mathrm{B}=1.46\pm0.04\,\mathrm{M}_\sun$, $M_\mathrm{C}=0.9\pm0.2\,\mathrm{M}_\sun$; $R_\mathrm{Aa}=0.82\pm0.02\,\mathrm{R}_\sun$, $R_\mathrm{Ab}=0.78\pm0.02\,\mathrm{R}_\sun$, $R_\mathrm{B}=1.7\pm0.1\,\mathrm{R}_\sun$, $R_\mathrm{C}=0.8\pm0.1\,\mathrm{R}_\sun$; and $T_\mathrm{Aa}=5400\pm50\,\mathrm{K}$, $T_\mathrm{Ab}=5200\pm50\,\mathrm{K}$, $T_\mathrm{B}=6930\pm70\,\mathrm{K}$ and $T_\mathrm{C}=5200\pm800\,\mathrm{K}$). Or, regarding the more accurate dimensionless quantities, these are as follows: $q_\mathrm{in}=0.95\pm0.01$, $q_\mathrm{mid}=0.82\pm0.01$, $q_\mathrm{out}=0.27\pm0.05$ and $r_\mathrm{Aa}=0.141\pm0.002$, $r_\mathrm{Ab}=0.133\pm0.002$, $r_\mathrm{B}=0.0103\pm0.0006$, $r_\mathrm{C}=0.00029\pm0.00004$, respectively.

As was mentioned above, the innermost pair in this system have the largest fractional radii among our sample.  But, even in this case, we do not see any tidally induced ellipsoidal variations (a proximity effect) in the out-of-eclipse sections of the light curve (see, again, the uppermost row of Fig.\,\ref{fig:lcs2}). Because the innermost pair, i.e., the EB, has a practically circular orbit ($e_\mathrm{in}=0.0004\pm0.0002$) we do not see any signal due tidally induced AM in the ETV curve (third row, right panel of Fig.\,\ref{fig:etvs}). We also note that the other two orbits in this system were found to be moderately eccentric with $e_\mathrm{mid}=0.22\pm0.01$ and $e_\mathrm{out}=0.28\pm0.06$.

The inner, triply eclipsing triple subsystem, similar to the other triples in our sample, was also found to be quite flat with $(i_\mathrm{mut})_\mathrm{A-B}=1\degr\pm1\degr$. This is far from true for the outermost orbit, as the mutual inclination between the innermost and the outermost orbit was found to be $(i_\mathrm{mut})_\mathrm{A-C}=27\degr\pm14\degr$--but, as was mentioned earlier, this latter value is not to be taken too seriously.

\subsubsection{TIC 379084660}
\label{Sect:discussion_TIC379084660}

In this system the only follow-up eclipse observation resulted in a much higher eclipse depth than the TESS light curve. Such a discrepancy (that cannot be explained by different photometric bands) may arise primarily from two different causes, that is, either (i) a large eclipse depth variation which is forced by a rapid inclination variation due to orbital plane precession or, (ii) different amounts of contaminating fluxes. In the current situation, the first reason can be safely ignored since the (primary) eclipses in both light curves appear to be total transits, that is, the disk of the smaller secondary star is located entirely in front of the disk of the larger primary star. This leads to U-shaped eclipses, and the eclipse depths depend only very weakly on the varying inclination. Moreover, the duration of the ground-based observed primary eclipse was similar to the durations of the TESS-observed ones, also implying that there are no significant varying inclinations. Therefore, we concluded that the TESS light curves are diluted substantially due to some contaminating light. Accordingly, we seek a consistent solution utilizing the correct third light contribution.

This is one of the two UCHT systems in our sample. As one can see nicely in the left panel of the bottom row of Fig.\,\ref{fig:etvs}, the ETV curve is clearly dynamically dominated, and it also shows low-amplitude, but rapid, dynamically dominated AM. The low amplitude is mainly due to the very small eccentricity of the EB, being $e_\mathrm{in}=0.0063\pm0.0004$. (Note, that the outer eccentricity was also found to be relatively low, at $e_\mathrm{out}=0.209\pm0.005$.) The true period of this AM (according to our numerical integration) is $P_\mathrm{apse}^\mathrm{obs}\approx8.76$\,yr, while the analytic value, which is calculated simply from the quadrupole-order approximation, is given in Table\,\ref{tab:syntheticfit_TIC379084660385933895} as $P_\mathrm{apse}=13.4\pm0.2$\,yr.  Consequently, the higher-order perturbative effects should also be taken into account for a quantitatively correct prediction.

The inner pair of this triple consists of two dwarf stars which are cooler and less massive than our Sun, while the more distant, third component was found to be the most massive and brightest, but somewhat more evolved than, the Sun. Quantitatively: $M_\mathrm{Aa}=0.71\pm0.01\,\mathrm{M}_\sun$, $M_\mathrm{Ab}=0.50\pm0.02\,\mathrm{M}_\sun$, $M_\mathrm{B}=0.91\pm0.03\,\mathrm{M}_\sun$; $R_\mathrm{Aa}=0.69\pm0.01\,\mathrm{R}_\sun$, $R_\mathrm{Ab}=0.50\pm0.02\,\mathrm{R}_\sun$, $R_\mathrm{B}=0.96\pm0.07\,\mathrm{R}_\sun$; and $T_\mathrm{Aa}=4500\pm100\,\mathrm{K}$, $T_\mathrm{Ab}=3540\pm80\,\mathrm{K}$, and $T_\mathrm{B}=5550\pm150\,\mathrm{K}$. Moreover, the dimensionless quantities are as follows: $q_\mathrm{in}=0.71\pm0.01$, $q_\mathrm{out}=0.75\pm0.02$ and $r_\mathrm{Aa}=0.0542\pm0.0005$, $r_\mathrm{Ab}=0.040\pm0.001$, $r_\mathrm{B}=0.0087\pm0.0005$, respectively.

This triple was found to be one of the two oldest ones in our sample, with an age of $8.5\pm1.0$\,Gyr. According to our results, the metallicity does not differ significantly from that of our Sun, being [M/H]$=0.06\pm0.09$. The distance of the system was found to be $d=730\pm40$\,pc, which differs by about $1-\sigma$ from the \textit{Gaia} DR3-based geometric distance: $d_\mathrm{DR3}=771\pm14$ of \citet{bailer-jonesetal21}. In this regard, however, we again call attention to the very large amount of  contaminated light ($\ell_4=63\pm4$\% in the TESS-band) that may strongly affect the determined photometric distance.

\subsubsection{TIC 385933895}
\label{Sect:discussion_TIC385933895}

This triple is another UCHT in our sample. It not only has the shortest outer period, with $P_\mathrm{out}=63\fd330\pm0\fd002$, but it is also the tightest system with $P_\mathrm{out}/P_\mathrm{in}\approx9.82$ among our eight triples.  We know of only 6 compact (i.e., $P_\mathrm{out}<1000$\,d) triple systems with a tightness parameter less than ten, and four of them are triply eclipsing.  Besides the current object, these are KIC\,7668648 \citep[$P_\mathrm{out}/P_\mathrm{in}\approx7.36$,][]{orosz23}, TIC\,242132789 \citep[$\approx8.13$,][]{rappaportetal22}, and TIC\,283846096 \citep[$\approx9.62$,][]{borkovitsetal25b}, while the two non-triply eclipsing very tight compact triples are KIC\,4079530 \citep[$\approx8.02$,][]{borkovitsetal25a} and the well-known $\lambda$~Tau \citep[$\approx8.36$,][]{ebbighausenstruve956}.

Since both the inner and the outer orbits are found to have only minor eccentricities ($e_\mathrm{in}=0.0161\pm0.0002$ and $e_\mathrm{out}=0.0447\pm0.0003$, respectively), this triple, despite its tightness, seems safely dynamically stable. In particular, it clearly satisfies the empirical stability criterion of \citet{mardlingaarseth01}, which suggests a stability-limit tightness ratio $P_{\rm out}/P_{\rm in} > 4.7$ for coplanar and circular orbits.  Note that coplanarity also holds for this system since our analysis finds $i_\mathrm{mut}=0\fdg2\pm0\fdg1$. On the other hand, we emphasize again that such a small outer eccentricity is quite unusual among triple star systems.

In this triple the third star is found to be the most massive, but the other two stars are also more massive and hotter than our Sun. The fundamental stellar parameters are as follows: $M_\mathrm{Aa}=1.39\pm0.02\,\mathrm{M}_\sun$, $M_\mathrm{Ab}=1.26\pm0.02\,\mathrm{M}_\sun$, $M_\mathrm{B}=1.86\pm0.04\,\mathrm{M}_\sun$; $R_\mathrm{Aa}=1.36\pm0.02\,\mathrm{R}_\sun$, $R_\mathrm{Ab}=1.21\pm0.02\,\mathrm{R}_\sun$, $R_\mathrm{B}=1.61\pm0.05\,\mathrm{R}_\sun$; and $T_\mathrm{Aa}=6940\pm60\,\mathrm{K}$, $T_\mathrm{Ab}=6560\pm60\,\mathrm{K}$, and $T_\mathrm{B}=8820\pm200\,\mathrm{K}$, while the dimensionless quantities are $q_\mathrm{in}=0.908\pm0.006$, $q_\mathrm{out}=0.702\pm0.008$ and $r_\mathrm{Aa}=0.0681\pm0.0009$, $r_\mathrm{Ab}=0.0607\pm0.0008$, $r_\mathrm{B}=0.0145\pm0.0004$, respectively.

We find that this system has a solar metallicty, to within the uncertainties, of [M/H]$=-0.02\pm0.05$. The photometrically inferred distance is found to be $d=1856_{-35}^{+46}$\,pc which agrees within $1.5\sigma$ with the Gaia DR3 derived (geometric) distance of $d_\mathrm{DR3}=1805\pm41$\,pc \citep{bailer-jonesetal21}.

\section{Summary, discussion, and conclusions}
\label{sec:discuss}

Interestingly, all of the eight investigated triple star systems are found to be nearly coplanar. For comparison, we considered the similar former analyses of 55 other triply eclipsing triples \citep[e.g.,][]{borkovitsetal19a,borkovitsetal20b,borkovitsetal22a,borkovitsetal22b,borkovitsetal25b,borkovitsetal26a,mitnyanetal20,rappaportetal22,rappaportetal23,rappaportetal24,rappaportetal26,powelletal22,czavalingaetal23,kostovetal24}.  We excluded those systems that either were supposed to be the inner triple subsystem of quadruple stellar systems, or some earlier mass exchange was inferred to have occurred.  Moreover, we also excluded those systems that contained red giant tertiaries. Among the remaining 33 triples, we find that the inner and outer orbital planes in these triples are also coplanar to within $\lesssim 5\degr$, with only one exception.  This is TIC\,47151245=V726 Sco, with $i_\mathrm{mut}=41\degr\pm1\degr$ \citep{rappaportetal23}. The question naturally arises as to whether this coplanarity is ubiquitous among the shortest outer period triples, or, if it is merely a selection effect. In the former case it might reflect some special formation mechanism(s) for compact triples, while in the latter case, it would simply indicate that compact triples can be most easily discovered when there are outer eclipses, and such extra eclipses are most likely when both the inner and outer orbits are nearly coplanar.

In order to answer this question, one should keep in mind that, due to the larger separation of the outer orbit, in most cases the outer orbit must viewed nearly on edge-on to observe third-body eclipsing events. (Naturally, this is valid only if we consider smaller, main sequence stars.)  For example, in the current sample of our eight systems, we found $89\fdg44\leqq i_\mathrm{out}\leqq90\fdg74$. For the inner EBs there is no such strict criterion on $i_{\rm in}$. In particular, we know of several EBs where $i_\mathrm{in}\lesssim75-80\degr$, which would allow even a mutual inclination of $i_\mathrm{mut}\sim15\degr$ or higher (in order to still have third body eclipses). Therefore, the relatively low number of such triply eclipsing triples that consist of only main sequence stars, and have significant non-zero mutual inclination angles, suggests that the flatness of the compact triples has some physical origin(s) (likely evolutionary), and is not simply a selection effect.

To further investigate this question, however, we collected a set of similarly compact triple stars where we do not observe third-body eclipses. Such compact triples in most cases can be discovered trough ETVs and/or EDVs or, even due to the rapid disappearance and/or reappearance of the eclipses. Because the vast majority of the previously investigated triply eclipsing triples have outer periods under 400 days, we concentrated on non-triply eclipsing triples limited to $P_{\rm out} \lesssim 400$ d\footnote{We understand that the short outer periods among the triply eclipsing triples are an observational selection effect, since the eclipse probability decreases quickly with $P_{\rm out}$ approximately as $\propto 1/P_{\rm out}^{2/3}$. We are also aware that some longer period triply eclipsing triple stars have been discovered. Currently the longest known outer period triply eclipsing triple is KIC\,6543764 with an outer period of $P_\mathrm{out}=1100\fd2\pm0\fd3$ \citep{masudaetal15}.}.

Unfortunately, no simple ETV studies can be used for such purposes. This is so because even if someone were to collect the $P_\mathrm{out}<400$\,d-period non-triply-eclipsing triples from ETV studies, one can imagine that these are (nearly) coplanar triples with a bit lower inclination that is still enough to produce regular EB eclipses, but too low for third-body eclipsing events. A good example of such a triple star system is TIC\,219885468 that is a non-triply eclipsing, tight and compact, triple star system, which was analyzed photodynamically in \citet{borkovitsmitnyan23}. These authors found a mutual inclination of $i_\mathrm{mut}=0\fdg2\pm0\fdg2$. Therefore, they concluded, that TIC\,219885468 is also practically a coplanar system but, simply, its outer inclination ($i_\mathrm{out}=88\fdg1\pm0\fdg2$) is insufficient to produce third-body eclipses.

Therefore, a better way is to find such short period triples that show signs of planar precession, i.e., that are certainly not coplanar. Planar precession, in general, causes variations in the (observed) inclination angle (i.e., $i_\mathrm{in}$) which usually manifests itself via EDVs. The problem, however, is that most times when we observe EDVs in an EB, we do not know the period of the non-coplanar, third-body.  We know of only a very few systems where the EDV is very clear, and we also know the outer period by other means. These are TIC\,167692429 \citet{borkovitsetal20a} with $P_\mathrm{out}=331\fd5\pm0\fd3$ and $i_\mathrm{mut}=27\fdg3\pm0\fdg3$ and, additionally, KIC\,8938628, where the eclipses completely disappeared between the \textit{Kepler} and TESS observations, but, from older ETV studies, we know that $P_\mathrm{out}=388\fd5\pm0\fd3$ and $i_\mathrm{mut}=13\degr\pm1\degr$ \citep{borkovitsetal15,borkovitsetal16}. We also mention the triple stars TICs\,236774836 and 420263614 for which the eclipses disappeared between the older and newer TESS sectors \citep{mitnyanetal24,borkovitsetal26b}. Finally, we should note the cases of V907~Sco \citep{zascheetal23} and HS~Hya \citep{vokrouhlickyzasche22} that are clearly inclined triples with outer periods less than 200\,days\footnote{There are some other systems, as well, where the EDVs are evident from the appearance or disappearance of the eclipses. However, in those cases, we either do not know the period of the outer orbit, or those triples violate our other conditions. This is the case, for example, with KIC\,7955301, where the eclipses also disappeared between the \textit{Kepler} and TESS era. The parameters of this triple system are quite robustly known, being $P_\mathrm{out}=209\fd76\pm0\fd02$ and $i_\mathrm{mut}=6\fdg2\pm0\fdg3$ \citep{gaulmeetal22}. The third component, however, in this triple is a red giant; therefore, we do not count it in our sample. Another case is the $P_\mathrm{out}=293.8$-d-period, long-known triple star IU~Aur \citep{mayer990}, for which the EDV is also well-documented. In this case, however, the inner EB is a semi-detached binary and, hence, a previous interaction between the binary members must have occurred. This excludes IU~Aur from our sample.}. 


In conclusion, the only thing we can say is that there are known short-outer-period, non-triply eclipsing, triple star systems where the third component certainly revolves on a mutually inclined orbit. Due to the small number of such triples, however, we are unable to make any rigorous statements, and cannot say whether the coplanarity, that is observed in most of the triply eclipsing triples, is only a selection effect or, whether it has evolutionary reasons. Intensive future investigations will be needed to resolve this issue.

Regarding the interesting orbital parameter of the outer eccentricity we find that, within our current sample, it ranges from 0.045 to 0.333.  The mean outer eccentricity is 0.176, while the median is 0.22. According to the \citet{mardlingaarseth01} stability criteria, 7 of the 8 systems in our sample could have had outer eccentricities in the range 0.45 to 0.79 before becoming dynamically unstable.  Only TIC 385933895 would be limited to $e_{\rm out}$ as low as 0.28 before becoming unstable.  On average, the 8 triples have only 1/3 the maximum eccentricity that they might have before become unstable.

In terms of the stellar parameters there are some revealing statistical facts.  The average inner mass ratio ($M_{\rm Ab}/M_{\rm Aa}$) for our sample of 8 triples is 0.85 and its median value is 0.92.  By contrast, the corresponding values for $q_{\rm out}$ ($M_{\rm B}/(M_{\rm Ab}+M_{\rm Aa})$) are 0.65 and 0.73, respectively.  Therefore, two things are clear even from this relatively small sample.  (i) Both mass ratios are of order unity, and (ii) the inner mass ratio is closer to unity than the outer mass ratio.  This is consistent with an evolutionary scenario wherein the masses grow by accretion onto the triple, and both mass ratios approach unity \citep[see, e.g.][]{tokovinin21}. In turn, this means that there is a tendency for the third body in these systems to be more massive than either star of the inner binary, and that is indeed the case for 5 of the 8 systems (with close to a tie in a sixth system).

The stellar masses in our sample tend to be low to intermediate mass stars.  Of the 24 stars in our sample, 23 of them have masses between 0.46 M$_\odot$ and 2.4 M$_\odot$.  Most of the stars have not evolved not very far from the ZAMS. We consider the quantity $R/M$ (where both $R$ and $M$ are expressed in solar units) as a crude proxy for how evolved the star is.  Twenty of the 24 stars have $R/M \lesssim 1$, so not very evolved.  Four of the stars have $1.05 < R/M < 1.15$, and these are just evolving off the main sequence.  Three of these four evolved stars are the tertiaries in the system.

When we compare our photodynamical distances for the 8 systems with respect to the Gaia distances \citep{bailer-jonesetal21} we find generally good agreement. The fitted slope to the Gaia--photodyamical distance comparison is $0.988 \pm 0.018$ with a correlation coefficient of 0.95. Six of the eight distances agree to within or, very close to their mutual 1-$\sigma$ uncertainties. One of our distances appears to be 6-$\sigma$ closer than the Gaia distance.  For TIC 232874003 we have a distance of $1184 \pm 66$ pc while Gaia gives $2220^{+2000}_{-700}$ pc.  The latter Gaia uncertainties are so large that the two distances formally agree, but the Gaia distance in this case seems wholly unreliable.   We also note that 4 of our 8 systems have significantly large Gaia RUWE\footnote{RUWE = `Renormalized Unit Weight Error'.} values ranging from 1.49 to 11.3.  The Gaia distances may be somewhat affected by these large RUWE's.  In fact, it is TIC 232874003 that has both a very large distance uncertainty and a huge RUWE of 11.3.

In the preceding discussion of distances, we have simply called attention to a couple of discrepancies, and listed some possible reasons for them. It is clear, however, that it will be very helpful in the future to incorporate the Gaia astrometric data into our photodynamical analyses after Gaia DR4.

Finally, we comment on two of systems which also have Gaia orbital solutions: TICs 220456364 and 358795247. In Table \ref{tbl:compare} we compare a selection of the orbital parameters from the Gaia solutions vs.~our photodynamical solutions. What we see is that all of the parameters agree to within 1-$\sigma$ of the less accurate Gaia solutions, with the exception that the two values for $e_{\rm out}$ differ by 1.5- and 2.1-$\sigma$, respectively.  The other thing we notice is that the photodynamical uncertainties range from factors of $\sim$ three to two orders of magnitude smaller than for the Gaia uncertainties.  We see nothing in this comparison which causes us to doubt the stated photodynamical accuracies. 

\begin{table}[ht]
\centering
\caption{Comparison of two Gaia orbital solutions with the photodynamical solutions}
\small
\begin{tabular}{lcc}
\hline
\hline
Source/Parameter & Photodynamical & Gaia\tablefootmark{a}  \\
\hline
TIC 220456364 &  & \\
$P_{\rm orb}$ [d] & $293.46 \pm 0.06$ & $295.0 \pm 1.6$ \\
$e_{\rm out}$ & $0.3332\pm 0.0004$ & $0.408 \pm 0.059$ \\
$\omega_{\rm out}$ [deg] & $338.5 \pm 0.2$ & $330 \pm 10$ \\
$i$ [deg] & $89.68 \pm 0.01$ & $89.4 \pm 1.8$ \\
\hline
TIC 358795247 &  &  \\
$P_{\rm orb}$ [d] & $133.23 \pm 0.07$ & $133.1 \pm 0.5$ \\
$e_{\rm out}$ & $0.226 \pm 0.008$ & $0.146 \pm 0.038$ \\
$\omega_{\rm out}$ [deg] & $251.0 \pm 0.5$ & $259 \pm 19$ \\
$K_{\rm out 1,out}$ [km s$^{-1}$ & $28.5 \pm 0.4$ & $29.4 \pm 1.4$ \\
\hline 
\label{tbl:compare}  
\end{tabular} 
\tablefoot{(a) Gaia's solution for TIC 220456364 is astrometric, while for TIC 358795247 it is spectroscopic.}
\end{table}

\section*{Data availability}

Tables H.1--H.8 are only available in electronic form at the CDS via anonymous ftp to cdsarc.u-strasbg.fr (130.79.128.5) or via \url{http://cdsweb.u-strasbg.fr/cgi-bin/qcat?J/A+A/}.

\begin{acknowledgements}
This project has received funding from the HUN-REN Hungarian Research Network.

T.\,B., T.\,M., I.\,B.\,B. and A.\,P. acknowledge the financial support of the Hungarian National Research, Development and Innovation Office -- NKFIH Grants K-147131 and K-138962.

V.\,B.\,K. is grateful for financial support from NASA grant 80NSSC22K0747.

Z.\,G., T.\,P., and Z.\,D. acknowledge the support of the ESA PRODEX projects Nos. 4000137122, 4000149202, and 4000149203 between the ELTE University and the European Space Agency, the support from SNN-147362 and the ADVANCED-153410 of the National Research, Development and Innovation Office (NKFIH, Hungary), and the support of the city of Szombathely. ZG was supported by the VEGA grant of the Slovak Academy of Sciences No. 2/0033/26 and by the Slovak Research and Development Agency, the contract Nos. APVV-24-0160 and APVV-25-0583. 

This paper makes extensive use of data collected by the \textit{TESS} mission. Funding for the \textit{TESS} mission is provided by the NASA Science Mission directorate. Some of the data presented in this paper were obtained from the Mikulski Archive for Space Telescopes (MAST). STScI is operated by the Association of Universities for Research in Astronomy, Inc., under NASA contract NAS5-26555. Support for MAST for non-HST data is provided by the NASA Office of Space Science via grant NNX09AF08G and by other grants and contracts.

Distances and other astrometric properties for all targets, and outer orbits for two of our sources,  were taken from the archives of European  Space Agency (ESA)  mission {\it Gaia}\footnote{\url{https://www.cosmos.esa.int/gaia}},  processed  by  the {\it   Gaia}   Data   Processing   and  Analysis   Consortium   (DPAC)\footnote{\url{https://www.cosmos.esa.int/web/gaia/dpac/consortium}}.  Funding for the DPAC  has been provided  by national  institutions, in  particular the institutions participating in the {\it Gaia} Multilateral Agreement.

Some of the SED fluxes and magnitudes were obtained with the Wide-field Infrared Survey Explorer, which is a joint project of the University of California, Los Angeles, and the Jet Propulsion Laboratory/California Institute of Technology, funded by the National Aeronautics and Space Administration. 

Additionally, some of the SED fluxes and magnitudes were obtained with the Two Micron All Sky Survey, which is a joint project of the University of Massachusetts and the Infrared Processing and Analysis Center/California Institute of Technology, funded by the National Aeronautics and Space Administration and the National Science Foundation.

We  used the  Simbad  service  operated by  the  Centre des  Donn\'ees Stellaires (Strasbourg,  France). 

This research has also made use of the VizieR catalogue access tool, CDS, Strasbourg, France (DOI : 10.26093/cds/vizier). The original description of the VizieR service was published in \citet{ochsenbein00}.

ZTF data are based on observations obtained with the Samuel Oschin Telescope 48-inch and the 60-inch Telescope at the Palomar Observatory as part of the Zwicky Transient Facility project. ZTF is supported by the National Science Foundation under Grants No. AST-1440341 and AST-2034437 and a collaboration including current partners Caltech, IPAC, the Oskar Klein Center at Stockholm University, the University of Maryland, University of California, Berkeley, the University of Wisconsin at Milwaukee, University of Warwick, Ruhr University, Cornell University, Northwestern University and Drexel University.

Some of the observations used here were obtained at the Smithsonian Astrophysical Observatory's Fred Lawrence Whipple Observatory Ridge Telescopes.

\end{acknowledgements}

\begin{appendix} 

\onecolumn

\section{Tabulated system parameters}
\label{app:inittables} 

Coordinates, catalogue passband magnitudes, and some other catalogue parameters of our eight target systems.

\begin{table*}[ht]
\hspace{-20px}
\footnotesize
\centering
\caption{Main properties of the eight triple systems from different catalogues}
\tiny
\scalebox{0.91}{\begin{tabular}{lcccccccc}
\hline
\hline
Parameter			       &   77177148	    & 220456364 	 &  232874003	    & 306690194            &  343763382        &    358795247	   &   379084660       &   385933895	    \\  
\hline
RA J2000 			       &  04:34:24.98	    & 04:57:46.63        &  16:41:06.64     & 06:31:00.25   	   & 18:14:41.19       & 18:54:32.69	   & 01:02:30.57       & 09:15:16.64	    \\
Dec J2000 		               & -41:04:33.55	    & -56:34:44.01	 &  -46:47:02.01    & 17:17:53.80   	   & -75:55:30.23      & 27:27:29.97	   & 71:31:03.36       & -56:25:15.50	    \\
$T^a$                                  & $12.38 \pm 0.01$   & $13.40 \pm 0.01$   & $11.52 \pm 0.01$ & $10.82 \pm 0.01$     & $11.07 \pm 0.01$  & $11.52 \pm 0.01$  & $14.00 \pm 0.01$  & $13.40 \pm 0.01$   \\
G$^b$ 				       & $12.92 \pm 0.00$   & $14.01 \pm 0.00$   & $11.87 \pm 0.00$ & $10.94 \pm 0.00$     & $11.53 \pm 0.00$  & $11.89 \pm 0.00$  & $14.66 \pm 0.00$  & $13.75 \pm 0.00$    \\ 
G$_{\rm BP}^b$ 		               & $13.36 \pm 0.00$   & $14.54 \pm 0.00$   & $12.06 \pm 0.00$ & $11.03 \pm 0.00$     & $11.88 \pm 0.00$  & $12.17 \pm 0.00$  & $15.24 \pm 0.00$  & $14.02 \pm 0.00$    \\
G$_{\rm RP}^b$ 		               & $12.31 \pm 0.00$   & $13.32 \pm 0.00$   & $11.45 \pm 0.00$ & $10.76 \pm 0.00$     & $11.00 \pm 0.00$  & $11.44 \pm 0.00$  & $13.93 \pm 0.00$  & $13.31 \pm 0.00$    \\
B$^a$                                  & $14.00 \pm 0.06$   & $15.25 \pm 0.02$   & $12.28 \pm 0.04$ & $11.29 \pm 0.23$     & $12.35 \pm 0.26$  & $12.56 \pm 0.28$  & $16.02 \pm 0.07$  & $14.34 \pm 0.00$    \\
V$^c$                                  & $13.16 \pm 0.06$   & $14.34 \pm 0.07$   & $11.89 \pm 0.02$ & $11.11 \pm 0.24$     & $11.71 \pm 0.02$  & $12.06 \pm 0.00$  & $14.97 \pm 0.10$  & $13.89 \pm 0.00$    \\       
J$^d$				       & $11.57 \pm 0.03$   & $12.50 \pm 0.02$   & $10.94 \pm 0.03$ & $10.51 \pm 0.02$     & $10.35 \pm 0.02$  & $10.92 \pm 0.02$  & $13.02 \pm 0.02$  & $12.73 \pm 0.02$    \\
H$^d$ 				       & $12.61 \pm 0.03$   & $11.95 \pm 0.03$   & $10.81 \pm 0.04$ & $10.52 \pm 0.02$     & $10.12 \pm 0.02$  & $10.68 \pm 0.02$  & $12.49 \pm 0.03$  & $12.52 \pm 0.02$     \\
K$^d$ 			               & $11.16 \pm 0.03$   & $11.84 \pm 0.03$   & $10.72 \pm 0.04$ & $10.47 \pm 0.02$     & $10.04 \pm 0.03$  & $10.62 \pm 0.02$  & $12.36 \pm 0.03$  & $12.53 \pm 0.03$    \\        
W1$^e$ 				       & $10.88 \pm 0.09$   & $11.80 \pm 0.02$   & $10.33 \pm 0.03$ & $10.43 \pm 0.02$     & $10.01 \pm 0.02$  & $10.53 \pm 0.02$  & $12.30 \pm 0.02$  & $12.43 \pm 0.02$     \\        
W2$^e$ 				       & $10.92 \pm 0.02$   & $11.83 \pm 0.02$   & $10.30 \pm 0.03$ & $10.46 \pm 0.02$     & $10.02 \pm 0.02$  & $10.56 \pm 0.02$  & $12.32 \pm 0.02$  & $12.46 \pm 0.02$    \\        
W3$^e$ 				       & $10.90 \pm 0.09$   & $11.87 \pm 0.15$   & $9.08 \pm NA$    & $10.47 \pm 0.12$     & $9.98 \pm 0.05$   & $10.64 \pm 0.09$  & $12.39 \pm 0.33$  & $12.01 \pm 0.24$    \\
$T_{\rm eff}$ [K]$^b$ 	               & $5028 \pm 17$      & $4705 \pm 16$	 & $10309 \pm 53$   & $10016 \pm 37$       & $5827 \pm 9$      & $6576 \pm 13$     & $4988 \pm 36$     &      NA       \\  
$T_{\rm eff}$ [K]$^a$ 	               & $5159 \pm 131$     & $4781 \pm 122$	 & $8589 \pm NA$    & $8947 \pm 197$       & $6093 \pm 132$    & $6361 \pm 133$    & $5112 \pm 122$    & $7908 \pm 122$        \\  
Radius [$R_\odot$]$^a$ 	               & $1.23 \pm 0.07$    & $1.11 \pm NA   $   & $2.20 \pm NA$    & $3.40 \pm 0.19$      & $2.16 \pm 0.10$   & $1.85 \pm 0.08$   & $1.34 \pm NA $    & $2.37 \pm NA$        \\
Distance [pc]$^f$		       & $411\pm 4$	    &  $483 \pm 6$	 & $2221 \pm 2175$  & $984 \pm 15$   	   & $482 \pm 4$       & $611\pm 12$	   & $771 \pm 14$      & $1805 \pm 41$  	\\  
$E(B-V)^a$                             & $0.02 \pm 0.01$    & $0.01 \pm 0.00$	 & NA               & $0.14 \pm 0.01$      & $0.13 \pm 0.01$   & $0.07 \pm 0.01$   & $0.20 \pm 0.01$   & $0.35 \pm 0.01$     \\ 
$\mu_\alpha$ [mas/yr]$^b$	       & $4.84 \pm 0.02$    & $14.27 \pm 0.04$   & $-0.85 \pm 0.25$ & $-1.33 \pm 0.02$     & $1.24 \pm 0.01$   & $1.97 \pm 0.02$   & $-10.82 \pm 0.02$ & $-6.00 \pm 0.01$    \\   
$\mu_\delta$ [mas/yr]$^b$  	       & $10.04 \pm 0.02$   & $ 8.77 \pm 0.04$   & $-1.09 \pm 0.23$ & $-2.09 \pm 0.02$     & $-30.86 \pm 0.01$ & $-2.57 \pm 0.03$  & $-8.45 \pm 0.02$  & $5.04 \pm 0.01$     \\  
RUWE$^{b,g}$                           & 1.49               & 2.29               & 11.25 	    & 1.05    		   &  0.91	       & 2.28		   & 1.15	       & 1.03			 \\
astr\_ex\_noise [mas]$^{b,h}$          &  0.16              &  0.31              &  1.91	    &  0.12	    	   &  0.11	       &  0.29  	   &  0.09	       &  0		  \\
astr\_ex\_noise\_sig$^{b,h}$           &  30                &   52               &  3697 	    &  20	    	   &   22	       &  149		   &  3.1	       &  0	    \\
$P_{\rm binary}$ [d]$^i$  & 3.3038             & 12.719             & 5.092      	    & 6.213	  	   & 5.8680	       & 1.2214 	   & 4.7588	       & 6.4508     \\
$P_{\rm triple}$ [d]$^i$  & 101.49             & 293.8              & 111.6     	    & 108.05	   	   & 149.33	       & 133.28 	   & 88.86	       & 63.33       \\
$P_\mathrm{triple}/P_\mathrm{binary}^j$& 30.68              & 23.10              & 21.92            & 17.39                & 25.45             & 109.12            & 18.67             & 9.82     \\
\hline
\label{tbl:mags8}
\end{tabular}}

\small
\textbf{Notes.} General: ``NA" and ellipses in this table indicate that the value is not available. (a) \textit{TESS} Input Catalog (TIC v8.2) \citep{TIC8}. (b) Gaia DR3 \citep{GaiaDR3}; the uncertainty in $T_{\rm eff}$ listed here is 1.5 times the geometric mean of the upper and lower error bars of \verb|teff_gspphot|. Magnitude uncertainties listed as 0.00 are $\lesssim 0.005$. (c) AAVSO Photometric All Sky Survey (APASS) DR9, \citep{APASS}, \url{http://vizier.u-strasbg.fr/viz-bin/VizieR?-source=II/336/apass9}. (d) 2MASS catalogue \citep{2MASS}.  (e) WISE point source catalogue \citep{WISE}. (f) \citet{bailer-jonesetal21}, geometric distances. (g) The Gaia renormalized unit weight error (RUWE) is the square root of the normalized $\chi^2$ of the astrometric fit to the along-scan observations. Values in excess of about unity are sometimes taken to be a sign of stellar multiplicity. (h) Abbreviations for \verb|astrometric_excess_noise| and \verb|astrometric_excess_noise_sig| (\citealt{lindegren21}; \url{https://gea.esac.esa.int/archive/documentation/GDR2/Gaia_archive/chap_datamodel/sec_dm_main_tables/ssec_dm_gaia_source.html}); these are a measure of ``the disagreement, expressed as an angle, between the observations of a source and the best-fitting standard astrometric model.'' Values of \verb|astrometric_excess_noise_sig| $\gtrsim 2$ are considered significant. (i) Binary and outer orbital periods from this work; given here for reference purposes. (j) This is essentially the tightness of the given system.
\end{table*}

\begin{table*}[ht]
\centering
\caption{TESS observation sectors for the triples}
\small
\begin{tabular}{lcc}
\hline
\hline
Object & Sectors observed & Third body events  \\
\hline
TIC 77177148 & S4--5,31--32,98 & 5,31,32,98 \\ 
TIC 220456364 & 2--7,9--10,12--13,27,29--33,35--37,39,63--69,87,90,93--98 & 4,12,33,36,65,68/69,87,90,97\\
TIC 232874003 & 12,39,66,93,103--104 & 12,39,104 \\
TIC 306690194\tablefootmark{a} & 43--45, 71--72, 1751 & 44,71 \\
TIC 343763382\tablefootmark{a} & 12--13,39,66,93,100--103 & 39,66,100,103 \\
TIC 358795247\tablefootmark{a} & 26,40,53--54,74,80--81 & 40,54,74,81 \\
TIC 379084660 & 18,24--25,52,58--59,78--79,85--86 & 18,52,58,78,85 \\
TIC 385933895 & 9--10,35,37,62--63,89--90 & 9,10,35,62,63,89,90 \\
\hline
\label{tbl:sectors}  
\end{tabular}
\tablefoot{(a) Ground-based RV data were obtained.}
\end{table*}


\FloatBarrier
\onecolumn


\section{Tabulated results of the photodynamical analyses}
\label{app:resulttables} 


In this appendix we tabulate the results of the joint photodynamical analyses.

\begin{table*}[ht]
\centering
\caption{Definitions of triple system parameters in Tables~\ref{tab:syntheticfit_TIC077177148220456364} -- \ref{tab:syntheticfit_TIC379084660385933895}}
\label{tbl:definitions}
\small
\begin{tabular}{lc}
\hline
\hline
Parameter$^a$ & Definition   \\
\hline
$t_0$ & Epoch time for osculating elements    \\
$P$ & Orbital period  \\ 
$a$ & Orbital semimajor axis  \\
$e$ & Orbital eccentricity \\
$\omega$ & Argument of periastron (of secondary) \\
$i$ & Orbital inclination angle \\
$\mathcal{T}_0^\mathrm{inf/sup}$ & Time of conjunction of the secondary$^b$ \\
$\tau$ & Time of periastron passage  \\
$\Omega$ & Longitude of the node relative to \\
& the node of the inner orbit \\
$i_{\rm mut}$ & Mutual inclination angle$^c$   \\
$q$ & Mass ratio (secondary/primary)  \\ 
$K_\mathrm{pri}$ & ``K'' velocity amplitude of primary \\
$K_\mathrm{sec}$ & ``K'' velocity amplitude of secondary \\
$R/a$ & Stellar radius divided by semimajor axis \\
$T_{\rm eff}/T_{\rm eff,Aa}$ & Temperature relative to EB primary \\
fractional flux  & Stellar contribution in the given band \\
$M$ & Stellar mass  \\
$R$ & Stellar radius   \\
$T_\mathrm{eff}$ & Stellar effective temperature  \\ 
$L_\mathrm{bol}$ & Stellar bolometric luminosity  \\
$M_\mathrm{bol}$ & Stellar absolute bolometric magnitude \\
$M_V$ & Stellar absolute visual magnitude \\
$\log g$ & log surface gravity (cgs units) \\
$[M/H]$ & log metallicity abundance to H, by mass \\
$E(B-V)$ & Colour excess in B-V bands  \\
extra light, $\ell_4$  & Contaminating flux in the given band   \\
$(M_V)_\mathrm{tot}$ & System absolute visual magnitude   \\ 
distance & Distance to the source  \\
\hline   
\end{tabular}

\tablefoot{(a) The units for the parameters are given in Tables~\ref{tab:syntheticfit_TIC077177148220456364}-- \ref{tab:syntheticfit_TIC379084660385933895}. (b) The superscript of ``inf/sup'' indicates inferior vs.~superior conjunctions. (By default we give inferior conjunctions. Superior conjunctions are indicated by asteriks.) (c) More explicitly, this is the angle between the orbital planes of the inner binary and the outer triple orbit.}

\end{table*} 

\begin{table*}
 \centering
\caption{Orbital and astrophysical parameters of TICs\,77177148 and 220456364 from the joint photodynamical light curve, ETV, SED and \texttt{PARSEC} isochrone solution. }
 \label{tab:syntheticfit_TIC077177148220456364}
\scalebox{0.91}{\begin{tabular}{@{}lllllll}
\hline
\hline
 & \multicolumn{3}{c}{TIC\,77177148} &  \multicolumn{3}{c}{TIC\,220456364}\\
\hline
\multicolumn{7}{c}{orbital elements} \\
\hline
   & \multicolumn{6}{c}{subsystem}  \\
   & \multicolumn{2}{c}{Aa--Ab} & A--B & \multicolumn{2}{c}{Aa--Ab} & A--B  \\
  \hline
  $t_0$ [BJD - 2400000]& \multicolumn{3}{c}{$58410.0$} & \multicolumn{3}{c}{$58350.0$} \\
  $P_\mathrm{ETV}^\mathrm{avg}$ [days] & \multicolumn{2}{c}{$3.303781496\pm0.000000006$} & $101.395\pm0.009$ & \multicolumn{2}{c}{$12.7187998\pm0.0000002$} & $293.46\pm0.06$ \\
  $P$ [days] & \multicolumn{2}{c}{$3.300754_{-0.000032}^{+0.000027}$} & $101.4875_{-0.0012}^{+0.0011}$ &  \multicolumn{2}{c}{$12.71266_{-0.00022}^{+0.00020}$} & $294.5554_{-0.0074}^{+0.0086}$ \\
  $a$ [R$_\odot$] & \multicolumn{2}{c}{$9.19_{-0.14}^{+0.20}$} & $111.7_{-1.3}^{+2.5}$ & \multicolumn{2}{c}{$22.55_{-0.29}^{+0.19}$} & $222.4_{-3.0}^{+2.0}$ \\
  $e$ & \multicolumn{2}{c}{$0.00287_{-0.00031}^{+0.00031}$} & $0.2441_{-0.0012}^{+0.0014}$ & \multicolumn{2}{c}{$0.00990_{-0.00017}^{+0.00017}$} & $0.33319_{-0.00037}^{+0.00042}$ \\
  $\omega$ [deg]& \multicolumn{2}{c}{$329.2_{-7.0}^{+7.5}$} & $131.32_{-0.43}^{+0.37}$ & \multicolumn{2}{c}{$122.84_{-1.10}^{+0.92}$} & $338.53_{-0.25}^{+0.21}$ \\ 
  $i$ [deg] & \multicolumn{2}{c}{$90.98_{-0.28}^{+0.24}$} & $89.60_{-0.02}^{+0.02}$ & \multicolumn{2}{c}{$89.45_{-0.07}^{+0.07}$} & $89.68_{-0.01}^{+0.01}$ \\
  $\mathcal{T}_0^\mathrm{inf/sup}$ [BJD - 2400000]& \multicolumn{2}{c}{$58414.0008_{-0.0003}^{+0.0003}$} & $58448.0065_{-0.0038}^{+0.0036}*$ & \multicolumn{2}{c}{$58356.3607_{-0.0005}^{+0.0005}$} & $58425.7402_{-0.0190}^{+0.0201}$\tablefootmark{*} \\
  $\tau$ [BJD - 2400000]& \multicolumn{2}{c}{$58411.241_{-0.064}^{+0.069}$} & $58353.621_{-0.102}^{+0.086}$ & \multicolumn{2}{c}{$58351.188_{-0.039}^{+0.033}$} & $58365.96_{-0.17}^{+0.15}$ \\
  $\Omega$ [deg] & \multicolumn{2}{c}{$0.0$} & $-1.95_{-0.46}^{+0.78}$ & \multicolumn{2}{c}{$0.0$} & $1.69_{-0.10}^{+0.10}$ \\
  $i_\mathrm{mut}$ [deg] & \multicolumn{3}{c}{$2.21_{-0.43}^{+0.43}$} & \multicolumn{3}{c}{$1.70_{-0.09}^{+0.10}$} \\
  $\varpi^\mathrm{dyn}$ [deg]& \multicolumn{2}{c}{$149.1_{-7.0}^{+7.5}$} & $311.32_{-0.43}^{+0.37}$ & \multicolumn{2}{c}{$302.83_{-1.10}^{+0.92}$} & $158.54_{-0.25}^{+0.21}$ \\
  $i^\mathrm{dyn}$ [deg] & \multicolumn{2}{c}{$1.98_{-0.38}^{+0.39}$} & $0.22_{-0.05}^{+0.05}$ & \multicolumn{2}{c}{$1.51_{-0.08}^{+0.09}$} & $0.19_{-0.01}^{+0.01}$ \\
  $\Omega^\mathrm{dyn}$ [deg] & \multicolumn{2}{c}{$241.8_{-17.8}^{+6.9}$} & $61.9_{-17.8}^{+6.9}$ & \multicolumn{2}{c}{$82.1_{-2.6}^{+2.8}$} & $262.1_{-2.6}^{+2.8}$ \\
  $i_\mathrm{inv}$ [deg] & \multicolumn{3}{c}{$89.99_{-0.03}^{+0.04}$} & \multicolumn{3}{c}{$89.66_{-0.01}^{+0.01}$} \\
  $\Omega_\mathrm{inv}$ [deg] & \multicolumn{3}{c}{$-1.75_{-0.41}^{+0.70}$} & \multicolumn{3}{c}{$1.50_{-0.08}^{+0.09}$} \\
  \hline
  mass ratio $[q=M_\mathrm{sec}/M_\mathrm{pri}]$ & \multicolumn{2}{c}{$0.941_{-0.014}^{+0.014}$} & $0.898_{-0.014}^{+0.018}$ & \multicolumn{2}{c}{$0.495_{-0.004}^{+0.004}$} & $0.786_{-0.010}^{+0.012}$ \\
  $K_\mathrm{pri}$ [km\,s$^{-1}$] & \multicolumn{2}{c}{$68.5_{-1.1}^{+1.1}$} & $27.20_{-0.27}^{+0.54}$ & \multicolumn{2}{c}{$29.74_{-0.44}^{+0.33}$} & $17.84_{-0.33}^{+0.27}$ \\ 
  $K_\mathrm{sec}$ [km\,s$^{-1}$] & \multicolumn{2}{c}{$72.6_{-1.4}^{+2.0}$} & $30.34_{-0.65}^{+0.66}$ & \multicolumn{2}{c}{$60.02_{-0.75}^{+0.53}$} & $22.66_{-0.24}^{+0.22}$ \\ 
  \hline
  \multicolumn{7}{c}{Apsidal and nodal motion related parameters} \\
  \hline
$P_\mathrm{apse}$ [year] & \multicolumn{2}{c}{$22.03_{-0.26}^{+0.21}$} & $194.2_{-1.0}^{+1.1}$ & \multicolumn{2}{c}{$47.49_{-0.41}^{+0.34}$} & $373.4_{-1.7}^{+1.8}$ \\ 
$P_\mathrm{apse}^\mathrm{dyn}$ [year] & \multicolumn{2}{c}{$10.41_{-0.11}^{+0.08}$} & $17.92_{-0.14}^{+0.10}$ & \multicolumn{2}{c}{$22.32_{-0.17}^{+0.15}$} & $37.84_{-0.22}^{+0.19}$ \\ 
$P_\mathrm{node}^\mathrm{dyn}$ [year] & \multicolumn{3}{c}{$19.74_{-0.14}^{+0.18}$} & \multicolumn{3}{c}{$42.11_{-0.25}^{+0.30}$} \\
$\Delta\omega_\mathrm{3b}$ [arcsec/cycle] & \multicolumn{2}{c}{$1123.3_{-9.1}^{+12.0}$} & $20098_{-116}^{+155}$ & \multicolumn{2}{c}{$2020_{-13}^{+16}$} & $27618_{-141}^{+164}$ \\ 
$\Delta\omega_\mathrm{GR}$ [arcsec/cycle] & \multicolumn{2}{c}{$0.860_{-0.025}^{+0.038}$} & $0.143_{-0.003}^{+0.006}$ & \multicolumn{2}{c}{$0.349_{-0.009}^{+0.006}$} & $0.0711_{-0.0019}^{+0.0013}$ \\ 
$\Delta\omega_\mathrm{tide}$ [arcsec/cycle] & \multicolumn{2}{c}{$0.577_{-0.086}^{+0.165}$} & $0.00056_{-0.00009}^{+0.00016}$ & \multicolumn{2}{c}{$0.0084_{-0.0008}^{+0.0007}$} & $0.000043_{-0.000005}^{+0.000006}$  \\ 
  \hline  
\multicolumn{7}{c}{stellar parameters} \\
\hline
   & Aa & Ab &  B & Aa & Ab &  B \\
  \hline
 \multicolumn{7}{c}{Relative quantities} \\
  \hline
 fractional radius [$R/a$]  & $0.0521_{-0.0019}^{+0.0029}$ & $0.0489_{-0.0015}^{+0.0022}$ & $0.0070_{-0.0001}^{+0.0002}$ & $0.0276_{-0.0006}^{+0.0004}$ & $0.0140_{-0.0004}^{+0.0002}$ & $0.0033_{-0.0001}^{+0.0001}$ \\
 temperature relative to $(T_\mathrm{eff})_\mathrm{Aa}$ & $1$ & $0.9782_{-0.0077}^{+0.0059}$ & $1.4981_{-0.0133}^{+0.0122}$ & $1$ & $0.7509_{-0.0138}^{+0.0164}$ & $1.1589_{-0.0133}^{+0.0141}$ \\
 fractional flux [in \textit{TESS}-band] & $0.0361_{-0.0016}^{+0.0017}$ & $0.0295_{-0.0017}^{+0.0014}$ & $0.4718_{-0.0188}^{+0.0221}$ & $0.2365_{-0.0105}^{+0.0114}$ & $0.0179_{-0.0016}^{+0.0023}$ & $0.6077_{-0.0379}^{+0.0522}$ \\
 \hline
 \multicolumn{7}{c}{Physical quantities} \\
  \hline 
 $M$ [M$_\odot$] & $0.493_{-0.023}^{+0.035}$ & $0.464_{-0.020}^{+0.028}$ & $0.857_{-0.025}^{+0.057}$ & $0.635_{-0.024}^{+0.017}$ & $0.315_{-0.013}^{+0.008}$ & $0.748_{-0.034}^{+0.024}$ \\
 $R$ [R$_\odot$] & $0.479_{-0.024}^{+0.038}$ & $0.449_{-0.020}^{+0.031}$ & $0.784_{-0.024}^{+0.045}$ & $0.622_{-0.021}^{+0.015}$ & $0.315_{-0.012}^{+0.008}$ & $0.723_{-0.032}^{+0.029}$ \\
 $T_\mathrm{eff}$ [K]& $3628_{-40}^{+74}$ & $3549_{-40}^{+65}$ & $5444_{-75}^{+93}$ & $4503_{-78}^{+87}$ & $3386_{-45}^{+58}$ & $5221_{-87}^{+94}$ \\
 $L_\mathrm{bol}$ [L$_\odot$] & $0.037_{-0.006}^{+0.004}$ & $0.030_{-0.004}^{+0.003}$ & $0.505_{-0.073}^{+0.045}$ & $0.142_{-0.016}^{+0.017}$ & $0.0116_{-0.0005}^{+0.0006}$ & $0.350_{-0.050}^{+0.056}$ \\
 $M_\mathrm{bol}$ & $8.33_{-0.12}^{+0.20}$ & $8.57_{-0.09}^{+0.16}$ & $5.51_{-0.09}^{+0.17}$ & $6.89_{-0.12}^{+0.13}$ & $9.61_{-0.05}^{+0.05}$ & $5.91_{-0.16}^{+0.17}$ \\
 $M_V           $ & $9.80_{-0.21}^{+0.21}$ & $10.18_{-0.21}^{+0.14}$ & $5.64_{-0.11}^{+0.19}$ & $7.48_{-0.18}^{+0.19}$ & $11.37_{-0.15}^{+0.14}$ & $6.11_{-0.19}^{+0.20}$ \\
 $\log g$ [dex] & $4.769_{-0.036}^{+0.023}$ & $4.798_{-0.032}^{+0.020}$ & $4.582_{-0.020}^{+0.014}$ & $4.653_{-0.010}^{+0.013}$ & $4.937_{-0.011}^{+0.017}$ & $4.592_{-0.023}^{+0.021}$ \\
 \hline
\multicolumn{7}{c}{Global system parameters} \\
  \hline
$\log$(age) [dex] &\multicolumn{3}{c}{$9.335_{-0.070}^{+0.120}$} & \multicolumn{3}{c}{$9.928_{-0.085}^{+0.045}$} \\
$[M/H]$  [dex]    &\multicolumn{3}{c}{$-0.162_{-0.065}^{+0.185}$} & \multicolumn{3}{c}{$-0.299_{-0.112}^{+0.081}$} \\
$E(B-V)$ [mag]    &\multicolumn{3}{c}{$0.039_{-0.025}^{+0.033}$} & \multicolumn{3}{c}{$0.077_{-0.032}^{+0.036}$} \\
extra light $\ell_4$ [in \textit{TESS}-band] & \multicolumn{3}{c}{$0.463_{-0.022}^{+0.019}$} & \multicolumn{3}{c}{$0.132_{-0.053}^{+0.049}$} \\
$(M_V)_\mathrm{tot}$  &\multicolumn{3}{c}{$5.60_{-0.11}^{+0.19}$} & \multicolumn{3}{c}{$5.83_{-0.18}^{+0.19}$} \\
distance [pc]           & \multicolumn{3}{c}{$308_{-14}^{+18}$} & \multicolumn{3}{c}{$456_{-22}^{+18}$} \\  
\hline
\end{tabular}}
\end{table*}

\begin{table*}
 \centering
\caption{Orbital and astrophysical parameters of TICs\,232874003 and 306690194 from the joint photodynamical light curve, ETV, SED and \texttt{PARSEC} isochrone solution. }
 \label{tab:syntheticfit_TIC232874003306690194}
\scalebox{0.91}{\begin{tabular}{@{}lllllll}
\hline
\hline
 & \multicolumn{3}{c}{TIC\,232874003} &  \multicolumn{3}{c}{TIC\,306690194}\\
\hline
\multicolumn{7}{c}{orbital elements} \\
\hline
   & \multicolumn{6}{c}{subsystem}  \\
   & \multicolumn{2}{c}{Aa--Ab} & A--B & \multicolumn{2}{c}{Aa--Ab} & A--B \\
  \hline
  $t_0$ [BJD - 2400000]& \multicolumn{3}{c}{$58624.0$} &  \multicolumn{3}{c}{$59474.0$} \\
  $P_\mathrm{ETV}^\mathrm{avg}$ [days] & \multicolumn{2}{c}{$5.09228805\pm0.00000002$} & $111.54\pm0.04$ & \multicolumn{2}{c}{$6.212704893\pm0.000000009$} & $108.395\pm0.005$ \\
  $P$ [days] & \multicolumn{2}{c}{$5.08921_{-0.00023}^{+0.00015}$} & $112.0055_{-0.0074}^{+0.0049}$ & \multicolumn{2}{c}{$6.211728_{-0.000068}^{+0.000077}$} & $108.7885_{-0.0010}^{+0.0011}$ \\
  $a$ [R$_\odot$] & \multicolumn{2}{c}{$20.70_{-0.72}^{+0.47}$} & $179.9_{-5.5}^{+4.7}$ & \multicolumn{2}{c}{$23.38_{-0.19}^{+0.11}$} & $174.8_{-1.4}^{+0.8}$ \\
  $e$ & \multicolumn{2}{c}{$0.001259_{-0.000062}^{+0.000085}$} & $0.0485_{-0.0026}^{+0.0026}$ & \multicolumn{2}{c}{$0.002226_{-0.000093}^{+0.000095}$} & $0.2437_{-0.0010}^{+0.0011}$ \\
  $\omega$ [deg]& \multicolumn{2}{c}{$95.7_{-2.4}^{+2.6}$} & $237.6_{-1.5}^{+1.4}$ & \multicolumn{2}{c}{$232.4_{-1.4}^{+1.3}$} & $275.63_{-0.12}^{+0.16}$ \\ 
  $i$ [deg] & \multicolumn{2}{c}{$88.12_{-0.12}^{+0.15}$} & $89.29_{-0.04}^{+0.02}$ & \multicolumn{2}{c}{$89.48_{-0.06}^{+0.06}$} & $89.53_{-0.01}^{+0.01}$ \\
  $\mathcal{T}_0^\mathrm{inf/sup}$ [BJD - 2400000] & \multicolumn{2}{c}{$58627.6856_{-0.0001}^{+0.0001}$} & $58637.3652_{-0.0106}^{+0.0104}$ & \multicolumn{2}{c}{$59475.5427_{-0.0001}^{+0.0001}$} & $59503.1643_{-0.0004}^{+0.0007}$\tablefootmark{*} \\
  $\tau$ [BJD - 2400000] & \multicolumn{2}{c}{$58625.223_{-0.034}^{+0.036}$} & $58628.19_{-0.47}^{+0.44}$ & \multicolumn{2}{c}{$59474.896_{-0.024}^{+0.022}$} & $59451.482_{-0.053}^{+0.077}$ \\
  $\Omega$ [deg] & \multicolumn{2}{c}{$0.0$} & $1.99_{-0.10}^{+0.09}$ & \multicolumn{2}{c}{$0.0$} & $0.02_{-0.09}^{+0.10}$ \\
  $i_\mathrm{mut}$ [deg] & \multicolumn{3}{c}{$2.31_{-0.17}^{ +0.13}$}  & \multicolumn{3}{c}{$0.10_{-0.05}^{+0.07}$} \\
  $\varpi^\mathrm{dyn}$ [deg] & \multicolumn{2}{c}{$275.7_{-2.4}^{+2.6}$} & $57.6_{-1.5}^{+1.4}$ & \multicolumn{2}{c}{$52.4_{-1.4}^{+1.3}$} & $95.63_{-0.12}^{+0.16}$ \\
  $i^\mathrm{dyn}$ [deg] & \multicolumn{2}{c}{$1.82_{-0.12}^{+0.10}$} & $0.49_{-0.06}^{+0.04}$ & \multicolumn{2}{c}{$0.08_{-0.04}^{+0.05}$} & $0.02_{-0.01}^{+0.02}$ \\
  $\Omega^\mathrm{dyn}$ [deg] & \multicolumn{2}{c}{$59.5_{-2.0}^{+3.0}$} & $239.5_{-2.0}^{+3.0}$ & \multicolumn{2}{c}{$4_{-92}^{+61}$} & $184_{-92}^{+61}$ \\
  $i_\mathrm{inv}$ [deg] & \multicolumn{3}{c}{$89.04_{-0.03}^{+0.02}$} & \multicolumn{3}{c}{$89.52_{-0.02}^{+0.02}$} \\
  $\Omega_\mathrm{inv}$ [deg] & \multicolumn{3}{c}{$1.56_{-0.07}^{+0.06}$} & \multicolumn{3}{c}{$0.01_{-0.07}^{+0.07}$} \\
  \hline
  mass ratio $[q=M_\mathrm{sec}/M_\mathrm{pri}]$ & \multicolumn{2}{c}{$0.934_{-0.007}^{+0.007}$} & $0.364_{-0.016}^{+0.026}$ & \multicolumn{2}{c}{$0.975_{-0.004}^{+0.004}$} & $0.362_{-0.003}^{+0.003}$ \\
  $K_\mathrm{pri}$ [km\,s$^{-1}$] & \multicolumn{2}{c}{$99.2_{-3.4}^{+2.5}$} & $21.6_{-0.6}^{+1.2}$ & \multicolumn{2}{c}{$93.98_{-0.67}^{+0.48}$} & $22.27_{-0.18}^{+0.17}$ \\ 
  $K_\mathrm{sec}$ [km\,s$^{-1}$] & \multicolumn{2}{c}{$106.2_{-3.3}^{+2.6}$} & $59.6_{-2.2}^{+1.8}$ & \multicolumn{2}{c}{$96.48_{-0.88}^{+0.50}$} & $61.54_{-0.51}^{+0.33}$ \\ 
  $V_\gamma$ [km\,s$^{-1}$] & \multicolumn{3}{c}{$-$} & & & $6.699_{-0.035}^{+0.034}$ \\ 
  \hline
  \multicolumn{7}{c}{Apsidal and nodal motion related parameters} \\
  \hline
$P_\mathrm{apse}$ [year] & \multicolumn{2}{c}{$32.8_{-1.6}^{+1.1}$} & $123.9_{-1.0}^{+1.6}$ & \multicolumn{2}{c}{$23.76_{-0.13}^{+0.14}$} & $78.56_{-0.10}^{+0.10}$ \\ 
$P_\mathrm{apse}^\mathrm{dyn}$ [year] & \multicolumn{2}{c}{$14.65_{-0.60}^{+0.42}$} & $21.80_{-0.60}^{+0.41}$ & \multicolumn{2}{c}{$10.34_{-0.05}^{+0.05}$} & $14.84_{-0.05}^{+0.05}$ \\ 
$P_\mathrm{node}^\mathrm{dyn}$ [year] & \multicolumn{3}{c}{$26.46_{-0.65}^{+0.94}$} & \multicolumn{3}{c}{$18.30_{-0.08}^{+0.08}$} \\
$\Delta\omega_\mathrm{3b}$ [arcsec/cycle] & \multicolumn{2}{c}{$1219_{-34}^{+53}$} & $18228_{-333}^{+515}$ & \multicolumn{2}{c}{$2128_{-11}^{+10}$} & $26005_{-88}^{+85}$ \\ 
$\Delta\omega_\mathrm{GR}$ [arcsec/cycle] & \multicolumn{2}{c}{$1.83_{-0.13}^{+0.08}$} & $0.287_{-0.017}^{+0.015}$ & \multicolumn{2}{c}{$1.570_{-0.026}^{+0.015}$} & $0.304_{-0.005}^{+0.003}$ \\ 
$\Delta\omega_\mathrm{tide}$ [arcsec/cycle] & \multicolumn{2}{c}{$12.65_{-0.21}^{+0.22}$} & $0.0208_{-0.0003}^{+0.0004}$ & \multicolumn{2}{c}{$2.363_{-0.042}^{+0.045}$} & $0.0058_{-0.0001}^{+0.0001}$ \\ 
  \hline  
\multicolumn{7}{c}{Stellar parameters} \\
\hline
   & Aa & Ab &  B & Aa & Ab & B \\
  \hline
 \multicolumn{7}{c}{Relative quantities} \\
  \hline
 fractional radius [$R/a$]  & $0.1129_{-0.0012}^{+0.0013}$ & $0.1014_{-0.0015}^{+0.0013}$ & $0.0088_{-0.0002}^{+0.0003}$ & $0.0776_{-0.0003}^{+0.0003}$ & $0.0760_{-0.0003}^{+0.0003}$ & $0.00851_{-0.00009}^{+0.00007}$ \\
 temperature relative to $(T_\mathrm{eff})_\mathrm{Aa}$ & $1$ & $0.9713_{-0.0034}^{+0.0039}$ & $0.8105_{-0.0312}^{+0.0455}$ & $1$ & $0.9836_{-0.0024}^{+0.0023}$ & $0.7732_{-0.0063}^{+0.0056}$ \\
 fractional flux [in TESS-band] & $0.4500_{-0.0095}^{+0.0107}$ & $0.3429_{-0.0107}^{+0.0106}$ & $0.1261_{-0.0124}^{+0.0321}$ & $0.4287_{-0.0035}^{+0.0032}$ & $0.3997_{-0.0031}^{+0.0031}$ & $0.1614_{-0.0051}^{+0.0045}$ \\
 \hline
 \multicolumn{7}{c}{Physical quantities} \\
  \hline 
 $M$ [M$_\odot$] & $2.370_{-0.235}^{+0.170}$ & $2.213_{-0.226}^{+0.159}$ & $1.660_{-0.121}^{+0.140}$ & $2.249_{-0.058}^{+0.032}$ & $2.191_{-0.051}^{+0.031}$ & $1.605_{-0.035}^{+0.025}$ \\
 $R$ [R$_\odot$] & $2.328_{-0.048}^{+0.054}$ & $2.091_{-0.087}^{+0.072}$ & $1.574_{-0.052}^{+0.087}$ & $1.813_{-0.016}^{+0.012}$ & $1.776_{-0.013}^{+0.012}$ & $1.486_{-0.015}^{+0.009}$ \\
 $T_\mathrm{eff}$ [K] & $9581_{-529}^{+589}$ & $9307_{-485}^{+539}$ & $7779_{-421}^{+561}$ & $9929_{-303}^{+161}$ & $9761_{-290}^{+157}$ & $7674_{-270}^{+151}$ \\
 $L_\mathrm{bol}$ [L$_\odot$] & $41_{-10}^{+14}$ & $30.0_{-8.4}^{+9.9}$ & $8.2_{-2.1}^{+3.3}$ & $28.7_{-3.7}^{+2.1}$ & $25.7_{-3.2}^{+1.9}$ & $6.86_{-0.88}^{+0.52}$ \\
 $M_\mathrm{bol}$ & $0.74_{-0.31}^{+0.29}$ & $1.08_{-0.31}^{+0.36}$ & $2.48_{-0.37}^{+0.32}$ & $1.12_{-0.08}^{+0.15}$ & $1.24_{-0.08}^{+0.14}$ & $2.68_{-0.08}^{+0.15}$ \\
 $M_V           $ & $0.88_{-0.19}^{+0.20}$ & $1.17_{-0.20}^{+0.29}$ & $2.43_{-0.34}^{+0.32}$ & $1.34_{-0.04}^{+0.09}$ & $1.43_{-0.04}^{+0.08}$ & $2.63_{-0.08}^{+0.15}$ \\
 $\log g$ [dex] & $4.073_{-0.021}^{+0.017}$ & $4.138_{-0.008}^{+0.008}$ & $4.258_{-0.012}^{+0.010}$ & $4.271_{-0.005}^{+0.004}$ & $4.278_{-0.006}^{+0.004}$ & $4.299_{-0.012}^{+0.009}$ \\
 \hline
\multicolumn{7}{c}{Global system parameters} \\
  \hline
$\log$(age) [dex] &\multicolumn{3}{c}{$8.597_{-0.099}^{+0.159}$} &\multicolumn{3}{c}{$8.205_{-0.033}^{+0.061}$} \\
$[M/H]$  [dex]    &\multicolumn{3}{c}{$0.070_{-0.084}^{+0.059}$} &\multicolumn{3}{c}{$0.064_{-0.043}^{+0.035}$} \\
$E(B-V)$ [mag]    &\multicolumn{3}{c}{$0.532_{-0.044}^{+0.038}$} &\multicolumn{3}{c}{$0.215_{-0.026}^{+0.015}$} \\
extra light $\ell_4$ [in TESS-band] &\multicolumn{3}{c}{$0.077_{-0.027}^{+0.015}$} & \multicolumn{3}{c}{$0.010_{-0.006}^{+0.009}$} \\
$(M_V)_\mathrm{tot}$  &\multicolumn{3}{c}{$0.12_{-0.21}^{+0.24}$} &\multicolumn{3}{c}{$0.47_{-0.05}^{+0.10}$} \\
distance [pc]       &\multicolumn{3}{c}{$1184_{-66}^{+66}$} & \multicolumn{3}{c}{$990_{-15}^{+11}$} \\  
\hline
\end{tabular}}
\end{table*}

\begin{table*}
 \centering
\caption{Orbital and astrophysical parameters of TICs\,343763382 and 358795247 from the joint photodynamical light curve, ETV, SED and \texttt{PARSEC} isochrone solution. }
 \label{tab:syntheticfit_TIC343763382358795247}
\scalebox{0.81}{\begin{tabular}{@{}llllllll}
\hline
 & \multicolumn{3}{c}{TIC\,343763382} &  \multicolumn{4}{c}{TIC\,358795247}\\
\hline
\multicolumn{8}{c}{orbital elements} \\
\hline
   & \multicolumn{7}{c}{subsystem}  \\
   & \multicolumn{2}{c}{Aa--Ab} & A--B & \multicolumn{2}{c}{Aa--Ab} & A--B & AB--C \\
  \hline
  $t_0$ [BJD - 2400000]& \multicolumn{3}{c}{$58624.0$} & \multicolumn{4}{c}{$59010.0$} \\
  $P_\mathrm{ETV}^\mathrm{avg}$ [days] & \multicolumn{2}{c}{$5.865596454\pm0.000000001$} & $149.16\pm0.02$ & \multicolumn{2}{c}{$1.221463308\pm0.000000003$} & $133.23\pm0.07$ & $8709\pm106$ \\
  $P$ [days] & \multicolumn{2}{c}{$5.8649203_{-0.0000080}^{+0.0000083}$} & $149.6053_{-0.0014}^{+0.0016}$ &  \multicolumn{2}{c}{$1.2213424_{-0.0000026}^{+0.0000025}$} & $133.277_{-0.014}^{+0.028}$ & $8931_{-2244}^{+975}$ \\
  $a$ [R$_\odot$] & \multicolumn{2}{c}{$18.34_{-0.10}^{+0.11}$} & $183.0_{-1.1}^{+0.9}$ & \multicolumn{2}{c}{$5.830_{-0.030}^{+0.031}$} & $162.6_{-1.2}^{+1.0}$ & $2912_{-526}^{+207}$ \\
  $e$ & \multicolumn{2}{c}{$0.0016401_{-0.0000085}^{+0.0000088}$} & $0.05747_{-0.00029}^{+0.00032}$ & \multicolumn{2}{c}{$0.00036_{-0.00015}^{+0.00025}$} & $0.2262_{-0.0112}^{+0.0045}$ & $0.282_{-0.045}^{+0.062}$ \\
  $\omega$ [deg]& \multicolumn{2}{c}{$312.56_{-0.47}^{+0.44}$} & $309.24_{-0.40}^{+0.38}$ & \multicolumn{2}{c}{$210_{-97}^{+26}$} & $251.04_{-0.58}^{+0.35}$ & $347.9_{-7.1}^{+7.8}$ \\ 
  $i$ [deg] & \multicolumn{2}{c}{$89.17_{-0.17}^{+0.16}$} & $89.71_{-0.01}^{+0.01}$ & \multicolumn{2}{c}{$89.02_{-0.47}^{+1.55}$} & $89.36_{-0.07}^{+0.15}$ & $79_{-16}^{+11}$ \\
  $\mathcal{T}_0^\mathrm{inf/sup}$ [BJD - 2400000]& \multicolumn{2}{c}{$58630.2007_{-0.0004}^{+0.0004}$} & $59371.3081_{-0.0077}^{+0.0082}$ & \multicolumn{2}{c}{$59011.4358_{-0.0071}^{+0.0018}$} & $59396.575_{-0.078}^{+0.039}$ & ... \\
  $\tau$ [BJD - 2400000]& \multicolumn{2}{c}{$58625.0287_{-0.0076}^{+0.0072}$} & $59236.34_{-0.15}^{+0.15}$ & \multicolumn{2}{c}{$59011.23_{-0.33}^{+0.09}$} & $59319.23_{-0.23}^{+0.19}$ & $58605_{-197}^{+459}$ \\
  $\Omega$ [deg] & \multicolumn{2}{c}{$0.0$} & $0.15_{-0.12}^{+0.12}$ & \multicolumn{2}{c}{$0.0$} & $0.47_{-1.36}^{+1.46}$ & $-11_{-17}^{+42}$ \\
  $i_\mathrm{mut}$ [deg] & \multicolumn{3}{c}{$0.56_{-0.18}^{+0.20}$} & \multicolumn{3}{c}{$1.27_{-0.53}^{+1.23}$} & $24_{-11}^{+17}$ \\
  $\varpi^\mathrm{dyn}$ [deg]& \multicolumn{2}{c}{$132.56_{-0.47}^{+0.44}$} & $129.24_{-0.40}^{+0.38}$ & \multicolumn{2}{c}{$28_{-95}^{+26}$} & $90.1_{-1.4}^{+1.9}$ & $71.4_{-4.3}^{+1.0}$ \\
  $i^\mathrm{dyn}$ [deg] & \multicolumn{2}{c}{$0.48_{-0.15}^{+0.17}$} & $0.09_{-0.03}^{+0.03}$ & \multicolumn{2}{c}{$19_{-9}^{+13}$} & $19_{-9}^{+13}$ & $5.2_{-2.7}^{+5.5}$ \\
  $\Omega^\mathrm{dyn}$ [deg] & \multicolumn{2}{c}{$16.1_{-10.8}^{+7.3}$} & $196.1_{-10.8}^{+7.3}$ & \multicolumn{2}{c}{$234_{-102}^{+39}$} & $235_{-105}^{+37}$ & $55_{-105}^{+37}$ \\
  $i_\mathrm{inv}$ [deg] & \multicolumn{3}{c}{$89.62_{-0.02}^{+0.03}$} & \multicolumn{4}{c}{$81_{-12}^{+9}$} \\
  $\Omega_\mathrm{inv}$ [deg] & \multicolumn{3}{c}{$0.13_{-0.10}^{+0.10}$} & \multicolumn{4}{c}{$-9_{-13}^{+30}$} \\
  \hline
  mass ratio $[q=M_\mathrm{sec}/M_\mathrm{pri}]$ & \multicolumn{2}{c}{$0.890_{-0.003}^{+0.003}$} & $0.524_{-0.006}^{+0.005}$ & \multicolumn{2}{c}{$0.947_{-0.009}^{+0.010}$} & $0.818_{-0.012}^{+0.014}$ & $0.292_{-0.075}^{+0.023}$ \\
  $K_\mathrm{pri}$ [km\,s$^{-1}$] & \multicolumn{2}{c}{$74.54_{-0.43}^{+0.48}$} & $21.33_{-0.18}^{+0.19}$ & \multicolumn{2}{c}{$117.48_{-0.82}^{+0.86}$} & $28.54_{-0.44}^{+0.33}$ & $3.82_{-0.78}^{+0.18}$ \\ 
  $K_\mathrm{sec}$ [km\,s$^{-1}$] & \multicolumn{2}{c}{$83.75_{-0.46}^{+0.46}$} & $40.62_{-0.17}^{+0.33}$ & \multicolumn{2}{c}{$124.04_{-0.84}^{+1.01}$} & $34.83_{-0.21}^{+0.19}$ & $13.05_{-0.55}^{+1.10}$ \\ 
  $V_\gamma$ [km\,s$^{-1}$]       & & & $-7.567_{-0.018}^{+0.023}$ & & & & $9.55_{-0.14}^{+0.15}$ \\
  \hline
  \multicolumn{8}{c}{Apsidal and nodal motion related parameters} \\
  \hline
$P_\mathrm{apse}$ [year] & \multicolumn{2}{c}{$40.04_{-0.22}^{+0.27}$} & $216.63_{-0.52}^{+0.41}$ & \multicolumn{2}{c}{$33.9_{-1.3}^{+1.4}$} & $346.9_{-5.3}^{+2.0}$ &  $39984_{-15992}^{+24076}$ \\ 
$P_\mathrm{apse}^\mathrm{dyn}$ [year] & \multicolumn{2}{c}{$18.38_{-0.09}^{+0.11}$} & $29.36_{-0.10}^{+0.13}$ & \multicolumn{2}{c}{$25.40_{-0.75}^{+0.79}$} & $94.32_{-0.80}^{+0.93}$ & $6239_{-1742}^{+1799}$ \\ 
$P_\mathrm{node}^\mathrm{dyn}$ [year] & \multicolumn{3}{c}{$33.96_{-0.18}^{+0.15}$} & \multicolumn{3}{c}{$95.7_{-3.9}^{+9.6}$} & $7215_{-1701}^{+1902}$ \\
$\Delta\omega_\mathrm{3b}$ [arcsec/cycle] & \multicolumn{2}{c}{$1129.2_{-6.8}^{+5.6}$} & $18079_{-78}^{+64}$ & \multicolumn{2}{c}{$82.45_{-0.50}^{+0.44}$} & $5014_{-49}^{+43}$ & $50000_{-1015}^{+955}$ \\ 
$\Delta\omega_\mathrm{GR}$ [arcsec/cycle] & \multicolumn{2}{c}{$1.084_{-0.011}^{+0.013}$} & $0.166_{-0.002}^{+0.002}$ & \multicolumn{2}{c}{$2.525_{-0.026}^{+0.027}$} & $0.173_{-0.002}^{+0.002}$ & $0.0130_{-0.0008}^{+0.0022}$ \\ 
$\Delta\omega_\mathrm{tide}$ [arcsec/cycle] & \multicolumn{2}{c}{$2.192_{-0.052}^{+0.055}$} & $0.00306_{-0.00008}^{+0.00008}$ & \multicolumn{2}{c}{$85.6_{-4.8}^{+5.0}$} & $0.0146_{-0.0008}^{+0.0009}$  & $0.00005_{-0.00001}^{+0.00002}$ \\ 
  \hline  
\multicolumn{8}{c}{stellar parameters} \\
\hline
   & Aa & Ab &  B & Aa & Ab &  B  & C \\
  \hline
 \multicolumn{8}{c}{Relative quantities} \\
  \hline
 fractional radius [$R/a$]  & $0.0715_{-0.0005}^{+0.0005}$ & $0.0591_{-0.0004}^{+0.0004}$ & $0.0071_{-0.0001}^{+0.0001}$ & $0.1410_{-0.0018}^{+0.0019}$ & $0.1331_{-0.0016}^{+0.0015}$ & $0.0103_{-0.0006}^{+0.0005}$ & $0.00029_{-0.00004}^{+0.00002}$ \\
 temperature relative to $(T_\mathrm{eff})_\mathrm{Aa}$ & $1$ & $0.9436_{-0.0018}^{+0.0018}$ & $0.9963_{-0.0048}^{+0.0041}$ & $1$ & $0.9620_{-0.0062}^{+0.0071}$ & $1.2816_{-0.0119}^{+0.0098}$ & $1.0101_{-0.2008}^{+0.0495}$ \\
 fractional flux [in \textit{TESS}-band] & $0.3846_{-0.0091}^{+0.0104}$ & $0.2163_{-0.0026}^{+0.0032}$ & $0.3698_{-0.0183}^{+0.0151}$ & $0.0809_{-0.0027}^{+0.0025}$ & $0.0619_{-0.0022}^{+0.0022}$ & $0.7527_{-0.0508}^{+0.0712}$ & $0.0955_{-0.0738}^{+0.0372}$ \\
 fractional flux [in Sloan $g'$-band] & $...$ & $...$ & $...$ & $0.0537_{-0.0027}^{+0.0026}$ & $0.0377_{-0.0021}^{+0.0023}$ & $0.8201_{-0.0555}^{+0.0671}$ & $0.0641_{-0.0557}^{+0.0386}$ \\
 fractional flux [in Sloan $r'$-band] & $...$ & $...$ & $...$ & $0.0710_{-0.0030}^{+0.0030}$ & $0.0530_{-0.0023}^{+0.0024}$ & $0.7664_{-0.0550}^{+0.0728}$ & $0.0850_{-0.0688}^{+0.0357}$ \\
  \hline
 \multicolumn{8}{c}{Physical quantities} \\
  \hline 
 $M$ [M$_\odot$] & $1.272_{-0.020}^{+0.022}$ & $1.132_{-0.018}^{+0.021}$ & $1.264_{-0.025}^{+0.021}$ & $0.914_{-0.014}^{+0.016}$ & $0.866_{-0.014}^{+0.014}$ & $1.461_{-0.043}^{+0.035}$ & $0.946_{-0.236}^{+0.063}$ \\
 $R$ [R$_\odot$] & $1.313_{-0.008}^{+0.009}$ & $1.084_{-0.012}^{+0.013}$ & $1.294_{-0.023}^{+0.021}$ & $0.822_{-0.014}^{+0.015}$ & $0.776_{-0.013}^{+0.012}$ & $1.676_{-0.113}^{+0.093}$ & $0.852_{-0.193}^{+0.071}$ \\
 $T_\mathrm{eff}$ [K]& $6292_{-34}^{+42}$ & $5937_{-30}^{+32}$ & $6264_{-38}^{+52}$ & $5408_{-51}^{+53}$ & $5205_{-52}^{+49}$ & $6930_{-66}^{+64}$ & $5528_{1093}^{+246}$ \\
 $L_\mathrm{bol}$ [L$_\odot$] & $2.427_{-0.067}^{+0.068}$ & $1.312_{-0.052}^{+0.054}$ & $2.321_{-0.129}^{+0.132}$ & $0.519_{-0.036}^{+0.039}$ & $0.397_{-0.028}^{+0.028}$ & $5.83_{-0.84}^{+0.71}$ & $0.61_{-0.46}^{+0.24}$\\
 $M_\mathrm{bol}$ & $3.81_{-0.03}^{+0.03}$ & $4.47_{-0.04}^{+0.04}$ & $3.86_{-0.06}^{+0.06}$ & $5.48_{-0.08}^{+0.08}$ & $5.77_{-0.07}^{+0.08}$ & $2.86_{-0.12}^{+0.17}$ & $5.30_{-0.36}^{+1.52}$ \\
 $M_V           $ & $3.79_{-0.03}^{+0.03}$ & $4.49_{-0.05}^{+0.05}$ & $3.84_{-0.06}^{+0.07}$ & $5.62_{-0.09}^{+0.09}$ & $5.98_{-0.09}^{+0.10}$ & $2.82_{-0.13}^{+0.17}$ & $5.41_{-0.40}^{+2.09}$ \\
 $\log g$ [dex] & $4.306_{-0.008}^{+0.007}$ & $4.421_{-0.004}^{+0.004}$ & $4.315_{-0.012}^{+0.011}$ & $4.568_{-0.009}^{+0.009}$ & $4.594_{-0.007}^{+0.008}$ & $4.152_{-0.038}^{+0.049}$ & $4.552_{-0.041}^{+0.098}$ \\
 \hline
\multicolumn{8}{c}{Global system parameters} \\
  \hline
$\log$(age) [dex] &\multicolumn{3}{c}{$9.180_{-0.083}^{+0.100}$} & \multicolumn{4}{c}{$9.156_{-0.075}^{+0.040}$} \\
$[M/H]$  [dex]    &\multicolumn{3}{c}{$0.245_{-0.040}^{+0.034}$} & \multicolumn{4}{c}{$0.043_{-0.031}^{+0.025}$} \\
$E(B-V)$ [mag]    &\multicolumn{3}{c}{$0.161_{-0.009}^{+0.011}$} & \multicolumn{4}{c}{$0.977_{-0.023}^{+0.029}$} \\
extra light $\ell_4$ [in \textit{TESS}-band] & \multicolumn{3}{c}{$0.027_{-0.018}^{+0.024}$} & \multicolumn{4}{c}{$0.013_{-0.009}^{+0.020}$} \\
extra light $\ell_4$ [in Sloan $g'$-band] & \multicolumn{3}{c}{$...$} & \multicolumn{4}{c}{$0.024_{-0.018}^{+0.026}$} \\
extra light $\ell_4$ [in Sloan $r'$-band] & \multicolumn{3}{c}{$...$} & \multicolumn{4}{c}{$0.025_{-0.015}^{+0.025}$} \\
$(M_V)_\mathrm{tot}$  &\multicolumn{3}{c}{$2.80_{-0.04}^{+0.04}$} & \multicolumn{4}{c}{$2.62_{-0.09}^{+0.11}$} \\
distance [pc]           & \multicolumn{3}{c}{$491.5_{-5.1}^{+5.3}$} & \multicolumn{4}{c}{$644_{-20}^{+20}$} \\  
\hline
\end{tabular}}
\end{table*}

\begin{table*}
 \centering
\caption{Orbital and astrophysical parameters of TICs\,379084660 and 385933895 from the joint photodynamical light curve, ETV, SED and \texttt{PARSEC} isochrone solution. }
 \label{tab:syntheticfit_TIC379084660385933895}
\scalebox{0.91}{\begin{tabular}{@{}lllllll}
\hline
 & \multicolumn{3}{c}{TIC\,379084660} &  \multicolumn{3}{c}{TIC\,385933895}\\
\hline
\multicolumn{7}{c}{orbital elements} \\
\hline
   & \multicolumn{6}{c}{subsystem}  \\
   & \multicolumn{2}{c}{Aa--Ab} & A--B & \multicolumn{2}{c}{Aa--Ab} & A--B  \\
  \hline
  $t_0$ [BJD - 2400000]& \multicolumn{3}{c}{$58790.0$} & \multicolumn{3}{c}{$58542.0$} \\
  $P_\mathrm{ETV}^\mathrm{avg}$ [days] & \multicolumn{2}{c}{$4.75882984\pm0.00000001$} & $88.982\pm0.004$ & \multicolumn{2}{c}{$6.45080413\pm0.00000007$} & $63.330\pm0.002$ \\
  $P$ [days] & \multicolumn{2}{c}{$4.74098_{-0.00034}^{+0.00042}$} & $89.2621_{-0.0084}^{+0.0077}$ &  \multicolumn{2}{c}{$6.33503_{-0.00072}^{+0.00083}$} & $64.2543_{-0.0031}^{+0.0034}$ \\
  $a$ [R$_\odot$] & \multicolumn{2}{c}{$12.68_{-0.12}^{+0.09}$} & $108.24_{-1.19}^{+0.88}$ & \multicolumn{2}{c}{$19.955_{-0.084}^{+0.093}$} & $111.64_{-0.50}^{+0.64}$ \\
  $e$ & \multicolumn{2}{c}{$0.00643_{-0.00054}^{+0.00031}$} & $0.2094_{-0.0050}^{+0.0044}$ & \multicolumn{2}{c}{$0.01614_{-0.00016}^{+0.00017}$} & $0.04466_{-0.00028}^{+0.00028}$ \\
  $\omega$ [deg]& \multicolumn{2}{c}{$139.3_{-2.2}^{+2.6}$} & $116.61_{-0.63}^{+0.85}$ & \multicolumn{2}{c}{$331.97_{-0.40}^{+0.40}$} & $226.77_{-0.24}^{+0.25}$ \\ 
  $i$ [deg] & \multicolumn{2}{c}{$90.08_{-0.23}^{+0.30}$} & $90.22_{-0.19}^{+0.10}$ & \multicolumn{2}{c}{$89.67_{-0.09}^{+0.10}$} & $89.44_{-0.02}^{+0.02}$ \\
  $\mathcal{T}_0^\mathrm{inf/sup}$ [BJD - 2400000]& \multicolumn{2}{c}{$58794.1165_{-0.0005}^{+0.0005}$} & {$58804.9110_{-0.0067}^{+0.0065}$}* & \multicolumn{2}{c}{$58543.6384.3937_{-0.0003}^{+0.0003}$} & $58545.4103_{-0.0042}^{+0.0034}$ \\
  $\tau$ [BJD - 2400000]& \multicolumn{2}{c}{$58792.402_{-0.029}^{+0.034}$} & $58719.92_{-0.15}^{+0.19}$ & \multicolumn{2}{c}{$58538.3665_{-0.0069}^{+0.0069}$} & $58538.305_{-0.044}^{+0.045}$ \\
  $\Omega$ [deg] & \multicolumn{2}{c}{$0.0$} & $-0.13_{-0.50}^{+0.76}$ & \multicolumn{2}{c}{$0.0$} & $-0.01_{-0.09}^{+0.08}$ \\
  $i_\mathrm{mut}$ [deg] & \multicolumn{3}{c}{$0.61_{-0.28}^{+0.32}$} & \multicolumn{3}{c}{$0.24_{-0.09}^{+0.11}$} \\
  $\varpi^\mathrm{dyn}$ [deg]& \multicolumn{2}{c}{$319.3_{-2.2}^{+2.6}$} & $296.61_{-0.63}^{+0.85}$ & \multicolumn{2}{c}{$151.97_{-0.40}^{+0.40}$} & $46.77_{-0.24}^{+0.25}$ \\
  $i^\mathrm{dyn}$ [deg] & \multicolumn{2}{c}{$0.53_{-0.24}^{+0.27}$} & $0.08_{-0.04}^{+0.04}$ & \multicolumn{2}{c}{$0.20_{-0.07}^{+0.09}$} & $0.04_{-0.01}^{+0.02}$ \\
  $\Omega^\mathrm{dyn}$ [deg] & \multicolumn{2}{c}{$322_{-69}^{+130}$} & $142_{-69}^{+130}$ & \multicolumn{2}{c}{$181_{-28}^{+17}$} & $1_{-28}^{+17}$ \\
  $i_\mathrm{inv}$ [deg] & \multicolumn{3}{c}{$90.20_{-0.13}^{+0.07}$} & \multicolumn{3}{c}{$89.48_{-0.02}^{+0.02}$} \\
  $\Omega_\mathrm{inv}$ [deg] & \multicolumn{3}{c}{$-0.12_{-0.43}^{+0.66}$} & \multicolumn{3}{c}{$-0.00_{-0.07}^{+0.07}$} \\
  \hline
  mass ratio $[q=M_\mathrm{sec}/M_\mathrm{pri}]$ & \multicolumn{2}{c}{$0.707_{-0.011}^{+0.016}$} & $0.749_{-0.019}^{+0.016}$ & \multicolumn{2}{c}{$0.908_{-0.006}^{+0.006}$} & $0.703_{-0.009}^{+0.007}$ \\
  $K_\mathrm{pri}$ [km\,s$^{-1}$] & \multicolumn{2}{c}{$56.07_{-0.98}^{+0.99}$} & $26.89_{-0.57}^{+0.44}$ & \multicolumn{2}{c}{$75.90_{-0.44}^{+0.43}$} & $36.33_{-0.36}^{+0.41}$ \\ 
  $K_\mathrm{sec}$ [km\,s$^{-1}$] & \multicolumn{2}{c}{$79.23_{-0.72}^{+0.72}$} & $35.86_{-0.42}^{+0.33}$ & \multicolumn{2}{c}{$83.58_{-0.53}^{+0.47}$} & $51.71_{-0.28}^{+0.25}$ \\ 
  \hline
  \multicolumn{7}{c}{Apsidal and nodal motion related parameters} \\
  \hline
$P_\mathrm{apse}$ [year] & \multicolumn{2}{c}{$13.39_{-0.14}^{+0.18}$} & $89.19_{-0.90}^{0.95}$ & \multicolumn{2}{c}{$5.74_{-0.04}^{+0.04}$} & $29.32_{-0.10}^{+0.09}$ \\ 
$P_\mathrm{apse}^\mathrm{dyn}$ [year] & \multicolumn{2}{c}{$6.23_{-0.06}^{+0.07}$} & $10.30_{-0.06}^{+0.09}$ & \multicolumn{2}{c}{$2.62_{-0.01}^{+0.02}$} & $4.13_{-0.02}^{+0.02}$ \\ 
$P_\mathrm{node}^\mathrm{dyn}$ [year] & \multicolumn{3}{c}{$11.65_{-0.12}^{+0.09}$} & \multicolumn{3}{c}{$4.80_{-0.03}^{+0.02}$} \\
$\Delta\omega_\mathrm{3b}$ [arcsec/cycle] & \multicolumn{2}{c}{$2700_{-32}^{+25}$} & $30746_{-258}^{+191}$ & \multicolumn{2}{c}{$8592_{-55}^{+47}$} & $55245_{-243}^{+210}$ \\ 
$\Delta\omega_\mathrm{GR}$ [arcsec/cycle] & \multicolumn{2}{c}{$0.793_{-0.015}^{+0.012}$} & $0.170_{-0.004}^{+0.003}$ & \multicolumn{2}{c}{$1.100_{-0.009}^{+0.010}$} & $0.335_{-0.003}^{+0.004}$ \\ 
$\Delta\omega_\mathrm{tide}$ [arcsec/cycle] & \multicolumn{2}{c}{$0.419_{-0.028}^{+0.022}$} & $0.0026_{-0.0005}^{+0.0004}$ & \multicolumn{2}{c}{$1.98_{-0.10}^{+0.13}$} & $0.0088_{-0.0005}^{+0.0006}$  \\ 
  \hline  
\multicolumn{7}{c}{stellar parameters} \\
\hline
   & Aa & Ab &  B & Aa & Ab &  B \\
  \hline
 \multicolumn{7}{c}{Relative quantities} \\
  \hline
 fractional radius [$R/a$]  & $0.0542_{-0.0005}^{+0.0004}$ & $0.0395_{-0.0011}^{+0.0011}$ & $0.0088_{-0.0006}^{+0.0004}$ & $0.0681_{-0.0009}^{+0.0009}$ & $0.0607_{-0.0007}^{+0.0008}$ & $0.0145_{-0.0003}^{+0.0004}$ \\
 temperature relative to $(T_\mathrm{eff})_\mathrm{Aa}$ & $1$ & $0.7782_{-0.0092}^{+0.0113}$ & $1.2230_{-0.0191}^{+0.0185}$ & $1$ & $0.9456_{-0.0037}^{+0.0039}$ & $1.2760_{-0.0188}^{+0.0199}$ \\
 fractional flux [in \textit{TESS}-band] & $0.0656_{-0.0039}^{+0.0038}$ & $0.0134_{-0.0008}^{+0.0010}$ & $0.2952_{-0.0458}^{+0.0382}$ & $0.2266_{-0.0046}^{+0.0046}$ & $0.1519_{-0.0032}^{+0.0032}$ & $0.6063_{-0.0095}^{+0.0120}$ \\
 fractional flux [in Sloan $r'$-band] & $0.1523_{-0.0162}^{+0.0249}$ & $0.0244_{-0.0029}^{+0.0047}$ & $0.7969_{-0.0289}^{+0.0217}$ & $...$ & $...$ & $...$ \\
 \hline
 \multicolumn{7}{c}{Physical Quantities} \\
  \hline 
 $M$ [M$_\odot$] & $0.713_{-0.018}^{+0.014}$ & $0.504_{-0.018}^{+0.015}$ & $0.914_{-0.036}^{+0.031}$ & $1.391_{-0.019}^{+0.020}$ & $1.263_{-0.017}^{+0.018}$ & $1.864_{-0.032}^{+0.044}$ \\
 $R$ [R$_\odot$] & $0.688_{-0.011}^{+0.008}$ & $0.501_{-0.019}^{+0.017}$ & $0.957_{-0.074}^{+0.044}$ & $1.358_{-0.022}^{+0.024}$ & $1.210_{-0.019}^{+0.021}$ & $1.612_{-0.041}^{+0.052}$ \\
 $T_\mathrm{eff}$ [K]& $4556_{-154}^{+95}$ & $3539_{-75}^{+64}$ & $5583_{-202}^{+98}$ & $6938_{-59}^{+62}$ & $6562_{-50}^{+54}$ & $8820_{-162}^{+210}$ \\
 $L_\mathrm{bol}$ [L$_\odot$] & $0.184_{-0.025}^{+0.021}$ & $0.035_{-0.005}^{+0.005}$ & $0.80_{-0.21}^{+0.12}$ & $3.85_{-0.20}^{+0.20}$ & $2.44_{-0.13}^{+0.13}$ & $14.3_{-1.0}^{+1.1}$ \\
 $M_\mathrm{bol}$ & $6.62_{-0.12}^{+0.16}$ & $8.41_{-0.14}^{+0.15}$ & $5.01_{-0.15}^{+0.32}$ & $3.31_{-0.06}^{+0.06}$ & $3.80_{-0.06}^{+0.06}$ & $1.88_{-0.08}^{+0.08}$ \\
 $M_V           $ & $7.19_{-0.19}^{+0.30}$ & $10.07_{-0.25}^{+0.30}$ & $5.10_{-0.17}^{+0.37}$ & $3.28_{-0.06}^{+0.06}$ & $3.79_{-0.06}^{+0.06}$ & $1.91_{-0.07}^{+0.06}$ \\
 $\log g$ [dex] & $4.615_{-0.005}^{+0.008}$ & $4.740_{-0.016}^{+0.017}$ & $4.440_{-0.036}^{+0.047}$ & $4.314_{-0.010}^{+0.009}$ & $4.372_{-0.009}^{+0.009}$ & $4.289_{-0.022}^{+0.024}$ \\
 \hline
\multicolumn{7}{c}{Global system parameters} \\
  \hline
$\log$(age) [dex] &\multicolumn{3}{c}{$9.926_{-0.057}^{+0.049}$} & \multicolumn{3}{c}{$8.450_{-0.149}^{+0.084}$} \\
$[M/H]$  [dex]    &\multicolumn{3}{c}{$0.012_{-0.041}^{+0.152}$} & \multicolumn{3}{c}{$-0.017_{-0.050}^{+0.053}$} \\
$E(B-V)$ [mag]    &\multicolumn{3}{c}{$0.264_{-0.079}^{+0.037}$} & \multicolumn{3}{c}{$0.384_{-0.015}^{+0.016}$} \\
extra light $\ell_4$ [in \textit{TESS}-band] & \multicolumn{3}{c}{$0.624_{-0.038}^{+0.048}$} & \multicolumn{3}{c}{$0.014_{-0.009}^{+0.010}$} \\
extra light $\ell_4$ [in Sloan $r'$-band] & \multicolumn{3}{c}{$0.024_{-0.016}^{+0.019}$} & \multicolumn{3}{c}{$...$} \\
$(M_V)_\mathrm{tot}$  &\multicolumn{3}{c}{$4.94_{-0.17}^{+0.34}$} & \multicolumn{3}{c}{$1.50_{-0.06}^{+0.06}$} \\
distance [pc]           & \multicolumn{3}{c}{$740_{-47}^{+27}$} & \multicolumn{3}{c}{$1856_{-35}^{+46}$} \\  
\hline
\end{tabular}}
\end{table*}

\FloatBarrier
\onecolumn


\section{Outer orbit phase-curves from the archival photometric data}
\label{app:phasecurves}

In this appendix we plot the outer-orbit folded lightcurves that were made from ATLAS, ASAS-SN and ZTF archival data, as was described earler in Sect.~\ref{sec:archivaldata}.

\begin{figure*}[h]
\centering
\includegraphics[width=0.45\textwidth]{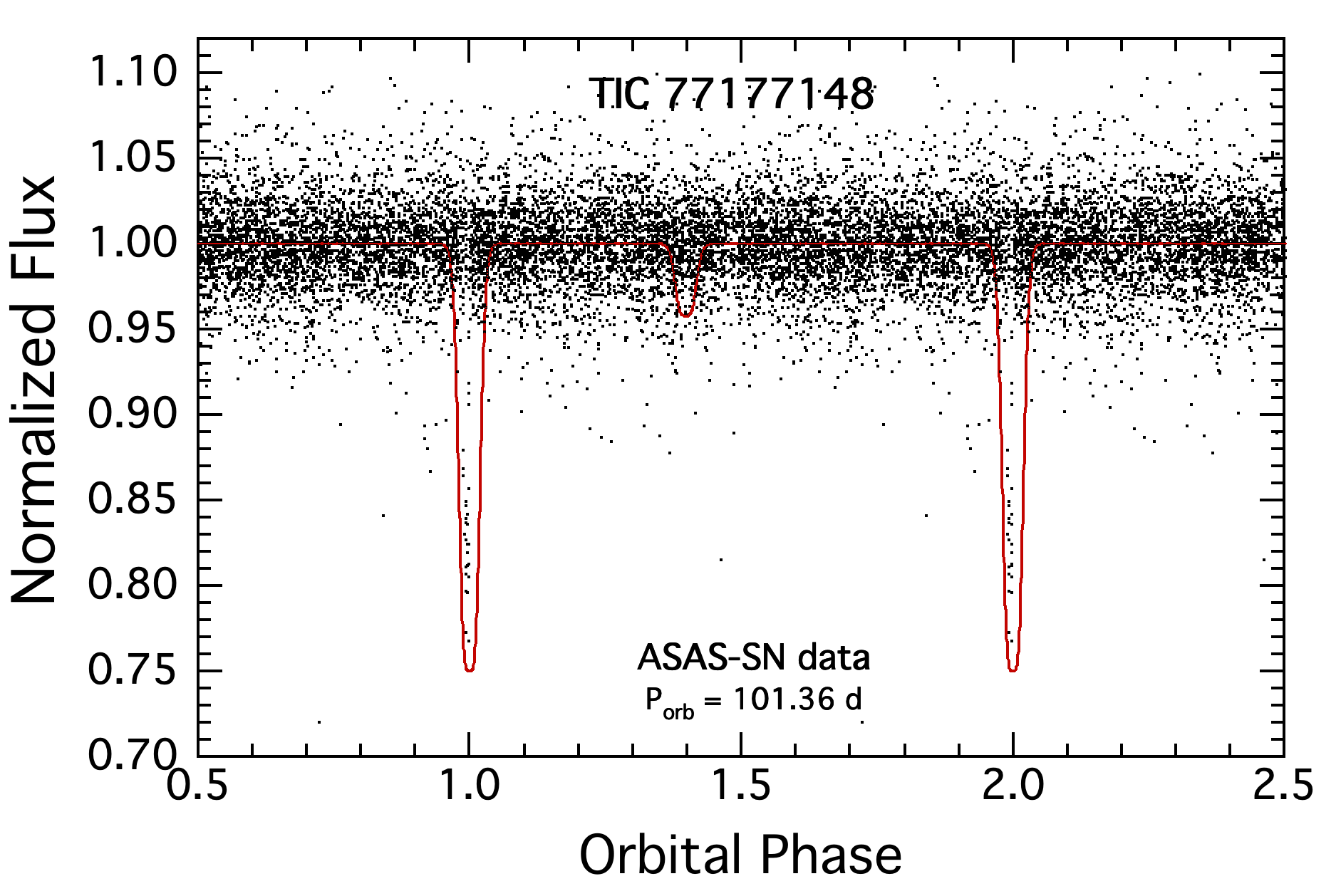}
\includegraphics[width=0.45\textwidth]{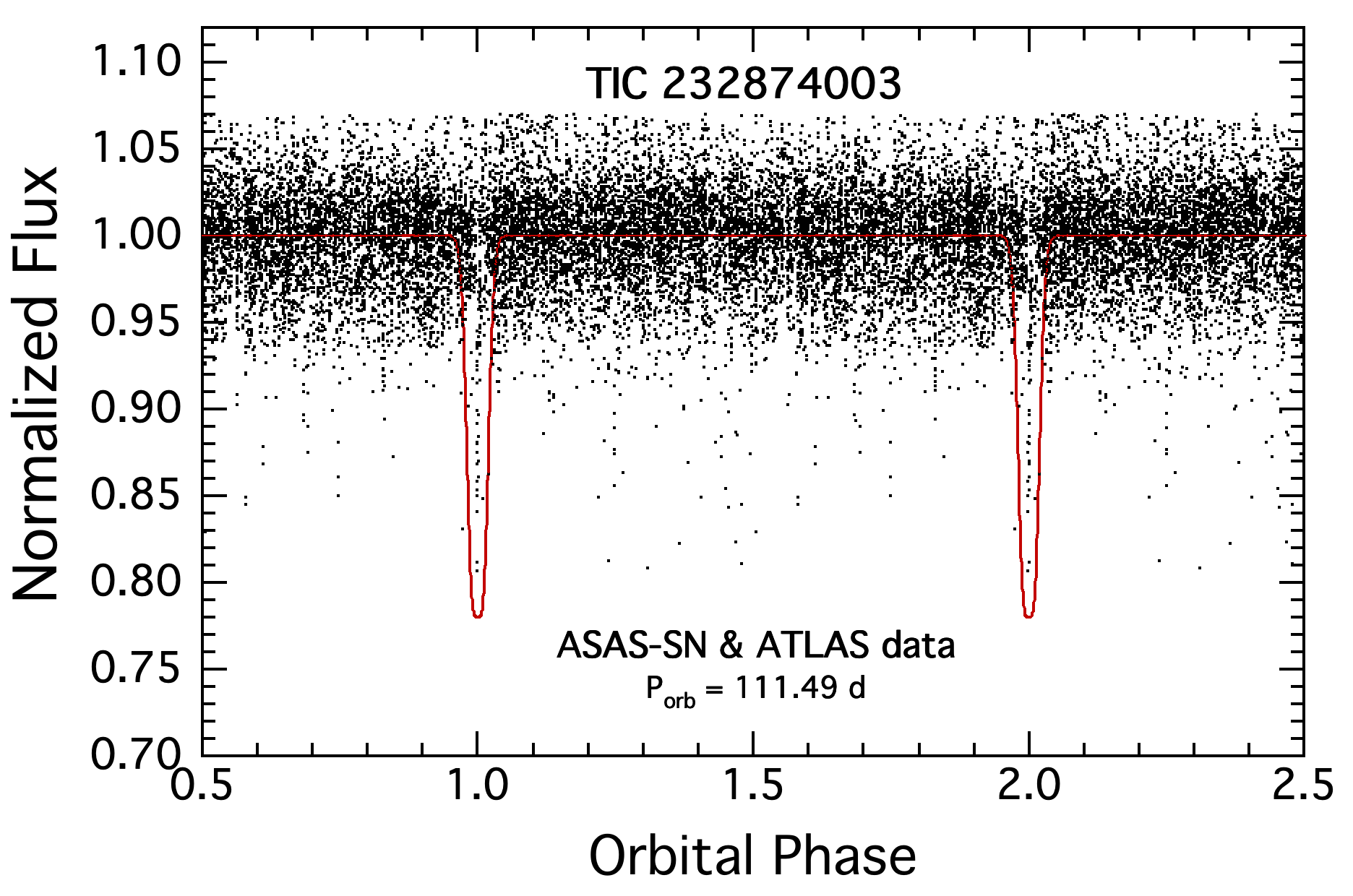}
\includegraphics[width=0.46\textwidth]{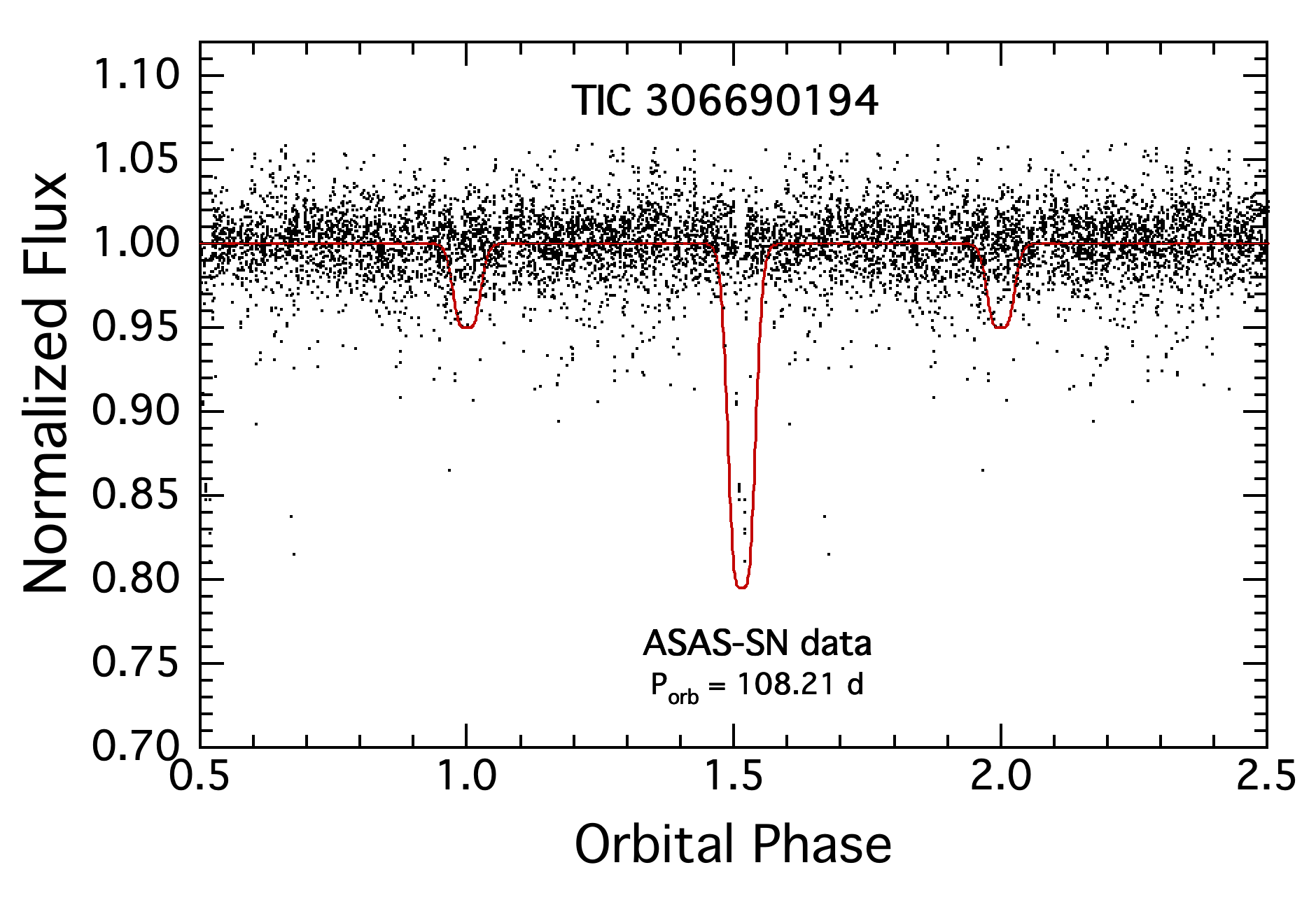}
\includegraphics[width=0.46\textwidth]{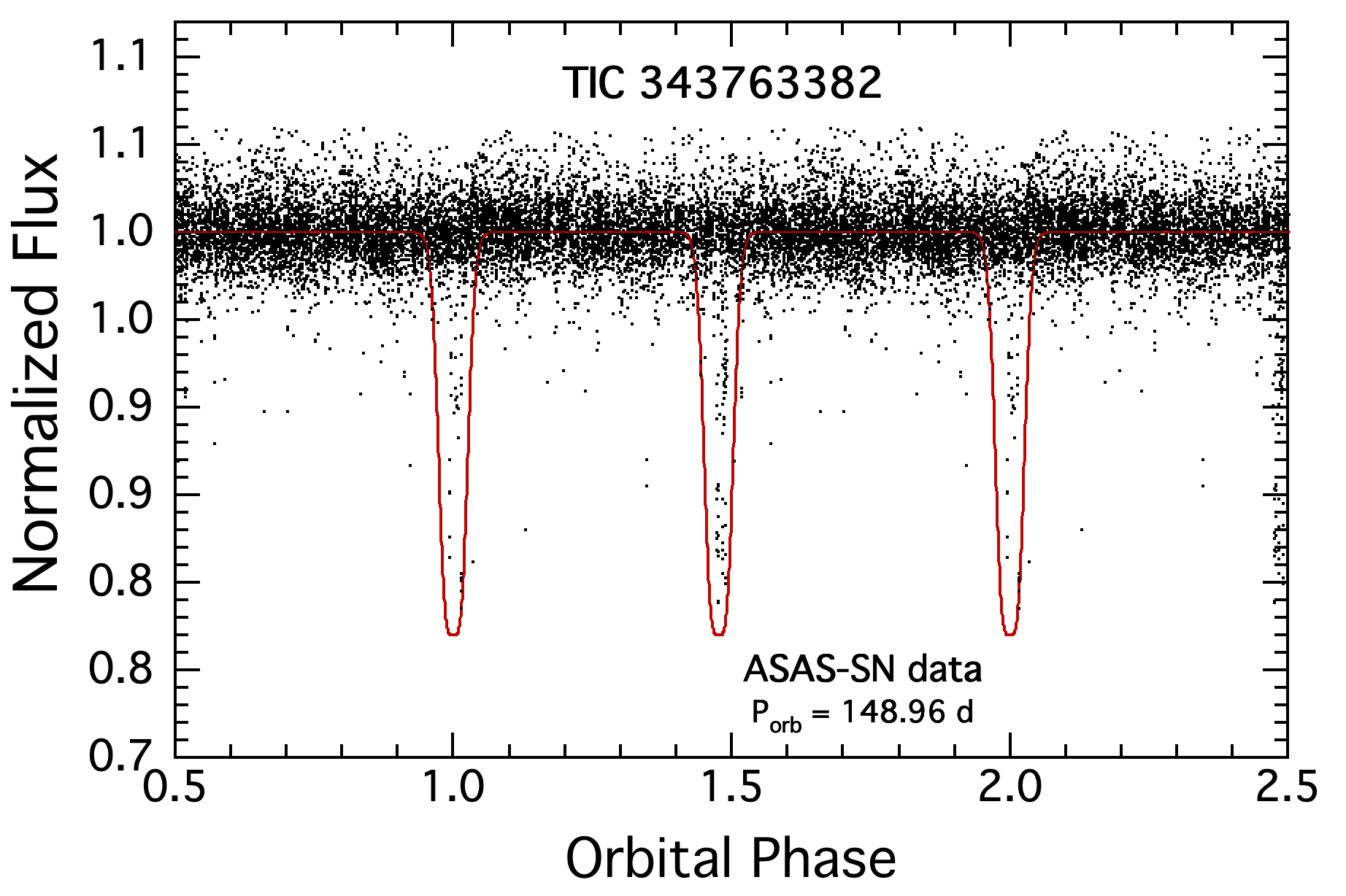}
\includegraphics[width=0.45\textwidth]{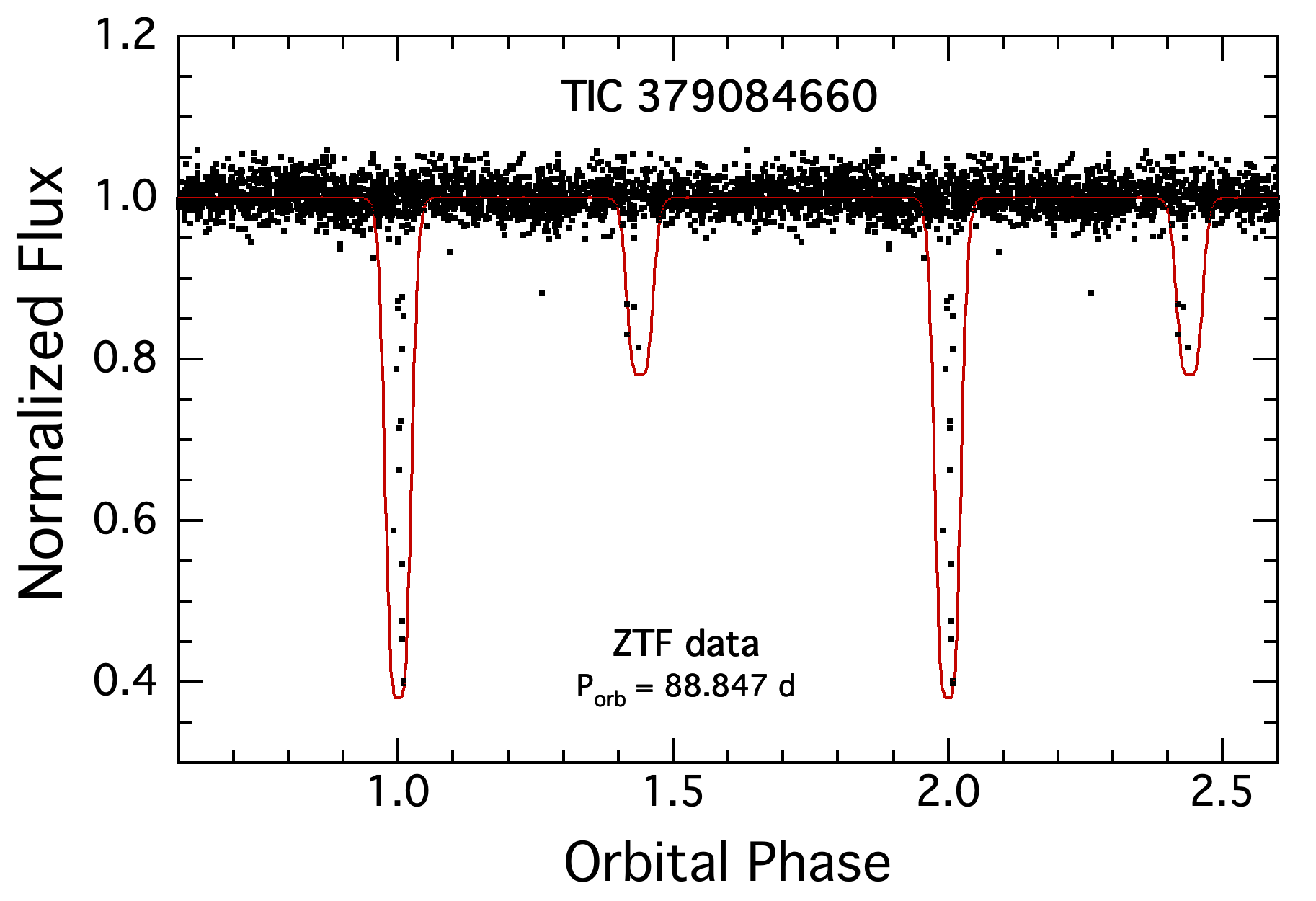}
\includegraphics[width=0.46\textwidth]{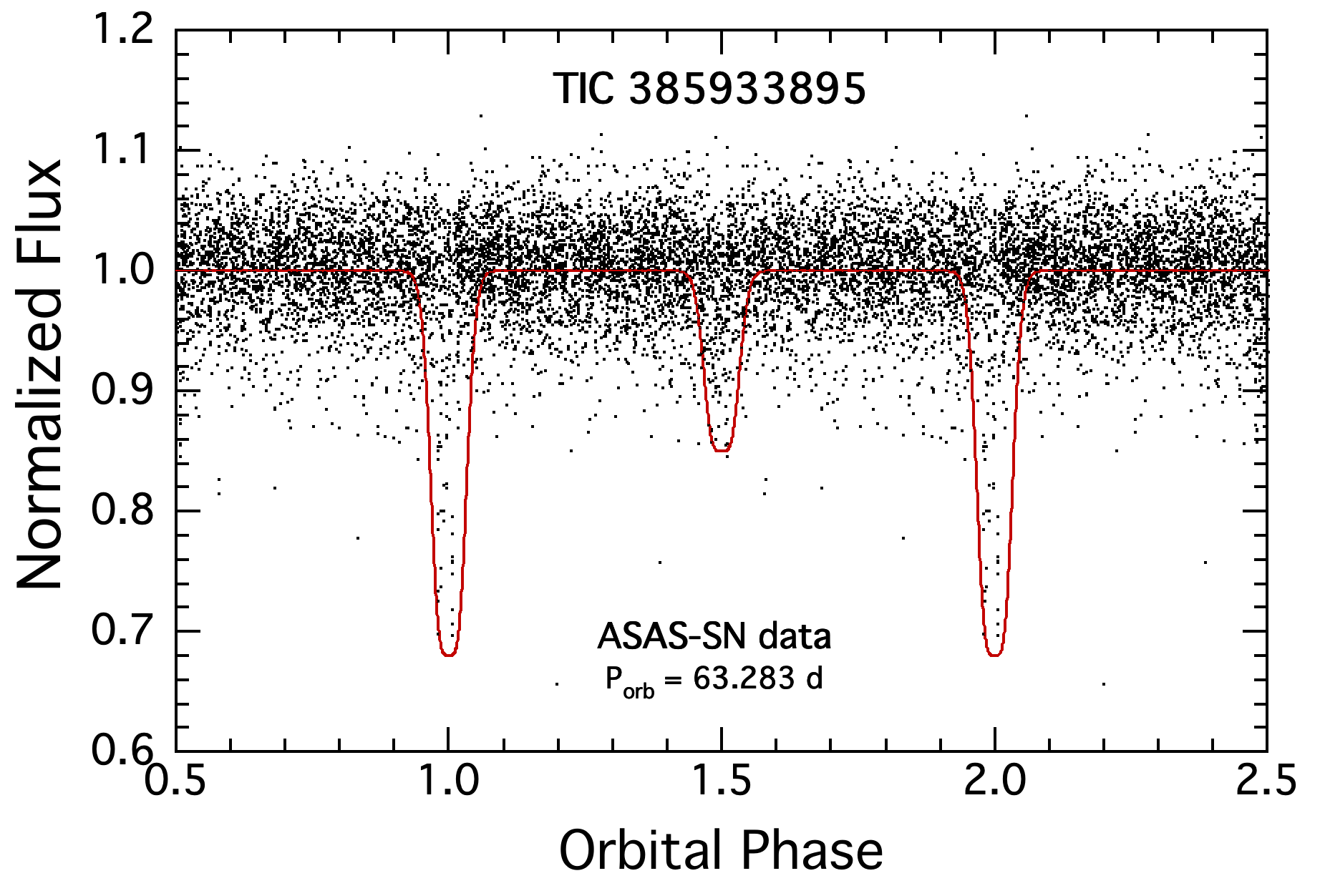}
\caption{Folds of the outer orbits made with ATLAS, ASAS-SN, and ZTF archival data (see text for references). Only the 6 of our 8 objects for which the outer orbital period could be found independently and unambiguously with archival data are shown here. The ${\rm sech} (x^2)$ function shown as a red curve merely outlines and highlights the regions where third-body outer eclipses are either detected in the archival or expected (based on TESS data).  They are not fits.  Orbital phase 1.0 corresponds to the time of a TESS outer eclipse. The outer eclipsing period, determined from a BLS transform of the archival data, is written in each panel.}
\label{fig:outer_folds}
\end{figure*}

\FloatBarrier

\onecolumn
\section{Light curve sections around the third-body eclipses}
\label{app:lightcurve}

In this appendix we plot characteristic light curve sections of seven of the eight investigated systems.

\begin{figure*}[ht]
   \centering
\includegraphics[width=0.30\textwidth]{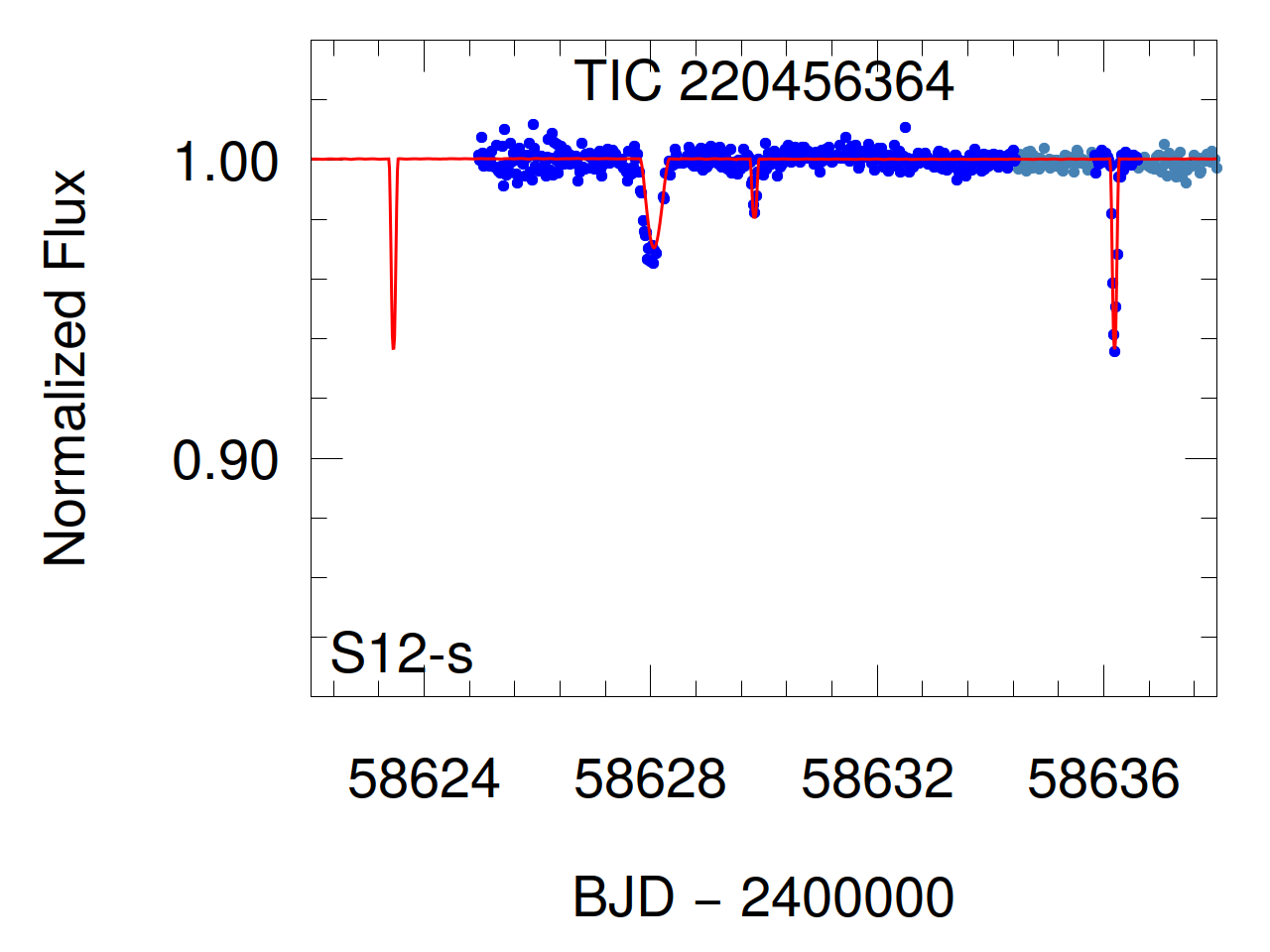}\includegraphics[width=0.30\textwidth]{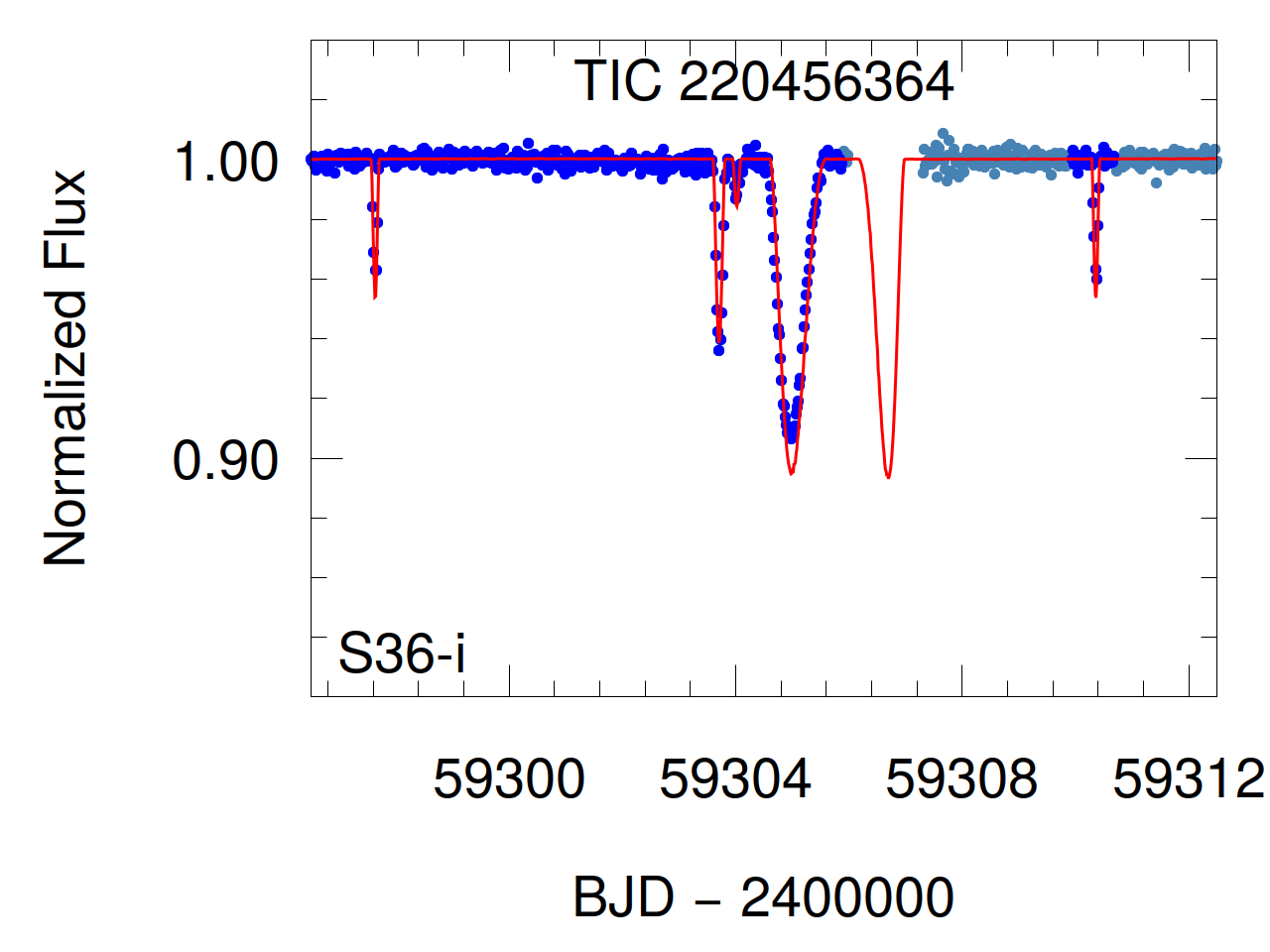}\includegraphics[width=0.30\textwidth]{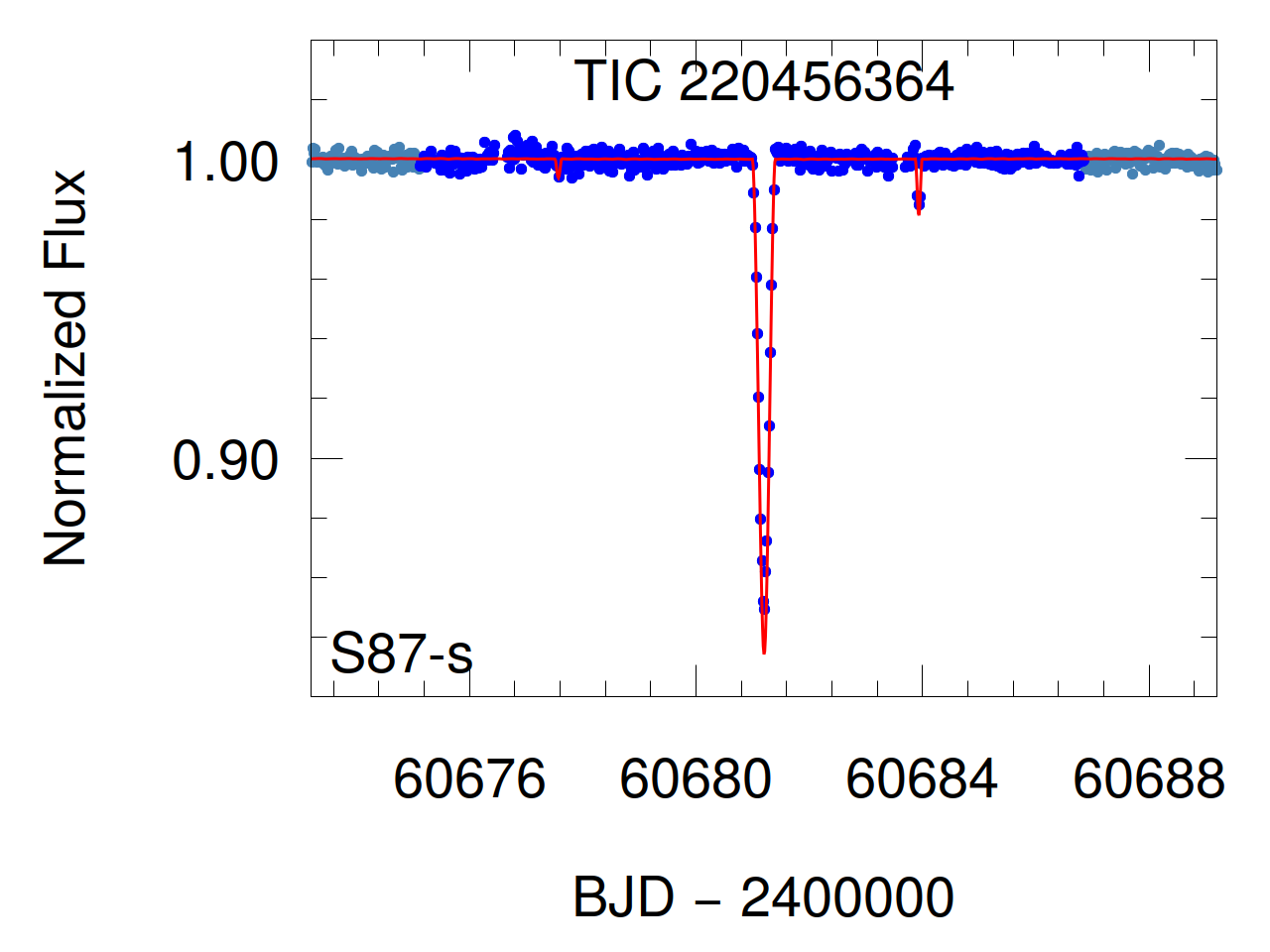}
\includegraphics[width=0.30\textwidth]{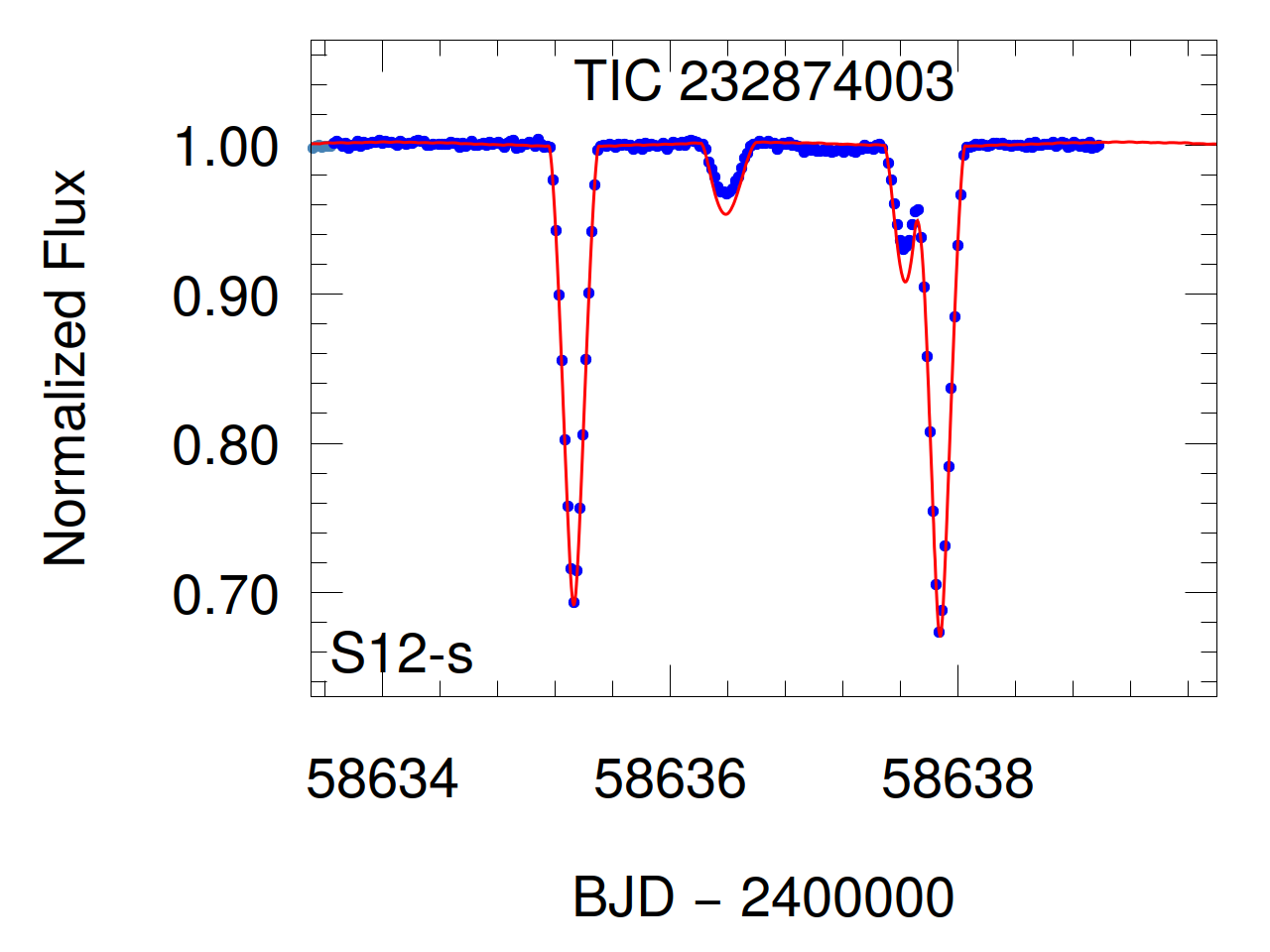}\includegraphics[width=0.30\textwidth]{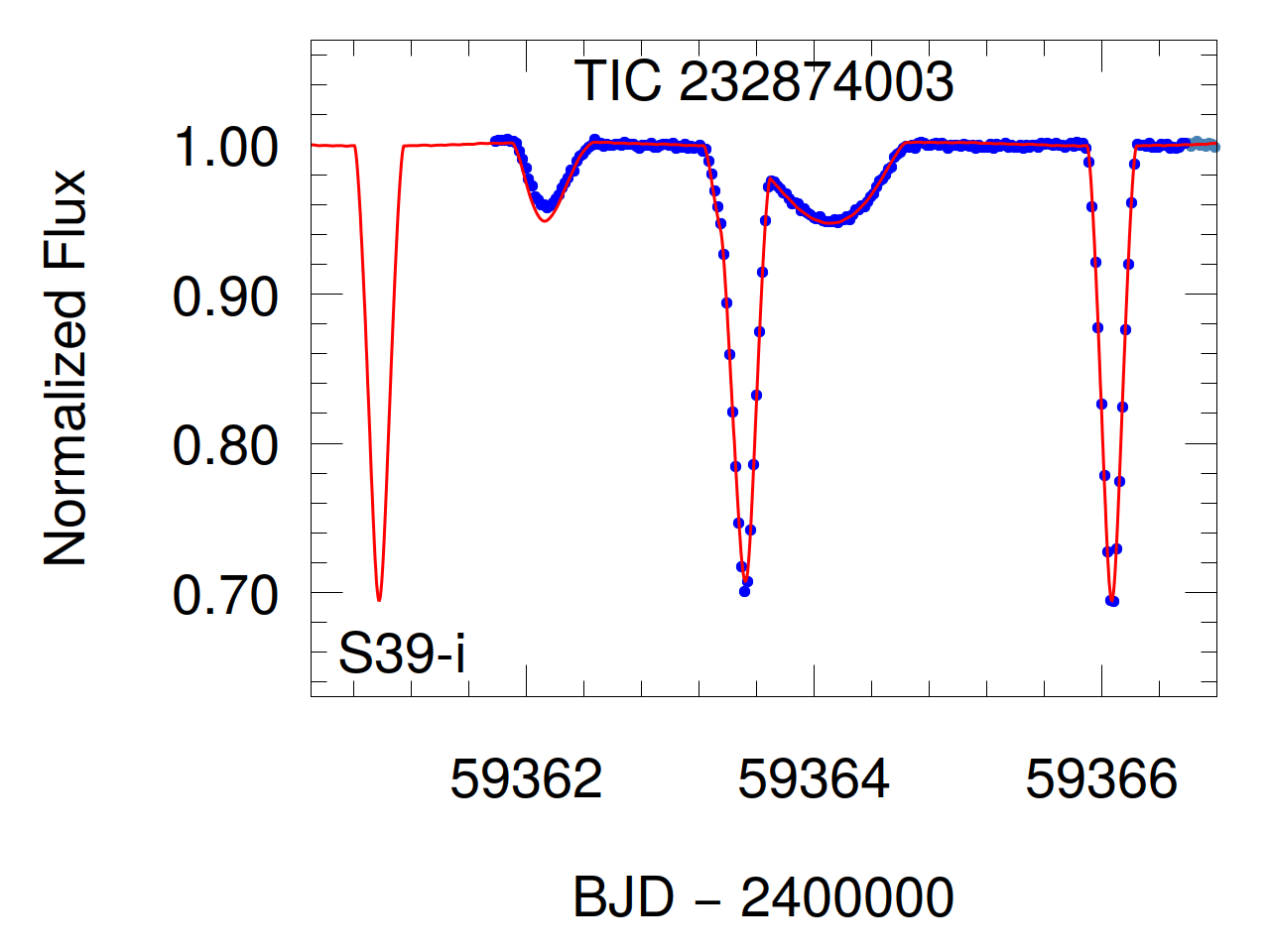}\includegraphics[width=0.30\textwidth]{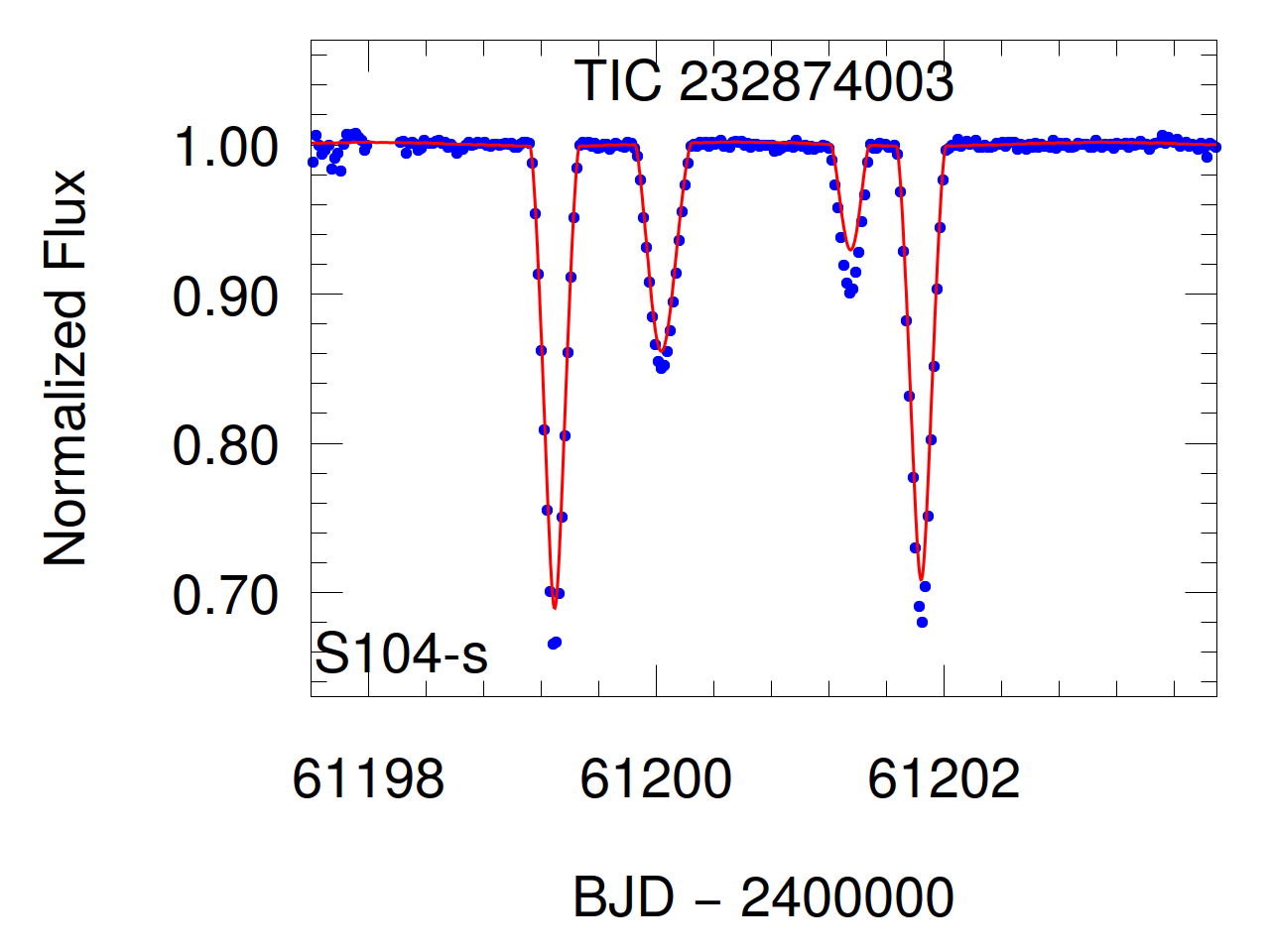}
\includegraphics[width=0.30\textwidth]{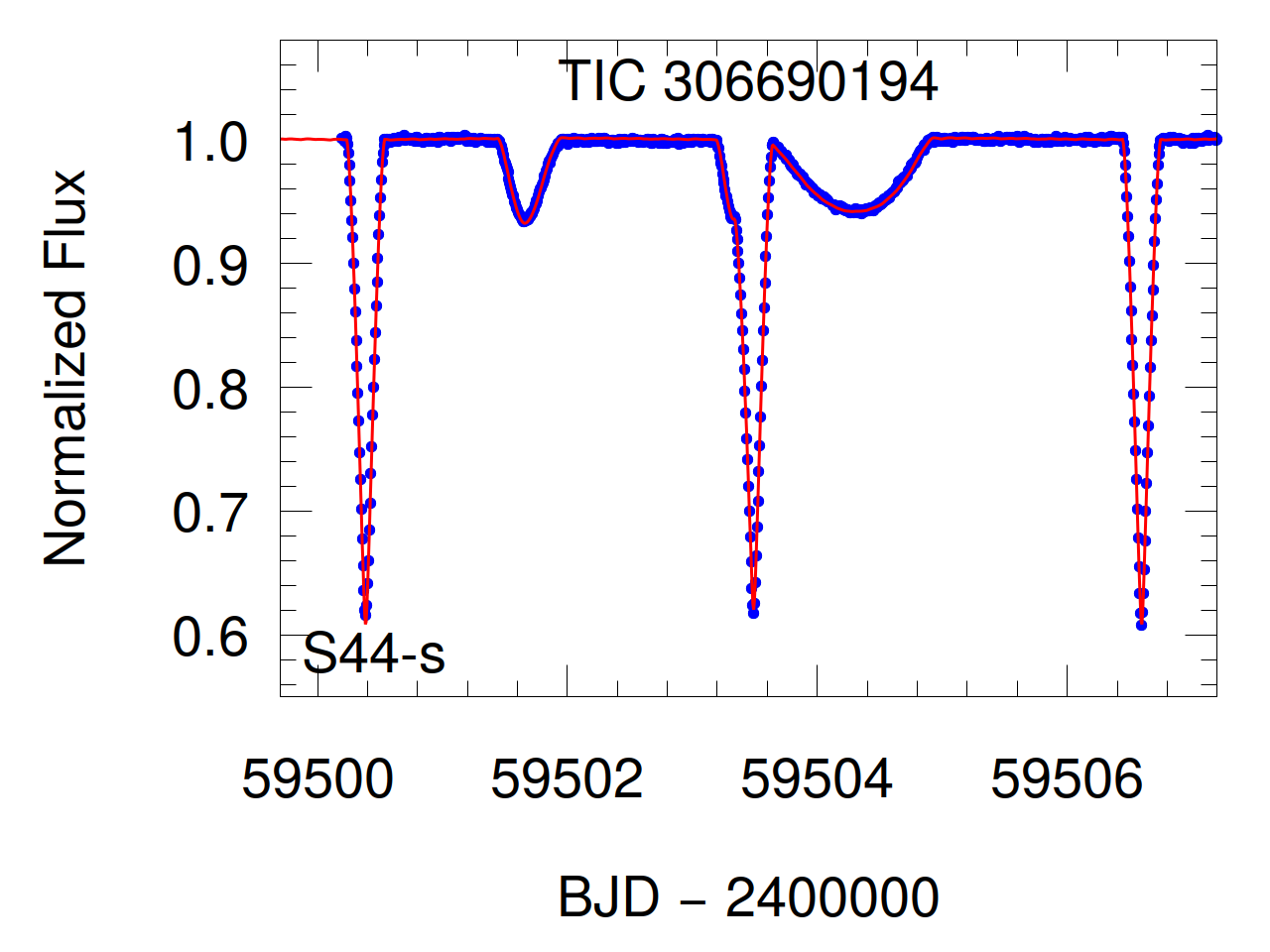}\includegraphics[width=0.30\textwidth]{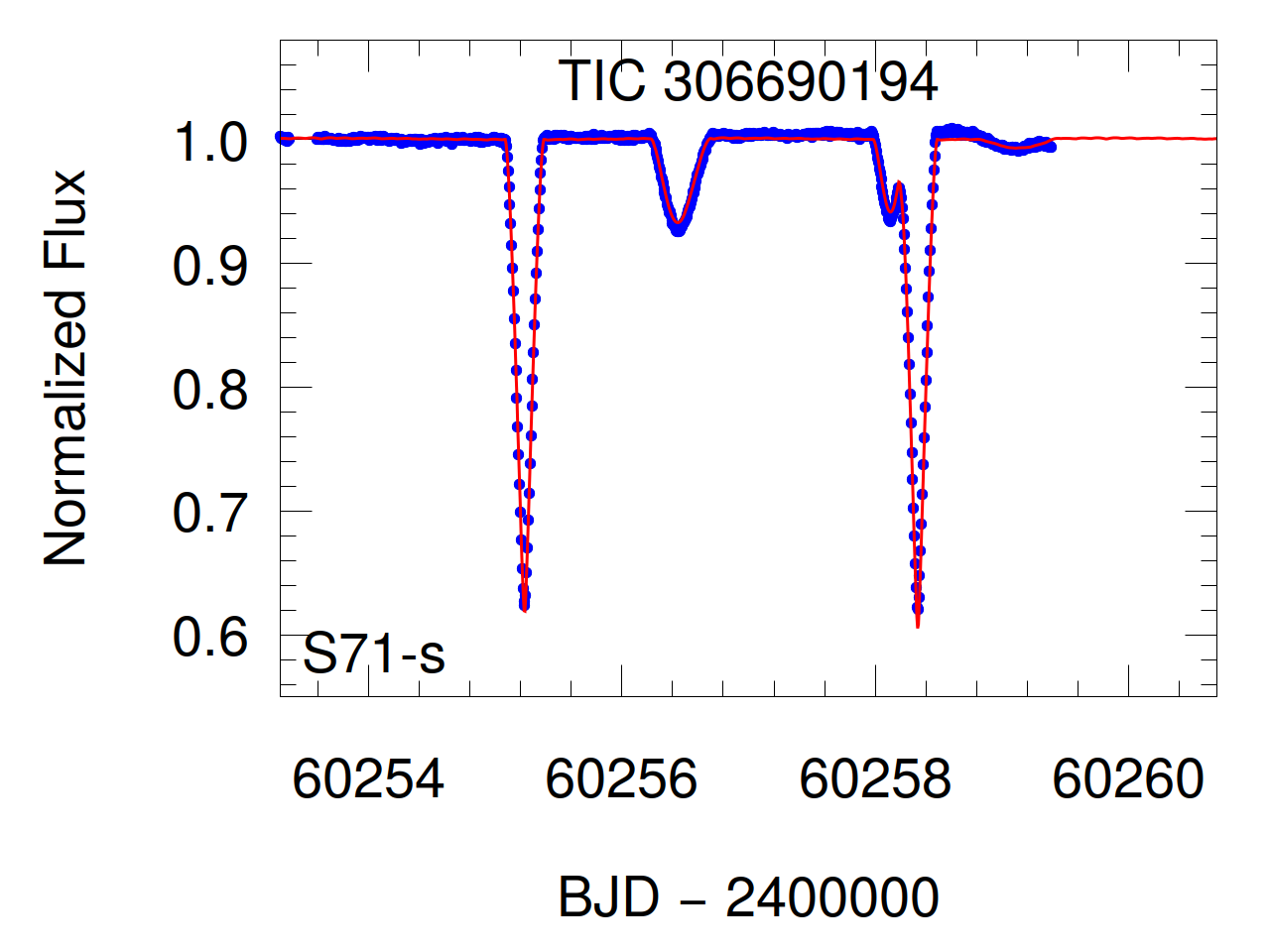}
\includegraphics[width=0.32\textwidth]{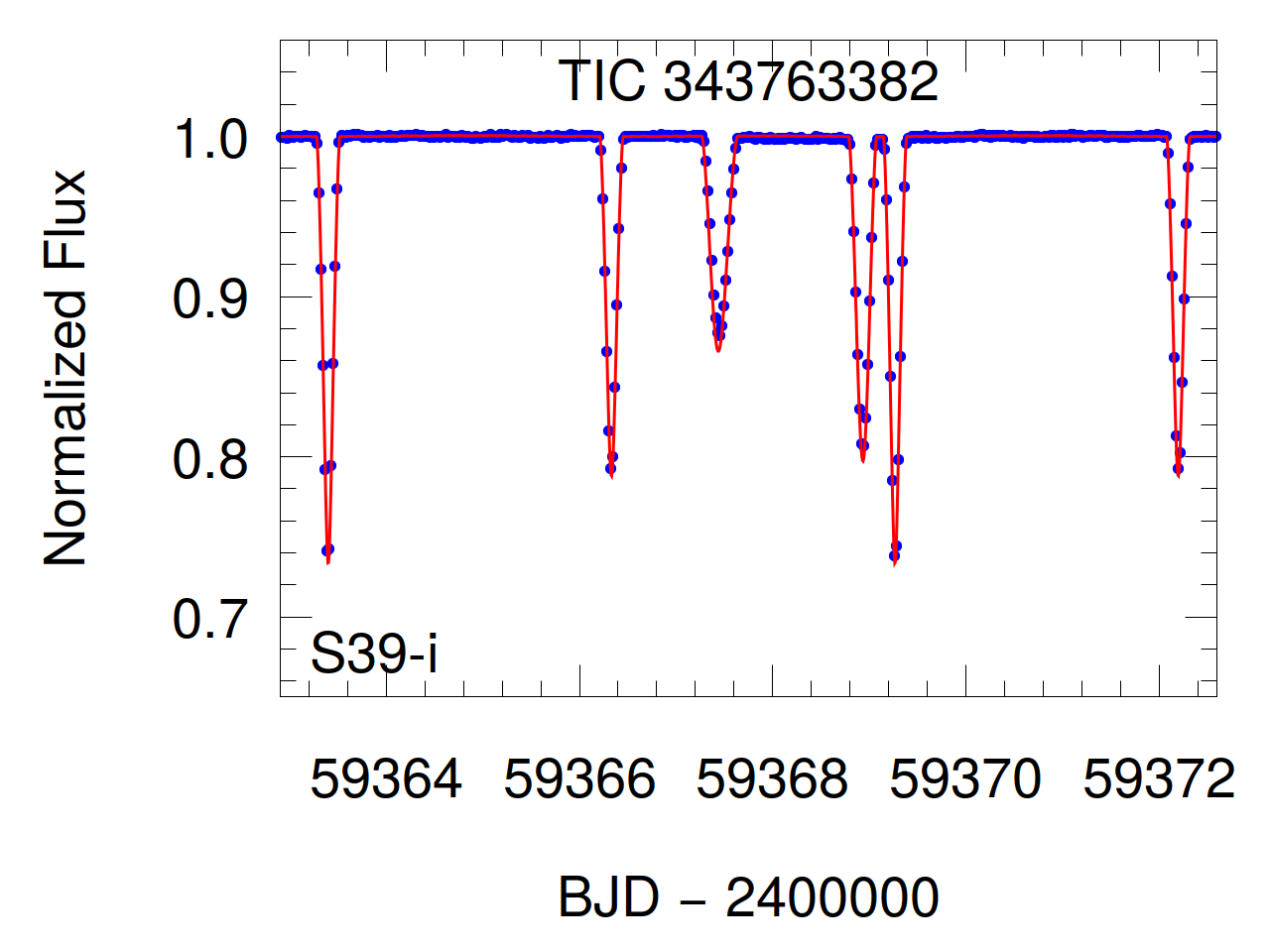}\includegraphics[width=0.32\textwidth]{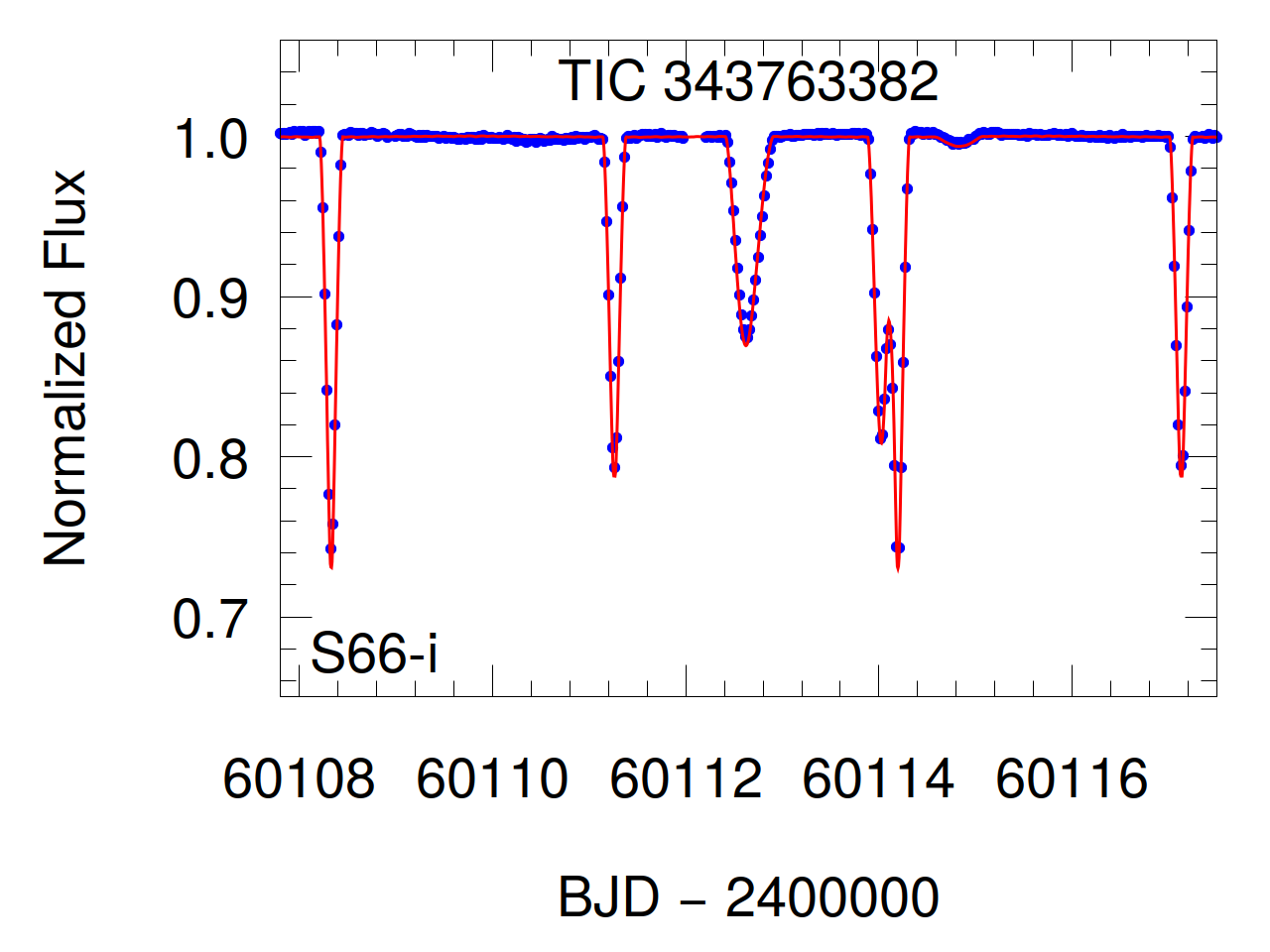}\includegraphics[width=0.32\textwidth]{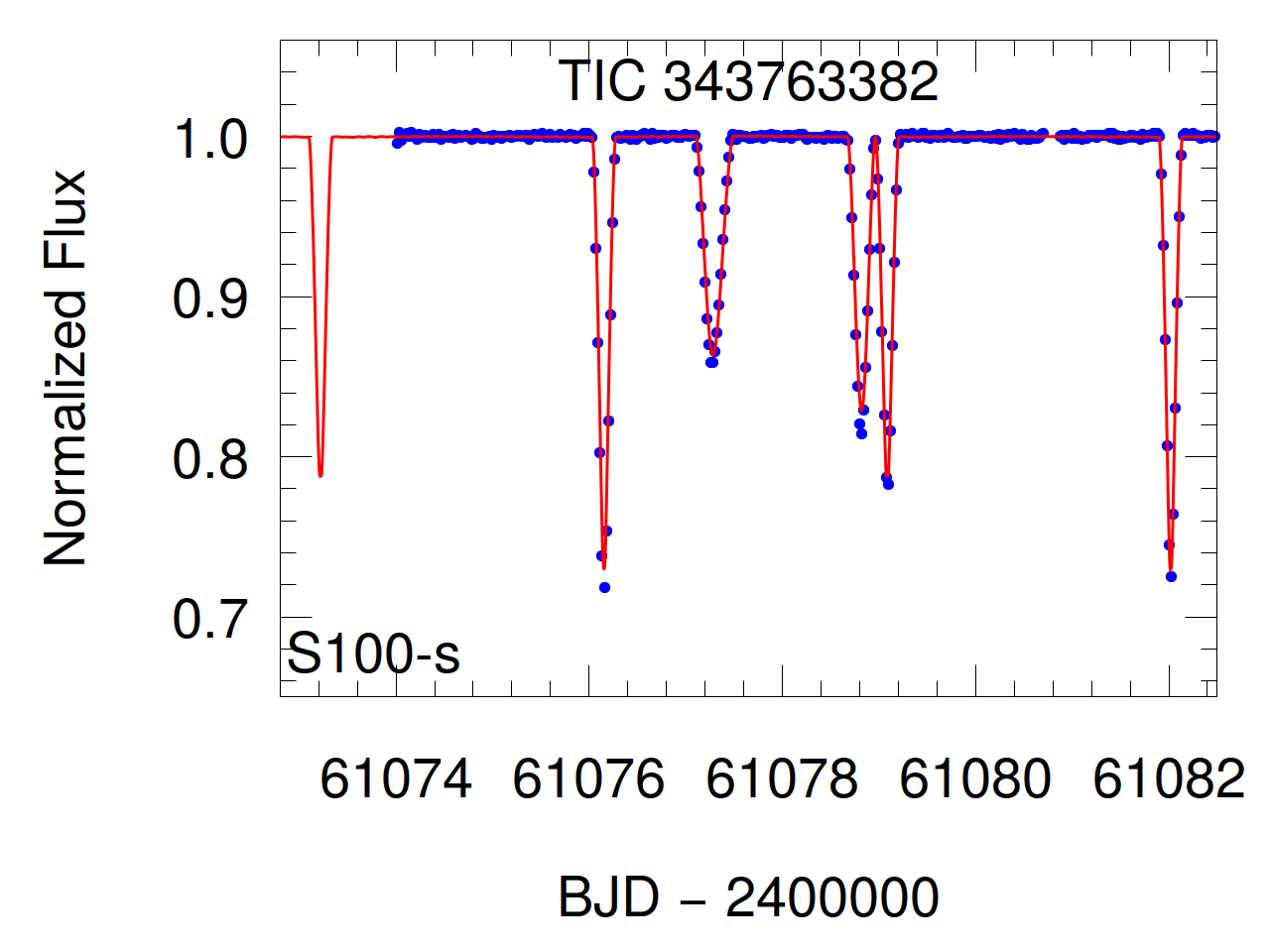}   
   \caption{Light curves (blue points) and model fits (smooth red curves) near the third-body eclipses of four targets. {\it From top to bottom panels:} 220456364, 232874003, 306690194 and, 343763382. Dark/pale blue points are for those light curve sections which were used/not-used for the photodynamical solution, after the removal of the likely effects of stellar activity. The sector numbers are indicated in the lower left corner of each panel. Letters `i' or `s' after the sector numbers refer to the inferior or superior conjunction of the third star, respectively.}
   \label{fig:lcs1}
   \end{figure*}  
   
\begin{figure*}[ht]
   \centering
\includegraphics[width=0.32\textwidth]{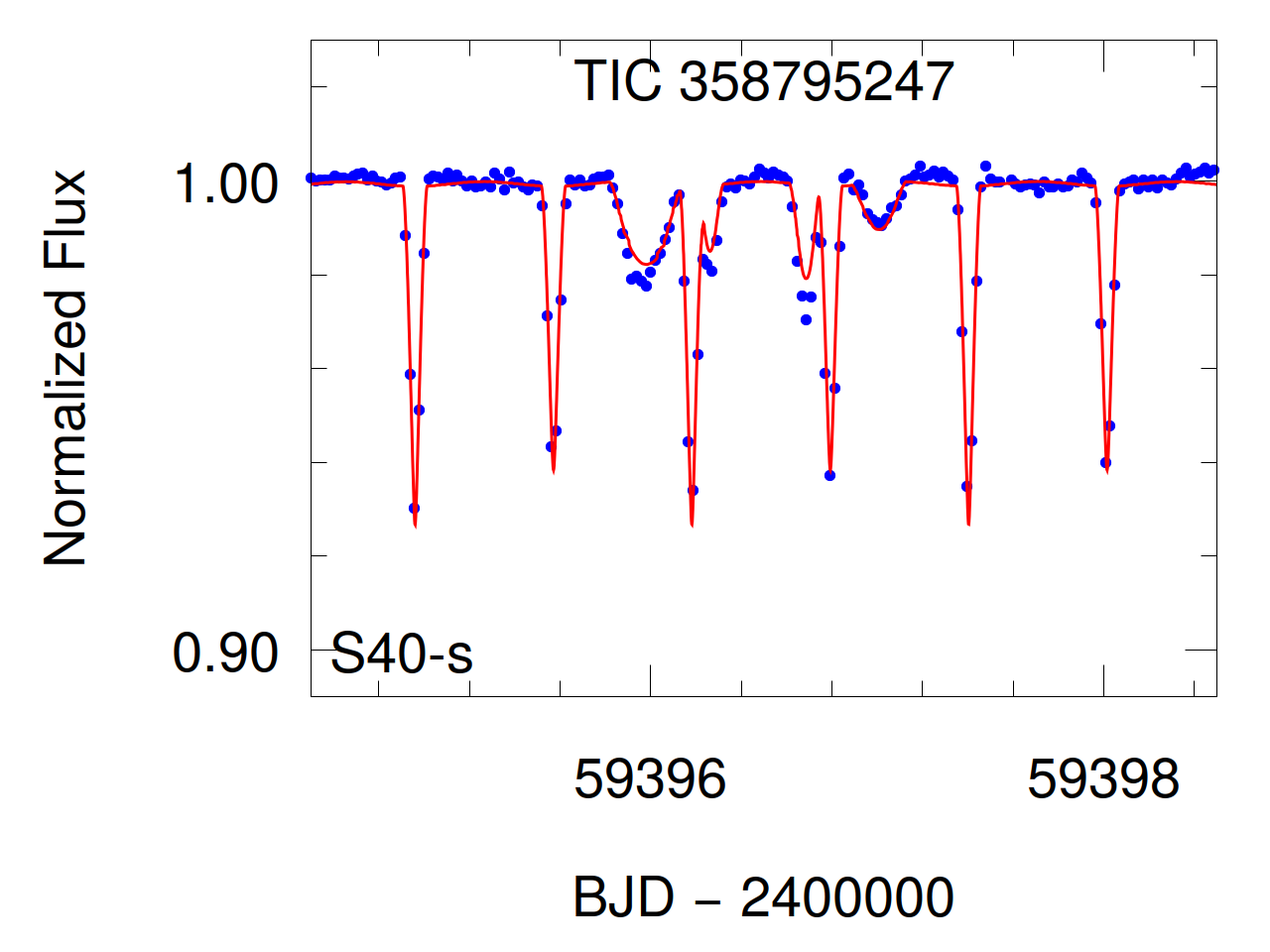}\includegraphics[width=0.32\textwidth]{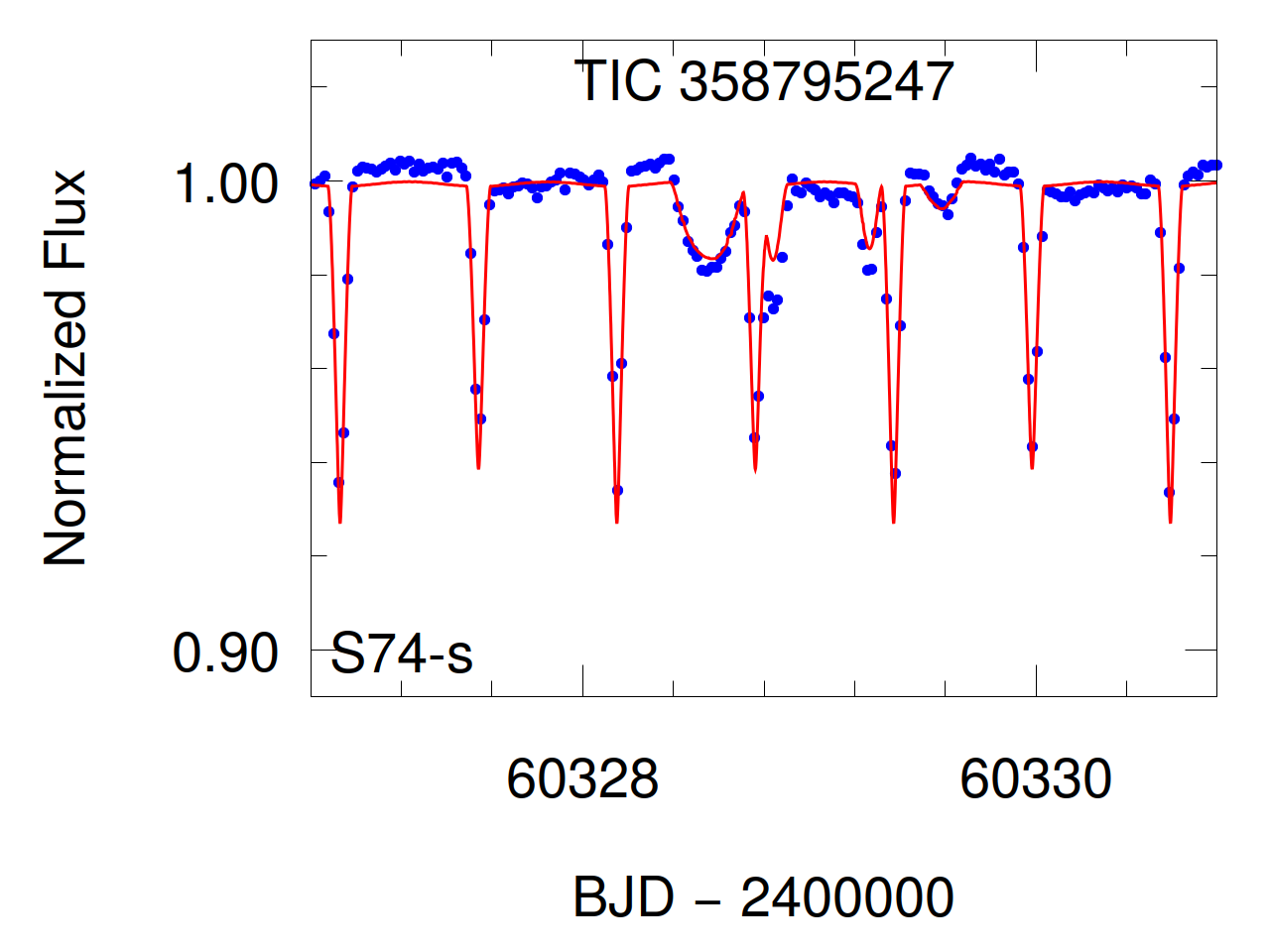}\includegraphics[width=0.32\textwidth]{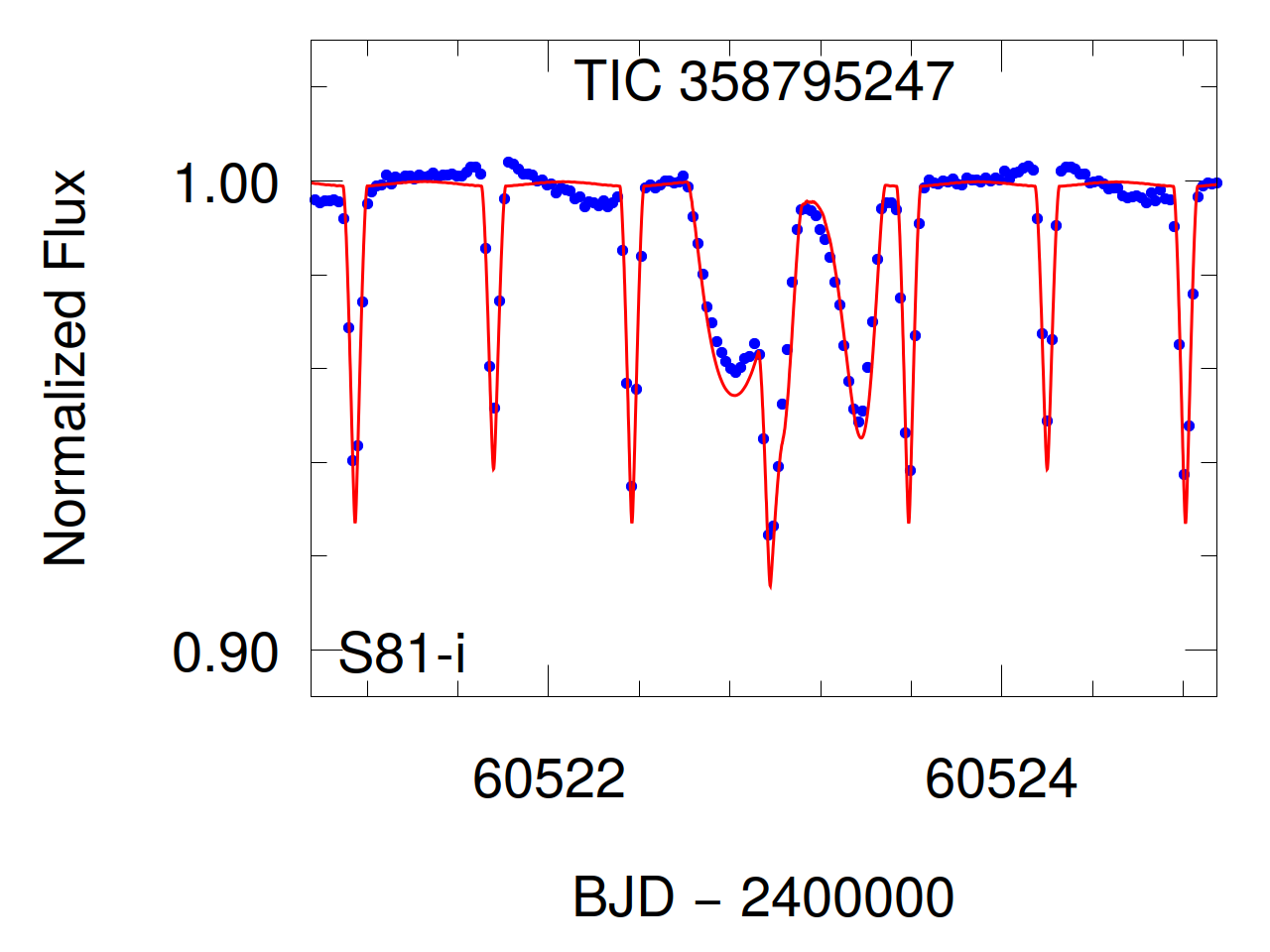}   
\includegraphics[width=0.32\textwidth]{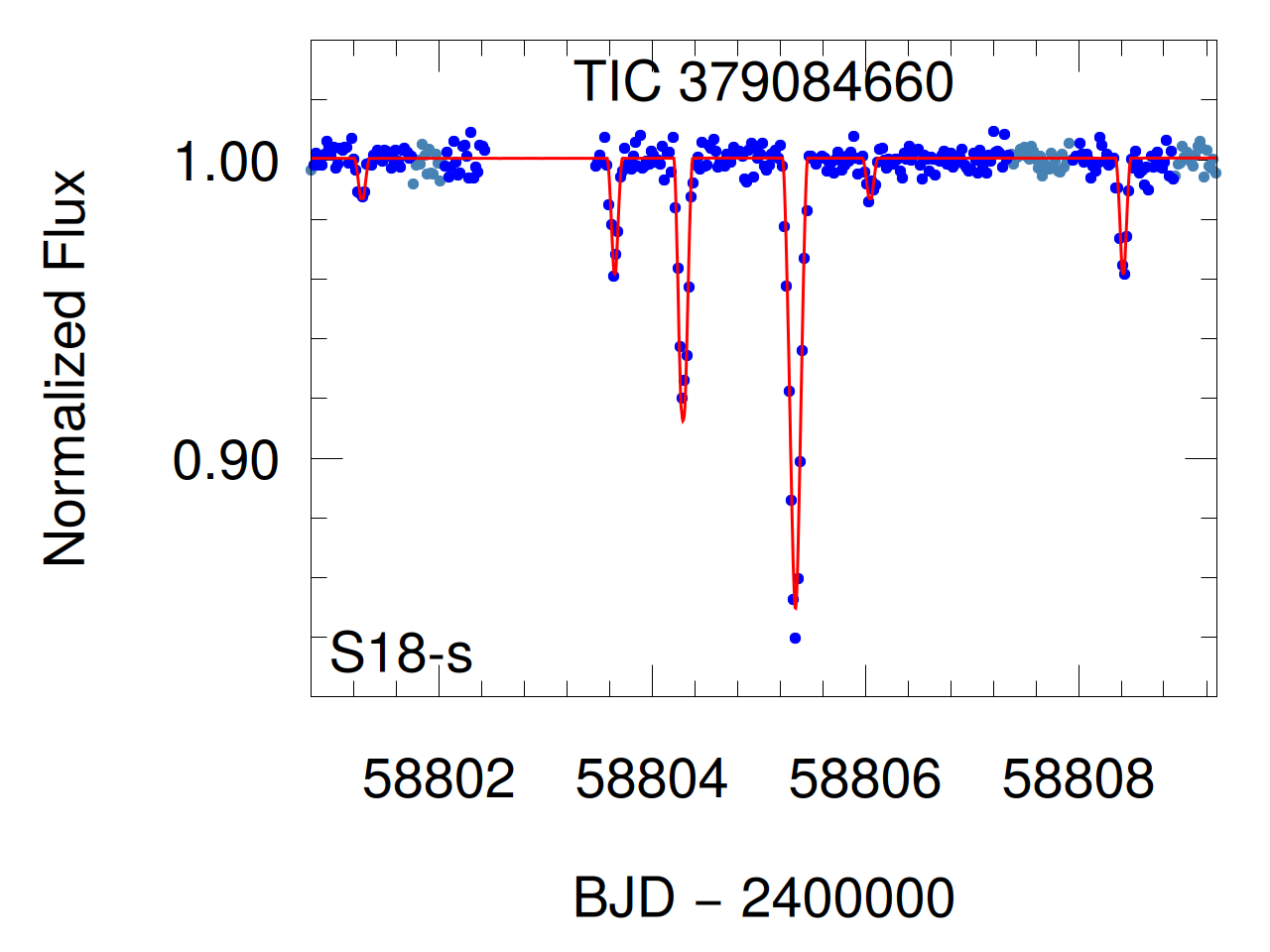}\includegraphics[width=0.32\textwidth]{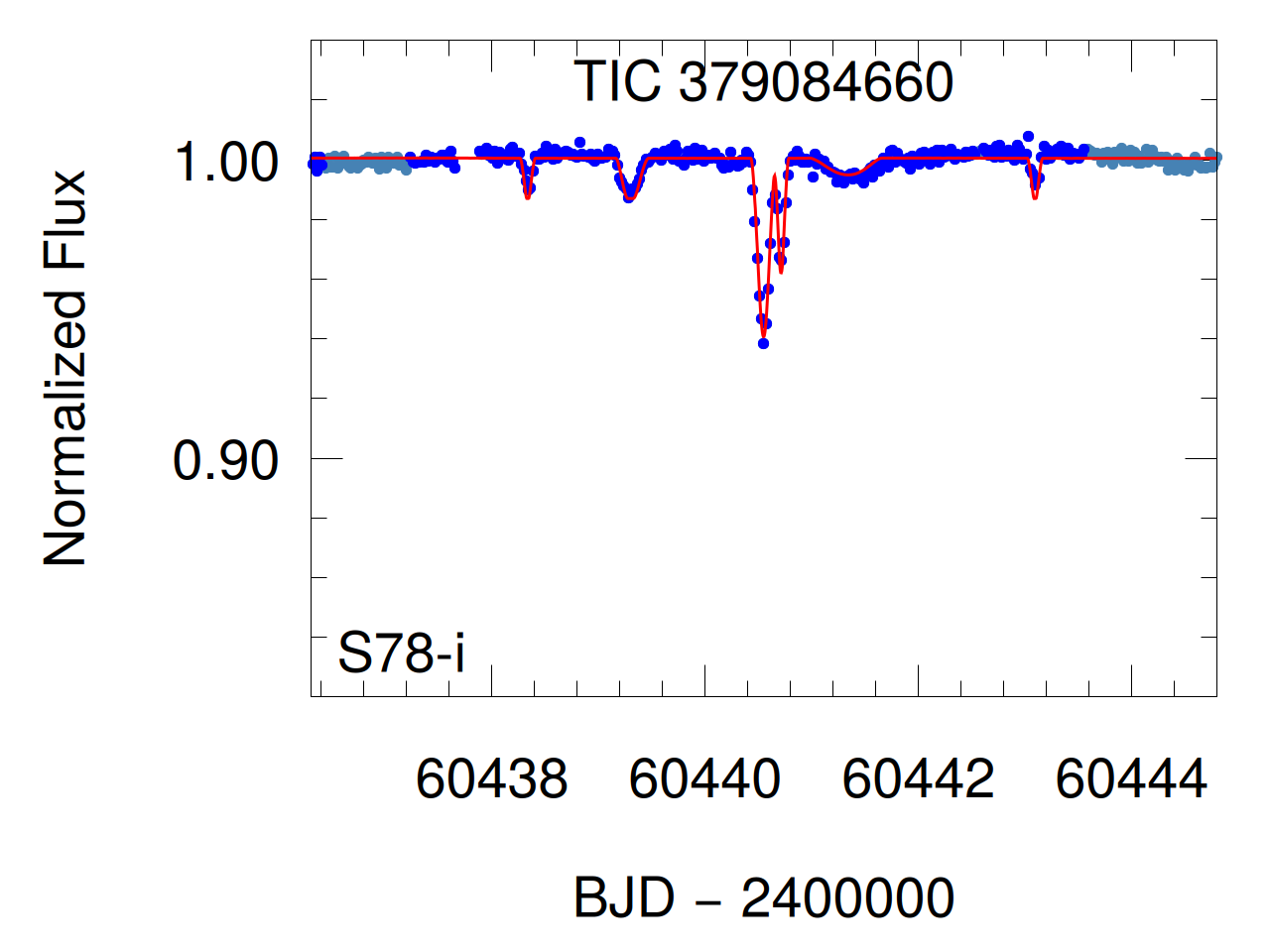}\includegraphics[width=0.32\textwidth]{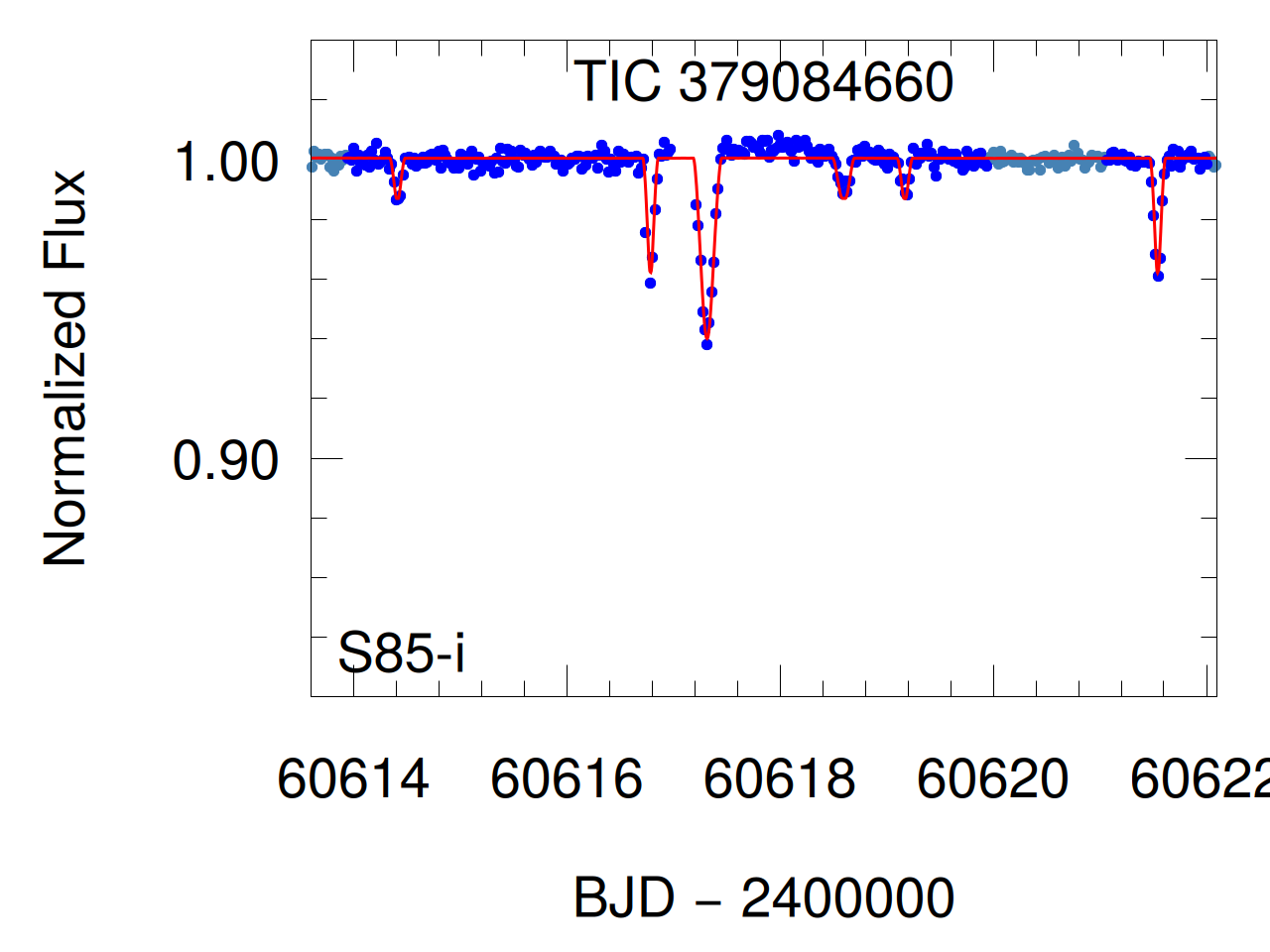}
\includegraphics[width=0.32\textwidth]{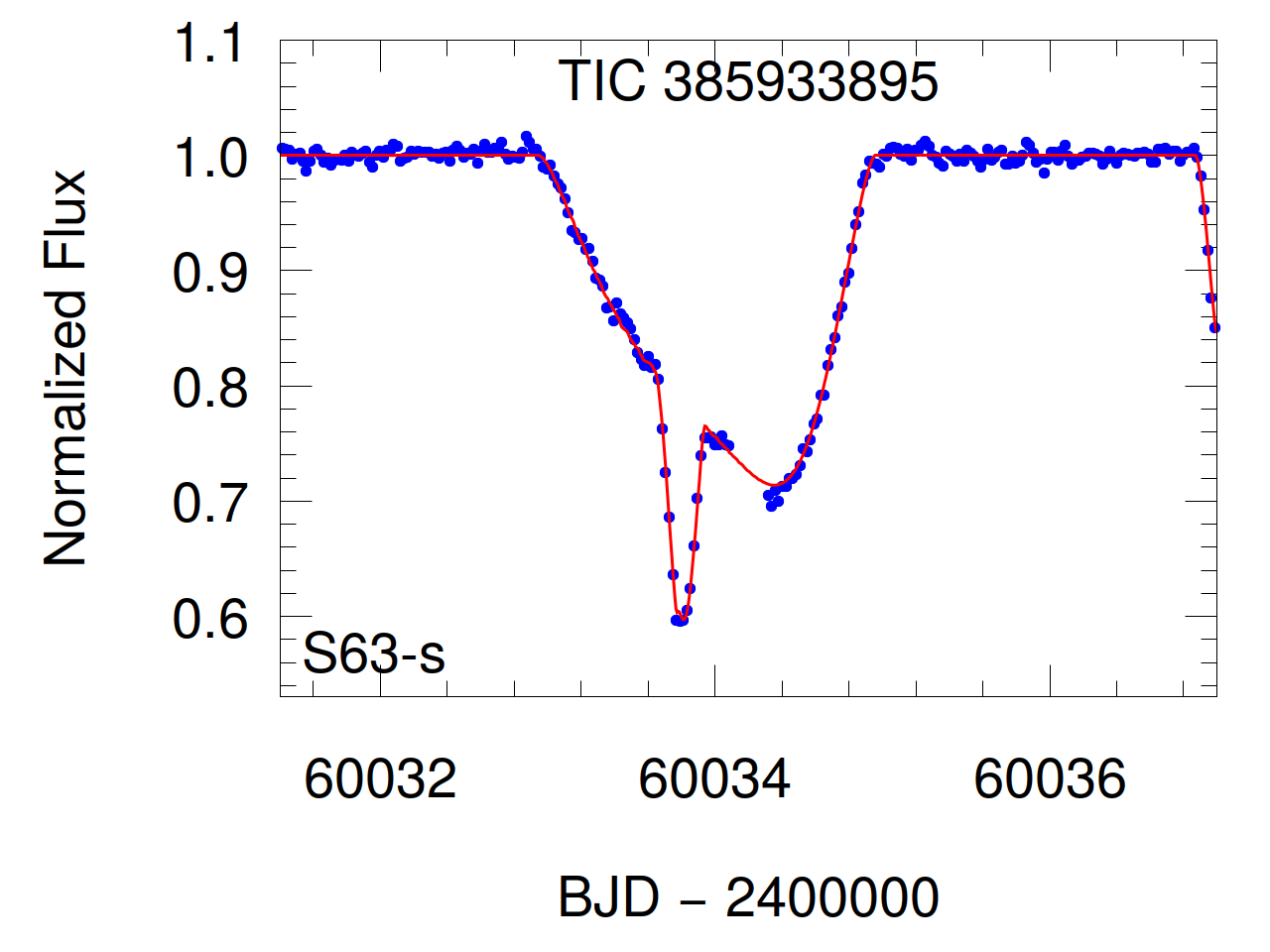}\includegraphics[width=0.32\textwidth]{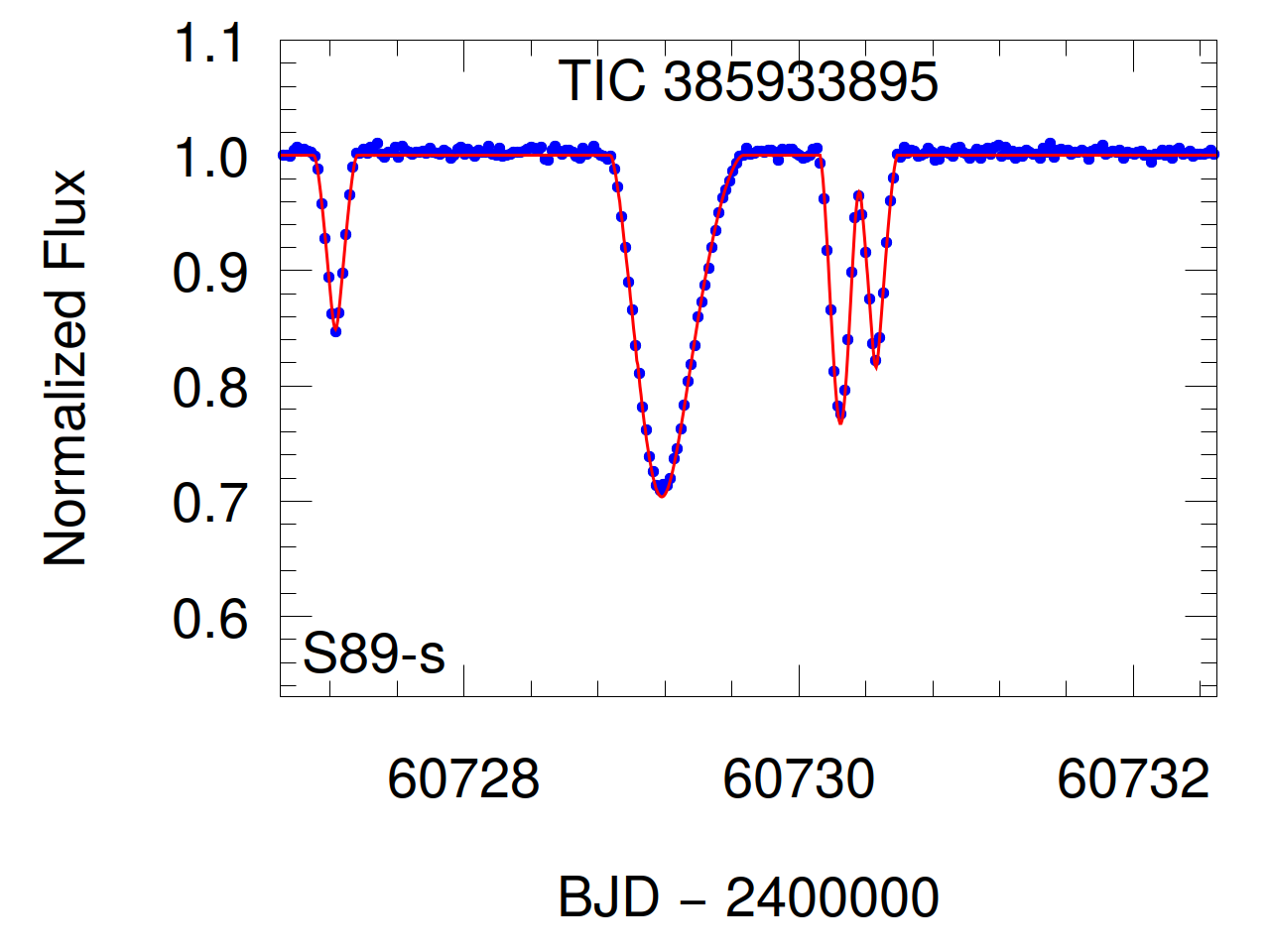}\includegraphics[width=0.32\textwidth]{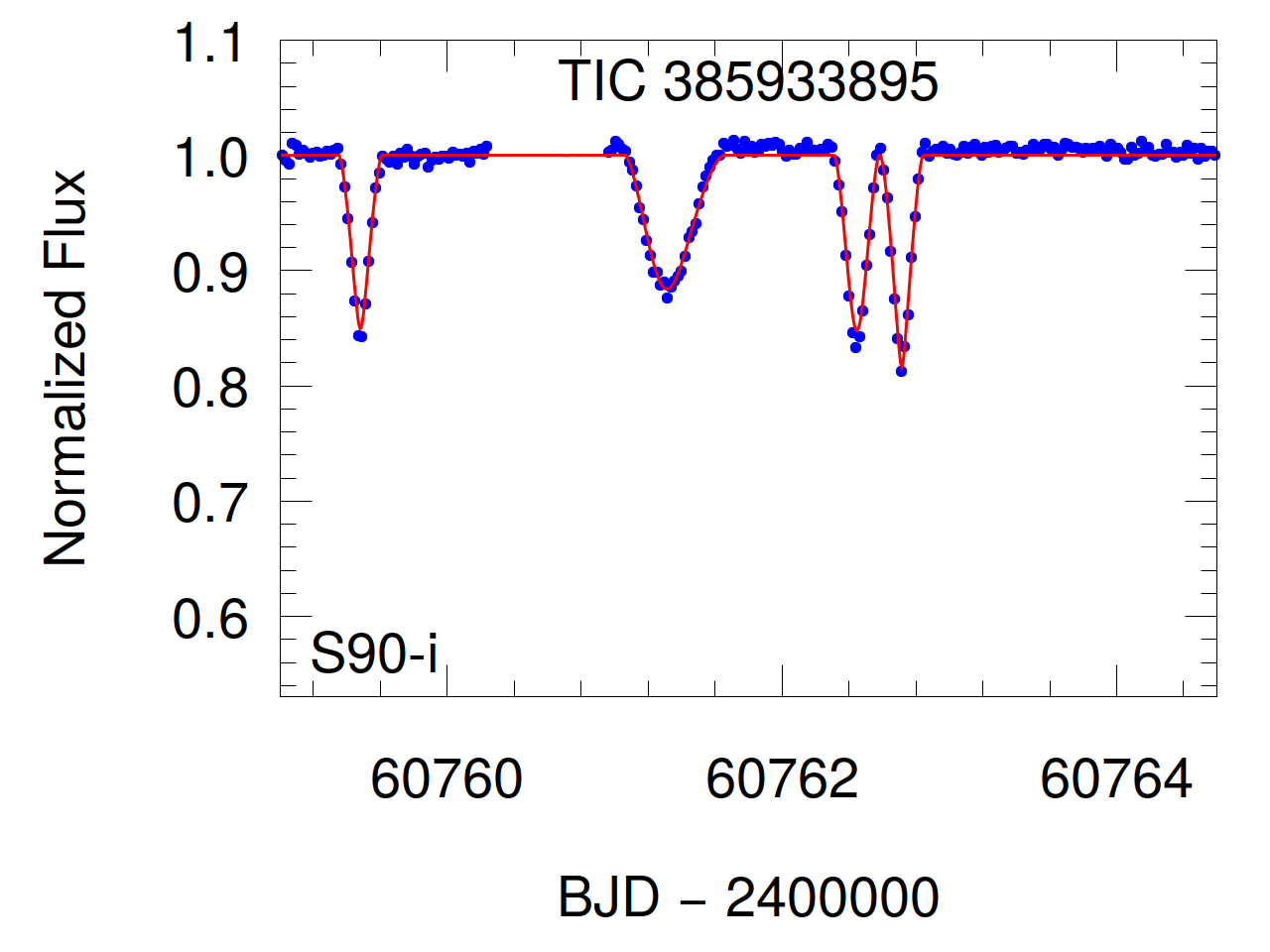}
   \caption{Light curves (blue points) and model fits (smooth red curves) near the third-body eclipses of the last three targets. {\it From top to bottom panels:} 358795247, 379084660, 385933895. Dark/pale blue points are for those light curve sections which were used/not-used for the photodynamical solution, after the removal of the likely effects of stellar activity. The sector numbers are indicated in the lower left corner of each panel. Letters `i' or `s' after the sector numbers refer to the inferior or superior conjunction of the third star, respectively.}
   \label{fig:lcs2}
   \end{figure*}  
   
\FloatBarrier

\onecolumn
\section{Radial velocity curve plots in the time and the phase domains}
\label{app:RVplots}

In this appendix we plot the RV curves both in the time and the phase domains for the three systems where RV measurements (tabulated in the forthcoming appendix) are available. In the case of TIC\,306690194, where only the two inner components, that is, stars Aa and Ab {yielded detectable lines}, besides the time-domain plot of the data, we give only one phase-folded plot, with the period of the inner binary. Oppositely, in the case of TIC\,343763382, where all three stars were observed, besides the formerly mentioned two plots, we also show a second phase-plot that was made using the period of the outer orbit. Finally, in the case of the (2+1)+1-type quadruple system, TIC\,358795247, where the RVs of all four components were obtained, we present a fourth plot as well that shows the observations, as well as the model RV curve, folded with the period of the outermost orbit.

\begin{figure*}[ht]
   \centering
\includegraphics[width=0.49\textwidth]{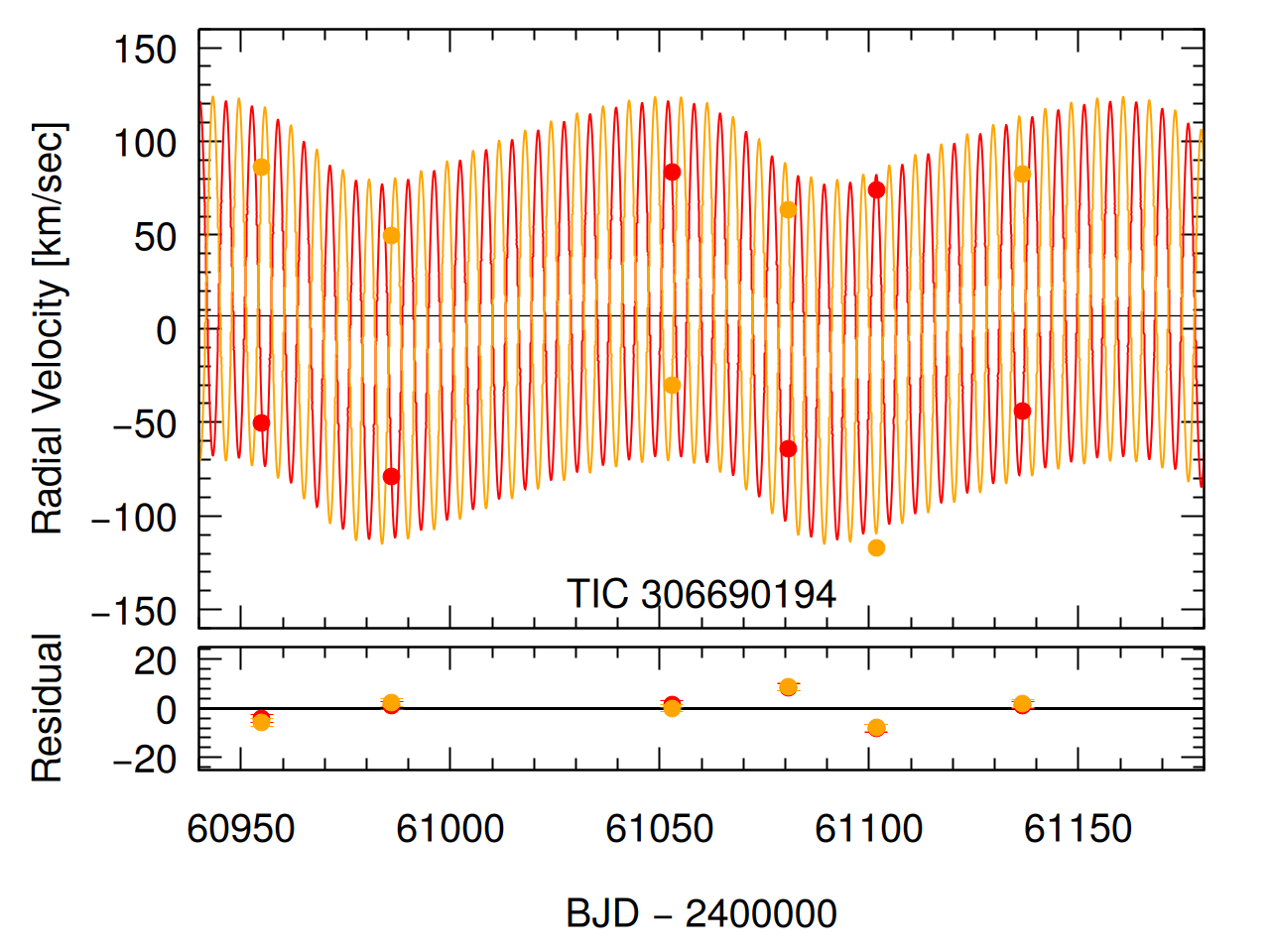}\includegraphics[width=0.49\textwidth]{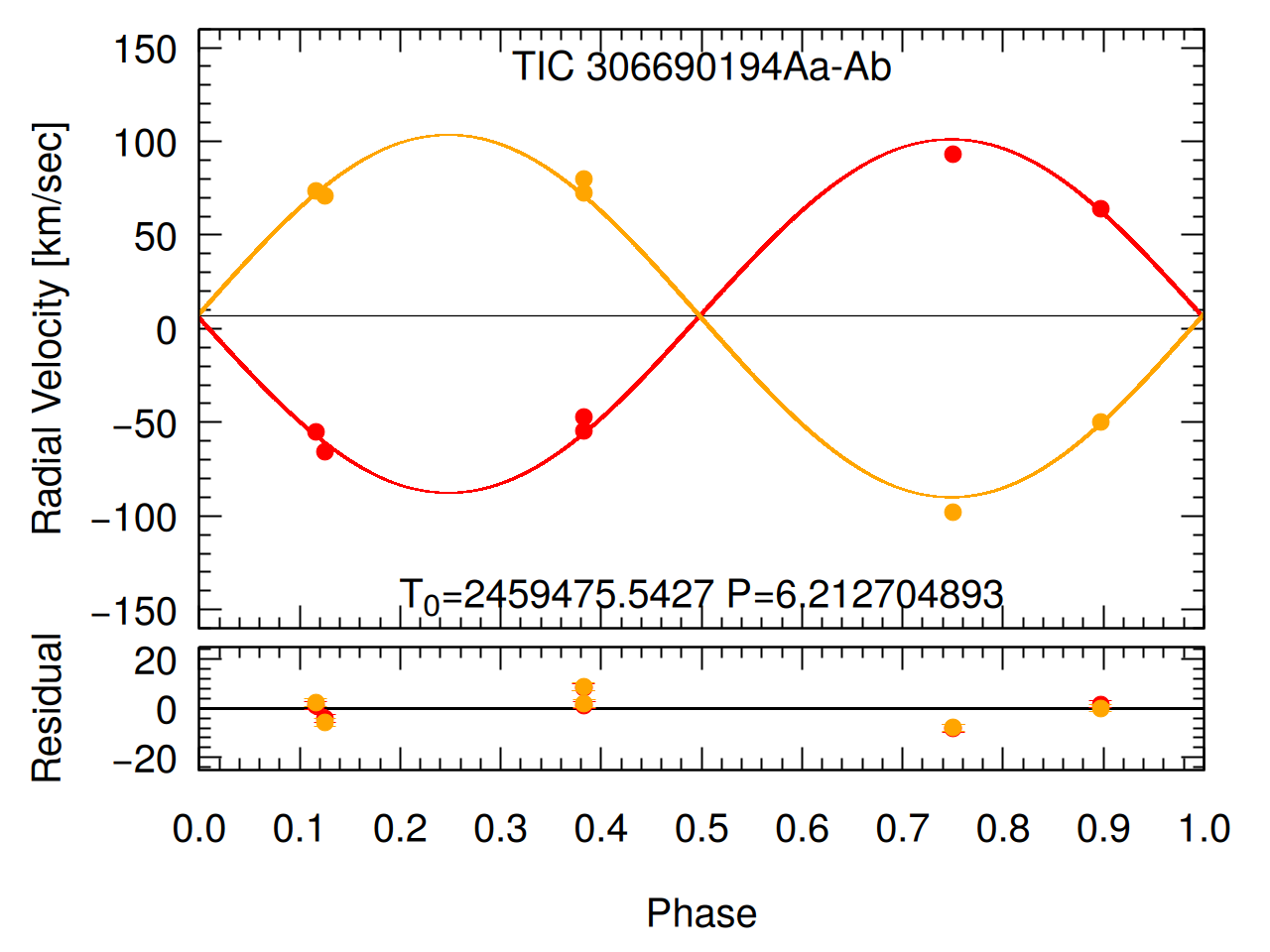}
\caption{The observed RV points (red circles for star Aa and orange ones for Ab, respectively), together with the best-fit spectro-photodynamical solution. {\it Left}: in the time domain; {\it right}: phased according to the period of the inner binary orbit. (Note, in this latter case, we removed the modelled effect of the motion in the outer orbit. Note also that the lower panels show the residuals from the fit, where the a priori estimated observational errors are also denoted.)}
\label{fig:T306690194RV}
\end{figure*}

\begin{figure*}[ht]
\includegraphics[width=0.33\textwidth]{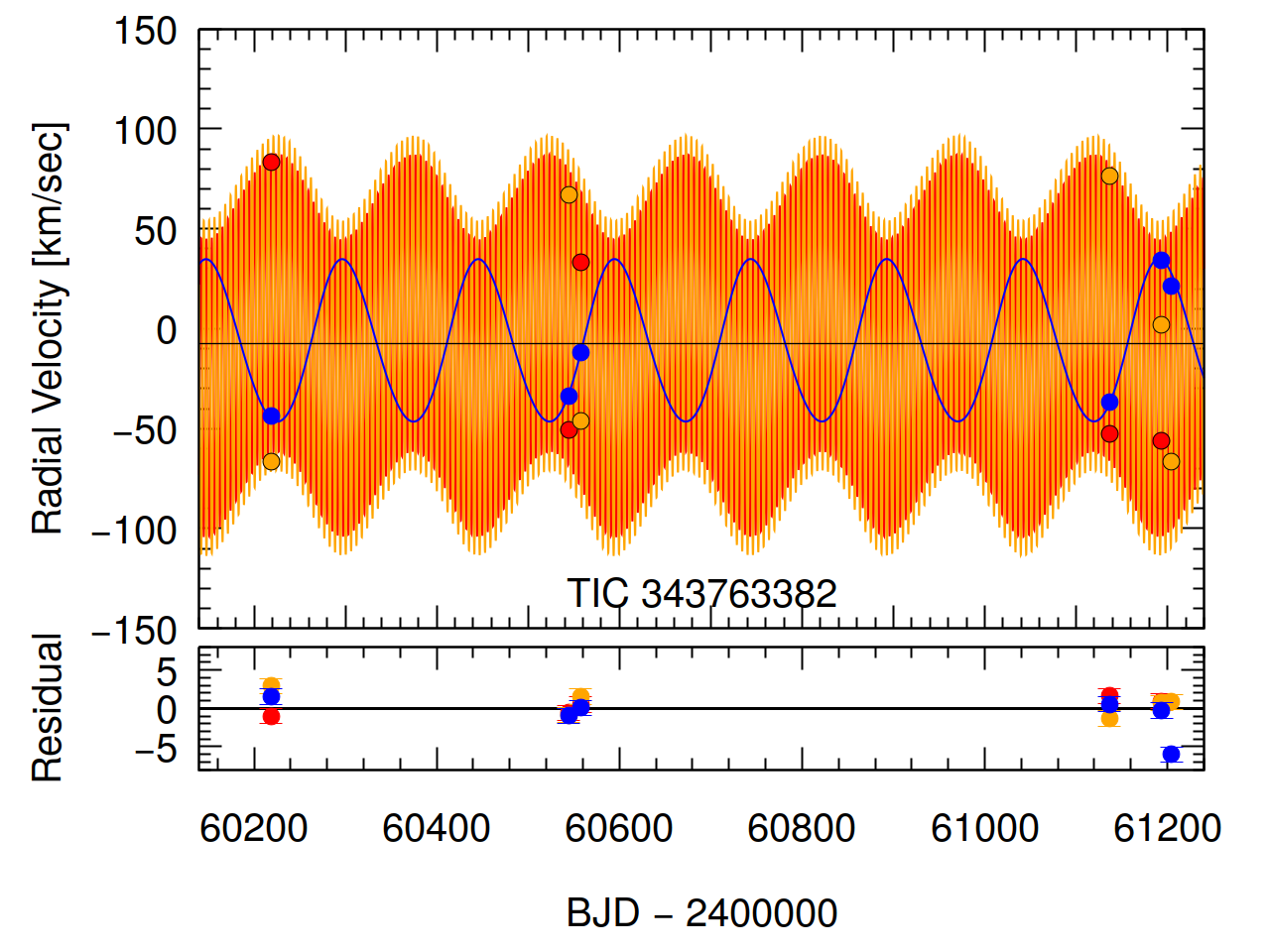}\includegraphics[width=0.33\textwidth]{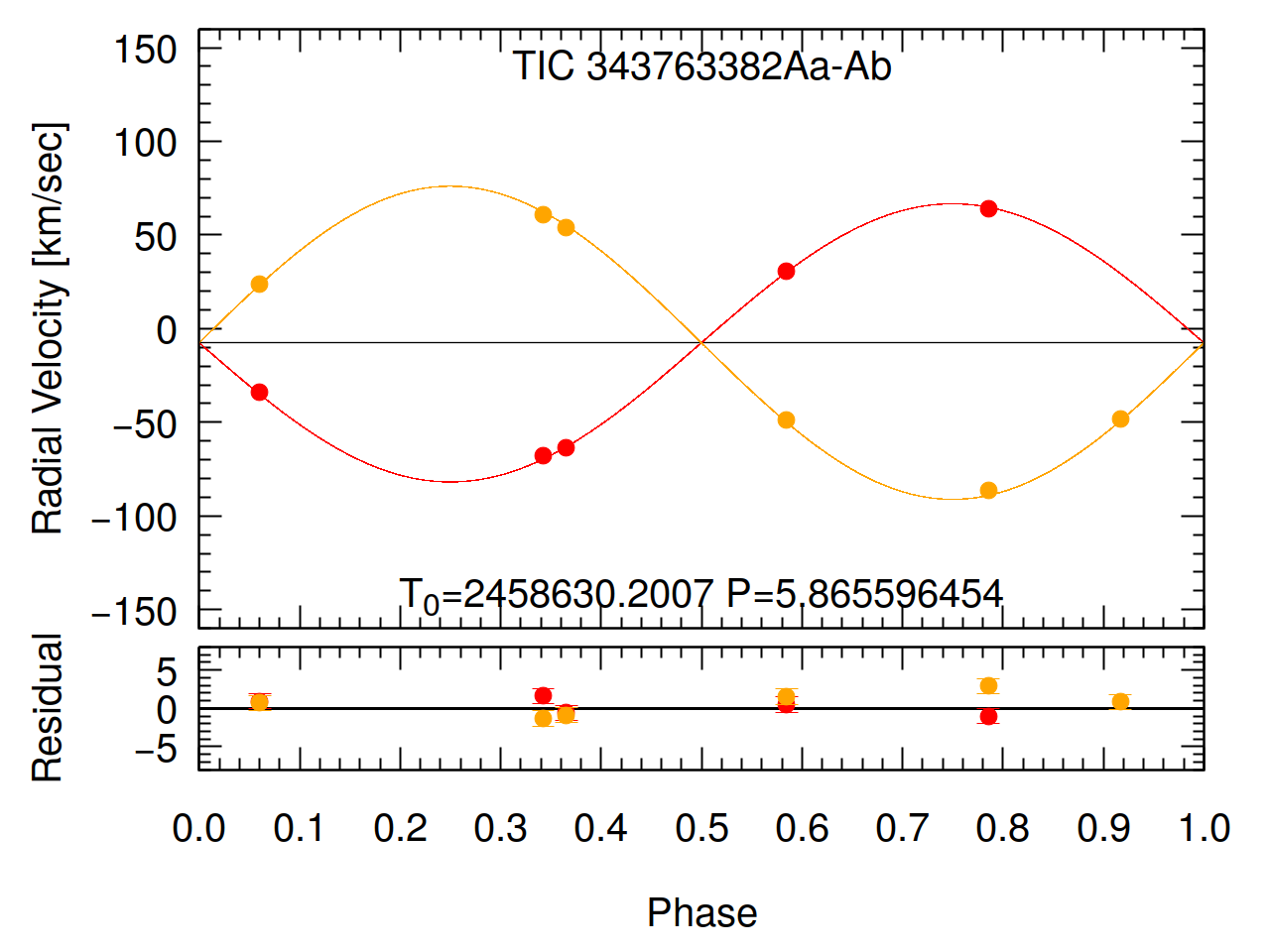}\includegraphics[width=0.33\textwidth]{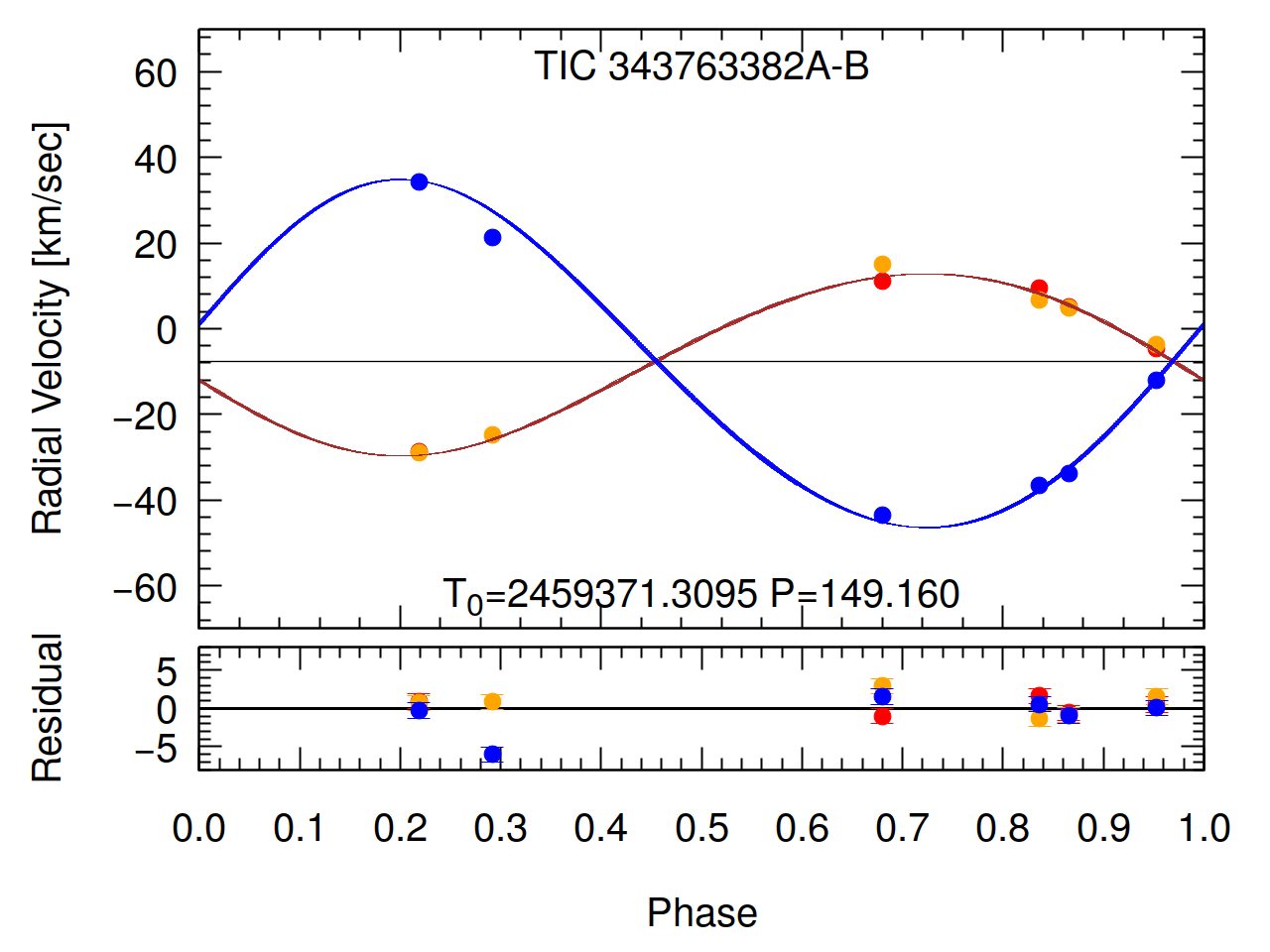}
\caption{The observed RV points (red circles for star Aa, orange ones for Ab, and blue circles for B, respectively), together with the best-fit spectro-photodynamical solution. {\it Left}: in the time domain; {\it middle}: phased according to the period of the inner binary orbit; {\it right}: phased according to the period of the outer orbit. (Note, in these latter two cases we removed the modelled effect of the motion in the other orbit. Note also, that the lower panels show the residuals from the fit, where the a priori estimated observational errors are also denoted.)}
\label{fig:T343763382RV}
\end{figure*}

\begin{figure*}[ht]
\includegraphics[width=0.49\textwidth]{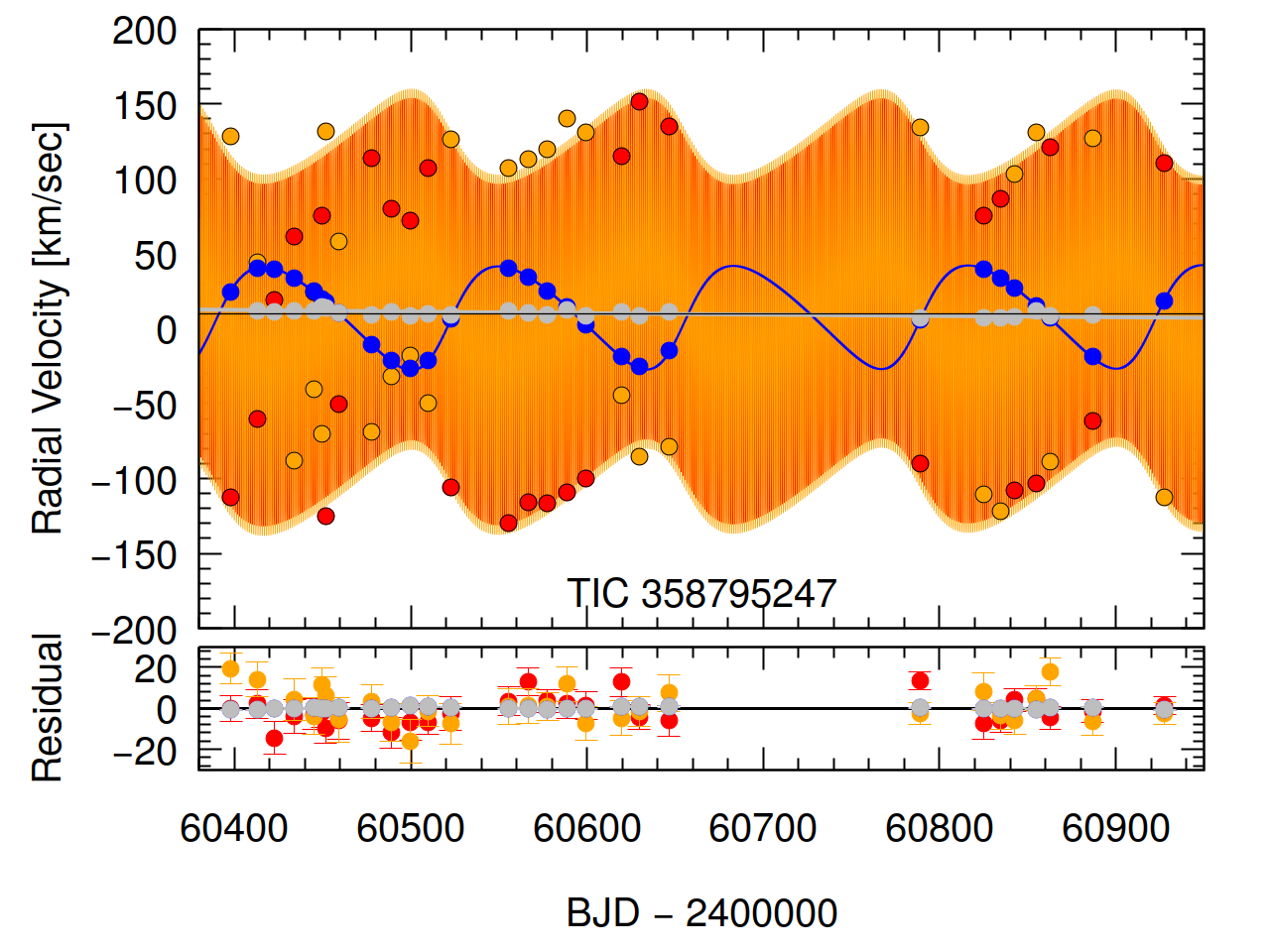}\includegraphics[width=0.49\textwidth]{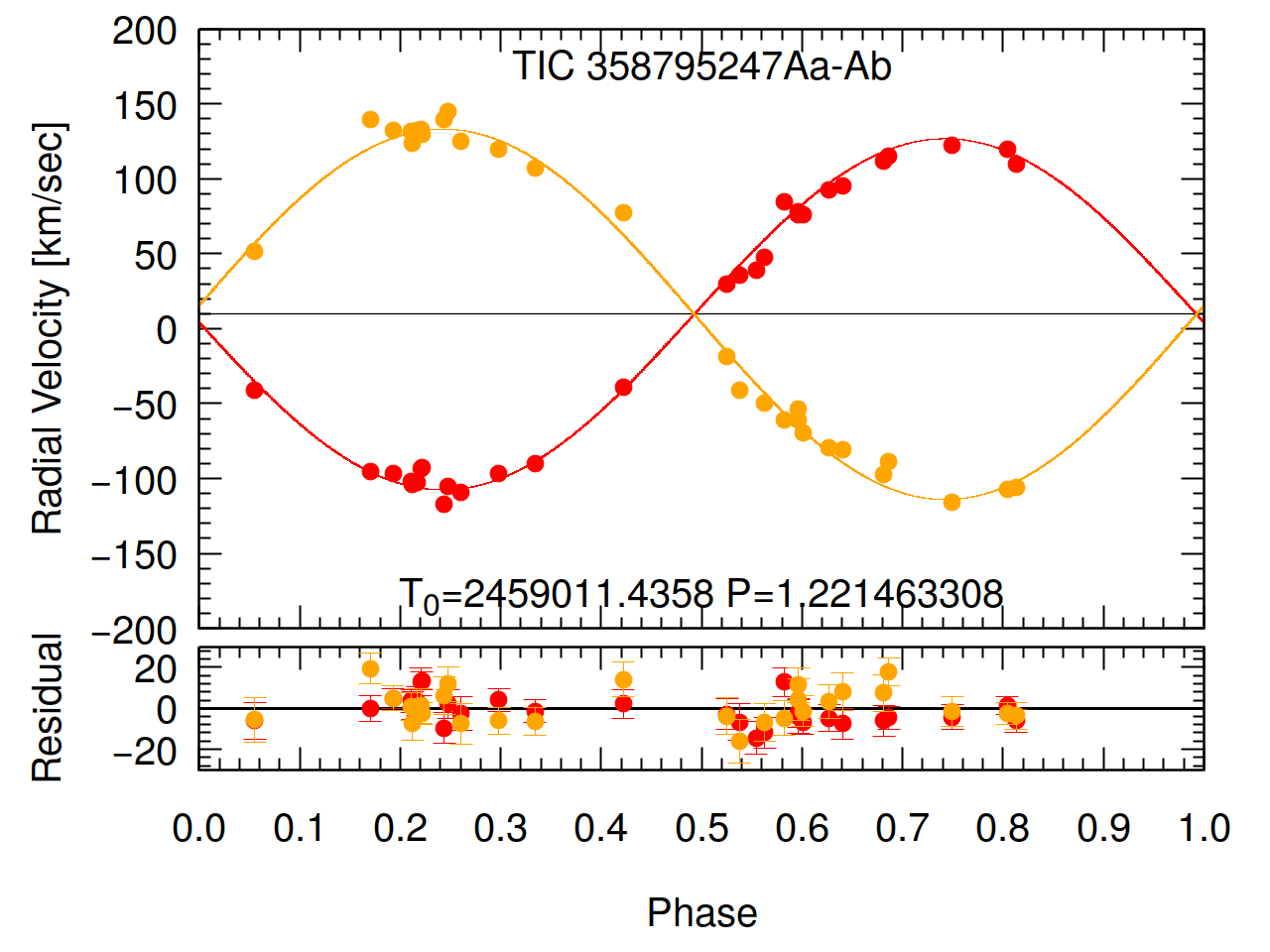}
\includegraphics[width=0.49\textwidth]{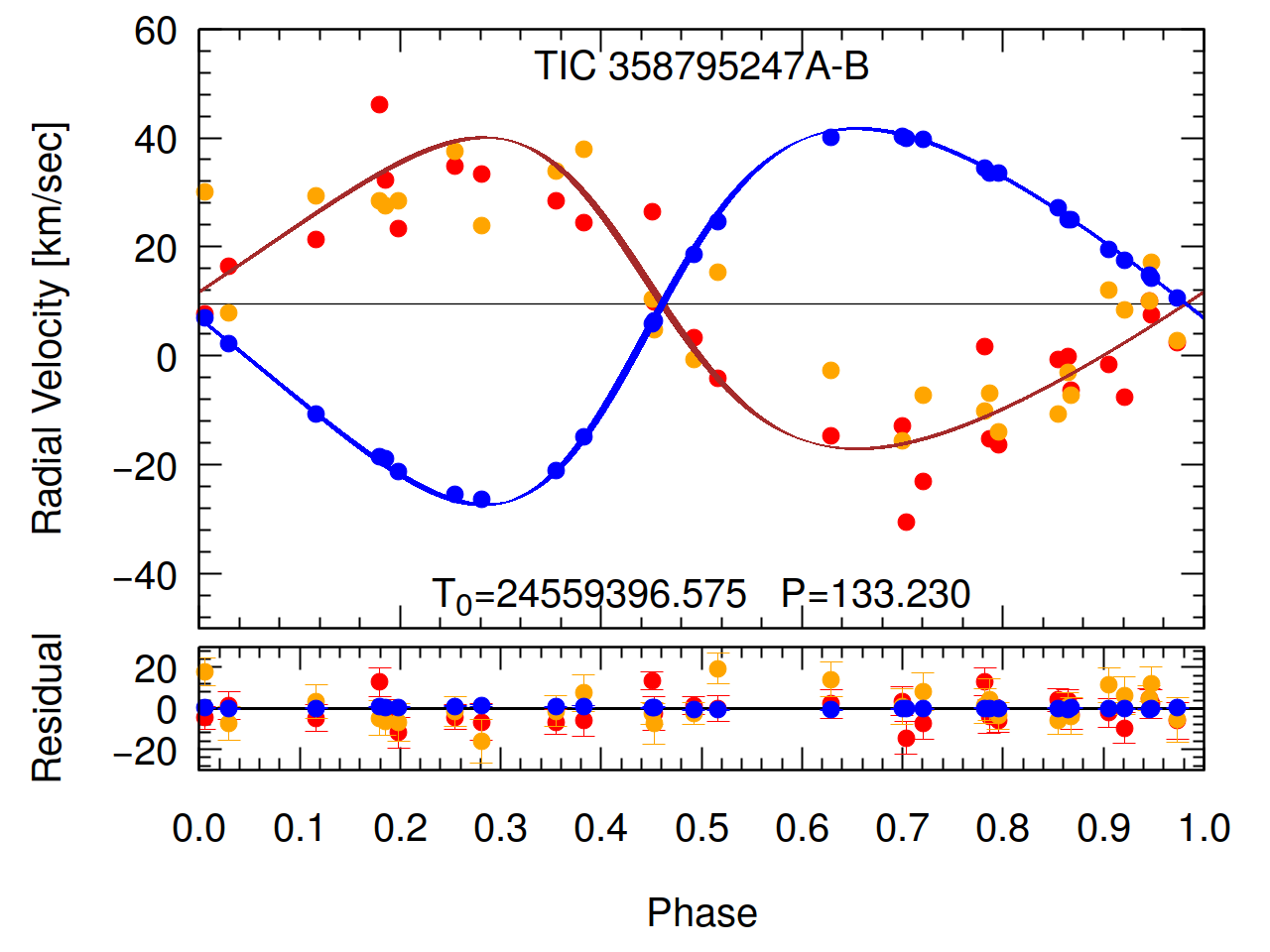}\includegraphics[width=0.49\textwidth]{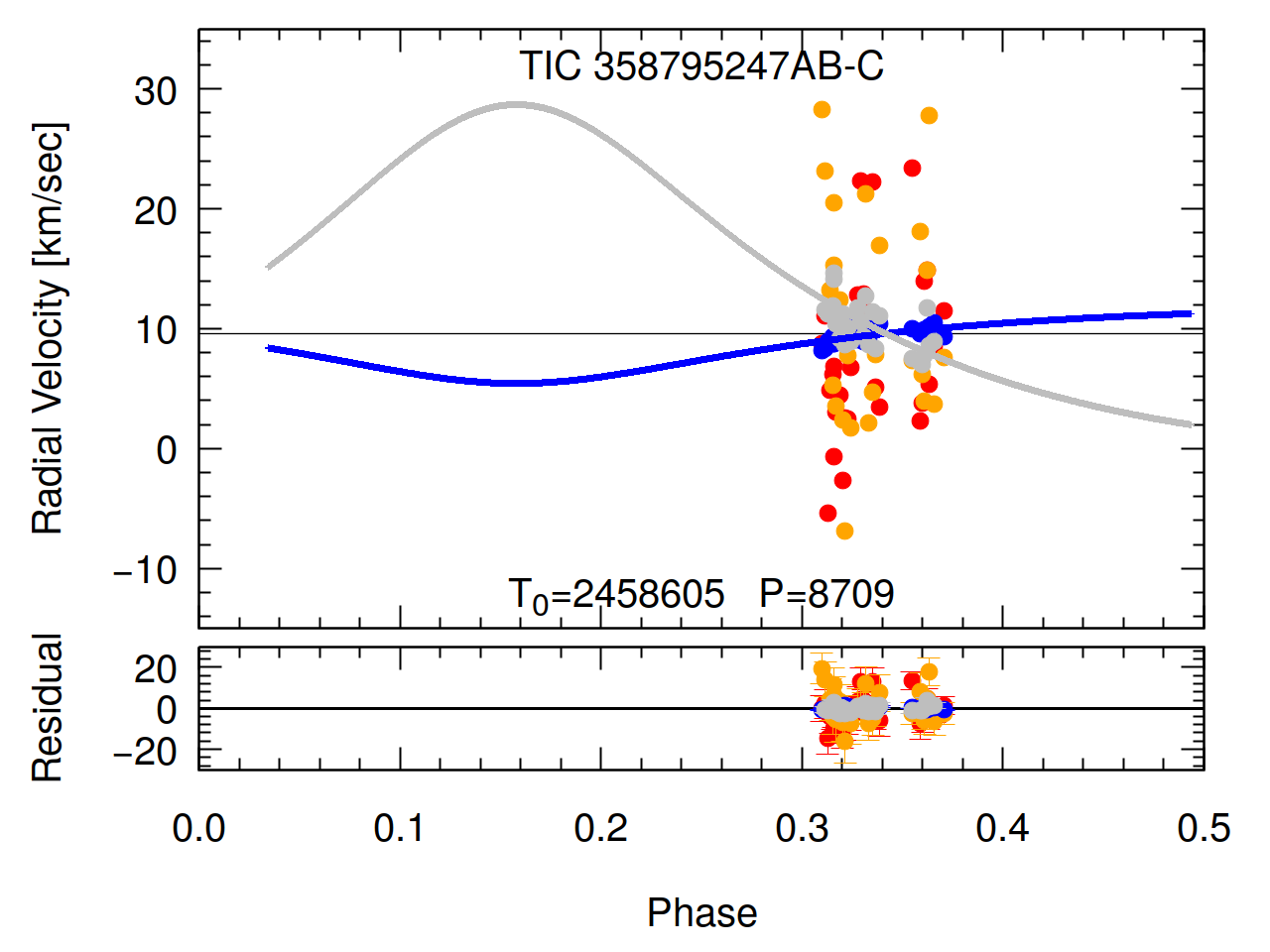}
\caption{The observed RV points (with red, orange, blue and gray circles for stars Aa, Ab, B, and C, respectively), together with the best-fit spectro-photodynamical solution. {\it Upper left}: in the time domain; {\it upper right}: phased according to the period of the inner binary orbit; {\it lower left}: phased according to the period of the middle orbit;  {\it lower right}: phased according to the period of the outer orbit. (Note, in the cases of the phase-folded plots, we removed the  modelled effects of the motions of the other orbits. Note also that the lower panels show the residuals from the fits, where the a priori estimated observational errors are also denoted.)}
\label{fig:T358795247RV}
\end{figure*}

\FloatBarrier
\onecolumn
\section{Use of the triple-lined spectroscopic triple, TIC\,343763382, to illustrate how stellar evolution and SED models may be used as proxies, in the absence of RV measurements}
\label{app:proxy}

In \citet{borkovitsetal22a} we carried out (spectro-)photodynamical analyses for TIC\,193993801 in four different ways. These were, as follows: (i) an astrophysical model-independent way without RV curves (denoted as `MIN'), that is, when only light and ETV curves were available; (ii) another, also astrophysical model-independent way where, however, RV data were available (`MIR'), that is, light curves, ETV curves and RV curves were analysed simultaneously; (iii) an astrophysical model-dependent method without RV curves (`MDN'), where the light curves, ETV curves, and net SED data were analysed simultaneously, together with built-in \texttt{PARSEC} stellar isochrones; (iv) and, finally, we also applied such an astrophysical model-dependent method, where the RV curves were also added to the previous data (`MDR' method) and, all these data were analysed simultaneously. In that paper we concluded that, `in the absence of reliable RV measurements, the MDN model can be used as a `tolerably good substitute'.  On the other hand, it was found that an MDN solution led to lower accuracies, that is, an accuracy of 5-10 percent, instead of 1-2 percent. Comparing, however, the MIR and the MDR models (that is, the two where RVs are included either without or with the use of pre-built \texttt{PARSEC} iscohrone tables and net SED data), we found that a careful inclusion of these latter theoretical data did not significantly alter the model independent (MIR) results. 

In the current appendix we again demonstrate this finding for the triple-lined triple star system TIC\,343763382. In the \citet{borkovitsetal22a} paper, however, we relaxed the strictly coeval condition for the three stars in the investigated triple system, TIC\,193993801, allowing slightly different ages for the three members of the triple system, in the MDR method. (Allowing different evolutionary ages for the three stars, might be allowable with sequential star formation from a given pre-stellar disk.) Here, however, we keep the strict coevality of the three components of the investigated triples, and investigate how the former findings change when we require this condition. 

We chose TIC\,343763382 because we have the necessary information to obtain both the model-independent MIR solution and the (astrophysical-)model-dependent MDN and MDR solutions. Naturally, the two different kinds of solutions require different variables to be adjusted. Moreover, in the model-independent MIR case, we need some a priori information, which is to be set independently. This is the temperature of one of the stars. Therefore, we should set the effective temperature of star Aa ($T_\mathrm{Aa}$) using some external information. Then, the temperatures of components Ab and B can be set relative to star Aa. (Therefore, in the MIR model the [logarithms of the] temperature ratios should be adjusted, instead of constraining the temperature itself from the built-in \texttt{PARSEC}-tables, as these are given in the MDN and MDR cases.) Furthermore, instead of the physical stellar radii, which are constrained from the same tables in the MDN and MDR models, in the MIR case, these should be replaced with the (adjusted) fractional radii of each star or, as was found to be more useful, with the sum and the ratio of the fractional radii in the case of the inner pair. These different sets of the variables which should be constrained or adjusted in the model-independent and model-dependent-runs are also described e.g., in Section 3 of \citet{borkovitsetal22a}.

In what follows, in Table\,\ref{tab:MDRvsMIR} we list simply the main astrophysical parameters as they were obtained from the three methods: MDR, MDN and MIR.  For five of our eight triples we applied the MDN method, while for the remaining three (where RVs were available), we used the MDR method. Here we demonstrate that the three methods give similar results within a few percent for this one illustrative extensively investigated triple system, TIC\,343763382. Interestingly, however, in this well studied case, the MDN masses are closer to the MIR ones than the masses obtained with the MDR method, though even these latter ones differ only slightly, within 2-3 percent. Therefore, we again conclude that the MDN method can be used as a satisfactory proxy for the stellar masses in those cases where RV data are not available.

\begin{table}[ht]
\centering
\caption{Comparison of the main astrophysical parameters obtained with different methods}
\small
\begin{tabular}{lccc}
\hline
Parameter & MDR & MIR & MDN\\
\hline
$M_\mathrm{Aa}$\tablefootmark{a}&$1.272_{-0.020}^{+0.022}$&$1.292_{-0.020}^{+0.012}$&$1.298_{-0.026}^{+0.019}$\\
$M_\mathrm{Ab}$\tablefootmark{b}&$1.132_{-0.018}^{+0.021}$&$1.161_{-0.018}^{+0.013}$&$1.153_{-0.024}^{+0.020}$\\
$M_\mathrm{B}$\tablefootmark{b}&$1.264_{-0.025}^{+0.021}$&$1.287_{-0.027}^{+0.024}$&$1.273_{-0.025}^{+0.034}$\\
$R_\mathrm{Aa}$&$1.313_{-0.008}^{+0.009}$\tablefootmark{c}&$1.346_{-0.017}^{+0.010}$\tablefootmark{b}&$1.324_{-0.011}^{+0.010}$\tablefootmark{c}\\
$R_\mathrm{Ab}$&$1.084_{-0.012}^{+0.013}$\tablefootmark{c}&$1.047_{-0.011}^{+0.007}$\tablefootmark{b}&$1.100_{-0.019}^{+0.015}$\tablefootmark{c}\\
$R_\mathrm{B}$&$1.294_{-0.023}^{+0.021}$\tablefootmark{c}&$1.202_{-0.029}^{+0.030}$\tablefootmark{b}&$1.289_{-0.028}^{+0.035}$\tablefootmark{c}\\
$T_\mathrm{Aa}$&$6292_{-34}^{+42}$\tablefootmark{c}&$6293_{-4}^{+2}$\tablefootmark{d}&$6402_{-47}^{+34}$\tablefootmark{c}\\
$T_\mathrm{Ab}$&$5937_{-30}^{+32}$\tablefootmark{c}&$5937_{-15}^{+16}$\tablefootmark{b}&$6030_{-40}^{+33}$\tablefootmark{c}\\
$T_\mathrm{B}$&$6264_{-38}^{+52}$\tablefootmark{c}&$6249_{-86}^{+88}$\tablefootmark{b}&$6350_{-54}^{+59}$\tablefootmark{c}\\
$r_\mathrm{Aa}$\tablefootmark{b}&$0.0715_{-0.0005}^{+0.0005}$&$0.0728_{-0.0006}^{+0.0005}$&$0.0718_{-0.0005}^{+0.0004}$\\
$r_\mathrm{Ab}$\tablefootmark{b}&$0.0591_{-0.0004}^{+0.0004}$&$0.0567_{-0.0004}^{+0.0003}$&$0.0596_{-0.0006}^{+0.0005}$\\
$r_\mathrm{B}$&$0.0071_{-0.0001}^{+0.0001}$\tablefootmark{b}&$0.0065_{-0.0002}^{+0.0002}$\tablefootmark{a}&$0.0070_{-0.0002}^{+0.0001}$\tablefootmark{b}\\
$T_\mathrm{Ab}/T_\mathrm{Aa}$\tablefootmark{b}&$0.9436_{-0.0018}^{0.0018}$&$0.9436_{-0.0023}^{+0.0024}$&$0.9421_{-0.0019}^{+0.0019}$\\
$T_\mathrm{B}/T_\mathrm{Aa}$\tablefootmark{b}&$0.9963_{-0.0048}^{+0.0041}$&$0.9936_{-0.0141}^{+0.0134}$&$0.9929_{-0.0074}^{+0.0071}$\\
$q_\mathrm{in}$\tablefootmark{a}&$0.890_{-0.003}^{+0.003}$&$0.899_{-006}^{+0.005}$&$0.889_{-0.003}^{+0.004}$\\
$q_\mathrm{out}$\tablefootmark{a}&$0.525_{-0.005}^{+0.004}$&$0.526_{-0.008}^{+0.008}$&$0.522_{-0.008}^{+0.007}$\\
\hline
\label{tab:MDRvsMIR}  
\end{tabular}
\tablefoot{(a) Directly adjusted; (b) derived; (c) constrained; (d) directly adjusted using a normal prior around the a posteriori median $T_\mathrm{Aa}$ value of the MDR method.}  
\end{table}

\FloatBarrier
\onecolumn
\section{Radial velocity curves obtained for three of the eight triples}
\label{app:RVs}

In this appendix we tabulate the RV data for three of the eight triples. The spectra of TICs\,306690194 and 358795247 were taken with the TRES spectrograph, as was described in Sect.~\ref{Subsect: spectroscopy}. On the other hand, the spectra of TIC\,343763382 were taken with the CHIRON instrument (see also in Sect.~\ref{Subsect: spectroscopy}).

\begin{table}[ht]
\centering
\caption{TRES Radial Velocity Measurements of TIC\,306690194\tablefootmark{a,b}}
\small
\begin{tabular}{lcc}
\hline
Time\tablefootmark{b} & RV Aa & RV Ab \\
\hline
   60954.9434 &    $-50.26$  &   86.05  \\
   60985.9504 &    $-78.84$  &   49.60  \\
   61052.9301 &      83.78   & $-29.96$ \\
   61080.7971 &    $-63.82$  &   63.42  \\
   61101.7214 &      73.83   &$-117.16$ \\
   61136.7154 &    $-43.90$  &   82.76  \\
\hline
\label{tbl:306690194_RVs}  
\end{tabular}
\tablefoot{(a) All radial velocities are in km s$^{-1}$. (b) The RV errors were set to be 1.5 km s$^{-1}$. (c) BJD - 2,400,000.}  
\end{table}
 
\begin{table}[ht]
\centering
\caption{CHIRON Radial Velocity Measurements of TIC\,343763382\tablefootmark{a}}
\small
\begin{tabular}{lcccccc}
\hline
Time\tablefootmark{b} & RV Aa & Err RV Aa & RV Ab & Err RV Ab & RV B & Err RV B \\
\hline
   60218.5200 &    83.371  & 7.546  &   $-66.514$  & 8.203 & $-43.538$ & 6.848 \\
   60544.5291 &  $-50.452$ & 6.397  &     67.186   & 6.955 & $-33.734$ & 7.594 \\
   60557.5445 &    32.996  & 7.712  &   $-46.058$  & 6.440 & $-12.083$ & 6.310 \\
   61136.8190 &  $-52.481$ & 7.911  &     76.474   & 7.270 & $-36.657$ & 6.552 \\
   61193.8197 &  $-56.033$ & 7.128  &      1.952   & 6.750 &   34.196  & 6.259 \\
   61204.7106 &     ...    &  ...   &   $-66.595$  & 5.880 &   21.180  &12.624 \\
\hline
\label{tbl:343763382_RVs}  
\end{tabular}
\tablefoot{(a) All radial velocities are in km s$^{-1}$. (b) BJD - 2,400,000.}  
\end{table}

\begin{table}[ht]
\centering
\caption{TRES Radial Velocity Measurements of TIC\,358795247\tablefootmark{a}}
\small
\begin{tabular}{lcccccccc}
\hline
Time\tablefootmark{b} & RV Aa & Err RV Aa & RV Ab & Err RV Ab & RV B & Err RV B & RV C & Err RV C \\
\hline
   60398.0048 & $-112.47$ & 6.21  &   128.33   &   7.55 &   24.56  & 0.25 & .... & ... \\
   60412.9698 &  $-60.01$ & 7.03  &    44.05   &   8.54 &   40.08  & 0.28 & 11.6 & 1.5 \\
   60422.9032 &    19.44  & 7.87  &    ....    &   ...  &   39.89  & 0.31 & 11.2 & 1.5 \\
   60433.9476 &    61.20  & 8.23  &  $-88.31$  &  10.01 &   33.55  & 0.33 & 11.7 & 1.5 \\
   60444.8537 &    24.61  & 7.49  &  $-40.69$  &   9.10 &   24.92  & 0.30 & 11.9 & 1.5 \\
   60449.8265 &    75.23  & 6.98  &  $-70.02$  &   8.48 &   19.57  & 0.28 & 14.6 & 1.5 \\
   60451.8391 & $-124.98$ & 7.52  &   131.51   &   9.14 &   17.56  & 0.30 & 14.1 & 1.5 \\
   60458.9376 &  $-50.62$ & 8.92  &    57.94   &  10.84 &   10.66  & 0.35 & 10.4 & 1.5 \\
   60477.9576 &   113.81  & 6.69  &  $-69.10$  &   8.13 & $-10.74$ & 0.26 &  9.2 & 1.5 \\
   60488.8722 &    79.79  & 7.54  &  $-31.88$  &   9.17 & $-21.28$ & 0.30 & 11.2 & 1.5 \\
   60499.8348 &    72.36  & 8.81  &  $-17.98$  &  10.71 & $-26.35$ & 0.35 &  8.7 & 1.5 \\
   60509.6839 &   106.83  & 6.17  &  $-49.43$  &   7.50 & $-21.04$ & 0.24 & 10.1 & 1.5 \\ 
   60522.7041 & $-106.19$ & 8.14  &   126.53   &   9.89 &    6.48  & 0.32 &  8.9 & 1.5 \\
   60555.6307 & $-129.73$ & 7.13  &   107.32   &   8.67 &   40.26  & 0.28 & 11.7 & 1.5 \\
   60566.6282 & $-115.52$ & 6.89  &   113.10   &   8.38 &   34.47  & 0.27 & 10.6 & 1.5 \\
   60577.6100 & $-116.75$ & 5.57  &   119.60   &   6.77 &   24.99  & 0.22 &  9.3 & 1.5 \\
   60588.6475 & $-109.46$ & 6.96  &   140.21   &   8.46 &   14.23  & 0.28 & 12.7 & 1.5 \\
   60599.5972 & $-100.13$ & 6.67  &   130.76   &   8.11 &    2.27  & 0.26 &  8.7 & 1.5 \\
   60619.5922 &   114.89  & 7.06  &  $-44.25$  &   8.58 & $-18.55$ & 0.28 & 11.4 & 1.5 \\
   60629.5689 &   151.55  & 6.00  &  $-85.59$  &   7.30 & $-25.41$ & 0.24 &  8.3 & 1.5 \\
   60646.5858 &   135.20  & 7.43  &  $-79.10$  &   9.03 & $-14.84$ & 0.29 & 11.1 & 1.5 \\
   60788.9355 &  $-90.06$ & 4.34  &   133.96   &   5.27 &    5.93  & 0.17 &  7.5 & 1.5 \\
   60824.8696 &    75.34  & 7.60  & $-110.22$  &   9.24 &   39.75  & 0.30 &  7.2 & 1.5 \\
   60834.8521 &    86.71  & 5.39  & $-121.93$  &   6.55 &   33.55  & 0.21 &  7.0 & 1.5 \\
   60842.7736 & $-107.67$ & 5.38  &   102.87   &   6.53 &   27.14  & 0.21 &  7.8 & 1.5 \\
   60854.8598 & $-103.56$ & 4.86  &   130.72   &   5.91 &   14.87  & 0.19 & 11.7 & 1.5 \\
   60862.7912 &   120.87  & 5.66  &  $-88.53$  &   6.88 &    6.97  & 0.22 &  8.6 & 1.5 \\
   60886.7901 &  $-61.25$ & 5.62  &   127.11   &   6.83 & $-18.84$ & 0.22 &  8.9 & 1.5 \\
   60927.6728 &   110.08  & 4.53  & $-112.29$  &   5.51 &   18.67  & 0.18 & .... & ... \\
\hline
\label{tbl:358795247_RVs}  
\end{tabular}
\tablefoot{(a) All radial velocities are in km s$^{-1}$. (b) BJD - 2,400,000.}  
\end{table}

\FloatBarrier
\onecolumn
\section{Eclipse times of the inner EBs of the eight triples}
\label{app:ToMs}

In this appendix we tabulate the times of the individual primary and secondary eclipses of the inner EBs of the triples considered in this study.  These naturally include mostly eclipses from TESS, plus a few that were observed from the ground (Tables~\ref{Tab:TIC_077177148_ToM}--\ref{Tab:TIC_385933895_ToM}). In the final, A\&A version, we present only the list of times of minima for TIC 77177148. The other seven tables, together with this sample table, are available at the CDS, in machine readable form. Here, in the arXiv version, however, all the eight tables are published.

\begin{table*}[ht]
\caption{Times of minima of TIC\,077177148}
 \label{Tab:TIC_077177148_ToM}
\scalebox{0.88}{\begin{tabular}{@{}lrllrllrllrl}
\hline
BJD & Cycle  & std. dev. & BJD & Cycle  & std. dev. & BJD & Cycle  & std. dev. & BJD & Cycle  & std. dev. \\ 
$-2\,400\,000$ & no. &   \multicolumn{1}{c}{$(d)$} & $-2\,400\,000$ & no. &   \multicolumn{1}{c}{$(d)$} & $-2\,400\,000$ & no. &   \multicolumn{1}{c}{$(d)$} & $-2\,400\,000$ & no. &   \multicolumn{1}{c}{$(d)$} \\ 
\hline
58412.34848 &    -0.5 & 0.00082 & 58458.60343 &    13.5 & 0.00162 & 59183.79210 &   233.0 & 0.00039 & 61010.77794 &   786.0 & 0.00034  \\ 
58414.00637 &     0.0 & 0.00144 & 58460.26168 &    14.0 & 0.00079 & 59185.43637 &   233.5 & 0.00063 & 61012.43406 &   786.5 & 0.00038  \\ 
58415.65273 &     0.5 & 0.00094 & 58461.90917 &    14.5 & 0.00105 & 59187.09642 &   234.0 & 0.00050 & 61014.08262 &   787.0 & 0.00035  \\ 
58417.30755 &     1.0 & 0.00105 & 58463.56403 &    15.0 & 0.02012 & 59188.74227 &   234.5 & 0.00041 & 61015.73554 &   787.5 & 0.00040  \\ 
58422.25304 &     2.5 & 0.00073 & 59145.77758 &   221.5 & 0.00037 & 59190.40085 &   235.0 & 0.00050 & 61017.38534 &   788.0 & 0.00035  \\ 
58425.55885 &     3.5 & 0.00263 & 59147.43491 &   222.0 & 0.00044 & 59192.04479 &   235.5 & 0.00054 & 61019.04095 &   788.5 & 0.00059  \\ 
58427.21510 &     4.0 & 0.00229 & 59149.08167 &   222.5 & 0.00038 & 59193.70619 &   236.0 & 0.00045 & 61020.68852 &   789.0 & 0.00030  \\ 
58428.85990 &     4.5 & 0.00092 & 59150.74136 &   223.0 & 0.00037 & 59195.35120 &   236.5 & 0.00054 & 61022.34392 &   789.5 & 0.00031  \\ 
58430.51885 &     5.0 & 0.00095 & 59152.38832 &   223.5 & 0.00066 & 59197.00635 &   237.0 & 0.00043 & 61023.99486 &   790.0 & 0.00028  \\ 
58432.16532 &     5.5 & 0.00229 & 59154.04388 &   224.0 & 0.00042 & 59198.65340 &   237.5 & 0.00046 & 61025.64638 &   790.5 & 0.00037  \\ 
58433.82168 &     6.0 & 0.00116 & 59155.69160 &   224.5 & 0.00039 & 60989.29598 &   779.5 & 0.00054 & 61027.29635 &   791.0 & 0.00031  \\ 
58435.46753 &     6.5 & 0.00081 & 59158.99625 &   225.5 & 0.00049 & 60990.94922 &   780.0 & 0.00050 & 61028.95063 &   791.5 & 0.00032  \\ 
58438.77218 &     7.5 & 0.00152 & 59160.65607 &   226.0 & 0.00044 & 60992.60429 &   780.5 & 0.00039 & 61030.59970 &   792.0 & 0.00030  \\ 
58440.42913 &     8.0 & 0.00095 & 59162.30292 &   226.5 & 0.00068 & 60994.25591 &   781.0 & 0.00050 & 61032.25236 &   792.5 & 0.00032  \\ 
58442.07575 &     8.5 & 0.00080 & 59163.96143 &   227.0 & 0.00054 & 60995.91176 &   781.5 & 0.00033 & 61033.90031 &   793.0 & 0.00028  \\ 
58443.73306 &     9.0 & 0.00059 & 59165.61046 &   227.5 & 0.00041 & 60997.56200 &   782.0 & 0.00038 & 61035.55620 &   793.5 & 0.00037  \\ 
58445.38136 &     9.5 & 0.00071 & 59167.27026 &   228.0 & 0.00058 & 60999.21739 &   782.5 & 0.00053 & 61037.20523 &   794.0 & 0.00035  \\ 
58447.04026 &    10.0 & 0.00240 & 59168.91489 &   228.5 & 0.00044 & 61000.86429 &   783.0 & 0.00048 & 61038.85905 &   794.5 & 0.00120  \\ 
58448.68477 &    10.5 & 0.00103 & 59175.52403 &   230.5 & 0.00034 & 61002.52014 &   783.5 & 0.00038 & 61040.50766 &   795.0 & 0.00034  \\ 
58451.99390 &    11.5 & 0.00089 & 59177.18487 &   231.0 & 0.00049 & 61004.16936 &   784.0 & 0.00057 & 61042.16011 &   795.5 & 0.00030  \\ 
58453.65096 &    12.0 & 0.00078 & 59178.83038 &   231.5 & 0.00037 & 61005.82529 &   784.5 & 0.00050 & 61043.81094 &   796.0 & 0.00030  \\ 
58455.29824 &    12.5 & 0.00105 & 59180.49015 &   232.0 & 0.00053 & 61007.47436 &   785.0 & 0.00037 & 61045.46189 &   796.5 & 0.00034  \\ 
58456.95770 &    13.0 & 0.00104 & 59182.13364 &   232.5 & 0.00054 & 61009.13008 &   785.5 & 0.00049  \\ 
\hline
\end{tabular}}
\textit{Notes}. Integer and half-integer cycle numbers denote primary and secondary eclipses, respectively.
\end{table*}

\begin{table*}
\caption{Times of minima of TIC\,220456364}
 \label{Tab:TIC_220456364_ToM}
\scalebox{0.88}{\begin{tabular}{@{}lrllrllrllrl}
\hline
BJD & Cycle  & std. dev. & BJD & Cycle  & std. dev. & BJD & Cycle  & std. dev. & BJD & Cycle  & std. dev. \\ 
$-2\,400\,000$ & no. &   \multicolumn{1}{c}{$(d)$} & $-2\,400\,000$ & no. &   \multicolumn{1}{c}{$(d)$} & $-2\,400\,000$ & no. &   \multicolumn{1}{c}{$(d)$} & $-2\,400\,000$ & no. &   \multicolumn{1}{c}{$(d)$} \\ 
\hline
58356.39560 &     0.0 & 0.00112 & 58642.56800 &    22.5 & 0.00092 & 59278.57334 &    72.5 & 0.00117 & 60149.79298 &   141.0 & 0.00087  \\ 
58362.77851 &     0.5 & 0.00084 & 58648.92738 &    23.0 & 0.00093 & 59284.91310 &    73.0 & 0.00062 & 60156.17737 &   141.5 & 0.00201  \\ 
58369.12751 &     1.0 & 0.00124 & 58655.31307 &    23.5 & 0.00080 & 59291.29801 &    73.5 & 0.00092 & 60162.51689 &   142.0 & 0.00071  \\ 
58375.50674 &     1.5 & 0.00098 & 58661.65837 &    24.0 & 0.00142 & 59297.63710 &    74.0 & 0.00080 & 60187.96634 &   144.0 & 0.00087  \\ 
58388.23876 &     2.5 & 0.00052 & 58674.38921 &    25.0 & 0.00109 & 59310.35906 &    75.0 & 0.00068 & 60194.29334 &   144.5 & 0.00529  \\ 
58394.58949 &     3.0 & 0.00085 & 58680.77863 &    25.5 & 0.00136 & 59316.73276 &    75.5 & 0.00099 & 60200.68582 &   145.0 & 0.00069  \\ 
58400.96524 &     3.5 & 0.00131 & 59036.91175 &    53.5 & 0.00074 & 59323.07379 &    76.0 & 0.00076 & 60207.03864 &   145.5 & 0.00267  \\ 
58413.68756 &     4.5 & 0.00121 & 59043.25402 &    54.0 & 0.00069 & 59329.44921 &    76.5 & 0.00088 & 60664.85715 &   181.5 & 0.00301  \\ 
58426.40762 &     5.5 & 0.00067 & 59049.62761 &    54.5 & 0.00067 & 59367.58095 &    79.5 & 0.00112 & 60671.20052 &   182.0 & 0.00085  \\ 
58432.75907 &     6.0 & 0.00107 & 59055.96554 &    55.0 & 0.00071 & 59373.92319 &    80.0 & 0.00071 & 60677.58920 &   182.5 & 0.00247  \\ 
58439.12422 &     6.5 & 0.00064 & 59094.10019 &    58.0 & 0.00062 & 59380.30506 &    80.5 & 0.00101 & 60683.93295 &   183.0 & 0.00082  \\ 
58445.47361 &     7.0 & 0.00091 & 59100.46755 &    58.5 & 0.00158 & 59386.63396 &    81.0 & 0.00066 & 60747.58023 &   188.0 & 0.00131  \\ 
58451.83820 &     7.5 & 0.00073 & 59106.81301 &    59.0 & 0.00071 & 60016.21215 &   130.5 & 0.00117 & 60753.94613 &   188.5 & 0.00301  \\ 
58458.18784 &     8.0 & 0.00084 & 59113.18595 &    59.5 & 0.00094 & 60022.55447 &   131.0 & 0.00056 & 60766.68431 &   189.5 & 0.00257  \\ 
58470.89987 &     9.0 & 0.00094 & 59119.52308 &    60.0 & 0.00066 & 60028.92878 &   131.5 & 0.00145 & 60836.60436 &   195.0 & 0.00124  \\ 
58483.61107 &    10.0 & 0.00105 & 59125.89362 &    60.5 & 0.00088 & 60035.26309 &   132.0 & 0.00061 & 60849.32866 &   196.0 & 0.00117  \\ 
58489.96970 &    10.5 & 0.00124 & 59132.23375 &    61.0 & 0.00048 & 60041.64201 &   132.5 & 0.00139 & 60862.03989 &   197.0 & 0.00122  \\ 
58496.32339 &    11.0 & 0.00089 & 59138.60358 &    61.5 & 0.00044 & 60047.97818 &   133.0 & 0.00067 & 60874.75497 &   198.0 & 0.00185  \\ 
58502.68571 &    11.5 & 0.00124 & 59144.94608 &    62.0 & 0.00067 & 60054.34102 &   133.5 & 0.00126 & 60887.46506 &   199.0 & 0.00130  \\ 
58509.03509 &    12.0 & 0.00103 & 59151.31756 &    62.5 & 0.00067 & 60060.69489 &   134.0 & 0.00060 & 60900.17618 &   200.0 & 0.00117  \\ 
58515.39591 &    12.5 & 0.00091 & 59164.03225 &    63.5 & 0.00089 & 60067.07052 &   134.5 & 0.00159 & 60912.88626 &   201.0 & 0.00131  \\ 
58547.16831 &    15.0 & 0.00098 & 59176.74773 &    64.5 & 0.00062 & 60073.40821 &   135.0 & 0.00073 & 60925.59551 &   202.0 & 0.00094  \\ 
58553.52671 &    15.5 & 0.00120 & 59183.08651 &    65.0 & 0.00065 & 60079.78388 &   135.5 & 0.00116 & 60938.30616 &   203.0 & 0.00131  \\ 
58559.88129 &    16.0 & 0.00129 & 59189.45944 &    65.5 & 0.00090 & 60092.50769 &   136.5 & 0.00156 & 60951.03069 &   204.0 & 0.00115  \\ 
58566.24473 &    16.5 & 0.00189 & 59195.80447 &    66.0 & 0.00052 & 60098.85780 &   137.0 & 0.00078 & 60963.74829 &   205.0 & 0.00107  \\ 
58572.59376 &    17.0 & 0.00128 & 59202.18222 &    66.5 & 0.00093 & 60105.23661 &   137.5 & 0.00226 & 60976.46379 &   206.0 & 0.00138  \\ 
58578.95599 &    17.5 & 0.00238 & 59208.52429 &    67.0 & 0.00061 & 60111.60285 &   138.0 & 0.00071 & 60989.20496 &   207.0 & 0.00138  \\ 
58585.30769 &    18.0 & 0.00107 & 59221.25288 &    68.0 & 0.00060 & 60118.00712 &   138.5 & 0.00160 & 61001.94290 &   208.0 & 0.00111  \\ 
58591.66900 &    18.5 & 0.00122 & 59259.44726 &    71.0 & 0.00064 & 60130.73140 &   139.5 & 0.00097 & 61014.66912 &   209.0 & 0.00118  \\ 
58629.83070 &    21.5 & 0.00061 & 59265.84915 &    71.5 & 0.00093 & 60137.05643 &   140.0 & 0.00098 & 61027.39775 &   210.0 & 0.00132  \\ 
58636.18874 &    22.0 & 0.00098 & 59272.18145 &    72.0 & 0.00069 & 60143.45492 &   140.5 & 0.00105 & 61040.12905 &   211.0 & 0.00118  \\ 
\hline
\end{tabular}}
\textit{Notes}. Integer and half-integer cycle numbers denote primary and secondary eclipses, respectively.
\end{table*}

\begin{table*}
\caption{Times of minima of TIC\,232874003}
 \label{Tab:TIC_232874003_ToM}
\scalebox{0.88}{\begin{tabular}{@{}lrllrllrllrl}
\hline
BJD & Cycle  & std. dev. & BJD & Cycle  & std. dev. & BJD & Cycle  & std. dev. & BJD & Cycle  & std. dev. \\ 
$-2\,400\,000$ & no. &   \multicolumn{1}{c}{$(d)$} & $-2\,400\,000$ & no. &   \multicolumn{1}{c}{$(d)$} & $-2\,400\,000$ & no. &   \multicolumn{1}{c}{$(d)$} & $-2\,400\,000$ & no. &   \multicolumn{1}{c}{$(d)$} \\ 
\hline
58625.14466 &    -0.5 & 0.00095 & 59378.79966 &   147.5 & 0.00020 & 60830.10202 &   432.5 & 0.00015 & 61166.19351 &   498.5 & 0.00012  \\ 
58627.68765 &     0.0 & 0.00120 & 59381.34410 &   148.0 & 0.00029 & 60832.64843 &   433.0 & 0.00011 & 61168.73905 &   499.0 & 0.00012  \\ 
58630.23584 &     0.5 & 0.00092 & 59383.89197 &   148.5 & 0.00037 & 60835.19468 &   433.5 & 0.00013 & 61171.28482 &   499.5 & 0.00023  \\ 
58632.78150 &     1.0 & 0.00070 & 59386.43721 &   149.0 & 0.00050 & 60837.73960 &   434.0 & 0.00012 & 61173.83127 &   500.0 & 0.00013  \\ 
58635.32875 &     1.5 & 0.00058 & 59388.98543 &   149.5 & 0.00046 & 60840.28738 &   434.5 & 0.00011 & 61176.37781 &   500.5 & 0.00012  \\ 
58640.42219 &     2.5 & 0.00031 & 60099.36263 &   289.0 & 0.00014 & 60845.37874 &   435.5 & 0.00014 & 61178.92393 &   501.0 & 0.00017  \\ 
58642.96769 &     3.0 & 0.00037 & 60101.90929 &   289.5 & 0.00014 & 60847.92587 &   436.0 & 0.00011 & 61181.47072 &   501.5 & 0.00011  \\ 
58645.51539 &     3.5 & 0.00044 & 60104.45429 &   290.0 & 0.00012 & 60850.47193 &   436.5 & 0.00012 & 61184.01649 &   502.0 & 0.00012  \\ 
58648.06016 &     4.0 & 0.00045 & 60107.00125 &   290.5 & 0.00019 & 60853.01737 &   437.0 & 0.00012 & 61186.56337 &   502.5 & 0.00012  \\ 
58650.60742 &     4.5 & 0.00063 & 60109.54561 &   291.0 & 0.00015 & 61153.46180 &   496.0 & 0.00011 & 61189.10926 &   503.0 & 0.00011  \\ 
59366.06804 &   145.0 & 0.00020 & 60112.09276 &   291.5 & 0.00021 & 61156.00847 &   496.5 & 0.00012 & 61194.20220 &   504.0 & 0.00010  \\ 
59368.61469 &   145.5 & 0.00016 & 60114.63693 &   292.0 & 0.00012 & 61158.55448 &   497.0 & 0.00012 & 61196.74932 &   504.5 & 0.00014  \\ 
59371.15986 &   146.0 & 0.00014 & 60117.18370 &   292.5 & 0.00014 & 61161.10108 &   497.5 & 0.00011 & 61199.29544 &   505.0 & 0.00010  \\ 
59373.70752 &   146.5 & 0.00014 & 60119.72909 &   293.0 & 0.00015 & 61163.64671 &   498.0 & 0.00010 & 61201.84271 &   505.5 & 0.00012  \\ 
59376.25259 &   147.0 & 0.00012 & 60122.27514 &   293.5 & 0.00021  \\ 
\hline
\end{tabular}}
\textit{Notes}. Integer and half-integer cycle numbers denote primary and secondary eclipses, respectively.
\end{table*}

\begin{table*}
\caption{Times of minima of TIC\,306690194}
 \label{Tab:TIC_306690194_ToM}
\scalebox{0.88}{\begin{tabular}{@{}lrllrllrllrl}
\hline
BJD & Cycle  & std. dev. & BJD & Cycle  & std. dev. & BJD & Cycle  & std. dev. & BJD & Cycle  & std. dev. \\ 
$-2\,400\,000$ & no. &   \multicolumn{1}{c}{$(d)$} & $-2\,400\,000$ & no. &   \multicolumn{1}{c}{$(d)$} & $-2\,400\,000$ & no. &   \multicolumn{1}{c}{$(d)$} & $-2\,400\,000$ & no. &   \multicolumn{1}{c}{$(d)$} \\ 
\hline
59475.54341 &     0.0 & 0.00008 & 59515.91050 &     6.5 & 0.00005 & 60236.59855 &   122.5 & 0.00004 & 60280.07168 &   129.5 & 0.00004  \\ 
59478.64853 &     0.5 & 0.00007 & 59519.01619 &     7.0 & 0.00005 & 60242.80995 &   123.5 & 0.00003 & 60283.17751 &   130.0 & 0.00004  \\ 
59481.75494 &     1.0 & 0.00006 & 59522.12075 &     7.5 & 0.00006 & 60245.91598 &   124.0 & 0.00004 & 60646.63131$^a$&   188.5 & 0.00004  \\ 
59487.96564 &     2.0 & 0.00005 & 59528.33148 &     8.5 & 0.00006 & 60252.12605 &   125.0 & 0.00004 & 60671.48862$^a$&   192.5 & 0.00047  \\ 
59491.06997 &     2.5 & 0.00005 & 59531.43789 &     9.0 & 0.00005 & 60255.23089 &   125.5 & 0.00004 & 60730.48597$^a$&   202.0 & 0.00009  \\ 
59494.17587 &     3.0 & 0.00006 & 59534.54254 &     9.5 & 0.00005 & 60258.33591 &   126.0 & 0.00005 & 60755.35860$^a$&   206.0 & 0.00027  \\ 
59497.28028 &     3.5 & 0.00005 & 59537.64930 &    10.0 & 0.00005 & 60261.44121 &   126.5 & 0.00006 & 61019.38967$^a$&   248.5 & 0.00066  \\ 
59500.38619 &     4.0 & 0.00007 & 59540.75448 &    10.5 & 0.00006 & 60264.54673 &   127.0 & 0.00006 & 61062.86659 &   255.5 & 0.00004  \\ 
59503.48723 &     4.5 & 0.00016 & 59543.86131 &    11.0 & 0.00007 & 60267.65090 &   127.5 & 0.00004 & 61103.26994$^a$&   262.0 & 0.00007  \\ 
59506.59629 &     5.0 & 0.00005 & 59546.96706 &    11.5 & 0.00006 & 60270.75599 &   128.0 & 0.00004 & 61106.37623$^a$&   262.5 & 0.00004  \\ 
59509.70029 &     5.5 & 0.00005 & 59550.07423 &    12.0 & 0.00006 & 60276.96667 &   129.0 & 0.00004 & 61134.32331$^a$&   267.0 & 0.00007  \\ 
\hline
\end{tabular}}
\textit{Notes}. Integer and half-integer cycle numbers denote primary and secondary eclipses, respectively. The majority of the mid-times of minima were collected with TESS, with the following exceptions: $^a:$ observed by R.\,U.
\end{table*}

\begin{table*}
\caption{Times of minima of TIC\,343763382}
 \label{Tab:TIC_343763382_ToM}
\scalebox{0.88}{\begin{tabular}{@{}lrllrllrllrl}
\hline
BJD & Cycle  & std. dev. & BJD & Cycle  & std. dev. & BJD & Cycle  & std. dev. & BJD & Cycle  & std. dev. \\ 
$-2\,400\,000$ & no. &   \multicolumn{1}{c}{$(d)$} & $-2\,400\,000$ & no. &   \multicolumn{1}{c}{$(d)$} & $-2\,400\,000$ & no. &   \multicolumn{1}{c}{$(d)$} & $-2\,400\,000$ & no. &   \multicolumn{1}{c}{$(d)$} \\ 
\hline
58627.26566 &    -0.5 & 0.00019 & 59369.26571 &   126.0 & 0.00003 & 60844.46029 &   377.5 & 0.00003 & 61123.07718 &   425.0 & 0.00003  \\ 
58630.20198 &     0.0 & 0.00011 & 59372.19761 &   126.5 & 0.00002 & 60847.39455 &   378.0 & 0.00003 & 61126.00846 &   425.5 & 0.00010  \\ 
58633.13316 &     0.5 & 0.00005 & 59378.06404 &   127.5 & 0.00003 & 60850.32645 &   378.5 & 0.00003 & 61128.94264 &   426.0 & 0.00003  \\ 
58636.06795 &     1.0 & 0.00009 & 59380.99847 &   128.0 & 0.00004 & 60853.26068 &   379.0 & 0.00004 & 61131.87430 &   426.5 & 0.00003  \\ 
58641.93389 &     2.0 & 0.00020 & 59383.93001 &   128.5 & 0.00003 & 61076.15395 &   417.0 & 0.00003 & 61134.80807 &   427.0 & 0.00003  \\ 
58644.86508 &     2.5 & 0.00005 & 59386.86444 &   129.0 & 0.00004 & 61079.08561 &   417.5 & 0.00003 & 61137.73991 &   427.5 & 0.00004  \\ 
58647.79926 &     3.0 & 0.00008 & 60099.52927 &   250.5 & 0.00002 & 61082.01918 &   418.0 & 0.00003 & 61140.67375 &   428.0 & 0.00003  \\ 
58650.73014 &     3.5 & 0.00005 & 60102.46401 &   251.0 & 0.00003 & 61084.95111 &   418.5 & 0.00003 & 61143.60562 &   428.5 & 0.00004  \\ 
58656.59516 &     4.5 & 0.00004 & 60105.39527 &   251.5 & 0.00002 & 61087.88472 &   419.0 & 0.00002 & 61146.53962 &   429.0 & 0.00004  \\ 
58659.52961 &     5.0 & 0.00008 & 60108.33030 &   252.0 & 0.00003 & 61090.81653 &   419.5 & 0.00003 & 61149.47176 &   429.5 & 0.00004  \\ 
58662.46032 &     5.5 & 0.00003 & 60111.26175 &   252.5 & 0.00002 & 61096.68212 &   420.5 & 0.00003 & 61158.27226 &   431.0 & 0.00002  \\ 
58665.39438 &     6.0 & 0.00016 & 60117.12823 &   253.5 & 0.00002 & 61099.61567 &   421.0 & 0.00004 & 61161.20478 &   431.5 & 0.00003  \\ 
58671.25988 &     7.0 & 0.00009 & 60120.06274 &   254.0 & 0.00003 & 61102.54732 &   421.5 & 0.00003 & 61164.13883 &   432.0 & 0.00003  \\ 
58674.19066 &     7.5 & 0.00003 & 60122.99502 &   254.5 & 0.00003 & 61105.48135 &   422.0 & 0.00003 & 61167.07100 &   432.5 & 0.00003  \\ 
58677.12510 &     8.0 & 0.00006 & 60832.72878 &   375.5 & 0.00003 & 61108.41270 &   422.5 & 0.00003 & 61170.00452 &   433.0 & 0.00003  \\ 
58680.05597 &     8.5 & 0.00004 & 60835.66299 &   376.0 & 0.00003 & 61111.34663 &   423.0 & 0.00003 & 61172.93715 &   433.5 & 0.00003  \\ 
59363.39936 &   125.0 & 0.00003 & 60838.59455 &   376.5 & 0.00003 & 61117.21181 &   424.0 & 0.00003 & 61175.87095 &   434.0 & 0.00003  \\ 
59366.33103 &   125.5 & 0.00002 & 60841.52851 &   377.0 & 0.00004  \\ 
\hline
\end{tabular}}
\textit{Notes}. Integer and half-integer cycle numbers denote primary and secondary eclipses, respectively.
\end{table*}

\begin{table*}
\caption{Times of minima of TIC\,358795247}
 \label{Tab:TIC_358795247_ToM}
\scalebox{0.78}{\begin{tabular}{@{}lrllrllrllrl}
\hline
BJD & Cycle  & std. dev. & BJD & Cycle  & std. dev. & BJD & Cycle  & std. dev. & BJD & Cycle  & std. dev. \\ 
$-2\,400\,000$ & no. &   \multicolumn{1}{c}{$(d)$} & $-2\,400\,000$ & no. &   \multicolumn{1}{c}{$(d)$} & $-2\,400\,000$ & no. &   \multicolumn{1}{c}{$(d)$} & $-2\,400\,000$ & no. &   \multicolumn{1}{c}{$(d)$} \\ 
\hline
59010.81566 &    -0.5 & 0.00012 & 59415.11417 &   330.5 & 0.00011 & 59791.32058 &   638.5 & 0.00012 & 60502.21132 &  1220.5 & 0.00005  \\ 
59011.42730 &     0.0 & 0.00023 & 59415.72602 &   331.0 & 0.00009 & 59791.93170 &   639.0 & 0.00011 & 60502.82251 &  1221.0 & 0.00007  \\ 
59012.03669 &     0.5 & 0.00005 & 59416.33593 &   331.5 & 0.00011 & 59792.54268 &   639.5 & 0.00014 & 60503.43306 &  1221.5 & 0.00008  \\ 
59012.64806 &     1.0 & 0.00042 & 59416.94742 &   332.0 & 0.00007 & 59793.15237 &   640.0 & 0.00015 & 60504.04428 &  1222.0 & 0.00009  \\ 
59013.25878 &     1.5 & 0.00004 & 59417.55798 &   332.5 & 0.00008 & 59793.76359 &   640.5 & 0.00016 & 60504.65458 &  1222.5 & 0.00010  \\ 
59013.86925 &     2.0 & 0.00004 & 59418.16899 &   333.0 & 0.00009 & 59794.98519 &   641.5 & 0.00008 & 60505.26474 &  1223.0 & 0.00007  \\ 
59014.48070 &     2.5 & 0.00042 & 59418.77967 &   333.5 & 0.00005 & 59795.59583 &   642.0 & 0.00006 & 60505.87533 &  1223.5 & 0.00007  \\ 
59015.09145 &     3.0 & 0.00014 & 59744.29876 &   600.0 & 0.00007 & 60312.88328 &  1065.5 & 0.00008 & 60507.09828 &  1224.5 & 0.00007  \\ 
59015.70172 &     3.5 & 0.00032 & 59744.90923 &   600.5 & 0.00009 & 60313.49364 &  1066.0 & 0.00008 & 60507.70784 &  1225.0 & 0.00005  \\ 
59016.31171 &     4.0 & 0.00032 & 59745.52068 &   601.0 & 0.00007 & 60314.71517 &  1067.0 & 0.00009 & 60508.31986 &  1225.5 & 0.00006  \\ 
59016.92393 &     4.5 & 0.00003 & 59746.13042 &   601.5 & 0.00015 & 60315.32647 &  1067.5 & 0.00010 & 60508.92953 &  1226.0 & 0.00007  \\ 
59017.53441 &     5.0 & 0.00017 & 59746.74086 &   602.0 & 0.00009 & 60315.93687 &  1068.0 & 0.00007 & 60509.54122 &  1226.5 & 0.00009  \\ 
59018.14495 &     5.5 & 0.00029 & 59747.35294 &   602.5 & 0.00011 & 60316.54839 &  1068.5 & 0.00009 & 60510.15145 &  1227.0 & 0.00005  \\ 
59018.75585 &     6.0 & 0.00010 & 59747.96280 &   603.0 & 0.00008 & 60317.15814 &  1069.0 & 0.00005 & 60510.76236 &  1227.5 & 0.00007  \\ 
59019.36699 &     6.5 & 0.00031 & 59748.57371 &   603.5 & 0.00016 & 60317.76938 &  1069.5 & 0.00006 & 60511.37302 &  1228.0 & 0.00006  \\ 
59019.97664 &     7.0 & 0.00002 & 59749.79441 &   604.5 & 0.00015 & 60318.37998 &  1070.0 & 0.00006 & 60511.98424 &  1228.5 & 0.00007  \\ 
59020.58861 &     7.5 & 0.00010 & 59750.40614 &   605.0 & 0.00006 & 60318.99095 &  1070.5 & 0.00005 & 60512.59421 &  1229.0 & 0.00006  \\ 
59021.19893 &     8.0 & 0.00020 & 59751.01576 &   605.5 & 0.00007 & 60319.60085 &  1071.0 & 0.00006 & 60513.20589 &  1229.5 & 0.00007  \\ 
59021.80990 &     8.5 & 0.00035 & 59751.62679 &   606.0 & 0.00010 & 60320.21188 &  1071.5 & 0.00009 & 60513.81613 &  1230.0 & 0.00007  \\ 
59023.64192 &    10.0 & 0.00014 & 59752.23655 &   606.5 & 0.00006 & 60320.82261 &  1072.0 & 0.00005 & 60514.42696 &  1230.5 & 0.00015  \\ 
59024.25320 &    10.5 & 0.00006 & 59752.84854 &   607.0 & 0.00009 & 60321.43326 &  1072.5 & 0.00006 & 60515.03671 &  1231.0 & 0.00013  \\ 
59024.86409 &    11.0 & 0.00042 & 59753.45879 &   607.5 & 0.00006 & 60322.04365 &  1073.0 & 0.00009 & 60516.26027 &  1232.0 & 0.00008  \\ 
59025.47397 &    11.5 & 0.00034 & 59754.07011 &   608.0 & 0.00004 & 60322.65541 &  1073.5 & 0.00005 & 60516.87116 &  1232.5 & 0.00005  \\ 
59026.08419 &    12.0 & 0.00045 & 59754.67970 &   608.5 & 0.00006 & 60323.26533 &  1074.0 & 0.00007 & 60517.48116 &  1233.0 & 0.00009  \\ 
59026.69582 &    12.5 & 0.00016 & 59755.29100 &   609.0 & 0.00004 & 60323.87638 &  1074.5 & 0.00007 & 60518.09236 &  1233.5 & 0.00005  \\ 
59027.30605 &    13.0 & 0.00016 & 59755.90127 &   609.5 & 0.00009 & 60324.48640 &  1075.0 & 0.00005 & 60518.70268 &  1234.0 & 0.00007  \\ 
59027.91740 &    13.5 & 0.00048 & 59757.12282 &   610.5 & 0.00010 & 60325.09832 &  1075.5 & 0.00007 & 60519.92406 &  1235.0 & 0.00006  \\ 
59028.52860 &    14.0 & 0.00012 & 59757.73323 &   611.0 & 0.00009 & 60325.70829 &  1076.0 & 0.00007 & 60520.53556 &  1235.5 & 0.00006  \\ 
59029.74970 &    15.0 & 0.00003 & 59758.34361 &   611.5 & 0.00006 & 60326.31945 &  1076.5 & 0.00005 & 60521.14582 &  1236.0 & 0.00006  \\ 
59030.36057 &    15.5 & 0.00007 & 59758.95490 &   612.0 & 0.00012 & 60326.92958 &  1077.0 & 0.00006 & 60521.75658 &  1236.5 & 0.00006  \\ 
59030.96998 &    16.0 & 0.00041 & 59759.56470 &   612.5 & 0.00009 & 60328.15103 &  1078.0 & 0.00011 & 60522.36688 &  1237.0 & 0.00005  \\ 
59031.58231 &    16.5 & 0.00004 & 59760.17611 &   613.0 & 0.00008 & 60329.37319 &  1079.0 & 0.00007 & 60523.58824 &  1238.0 & 0.00006  \\ 
59032.19296 &    17.0 & 0.00013 & 59760.78703 &   613.5 & 0.00014 & 60329.98358 &  1079.5 & 0.00008 & 60524.20019 &  1238.5 & 0.00007  \\ 
59032.80420 &    17.5 & 0.00022 & 59761.39736 &   614.0 & 0.00013 & 60330.59377 &  1080.0 & 0.00007 & 60524.81011 &  1239.0 & 0.00005  \\ 
59033.41393 &    18.0 & 0.00041 & 59763.22913 &   615.5 & 0.00011 & 60331.20524 &  1080.5 & 0.00009 & 60525.42146 &  1239.5 & 0.00008  \\ 
59034.02535 &    18.5 & 0.00005 & 59763.83968 &   616.0 & 0.00005 & 60331.81544 &  1081.0 & 0.00006 & 60526.03115 &  1240.0 & 0.00005  \\ 
59034.63604 &    19.0 & 0.00003 & 59764.45096 &   616.5 & 0.00008 & 60332.42695 &  1081.5 & 0.00007 & 60526.64276 &  1240.5 & 0.00007  \\ 
59390.68513 &   310.5 & 0.00008 & 59765.06175 &   617.0 & 0.00007 & 60333.64853 &  1082.5 & 0.00008 & 60527.25287 &  1241.0 & 0.00005  \\ 
59391.29667 &   311.0 & 0.00011 & 59765.67216 &   617.5 & 0.00010 & 60334.25859 &  1083.0 & 0.00005 & 60527.86433 &  1241.5 & 0.00006  \\ 
59391.90716 &   311.5 & 0.00006 & 59766.28299 &   618.0 & 0.00007 & 60334.86995 &  1083.5 & 0.00006 & 60528.47446 &  1242.0 & 0.00005  \\ 
59392.51753 &   312.0 & 0.00012 & 59766.89309 &   618.5 & 0.00015 & 60335.47967 &  1084.0 & 0.00006 & 60529.08554 &  1242.5 & 0.00008  \\ 
59393.12836 &   312.5 & 0.00007 & 59767.50470 &   619.0 & 0.00010 & 60336.09119 &  1084.5 & 0.00006 & 60529.69615 &  1243.0 & 0.00006  \\ 
59393.73944 &   313.0 & 0.00005 & 59768.11492 &   619.5 & 0.00006 & 60336.70192 &  1085.0 & 0.00006 & 60530.30727 &  1243.5 & 0.00005  \\ 
59394.34973 &   313.5 & 0.00009 & 59768.72548 &   620.0 & 0.00007 & 60337.31246 &  1085.5 & 0.00008 & 60530.91746 &  1244.0 & 0.00006  \\ 
59394.96085 &   314.0 & 0.00009 & 59769.94693 &   621.0 & 0.00006 & 60337.92278 &  1086.0 & 0.00005 & 60531.52889 &  1244.5 & 0.00006  \\ 
59395.57168 &   314.5 & 0.00007 & 59770.55721 &   621.5 & 0.00009 & 60338.53446 &  1086.5 & 0.00009 & 60532.13903 &  1245.0 & 0.00005  \\ 
59396.18210 &   315.0 & 0.00016 & 59771.16815 &   622.0 & 0.00007 & 60339.14420 &  1087.0 & 0.00006 & 60532.75008 &  1245.5 & 0.00006  \\ 
59396.79303 &   315.5 & 0.00011 & 59771.77930 &   622.5 & 0.00007 & 60480.22393 &  1202.5 & 0.00006 & 60591.37549$^a$&  1293.5 & 0.00012  \\ 
59397.40292 &   316.0 & 0.00006 & 59772.38961 &   623.0 & 0.00008 & 60480.83399 &  1203.0 & 0.00004 & 60752.60930$^a$&  1425.5 & 0.00006  \\ 
59398.01460 &   316.5 & 0.00007 & 59773.00023 &   623.5 & 0.00006 & 60481.44463 &  1203.5 & 0.00006 & 60793.53244$^a$&  1459.0 & 0.00005  \\ 
59398.62534 &   317.0 & 0.00011 & 59773.61043 &   624.0 & 0.00006 & 60482.05534 &  1204.0 & 0.00006 & 60809.41230$^b$&  1472.0 & 0.00025  \\ 
59399.23578 &   317.5 & 0.00007 & 59774.22169 &   624.5 & 0.00006 & 60482.66595 &  1204.5 & 0.00004 & 60826.50882$^b$&  1486.0 & 0.00009  \\ 
59399.84623 &   318.0 & 0.00009 & 59774.83220 &   625.0 & 0.00005 & 60483.27675 &  1205.0 & 0.00004 & 60840.55507$^b$&  1497.5 & 0.00037  \\ 
59400.45679 &   318.5 & 0.00010 & 59775.44166 &   625.5 & 0.00007 & 60483.88765 &  1205.5 & 0.00006 & 60889.41491$^b$&  1537.5 & 0.00016  \\ 
59401.06822 &   319.0 & 0.00007 & 59776.05269 &   626.0 & 0.00087 & 60484.49857 &  1206.0 & 0.00006 & 60892.46827$^b$&  1540.0 & 0.00008  \\ 
59401.67865 &   319.5 & 0.00007 & 59776.66415 &   626.5 & 0.00007 & 60489.38398 &  1210.0 & 0.00007 & 60919.34450$^b$&  1562.0 & 0.00013  \\ 
59402.28878 &   320.0 & 0.00017 & 59777.27481 &   627.0 & 0.00011 & 60489.99542 &  1210.5 & 0.00007 & 60949.26605$^b$&  1586.5 & 0.00043  \\ 
59402.90052 &   320.5 & 0.00006 & 59777.88586 &   627.5 & 0.00013 & 60490.60656 &  1211.0 & 0.00011 & 61127.60332$^b$&  1732.5 & 0.00015  \\ 
59403.51106 &   321.0 & 0.00008 & 59778.49633 &   628.0 & 0.00008 & 60491.21702 &  1211.5 & 0.00011 & 61135.53633$^b$&  1739.0 & 0.00017  \\ 
59404.12123 &   321.5 & 0.00008 & 59779.10709 &   628.5 & 0.00012 & 60491.82789 &  1212.0 & 0.00006 & 61165.46568$^c$&  1763.5 & 0.00003  \\ 
59405.34289 &   322.5 & 0.00029 & 59779.71667 &   629.0 & 0.00012 & 60492.43900 &  1212.5 & 0.00007 & 61173.40417$^a$&  1770.0 & 0.00011  \\ 
59405.95400 &   323.0 & 0.00006 & 59780.93914 &   630.0 & 0.00009 & 60493.04929 &  1213.0 & 0.00006 & 61182.56785$^a$&  1777.5 & 0.00020  \\ 
59406.56444 &   323.5 & 0.00010 & 59781.54936 &   630.5 & 0.00008 & 60493.65997 &  1213.5 & 0.00005 & 61184.40002$^a$&  1779.0 & 0.00016  \\ 
59407.17550 &   324.0 & 0.00013 & 59782.15980 &   631.0 & 0.00005 & 60494.27063 &  1214.0 & 0.00005 & 61187.45381$^c$&  1781.5 & 0.00010  \\ 
59407.78658 &   324.5 & 0.00007 & 59783.99220 &   632.5 & 0.00008 & 60494.88190 &  1214.5 & 0.00005 & 61190.50757$^a$&  1784.0 & 0.00010  \\ 
59408.39709 &   325.0 & 0.00006 & 59784.60308 &   633.0 & 0.00010 & 60495.49225 &  1215.0 & 0.00007 & 61209.44053$^c$&  1799.5 & 0.00004  \\ 
59409.00737 &   325.5 & 0.00008 & 59785.21366 &   633.5 & 0.00007 & 60496.10360 &  1215.5 & 0.00007 & 61215.54324$^b$&  1804.5 & 0.00022  \\ 
59409.61835 &   326.0 & 0.00005 & 59785.82405 &   634.0 & 0.00008 & 60496.71358 &  1216.0 & 0.00005 & 61217.38008$^c$&  1806.0 & 0.00003  \\ 
59410.22843 &   326.5 & 0.00010 & 59786.43561 &   634.5 & 0.00007 & 60497.32492 &  1216.5 & 0.00007 & 61231.42341$^a$&  1817.5 & 0.00016  \\ 
59410.83921 &   327.0 & 0.00010 & 59787.04531 &   635.0 & 0.00006 & 60497.93572 &  1217.0 & 0.00007 & 61242.41727$^c$&  1826.5 & 0.00008  \\ 
59411.45066 &   327.5 & 0.00012 & 59787.65619 &   635.5 & 0.00009 & 60498.54667 &  1217.5 & 0.00007 & 61245.47094$^b$&  1829.0 & 0.00025  \\ 
59412.06166 &   328.0 & 0.00010 & 59788.26713 &   636.0 & 0.00010 & 60499.15702 &  1218.0 & 0.00005 & 61253.40961$^c$&  1835.5 & 0.00004  \\ 
59412.67176 &   328.5 & 0.00011 & 59788.87834 &   636.5 & 0.00013 & 60499.76875 &  1218.5 & 0.00007 & 61264.40338$^b$&  1844.5 & 0.00014  \\ 
59413.28275 &   329.0 & 0.00009 & 59789.48828 &   637.0 & 0.00008 & 60500.37858 &  1219.0 & 0.00007 & 61286.39093$^b$&  1862.5 & 0.00023  \\ 
59413.89282 &   329.5 & 0.00009 & 59790.09977 &   637.5 & 0.00005 & 60500.98942 &  1219.5 & 0.00006 & 61308.37740$^b$&  1880.5 & 0.00027  \\ 
59414.50481 &   330.0 & 0.00009 & 59790.71005 &   638.0 & 0.00010 & 60501.60022 &  1220.0 & 0.00007  \\ 
\hline
\end{tabular}}

\textit{Notes}. Integer and half-integer cycle numbers denote primary and secondary eclipses, respectively;
The majority of the mid-times of minima were collected with TESS, with the following exceptions: $^a:$ GAO80 minimum; $^b:$ Observed by R.U. $^c$: BAO80 minimum.
\end{table*}

\begin{table*}
\caption{Times of minima of TIC\,379084660}
 \label{Tab:TIC_379084660_ToM}
\scalebox{0.88}{\begin{tabular}{@{}lrllrllrllrl}
\hline
BJD & Cycle  & std. dev. & BJD & Cycle  & std. dev. & BJD & Cycle  & std. dev. & BJD & Cycle  & std. dev. \\ 
$-2\,400\,000$ & no. &   \multicolumn{1}{c}{$(d)$} & $-2\,400\,000$ & no. &   \multicolumn{1}{c}{$(d)$} & $-2\,400\,000$ & no. &   \multicolumn{1}{c}{$(d)$} & $-2\,400\,000$ & no. &   \multicolumn{1}{c}{$(d)$} \\ 
\hline
58791.75879 &    -0.5 & 0.00607 & 58994.02578 &    42.0 & 0.00132 & 59905.33792 &   233.5 & 0.00211 & 60471.63044 &   352.5 & 0.00222  \\ 
58794.13020 &     0.0 & 0.00127 & 58998.78371 &    43.0 & 0.00162 & 59912.48099 &   235.0 & 0.00078 & 60474.00234 &   353.0 & 0.00080  \\ 
58796.52039 &     0.5 & 0.00724 & 59001.16993 &    43.5 & 0.00314 & 59914.86650 &   235.5 & 0.00213 & 60476.38638 &   353.5 & 0.00289  \\ 
58798.88493 &     1.0 & 0.00164 & 59003.54333 &    44.0 & 0.00102 & 59917.23677 &   236.0 & 0.00078 & 60478.75835 &   354.0 & 0.00064  \\ 
58801.27265 &     1.5 & 0.00967 & 59005.92427 &    44.5 & 0.00390 & 59919.61872 &   236.5 & 0.00353 & 60612.03272 &   382.0 & 0.00085  \\ 
58803.64643 &     2.0 & 0.00214 & 59008.30469 &    45.0 & 0.00213 & 59921.99187 &   237.0 & 0.00081 & 60614.41809 &   382.5 & 0.00291  \\ 
58806.04705 &     2.5 & 0.01058 & 59719.75219 &   194.5 & 0.00806 & 59924.35773 &   237.5 & 0.00363 & 60619.17510 &   383.5 & 0.00325  \\ 
58808.42110 &     3.0 & 0.00161 & 59722.13125 &   195.0 & 0.00089 & 59926.74920 &   238.0 & 0.00093 & 60621.54561 &   384.0 & 0.00077  \\ 
58810.79457 &     3.5 & 0.08872 & 59724.51170 &   195.5 & 0.00388 & 59929.12938 &   238.5 & 0.00293 & 60623.93267 &   384.5 & 0.00431  \\ 
58813.18408 &     4.0 & 0.00480 & 59726.89037 &   196.0 & 0.00059 & 59931.50558 &   239.0 & 0.00083 & 60626.30300 &   385.0 & 0.00098  \\ 
58955.93677 &    34.0 & 0.00086 & 59729.26943 &   196.5 & 0.00193 & 59933.88187 &   239.5 & 0.00210 & 60628.67842 &   385.5 & 0.00397  \\ 
58958.31678 &    34.5 & 0.00893 & 59731.64877 &   197.0 & 0.00064 & 59936.25895 &   240.0 & 0.00075 & 60631.05592 &   386.0 & 0.00082  \\ 
58960.69030 &    35.0 & 0.00123 & 59734.02827 &   197.5 & 0.00197 & 60435.95465 &   345.0 & 0.00070 & 60633.44153 &   386.5 & 0.00183  \\ 
58963.08524 &    35.5 & 0.00354 & 59736.40184 &   198.0 & 0.00066 & 60438.34176 &   345.5 & 0.00214 & 60635.81403 &   387.0 & 0.00079  \\ 
58965.44943 &    36.0 & 0.00194 & 59738.78134 &   198.5 & 0.00184 & 60443.09991 &   346.5 & 0.00285 & 60638.20515 &   387.5 & 0.00526  \\ 
58967.83998 &    36.5 & 0.00303 & 59741.16200 &   199.0 & 0.00064 & 60447.85372 &   347.5 & 0.00232 & 60640.56478 &   388.0 & 0.00113  \\ 
58970.20497 &    37.0 & 0.00099 & 59883.92946 &   229.0 & 0.00085 & 60452.61059 &   348.5 & 0.00202 & 60645.32479 &   389.0 & 0.00086  \\ 
58972.59598 &    37.5 & 0.00490 & 59886.30007 &   229.5 & 0.00253 & 60454.98141 &   349.0 & 0.00088 & 60647.70878 &   389.5 & 0.00366  \\ 
58974.96290 &    38.0 & 0.00092 & 59888.69091 &   230.0 & 0.00078 & 60457.36100 &   349.5 & 0.00366 & 60650.08060 &   390.0 & 0.00096  \\ 
58977.34988 &    38.5 & 0.00395 & 59891.06162 &   230.5 & 0.00215 & 60459.73649 &   350.0 & 0.00093 & 60652.45310 &   390.5 & 0.00270  \\ 
58979.72824 &    39.0 & 0.00107 & 59893.44792 &   231.0 & 0.00084 & 60462.12160 &   350.5 & 0.00221 & 60654.84286 &   391.0 & 0.00716  \\ 
58984.49493 &    40.0 & 0.00122 & 59895.83279 &   231.5 & 0.00177 & 60464.49218 &   351.0 & 0.00072 & 60659.59384 &   392.0 & 0.00091  \\ 
58986.88767 &    40.5 & 0.00513 & 59898.21069 &   232.0 & 0.00085 & 60466.88240 &   351.5 & 0.00217 & 60661.98780 &   392.5 & 0.00235  \\ 
58989.26254 &    41.0 & 0.00130 & 59900.59098 &   232.5 & 0.00172 & 60469.24476 &   352.0 & 0.00077 & 61273.47928$^a$&   521.0 & 0.00026  \\ 
58991.65073 &    41.5 & 0.00604 & 59902.96578 &   233.0 & 0.00089  \\ 
\hline
\end{tabular}}
\textit{Notes}. Integer and half-integer cycle numbers denote primary and secondary eclipses, respectively. All but one eclipses observed with TESS. $^a$: The very last eclipse was observed with BAO80 instrument.
\end{table*}

\begin{table*}
\caption{Times of minima of TIC\,385933895}
 \label{Tab:TIC_385933895_ToM}
\scalebox{0.88}{\begin{tabular}{@{}lrllrllrllrl}
\hline
BJD & Cycle  & std. dev. & BJD & Cycle  & std. dev. & BJD & Cycle  & std. dev. & BJD & Cycle  & std. dev. \\ 
$-2\,400\,000$ & no. &   \multicolumn{1}{c}{$(d)$} & $-2\,400\,000$ & no. &   \multicolumn{1}{c}{$(d)$} & $-2\,400\,000$ & no. &   \multicolumn{1}{c}{$(d)$} & $-2\,400\,000$ & no. &   \multicolumn{1}{c}{$(d)$} \\ 
\hline
58543.63837 &     0.0 & 0.00434 & 59262.89868 &   111.5 & 0.00239 & 59998.28284 &   225.5 & 0.00063 & 60724.00452 &   338.0 & 0.00039  \\ 
58546.87155 &     0.5 & 0.00292 & 59272.57451 &   113.0 & 0.00282 & 60001.51981 &   226.0 & 0.00167 & 60727.23086 &   338.5 & 0.00045  \\ 
58550.10314 &     1.0 & 0.00153 & 59275.80175 &   113.5 & 0.00233 & 60004.75355 &   226.5 & 0.00054 & 60733.68919 &   339.5 & 0.00043  \\ 
58553.32951 &     1.5 & 0.00103 & 59279.03427 &   114.0 & 0.00194 & 60007.98541 &   227.0 & 0.00057 & 60736.91896 &   340.0 & 0.00039  \\ 
58559.77158 &     2.5 & 0.00164 & 59308.06829 &   118.5 & 0.00131 & 60011.21224 &   227.5 & 0.00080 & 60740.13915 &   340.5 & 0.00046  \\ 
58562.99041 &     3.0 & 0.00100 & 59311.29929 &   119.0 & 0.00070 & 60014.43272 &   228.0 & 0.00078 & 60743.36464 &   341.0 & 0.00043  \\ 
58566.21127 &     3.5 & 0.00100 & 59314.52336 &   119.5 & 0.00068 & 60017.65359 &   228.5 & 0.00101 & 60746.58110 &   341.5 & 0.00050  \\ 
58572.65917 &     4.5 & 0.00259 & 59317.74367 &   120.0 & 0.00045 & 60020.87079 &   229.0 & 0.00072 & 60749.80476 &   342.0 & 0.00038  \\ 
58579.11691 &     5.5 & 0.00083 & 59320.96239 &   120.5 & 0.00091 & 60024.09443 &   229.5 & 0.00070 & 60753.02409 &   342.5 & 0.00047  \\ 
58585.57639 &     6.5 & 0.00161 & 59324.18264 &   121.0 & 0.00056 & 60027.31884 &   230.0 & 0.00052 & 60756.25139 &   343.0 & 0.00042  \\ 
58588.80298 &     7.0 & 0.00073 & 59327.40371 &   121.5 & 0.00066 & 60030.54355 &   230.5 & 0.00053 & 60759.47924 &   343.5 & 0.00053  \\ 
58592.02299 &     7.5 & 0.00074 & 59330.62833 &   122.0 & 0.00046 & 60036.99764 &   231.5 & 0.00101 & 60765.94681 &   344.5 & 0.00052  \\ 
58595.24331 &     8.0 & 0.00100 & 59988.61612 &   224.0 & 0.00233 & 60040.23001 &   232.0 & 0.00116 & 60769.17771 &   345.0 & 0.00037  \\ 
59256.46164 &   110.5 & 0.00259 & 59991.82838 &   224.5 & 0.00077 & 60720.77741 &   337.5 & 0.00055 & 60772.40164 &   345.5 & 0.00058  \\ 
59259.67639 &   111.0 & 0.00235  \\ 
\hline
\end{tabular}}
\textit{Notes}. Integer and half-integer cycle numbers denote primary and secondary eclipses, respectively.
\end{table*}

\end{appendix}
\end{document}